\documentclass[11pt]{article}
\usepackage{jheppub}
\usepackage[cp1251]{inputenc}
\usepackage{amsmath}
\usepackage{amssymb}
\usepackage{array}
\usepackage{graphicx,xcolor,slashed}
\usepackage{enumerate}
\usepackage{ifpdf}
\usepackage{mathtools}
\usepackage[english]{babel}
\usepackage{url}
\usepackage{booktabs}
\usepackage{xspace}
\usepackage{xinttools}
\usepackage{mathrsfs}
\usepackage{tensor}
\usepackage{cancel}
\usepackage{cleveref}
\usepackage{subfig}
\notoc

\definecolor{darkgreen}{rgb}{0.0, 0.26, 0.15}
\definecolor{darkred}{rgb}{0.65,0.15,0}

\usepackage{float}

\hypersetup{bookmarksnumbered,bookmarksdepth=3}

\allowdisplaybreaks[1]

\DeclareFontFamily{U}{mathx}{\hyphenchar\font45}
\DeclareFontShape{U}{mathx}{m}{n}{
      <5> <6> <7> <8> <9> <10>
      <10.95> <12> <14.4> <17.28> <20.74> <24.88>
      mathx10
      }{}
\DeclareSymbolFont{mathx}{U}{mathx}{m}{n}
\DeclareFontSubstitution{U}{mathx}{m}{n}
\DeclareMathAccent{\widecheck}{0}{mathx}{"71}

\restylefloat{figure}
\definecolor{dgreen}{rgb}{0,0.70,0.30}
\definecolor{gold}{rgb}{0.85,.66,0}
\definecolor{purple}{rgb}{1.0,0.3,0.6}

\usepackage{tikz}
\usetikzlibrary{calc}
\usetikzlibrary{patterns}
\usetikzlibrary{decorations.pathreplacing}
\usetikzlibrary{decorations.markings}
\usetikzlibrary{decorations.pathmorphing}
\usetikzlibrary{positioning}
\usetikzlibrary{arrows}
\usetikzlibrary{arrows.meta}

\newcommand{\bea}{\begin{eqnarray}}
\newcommand{\eea}{\end{eqnarray}}
\def\beq{\begin{equation}}
\def\eeq{\end{equation}}

\let\Im\relax

\DeclareMathOperator{\Im}{Im}

\newcommand{\la}{\lambda}
\newcommand{\ep}{\epsilon}

\newcommand{\dd}{\mathrm{d}}
\newcommand{\te}{\textrm}
\newcommand{\ap}{{\alpha'}}

\newcommand{\RR}{\mathbb R}

\newcommand{\ZZ}{\mathbb Z}

\DeclareMathOperator{\coef}{coef}

\newcommand{\KN}{{\cal J}}

\newcommand{\tmax}{t_{\rm max}}

\newcommand{\afouruv}[1]{\ensuremath{
    \widehat A^{\te{1-loop}}_{\te{eff}}(1,2,3,4) \,\Big|^{#1}_{\rm UV}}}
\newcommand{\mfouruv}[1]{\ensuremath{
    \widehat M^{\te{1-loop}}_{4,{\rm eff}}   \, \Big|^{#1}_{\rm UV} }}

\newcommand{\eg}{{e.g.~}}
\newcommand{\ie}{{i.e.~}}

\crefname{equation}{}{}

\title{One-loop matrix elements of effective superstring interactions: $\alpha'$-expanding loop integrands}

\author[a]{Alex Edison,}
\author[a]{Max Guillen,}
\author[a,b]{Henrik Johansson,}
\author[a]{Oliver Schlotterer}
\author[a,c]{and Fei Teng}

\affiliation[a]{Department of Physics and Astronomy, Uppsala University, Box 516, 75120 Uppsala, Sweden}
\affiliation[b]{Nordita, Stockholm University and KTH Royal Institute of Technology, Hannes Alfv\'{e}ns v\"{a}g 12, 10691 Stockholm, Sweden}
\affiliation[c]{Institute for Gravitation and the Cosmos, Department of Physics, Pennsylvania State University, University Park, PA 16802, USA}

\emailAdd{alexander.edison@physics.uu.se}
\emailAdd{max.guillen@physics.uu.se}
\emailAdd{henrik.johansson@physics.uu.se}
\emailAdd{oliver.schlotterer@physics.uu.se}
\emailAdd{fei.teng@psu.edu}

\date{\today}

\abstract{In the low-energy effective action of string theories, non-abelian gauge interactions and supergravity are augmented by infinite towers of higher-mass-dimension operators. We propose a new method to construct one-loop matrix elements with insertions of operators $D^{2k} F^n$ and $D^{2k} R^n$ in the tree-level effective action of type-I and type-II superstrings. Inspired by ambitwistor string theories, our method is based on forward limits of moduli-space integrals using string tree-level amplitudes with two extra points, expanded in powers of the inverse string tension $\alpha'$. Similar to one-loop ambitwistor computations, intermediate steps feature non-standard linearized Feynman propagators which eventually recombine to conventional quadratic propagators. With linearized propagators the loop integrand of the matrix elements obey one-loop versions of the monodromy and KLT relations. 
We express a variety of four- and five-point examples in terms of quadratic propagators and 
formulate a criterion on the underlying genus-one correlation functions that should make this recombination
possible at all orders in $\alpha'$.
The ultraviolet divergences of the one-loop matrix elements are crosschecked against the non-separating degeneration of genus-one integrals in string amplitudes. Conversely, our results can be used as a constructive method to determine degenerations of elliptic multiple zeta values and modular graph forms at arbitrary weight.}

\preprint{UUITP--31/21 \\
\phantom{~} \hfill NORDITA 2021-065}  

\begin{document}

\maketitle{}

\newpage

\setcounter{page}{1}
\pagenumbering{roman}

\setcounter{tocdepth}{2}


\tableofcontents


\numberwithin{equation}{section}

 \newpage


\section{Introduction}
\label{sec:intro}

\setcounter{page}{1}
\pagenumbering{arabic}

Recent years have witnessed remarkable synergies between properties and
building blocks of massless string amplitudes and new structures of
field-theory amplitudes.  A central example is the double-copy
structure of perturbative gravity \cite{BCJ, loopBCJ, Bern:2019prr},
which is manifest in the Kawai--Lewellen--Tye open/closed string relations
\cite{Kawai:1985xq}, the chiral-splitting description of string
amplitudes \cite{DHoker:1988pdl, DHoker:1989cxq}, and kinematic numerators in multiloop field-theory amplitudes~\cite{loopBCJ, Carrasco:2011mn, Bern:2012uf}. Ambitwistor string theories
\cite{Mason:2013sva, Berkovits:2013xba, Adamo:2013tsa, Adamo:2015hoa} added important
facets to the elegant realization of double copy via string
worldsheets and extracted key insights on (possibly non-supersymmetric) loop integrands 
from forward limits of tree-level diagrams \cite{Geyer:2015bja, Baadsgaard:2015hia, He:2015yua, Geyer:2015jch, Cachazo:2015aol, Cardona:2016bpi, Cardona:2016wcr,  He:2016mzd, He:2017spx, Geyer:2017ela, Edison:2020uzf}.
  
The double-copy structures of string amplitudes have profound
implications on the effective field theories that govern the
low-energy dynamics of the gauge and gravity multiplets from the
massless string excitations.
For open (type-I) superstrings and closed (type-II) superstrings, the
leading orders in the low-energy effective actions at tree level are
schematically given by (see e.g.\
\cite{Tseytlin:1986ti, Gross:1986iv, Kitazawa:1987xj,
  Tseytlin:1997csa, Koerber:2002zb, Barreiro:2005hv, Oprisa:2005wu,
  Stieberger:2009rr, Schlotterer:2012ny, Barreiro:2012aw, Garousi:2020gio, Garousi:2020lof} and
references therein)
\begin{align}
S_{\rm eff}^{\rm open}&= \int \dd^D x \, {\rm Tr}\Big\{ F^2 + \ap^2 \zeta_2 F^4 
+ \ap^3 \zeta_3(D^2 F^4 {+}F^5)+ \ap^4 \zeta_4(D^4 F^4 {+}D^2F^5{+}F^6) + \ldots \Big\}
\notag
\\
S_{\rm eff}^{\rm closed}&= \int \dd^D x \,  \Big\{ R + \ap^3 \zeta_3 R^4 
+ \ap^5 \zeta_5(D^4 R^4 {+}D^2 R^5{+}R^6)+ \ldots \Big\}
\label{seffclosed}
\end{align}
with Riemann zeta values $\zeta_n = \sum_{k=1}^{\infty} k^{-n}$,
non-abelian gauge-field strengths $F$, Riemann or Ricci curvature $R$
and their respective covariant derivatives $D$.  The ellipses refer to
infinite power series in the inverse string tension $\alpha'$ that
organize the low-energy expansion\footnote{More precisely, low-energy-
or $\alpha'$-expansions refer to series expansion of string amplitudes
in the dimensionless Mandelstam invariants $\alpha' k_i \cdot k_j$
with external momenta $k_i$. The latter signal gauge- or diffeomorphism
covariant derivatives upon translation to the effective interactions in (\ref{seffclosed}).} 
as well as the supersymmetric completions, see \eg \cite{Green:1998by, Cederwall:2001td, Koerber:2001uu,
Collinucci:2002ac, Berkovits:2002ag, Drummond:2003ex,
Policastro:2006vt, Howe:2010nu, Cederwall:2011vy, Wang:2015jna}.  
From an effective-field-theory perspective, the infinitely many coefficients in (\ref{seffclosed}) are exactly determined by string perturbation theory at tree level. Furthermore, loop corrections to (\ref{seffclosed}) 
and non-perturbative effects are related to string dualities and
intricate properties of string vacua, which are beyond the scope of this work.

The elegant structures of superstring tree-level amplitudes involving gauge and gravity supermultiplets
expose a rich web of relations among the matrix elements of the
higher-mass-dimension operators in (\ref{seffclosed}). Moreover,
properties of bosonic- and heterotic-string amplitudes extend the
harmonious interplay of tree-level matrix elements to
non-supersymmetrizable effective operators including
$\alpha'{\rm Tr}(F^3),\alpha'R^2$ and $\alpha'^2R^3$
\cite{Broedel:2012rc, Huang:2016tag, Azevedo:2018dgo}. 

\begin{figure}
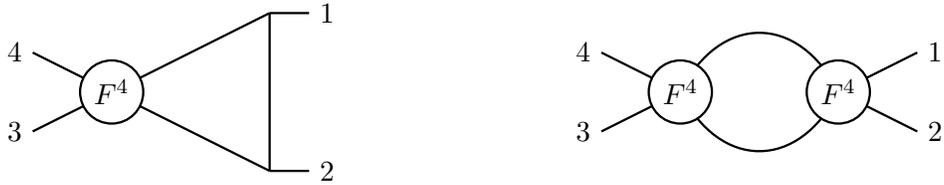

\begin{center}
\tikzpicture [scale=1.05, line width=0.30mm]
\draw(0,0)--(2,1);
\draw(0,0)--(2,-1);
\draw(2,-1)--(2,1);
\draw(2,1)--(2.5,1)node[right]{$1$};
\draw(2,-1)--(2.5,-1)node[right]{$2$};
\draw(0,0)--(-1,-0.5)node[left]{$3$};
\draw(0,0)--(-1,0.5)node[left]{$4$};
\draw [fill=white] (0,0)circle (0.4cm);
\draw(0,0)node{$F^4$};
\scope[xshift=7.2cm]
\draw(0,0) .. controls (0.5,1) and (1.5,1) .. (2,0);
\draw(0,0) .. controls (0.5,-1) and (1.5,-1) .. (2,0);
\draw(0,0)--(-1,-0.5)node[left]{$3$};
\draw(0,0)--(-1,0.5)node[left]{$4$};
\draw [fill=white] (0,0)circle (0.4cm);
\draw(0,0)node{$F^4$};
\draw(2,0)--(3,-0.5)node[right]{$2$};
\draw(2,0)--(3,0.5)node[right]{$1$};
\draw [fill=white] (2,0)circle (0.4cm);
\draw(2,0)node{$F^4$};
\endscope
\endtikzpicture
\end{center}
\caption{Contributions to the one-loop matrix elements with single- and double-insertion of the
operator $\alpha'^2 {\rm Tr}(F^4)$ in the open-string effective action (\ref{seffclosed}).}
\label{appetize}
\end{figure}

In this work, we initiate a systematic construction and structural
understanding of one-loop matrix elements involving the effective
tree-level interactions of open and closed superstrings as in
(\ref{seffclosed}).  More specifically, we use forward limits of
string tree amplitudes to propose expressions for, and relations among,
loop integrands involving insertions of the $D^{2k} F^n$ and
$D^{2k} R^n$ operators in tree-level effective actions.  Typical
diagrams in these loop integrands are drawn in Figure~\ref{appetize},
and we will give the explicit form of the kinematic factors for four and
five gauge or gravity states at various orders in $\alpha'$.  In fact,
the vertices and internal lines in Figure~\ref{appetize} represent the
full supersymmetry multiplets of the gluon and graviton, and our
results incorporate cancellations between bosonic and fermionic states
in the loop.

Our construction of one-loop matrix elements applies to
arbitrary multiplicities, spacetime dimensions and orders in $\alpha'$, and it bypasses
a variety of difficulties that would arise in other lines of attack.
First, a Feynman-diagrammatic computation in supersymmetry components involving 
different $n$-gon diagrams and the fermionic superpartners
of the operators in (\ref{seffclosed}) appears hopeless by its combinatorial complexity.
Second, superspace descriptions of $D^{2k} F^n$ and $D^{2k} R^n$ operators
are only available at low mass dimensions, but not for generic orders in $\alpha'$ that
are aimed for in this work.
Third, the $\alpha'$-expansion of one-loop string amplitudes in the literature 
usually proceeds by integrating out the loop momentum in early stages of the
computation. Hence, the desired momentum-space 
representations of one-loop matrix elements cannot be extracted from string-theory
references.

The one-loop matrix elements computed from our method determine the
non-analytic thresholds in the low-energy expansion of one-loop string
amplitudes, i.e.\ their branch cuts due to massless states in the
string loop.  However, the matrix elements of this work do not fix the
analytic contributions to one-loop string amplitudes, i.e.\ the terms
with power-series behaviour in the Mandelstam variables. 
Such analytic terms and the associated loop corrections to the superstring 
effective actions (\ref{seffclosed}) can be thought of as partially originating from purely massive loop integrals, which have no branch cuts below the ultraviolet (UV) cutoff of the effective field theory.
They are beyond the reach of the computations in this work
since all the propagators in the $\alpha'$-expansion of our one-loop matrix
elements (see e.g.\ Figure~\ref{appetize}) refer to massless states.

\subsection{Inspiration from ambitwistor strings}
\label{sec:1.1}

Our construction of one-loop matrix elements is inspired by
the ambitwistor-string formulae for loop integrands in gauge theories
and (super)gravity. 
Although ambitwistor theories only involve {\it closed}
strings, selected features of their one-loop amplitudes to be detailed below will be exported
to propose matrix elements of type-I {\it open} superstrings. In ambitwistor
string theories, scattering equations reduce the genus-one
worldsheets for one-loop integrands to nodal Riemann spheres
\cite{Geyer:2015bja, Geyer:2015jch} and introduce forward limits of
various tree-level building blocks \cite{He:2015yua,
Cachazo:2015aol} akin to the Q-cuts of \cite{Baadsgaard:2015twa}.\footnote{In fact, the 
role of forward limits in one-loop amplitudes was already understood to a great detail
through the Feynman tree theorem \cite{Feynman:1963ax}, see for instance \cite{Catani:2008xa} and
references therein.}
All the Feynman propagators in these loop
integrands can be derived from forward limits of (doubly-ordered) tree
amplitudes of bi-adjoint scalars. The polarization degrees of freedom
in turn are organized by a conformal field theory on the
ambitwistor-string worldsheet whose correlators can be freely
interchanged with those of conventional strings (see \cite{Gomez:2013wza, He:2017spx, Kalyanapuram:2021xow, Kalyanapuram:2021vjt} for a standard ambitwistor derivation, and \cite{Huang:2016bdd, Leite:2016fno, Li:2017emw, Casali:2016atr} for a more stringy interpretation in the context of the so-called chiral strings).

The tree amplitudes of bi-adjoint scalars also appear from the
$\alpha' \rightarrow 0$ limit of moduli-space integrals over disk and
sphere worldsheets in tree-level amplitudes of conventional strings
\cite{Cachazo:2013iea, Stieberger:2014hba}. The key idea in our
proposal is to also export higher orders in the $\alpha'$-expansion of
disk and sphere integrals into forward-limit formulae for one-loop
quantities. In other words, we convert the low-energy expansion of
string tree integrals into $\alpha'$-uplifts of the loop
integrands from ambitwistor strings. The resulting power series only
involve propagators and momentum invariants, i.e.\ the entire
$\alpha'$-dependence in our one-loop matrix elements of gauge and
gravity multiplets is carried by simple {\it scalar} ingredients.
Their dressing by polarization degrees of freedom follows the
thoroughly studied conformal-field-theory mechanisms that are common
to ambitwistor strings and conventional strings.

\subsection{Comparison with conventional strings}

In later sections, we will also motivate our proposal for one-loop matrix elements
from first principles of conventional string theories in the chiral-splitting formalism. 
Although we are not providing a mathematically rigorous derivation,
our proposal is supported by two kinds of consistency checks.
\begin{itemize}
\item The UV
divergences in the $\alpha'$-expansions of the one-loop matrix
elements in various spacetime dimensions are checked to
match expectations from string theory \cite{Metsaev:1987ju,Green:2010sp,
Pioline:2018pso}. Even though the one-loop amplitudes of 
open superstrings with gauge group $SO(32)$ and closed  superstrings are UV finite,
the effective-field-theory viewpoint introduces spurious UV divergences
order by order in $\alpha'$ which can be matched with degenerations 
of the string worldsheet.
\item The discontinuities of gravitational
four-point examples at the orders of $\alpha'^{\leq 8}$ are shown to
be consistent with the all-order results on the non-analytic sector of
the four-point closed-string amplitude
\cite{DHoker:2019blr}\footnote{Also see for instance
  \cite{DHoker:1994gnm, Green:1999pv, Green:2008uj, DHoker:2015gmr}
  for earlier work on the discontinuity structure at four
  points.}. 
 The method of this work should be a key steppingstone to
evaluate the non-analytic sector of open-string amplitudes and
($n{\geq} 5$)-point closed-string amplitudes at one loop, and 
similar techniques are hoped to be applicable to higher loops. 
\end{itemize}
Furthermore, the study of matrix elements with interactions between
gauge and gravity multiplets (such as $F^m R^n$) and the application
of our method to heterotic strings are relegated to the future. On the
one hand, the loop diagrams with $D^{2k}F^n$ insertions are closely
related to the cylinder diagram in one-loop open-string amplitudes
which also propagates gravitational states in the non-planar case. On
the other hand, the closed-string exchange in non-planar cylinders
leads to analytic dependence on the kinematic variables, whereas the
results of this work are only sensitive to the non-analytic parts of
one-loop string amplitudes. This is consistent with the absence of
gravitational states in the loop of the ambitwistor-string
prescription for gauge-theory amplitudes \cite{Geyer:2015bja} that
serves as a starting point for our proposal.
Indeed, the massless spectra of heterotic ambitwistor strings
and conventional strings are different \cite{Berkovits:2018jvm, Flores:2019khe}, also see \cite{Geyer:2015jch} for mismatches between
ambitwistor-string amplitudes and field-theory limits of
conventional heterotic strings.

\subsection{Linearized versus quadratic propagators}
\label{sec:1.2}

The forward limit we consider identifies the loop momentum $\ell$ with a
lightlike external momentum of a tree diagram, and consequently most of the inverse Feynman propagators
in ambitwistor-string amplitudes are linear in $\ell$. The traditional
form of inverse Feynman propagators quadratic in $\ell$ is often better suited for integration, for instance since the Feynman $i\epsilon$ prescription is the familiar one, and such integrals have been studied for many decades. The two types of propagators are related via
partial-fraction manipulations combined with shifts of loop momenta
\cite{Geyer:2015bja, Geyer:2015jch}.  The conversions from linearized
to quadratic propagators has been actively discussed in the recent
literature \cite{Gomez:2016cqb, Gomez:2017lhy, Gomez:2017cpe,
  Ahmadiniaz:2018nvr, Agerskov:2019ryp, Farrow:2020voh} and will also
be performed for the one-loop matrix elements in this work. While the
$\alpha'$-expansions of the sphere and disk integrals in our proposal
do not involve~$\ell^2$, all of our examples of one-loop matrix
elements will be presented after conversion to quadratic propagators.

In spite of its drawbacks in extracting UV properties and
relating to traditional Feynman vertices, the linearized-propagator form 
of loop integrands offers additional flexibility in manifesting the color-kinematics
duality and double-copy structures. At the time of writing, the linearized propagators 
derived from ambitwistor strings are the only framework where a one-loop KLT formula
is known in field theory \cite{He:2016mzd, He:2017spx}. Our gravitational one-loop 
matrix elements obey a stringy (i.e.\ $\alpha'$-dressed) version of the field-theory KLT relations 
for loop integrands in the references which follow from forward limits of their tree-level 
counterparts~\cite{Kawai:1985xq}. Similarly, our one-loop matrix elements of gauge multiplets enjoy an echo of the
monodromy relations among open-string tree-level amplitudes \cite{Plahte:1970wy, BjerrumBohr:2009rd, Stieberger:2009hq}.\footnote{It would be interesting to connect the monodromy relations of one-loop matrix
elements in this work with those of the full-fledged one-loop open-string amplitudes \cite{Tourkine:2016bak, 
Hohenegger:2017kqy, Ochirov:2017jby, Tourkine:2019ukp, Casali:2019ihm, Casali:2020knc, Stieberger:2021daa}.}
These relations among loop integrands are exact in $\alpha'$ and mix
the building blocks for the planar and non-planar sectors of the matrix elements.

The recombination of linearized propagators to quadratic ones is
well-known to be possible for the complete one-loop integrands in
ambitwistor-string formulae. As a refinement of this, we will propose
a criterion to identify smaller subsectors of the correlation
functions which allow for quadratic-propagator representations.  Our
criterion concerns the double-periodicity of correlators on a torus
worldsheet {\it prior to} its degeneration to the nodal Riemann sphere
\cite{Geyer:2015bja, Geyer:2015jch}. Translations of the torus
punctures around the homology $B$-cycle require a simultaneous shift
of loop momentum \cite{DHoker:1989cxq}. Functions of the punctures and
$\ell$ with an invariance of this combined homology action have been
studied under the name of generalized elliptic integrands
\cite{Mafra:2017ioj, Mafra:2018pll} or {\it homology invariants}
\cite{DHoker:2020prr}.

The Feynman integrals descending from the moduli-space integration
of homology invariants in ambitwistor string theories are conjectured
to always have a quadratic-propagator representation. Moreover, the
same recombination to quadratic propagators is claimed to be possible
for our $\alpha'$-uplifts of the ambitwistor-string integrals. We
will illustrate these statements at the five-point level, where the
chiral worldsheet correlators for the gauge multiplet are built from seven
homology invariants to be denoted by $E_{\ldots}$ that \emph{individually}
recombine to quadratic propagators.

\subsection{Scope of our method and results}
\label{sec:1.scope}

The loop integrands of the matrix elements in this work are universal
to any number of spacetime dimensions.  The examples discussed in this
work are tailored to type-II superstrings with 32 supercharges and the
gauge sector of type-I superstrings with 16 supercharges. Maximal supersymmetry
in spacetime dimensions $D<10$ can be attained by compactification of the initially
ten-dimensional theories on a $(10{-}D)$-torus. The one-loop matrix elements
in this work refer to the effective field theory of massless states in $D$ dimensions
and therefore exclude Kaluza-Klein and winding modes from their propagators.
One-loop string amplitudes, by contrast, depend on the number of dimensions and
radii of the compactification torus through their analytic part which is currently
inaccessible to our method.

Our method can also be applied to type-I and type-II string amplitudes with
reduced supersymmetry due to K3 or Calabi-Yau compactifications
\cite{Bianchi:2006nf, Bianchi:2015vsa, Berg:2016wux, Berg:2016fui}.
This follows from the fact that the associated ambitwistor-string
correlators of \cite{He:2017spx, Geyer:2017ela, Edison:2020uzf} have a
natural uplift to forward limits of disk and sphere integrals. Like
this, one can for instance obtain one-loop matrix elements of the
operator $R^2$ and its completion under 16 supercharges which plays a
prominent role for the UV structure of four-dimensional
${\cal N}=4$ supergravity \cite{Carrasco:2013ypa, Bern:2015xsa, 
Bern:2017puu, Bern:2017tuc}.

\subsection{Further lines of motivation}
\label{sec:1.3}

As another line of motivation, the one-loop matrix elements in this
work serve as testing grounds whether large numbers of loop momenta in
the numerator can be reconciled with the color-kinematics duality~\cite{BCJ,loopBCJ}.  While BCJ numerators are known up to four loops in maximal super-Yang-Mills (SYM) and supergravity~\cite{loopBCJ, Carrasco:2011mn, Bern:2012uf}, the corresponding four-point five-loop amplitude has so far resisted a manifest realization of color-kinematics duality; that is, with all kinematic Jacobi identities among cubic diagrams imposed and with local numerators. Instead, the five-loop integrand was obtained through a relaxed approach as a
generalized double copy~\cite{Bern:2017yxu, Bern:2017ucb}, and subsequently used to determine the UV behavior~\cite{Bern:2018jmv}. The exact cause of a putative five-loop obstruction is difficult to pin down, due to the complexity of a five-loop calculation. If lower-loop calculations encounter similar problems, then they would provide much simpler testing ground for overcoming the obstacles. For example, in \cite{Mogull:2015adi} such an obstacle was overcome at two loops in non-supersymmetric Yang-Mills and gravity through an increase of the powers of loop momenta. One may expect that such effects can also be observed at one loop by 
varying the numbers of loop-momentum factors in the numerators.  

The infinite number of operators in the open-string effective action which are know to be compatible with tree-level color-kinematics duality \cite{Broedel:2012rc, Stieberger:2014hba, Huang:2016tag}, generate helpful illustrative examples for probing this question. Already their one-loop matrix elements involve high
powers of loop momenta and might serve as toy examples to settle their
compatibility with kinematic Jacobi relations. For instance, the
moderately sized one-loop matrix elements in section \ref{sec:4} with
$\alpha'^3 \zeta_3 D^2F^4$ and $\alpha'^5 \zeta_5 D^6F^4$ insertions
furnish a manageable laboratory for manifestly color-kinematic dual
representations that bypasses the combinatorial complexity of the
five-loop problem in field theory.

Finally, the results of this work offer a new line of attack for a
challenging problem at the interface of string amplitudes, number
theory and algebraic geometry: intermediate steps in computing the
low-energy expansion of one-loop closed-string amplitudes introduce a
fascinating class of non-holomorphic modular forms dubbed {\it modular
  graph forms} \cite{DHoker:2015wxz, DHoker:2016mwo}. Many of their
properties and their relations have been well-understood on the basis
of their differential equations in the modular parameter $\tau$
\cite{DHoker:2015gmr, DHoker:2016mwo, DHoker:2016quv, Broedel:2018izr,
  Gerken:2019cxz, Gerken:2020yii} which still necessitate further
input on their behaviour at the cusp $\tau \rightarrow i
\infty$. In spite of steady progress in the mathematics and physics
literature \cite{Zerbini:2015rss, DHoker:2019xef, Panzertalk,
  Zagier:2019eus, Vanhove:2020qtt}, computing the cusp degeneration of
generic modular graph forms is still a daunting task.

As will be detailed in section \ref{sec:5}, the UV properties
of our one-loop matrix elements are in one-to-one correspondence with
the expansion of modular graph forms around the cusp. The Feynman
integrals at fixed order in the $\alpha'$-expansion of one-loop matrix
elements still carry information on any higher $\alpha'$-order through
their higher-dimensional UV divergences. In this way,
infinitely many contributions to degenerate modular graph forms can be
inferred from straightforward UV properties of triangle and
bubble integrals. In simple cases where the behaviour of modular graph
forms around the cusp is known, this connection has been used as a
check of the loop integrands in the matrix elements. At higher orders
in $\alpha'$, the logic can be turned around such as to use the matrix
element as a new method to infer mathematical information.

\subsection{Outline}
\label{sec:1.4}

The main body of this work begins with a review of various ideas from
ambitwistor-string amplitudes, $\alpha'$-expansions of genus-zero integrals
and superstring effective actions in section~\ref{sec:2}. Then, the key definitions
of one-loop matrix elements as well as their KLT- and monodromy relations can
be found in section~\ref{sec:3}, followed by detailed four- and five-point
examples in section~\ref{sec:4}. Their connection with superstring amplitudes
and the conjectural origin of quadratic propagators from homology invariance
is discussed in section \ref{stringysec}. In section \ref{sec:5}, we relate the 
UV structure of one-loop matrix elements to the non-separating degeneration
of genus-one string integrals. Section \ref{mainsec7} is dedicated to the discontinuities 
of one-loop matrix elements and superstring amplitudes. Finally, we conclude with
a brief summary and outlook in section~\ref{sec:out}. Several appendices complement
the material in the main text.


\section{Review}
\label{sec:2}


\subsection{One-loop amplitudes from ambitwistor strings}
\label{sec:2.1}

Our starting point is the prescription for one-loop SYM and
supergravity amplitudes from ambitwistor string theories. Both the RNS
\cite{Mason:2013sva, Adamo:2013tsa} and the pure-spinor formulation
\cite{Berkovits:2013xba, Adamo:2015hoa} of ambitwistor strings express
one-loop amplitudes via moduli-space integrals over punctured tori
that are localized on the solutions of the scattering equations. These
integrals were reduced to forward limits of tree-level building blocks
\cite{Geyer:2015bja, Geyer:2015jch} from the boundary of moduli space
where the torus degenerates to a nodal Riemann sphere.

More specifically, the ambitwistor prescription for $n$-point one-loop
amplitudes in $D$-dimensional theories with double-copy structure
$L \otimes R$ is given by\footnote{Note that the $D$-dimensional loop integrand in (\ref{rev.1}) can also be obtained from CHY representations of tree-level amplitudes in $D{+}1$ dimensions \cite{Cachazo:2015aol}.}
\beq
M^{\te{1-loop}}_{n, L \otimes R} = \int \frac{ \dd^D \ell}{\ell^2} \lim_{k_{\pm} \rightarrow \pm \ell} \int_{\mathbb C^{n-1}} \dd \mu_{n+2}^{\te{tree}} \,{\cal I}_L(\ell) \,{\cal I}_R(\ell)\, .
\label{rev.1}
\eeq
We define the volume element in loop-momentum space as
$\dd^D \ell = -i \dd\ell^0 \,\dd\ell^1 \ldots \dd\ell^{D-1}$
with a factor of $-i$ to simplify later formulae,
and use a spacetime metric of signature $(-,+,+,\ldots,+)$.  The
forward limit $k_{\pm} \rightarrow \pm \ell$ amounts to integrating
  over $(n{+}2)$-punctured Riemann spheres with the well-known measure
  from the CHY formulae for tree-level amplitudes in
  \cite{Cachazo:2013gna, Cachazo:2013hca, Cachazo:2013iea}
\beq
\dd \mu_{n+2}^{\te{tree}} = \dd \sigma_2 \, \dd \sigma_3 \, \ldots \, \dd \sigma_n \prod_{i=2}^n \delta \bigg( 2\sum^{n+2}_{j \neq i} \frac{ k_i \cdot k_j }{\sigma_{ij}} \bigg) \, , \ \ \ \ \ \ \sigma_{ij} = \sigma_i - \sigma_j
\label{rev.2}
\eeq
and identifying the momenta $k_{+}=k_{n+1}, \ k_-= k_{n+2}$ of the last two legs with (plus or minus) the 
$D$-dimensional loop momentum $\ell$. The half integrands ${\cal I}_{L,R}(\ell)$ in (\ref{rev.1}) 
depend on the punctures $\sigma_1,\sigma_2,\ldots,\sigma_n$, the internal and external momenta $\ell,k_j$ and
the polarization- or color degrees of freedom specific to the theory. In this work, we need the following half integrands 
for color ${\cal I}_{{\rm col}}$ and kinematics $ {\cal I}_{{\rm kin}}(\ell)$:
\begin{align}
\te{SYM}: \ \ \ &{\cal I}_L(\ell) {\cal I}_R(\ell) \rightarrow {\cal I}_{\rm col}
{\cal I}_{\rm kin}(\ell) \label{symsugra} \\
\te{supergravity}: \ \ \ &{\cal I}_L(\ell) {\cal I}_R(\ell)\rightarrow {\cal I}_{\rm kin}(\ell)
 \bar {\cal I}_{\rm kin}(\ell) \notag
 \end{align}
 As indicated by the bar notation,
the kinematic half integrands ${\cal I}_{\rm kin}(\ell)$ and $ \bar {\cal I}_{\rm kin}(\ell)$ of supergravity
may carry different amounts of supersymmetry. 

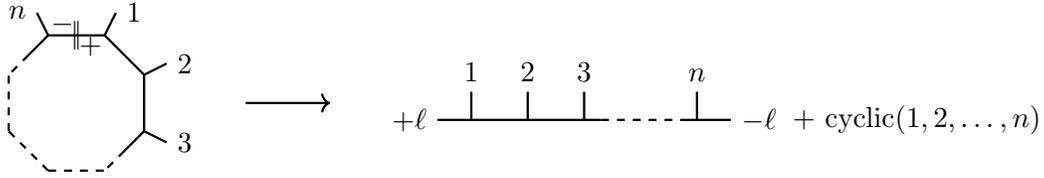
\begin{figure}
\begin{center}
\begin{tikzpicture} [scale=0.75, line width=0.30mm]
\begin{scope}[xshift=-0.8cm]
\draw (0.5,0)--(-0.5,0);
\draw (-0.5,0)--(-0.85,-0.35);
\draw [dashed](-0.85,-0.35)--(-1.2,-0.7);
\draw (0.5,0)--(1.2,-0.7);
\draw[dashed] (-1.2,-1.7)--(-1.2,-0.7);
\draw (1.2,-1.7)--(1.2,-0.7);
\draw (1.2,-1.7)--(0.85,-2.05);
\draw[dashed] (0.85,-2.05)--(0.5,-2.4);
\draw[dashed] (-0.5,-2.4)--(0.5,-2.4);
\draw[dashed] (-0.5,-2.4)--(-1.2,-1.7);
\draw (-0.5,0)--(-0.7,0.4)node[left]{$n$};
\draw (0.5,0)--(0.7,0.4)node[right]{$1$};
\draw (1.2,-0.7)--(1.6,-0.5)node[right]{$2$};
\draw (1.2,-1.7)--(1.6,-1.9)node[right]{$3$};
\draw (0,0) node{$| \! |$};
\draw (-0.25,0.2)node{$-$};
\draw (0.25,-0.2)node{$+$};
\end{scope}
\draw[-> ](2.2,-1.2)  -- (3.7,-1.2);
\begin{scope}[xshift=2.7cm, yshift=0.5cm]
\draw(11.4,-2)node{$+ \ {\rm cyclic}(1,2,\ldots,n)$};
\draw (2.9,-2)node[left]{$+\ell$} -- (5.8,-2);
\draw (7.2,-2) -- (8.1,-2)node[right]{$-\ell$};
\draw (3.5,-2) -- (3.5,-1.5)node[above]{$1$};
\draw (4.5,-2) -- (4.5,-1.5)node[above]{$2$};
\draw (5.5,-2) -- (5.5,-1.5)node[above]{$3$};
\draw[dashed] (5.8,-2) -- (7.2,-2);
\draw (7.5,-2) -- (7.5,-1.5)node[above]{$n$};
\end{scope}
\end{tikzpicture}
\caption{Left panel: $n$-gon diagram associated with the color factor $c_{12\ldots n}$ in (\ref{rev.5}).
Right panel: the partial-fraction decomposition to be reviewed in (\ref{ngonPF}) relates each $n$-gon
with a cyclic orbit of the depicted $(n{+}2)$-point tree-level diagram.\label{figcutting}}
\end{center}
\end{figure}

It will be convenient to express the half integrands in terms of Parke-Taylor factors
\beq
{\rm PT}(+,1,2,\ldots,n,-) = \frac{1}{\sigma_{+1} \sigma_{12} \sigma_{23}\ldots \sigma_{n-1,n} }\,,
\label{rev.3}
\eeq
which are already adapted to our ubiquitous choice of ${\rm SL}_2(\mathbb C)$-frame on the Riemann sphere
(implicitly used in (\ref{rev.2}) by the absence of $\dd \sigma_1$)
\beq
\sigma_1 = 1 \, , \ \ \ \ \ \ \sigma_+ =0 \, , \ \ \ \ \ \ \sigma_- \rightarrow \infty \, .
\label{rev.4}
\eeq
The half integrand for the color dependence of SYM is given in terms of structure constants
$f^{abc}$ of some gauge group ${\cal G}$ (with generators $t^a$, commutation
relations $[t^a, t^b]=  f^{abc} t^c$ and adjoint indices $a=1,2,\ldots,{\rm dim}({\cal G})$) 
\cite{Geyer:2015bja, Geyer:2015jch},
\begin{align}
{\cal I}_{n,{\rm col}} &= \sum_{\rho \in S_n} 
c_{\rho(12\ldots n) }  {\rm PT}(+,\rho(1,2,\ldots,n),-) 
\notag \\
c_{12\ldots n} &= f^{z a_1 b} f^{b a_2 c} f^{c a_3 d} \ldots f^{y a_{n} z}  \label{rev.5} \, .
\end{align}
The color tensors $c_{12\ldots n} $ arise from the $n$-gon diagrams in
Figure~\ref{figcutting}. By the Jacobi identities among contracted
structure constants $f^{e [ab} f^{c]de}=0$, they form a basis for
those of all other cubic diagrams in planar or non-planar one-loop SYM
amplitudes. By the cyclic invariance
$c_{12\ldots n} = c_{23\ldots n1} $, color ordered amplitudes
following from (\ref{rev.5}) involve a cyclic sum of Parke-Taylor
factors (\ref{rev.3}),
\beq
A^{\te{1-loop}}_{\rm SYM}(1,2,\ldots,n) = \int \frac{ \dd^D \ell}{\ell^2} \lim_{k_{\pm} \rightarrow \pm \ell} \int\limits_{\mathbb C^{n-1}} \dd \mu_{n+2}^{\te{tree}} \, {\cal I}_{n}(\ell) \! \!  \! \! \! \! \sum_{\gamma \in {\rm cyc}(1,2,\ldots,n)} \! \! \! \! \! \! {\rm PT}(+,\gamma(1,2,\ldots,n),-)  \, ,
\label{rev.6}
\eeq
where we drop the subscript $_{\rm kin}$ of
${\cal I}_{n}(\ell) ={\cal I}_{n,{\rm kin}}(\ell) $ here and below to
avoid cluttering. The Parke-Taylor decomposition (\ref{rev.5}) of the
color-half integrand can also be attained for its kinematic counterpart \cite{He:2017spx}
\begin{align}
{\cal I}_{n}(\ell) &= \sum_{\rho \in S_n} 
N_{+|\rho(12\ldots n)| -}(\ell)  {\rm PT}(+,\rho(1,2,\ldots,n),-) \,,
\label{rev.7}
\end{align}
with $n!$ independent kinematic numerators
$N_{+|\rho(12\ldots n)| -}(\ell) $ associated with $n$-gon
diagrams. More specifically, each $N_{+|12\ldots n| -}(\ell)$
corresponds to an $(n{+}2)$-point tree-level diagram in
Figure~\ref{figcutting} obtained from the partial-fraction
decomposition to be reviewed in (\ref{ngonPF}) below.

While the color factor (\ref{rev.5}) is a thoroughly tested prescription, the
kinematic half integrands (\ref{rev.7}) are derived from vertex
operators for SYM and supergravity multiplets. Their worldsheet correlation
functions on the torus can be imported from those of conventional
strings \cite{Tsuchiya:1988va, Broedel:2014vla, Mafra:2018qqe} in the
chiral-splitting formalism \cite{DHoker:1988pdl, DHoker:1989cxq}. The
degeneration of these torus correlators to the Riemann sphere yields
linear combinations of Parke-Taylor factors as in (\ref{rev.7}) after
repeated use of integration by parts or scattering equations
\cite{He:2017spx}. Alternatively, (\ref{rev.7}) can be obtained from
$(n{+}2)$-point correlators on the Riemann sphere by taking the
forward limit in pairs of bosons and fermions \cite{Edison:2020uzf},
also see \cite{Geyer:2015jch, Geyer:2017ela} for an earlier discussion of the bosonic
forward limits. In both approaches, worldsheet techniques yield
explicit expressions for the numerators
$N_{+|\rho(12\ldots n)| -}(\ell) $ in theories with variable amount of
supersymmetry.

For non-maximal supersymmetry the forward limit can be subtle since the collinear momenta $\ell$ and $-\ell$ may in principle introduce singularities corresponding to tadpole and external-bubble diagrams. However, they can often be removed with proper treatment such as the Minahaning procedure \cite{Minahan:1987ha} in the organization of string amplitudes with half-maximal supersymmetry in \cite{Berg:2016wux, Berg:2016fui}. Alternatively, see the approach of \cite{Johansson:2014zca} for identifying external bubbles that arise from 0/0 cancelations in half-maximally supersymmetric field theories. In non-supersymmetric cases, forward-limit divergences are regularized by the dropout of singular solutions of the scattering equations \cite{He:2015yua, Cachazo:2015aol} which has been used to obtain explicit numerators in \cite{Geyer:2017ela}.

For maximally supersymmetric SYM with 16 supercharges, the numerators
$N_{+|12\ldots n| -}(\ell) $ in (\ref{rev.7}) are well-known to be degree-$(n{-}4)$
polynomials in the loop momentum. Manifestly supersymmetric
expressions in pure-spinor superspace \cite{Berkovits:2000fe,
  Berkovits:2006ik} can be found in \cite{Mafra:2014gja,
  Bridges:2021ebs}.  For external gluons, the four- and five-point
examples to be discussed in section \ref{sec:4} are expressible in
terms of the standard $t_8$-tensor contracting polarization vectors.


\subsection{Linearized versus quadratic propagators}
\label{sec:2.2}

By the organization of the color and kinematic half integrands in (\ref{rev.5}) and (\ref{rev.7}), the 
$\sigma_j$-integrals in the one-loop amplitudes (\ref{rev.1}) boil down to those over pairs of
Parke-Taylor factors. From the early days of CHY formulae, such Parke-Taylor integrals are
given by doubly-partial amplitudes $m(\alpha|\beta)$ of bi-adjoint scalars \cite{Cachazo:2013iea},
and the scattering equations at one loop pick up their forward limits
\begin{align}
\lim_{k_{\pm} \rightarrow \pm \ell} \int \limits_{\mathbb C^{n-1}} \dd \mu_{n+2}^{\te{tree}}\, {\rm PT}\big(\alpha(+,1,2,\ldots,n,-)\big)  {\rm PT}\big(\beta(+,1,2,\ldots,n,-)\big) = \lim_{k_{\pm} \rightarrow \pm \ell}  m(\alpha |\beta)\, ,
\label{rev.11}
\end{align}
where $\alpha,\beta \in S_{n+2}$. Doubly partial amplitudes
are defined via tree-level amplitudes $M^{\rm tree}_{n,\phi^3}$ of bi-adjoint scalars $\Phi = \phi_{a,\tilde b} t^a \otimes \tilde t^{\tilde b}$
with cubic interaction ${\cal L}_{\rm int}\sim \phi_{a,\tilde a} \phi_{b,\tilde b} \phi_{c,\tilde c} f^{abc} \tilde f^{\tilde a\tilde b\tilde c}$ by color-decomposing w.r.t.\ traces of the two types of Lie-algebra generators 
$t^{a}$ and $\tilde t^{\tilde b}$,
\beq
M^{\rm tree}_{n,\phi^3} = \sum_{\alpha,\beta \in S_{n}/\mathbb Z_n} 
{\rm Tr}(t^{a_{\alpha(1)}} t^{a_{\alpha(2)}}\ldots t^{a_{\alpha(n)}})
{\rm Tr}(\tilde t^{\tilde b_{\beta(1)}}  \tilde t^{\tilde b_{\beta(2)}}\ldots \tilde t^{\tilde b_{\beta(n)}})
m(\alpha | \beta)\, .
\label{defmab}
\eeq
Accordingly, doubly-partial amplitudes depend on the external momenta $k_{j=1,2,\ldots,n}$ and $k_{\pm}$ through the 
inverse propagators
\beq
s_{ab} =2 k_a \cdot k_b \, , \ \ \ \ \ \
s_{a_1 a_2 \ldots a_p} =2 \sum_{1\leq i < j}^p k_{a_i} \cdot k_{a_j}\,,
\label{rev.12}
\eeq
of cubic diagrams that are compatible with the cyclic orderings of both $\alpha$ and $\beta$. Any
$m(\alpha|\beta)$ can be straightforwardly computed via graphical methods \cite{Cachazo:2013iea} or through the 
Berends-Giele recursion given in \cite{Mafra:2016ltu}. These methods yield the $\mathbb C^{n-1}$ integrals in the 
one-loop amplitudes~(\ref{rev.5}) and (\ref{rev.7}) as a rational function of $s_{a_1 \ldots a_p}$
with $a_i \in \{1,2,\ldots,n,+,-\}$.

By the vanishing of $s_{12\ldots n-1}$ and $s_{12\ldots n}$ in the momentum phase space of
$n$ massless particles subject to momentum conservation $\sum_{j=1}^n k_j=0$, individual
$m(\alpha|\beta)$ in one-loop amplitudes may introduce forward-limit divergences. In 
supersymmetric amplitudes, these divergences can be efficiently cancelled by first performing the permutation
sums over $m(\alpha|\beta)$ prescribed by ${\cal I}_{L}(\ell){\cal I}_{R}(\ell)$ before taking
the forward limit and imposing phase-space constraints. 

The inverse Feynman propagators in the loop integrands of scattering amplitudes
are of the form $(\ell+K)^2$ quadratic in $\ell$, with $K$ a sum of external momenta.
However, the SYM and supergravity amplitudes resulting from
(\ref{rev.1}), (\ref{rev.6}), (\ref{rev.7}) and (\ref{rev.11}),
\begin{align}
A^{\te{1-loop}}_{\rm SYM}(1,2,\ldots,n) &=  \int \frac{ \dd^D \ell}{\ell^2} \lim_{k_{\pm} \rightarrow \pm \ell} 
\sum_{\substack{\gamma \in {\rm cyc}(1,2,\ldots,n)  \\  \rho \in S_n}} m(+,\gamma,-|+,\rho,-) N_{+|\rho|-}(\ell)\,,
\label{rev.21} \\
M^{\te{1-loop}}_{n,{\rm SUGRA}} &=  \int \frac{ \dd^D \ell}{\ell^2} \lim_{k_{\pm} \rightarrow \pm \ell} 
\sum_{\gamma, \rho \in S_n} m(+,\gamma,-|+,\rho,-) N_{+|\gamma |-}(\ell) \bar N_{+|\rho|-}(\ell)\,,
\label{rev.22}
\end{align}
mostly feature linearized inverse propagators $2\ell \cdot K + K^2$ instead of $(\ell+K)^2$ from the
variables $s_{ij\ldots p,\pm}$ in doubly partial amplitudes. Using the notation
\beq
s_{a_1a_2\ldots a_p,\pm \ell} = s_{a_1a_2\ldots a_p} \pm 2 \ell\cdot k_{a_1a_2\ldots a_p}  \, , \ \ \ \ \ \ 
k_{a_1a_2\ldots a_p} = \sum_{i=1}^p k_{a_i}\,,
\label{rev.23}
\eeq
a representative four-point example of linearized propagators is
\beq
\sum_{\rho \in S_4} m(+,1,2,3,4,-|+,\rho(1,2,3,4),-) = \frac{1}{s_{1,\ell} s_{12,\ell} s_{123,\ell}} \, .
\label{rev.24}
\eeq
The denominator is obtained by manually dropping the $\ell^2$ in three of
the Feynman propagators on the left-hand side of the four-point box integral
\begin{align}
\int \frac{ \dd^D \ell  }{ \ell^2 (\ell{+}k_1)^2 (\ell {+} k_{12})^2 (\ell {+} k_{123})^2}
= \int \frac{\dd^D \ell }{ \ell^2} \bigg( \frac{1}{ s_{1,\ell} s_{12,\ell} s_{123,\ell} } + {\rm cyc}(1,2,3,4) \bigg)   \, .
\label{rev.25}
\end{align}
In passing to the right-hand side, however, partial-fraction manipulations and shifts of the
loop momenta $\ell \rightarrow \ell \pm k_i$ have been used to rewrite the box integral
in terms of linearized-propagator expressions as in (\ref{rev.25}). The same manipulations
apply to $n$-gons
\begin{align}
&\int {  {\rm d}^D \ell \over \ell^2 (\ell{+}k_1)^2 (\ell{+}k_{12})^2 \ldots (\ell {+} k_{12\ldots n-1})^2}
= \sum_{i=0}^{n-1} \int {   {\rm d}^D \ell \over (\ell {+} k_{12\ldots i})^2 } \prod^n_{j \neq i} {1\over (\ell {+} k_{12\ldots j})^2 - (\ell {+} k_{12\ldots i})^2 }\notag \\
& \ \ \ \ \ = \sum_{i=0}^{n-1} \int { {\rm d}^D \ell \over \ell^2}   \prod_{j=0}^{i-1} {1\over s_{j+1,j+2,\ldots,i,-\ell}} \prod_{j=i+1}^{n-1} {1\over s_{i+1,i+2,\ldots,j,\ell} } \label{ngonPF}\, ,
\end{align}
and their generalizations to massive corners with momenta $k_A = k_{a_1}+k_{a_2}+ \ldots +k_{a_p}$
for $A=a_1a_2\ldots a_p$, see (\ref{rev.23}). The example of the box integral in (\ref{rev.24}) and (\ref{rev.25})
illustrates that the recombination of linearized propagators to quadratic ones generically 
relies on the sum over cyclic permutations $\gamma$ in the color-ordered SYM amplitudes (\ref{rev.21}). Similarly, 
the quadratic-propagator representation of supergravity amplitudes is far from manifest in (\ref{rev.22})
and relies on the interplay between different terms in the permutation sum over $\gamma,\rho \in S_n$.

Consider a generic numerator $\mathfrak{N}(\ell)$ belonging to a $(p\leq n)$-gon diagram with quadratic propagators,
then after inserting it into the cyclic sum of \cref{ngonPF} each term will become a shifted version 
$\mathfrak{N}(\ell{-}k_{\ldots})$. 
For the box integral, we get
\begin{align}
\int \frac{ \mathfrak{N}(\ell)\, \dd^D \ell  }{ \ell^2 (\ell{+}k_1)^2 (\ell {+} k_{12})^2 (\ell {+} k_{123})^2}
&= \int \frac{ \dd^D \ell }{ \ell^2} \, \bigg( \frac{\mathfrak{N}_{+|1234|- }(\ell)}{ s_{1,\ell} s_{12,\ell} s_{123,\ell} } 
+\frac{\mathfrak{N}_{+|2341|- }(\ell)}{ s_{2,\ell} s_{23,\ell} s_{234,\ell} }   \label{shiftnums} \\
&\ \ \ \ \ \ \ \ \ \ \ \ \,
+\frac{\mathfrak{N}_{+|3412|- }(\ell)}{ s_{3,\ell} s_{34,\ell} s_{341,\ell} } 
+\frac{\mathfrak{N}_{+|4123|- }(\ell)}{ s_{4,\ell} s_{41,\ell} s_{412,\ell} } \bigg)   \, ,
\notag
\end{align}
where
\begin{align}
  & \mathfrak{N}_{+|1234|-}(\ell)=\mathfrak{N}(\ell)\,, &&\mathfrak{N}_{+|2341|-}(\ell)=\mathfrak{N}(\ell{-}k_1)\,, \\
  & \mathfrak{N}_{+|3412|-}(\ell)=\mathfrak{N}(\ell{-}k_{12})\,, &&\mathfrak{N}_{+|4123|-}(\ell)=\mathfrak{N}(\ell{-}k_{123})\,.\nonumber
\end{align}
Therefore, we can trivially transform a $p$-gon integrand in the linearized-propagator 
representation back to standard quadratic propagators if
\begin{align}\label{eq:Nshift}
\mathfrak{N}_{+|i+1,i+2,\ldots,p,1,\ldots,i|-}(\ell)=\mathfrak{N}_{+|123\ldots p|-}(\ell{-}k_{12\ldots i})\,,\qquad 1\leq i\leq p{-}1\,,
\end{align}
for example,
\begin{align}
\mathfrak{N}_{+|23\ldots p1|-}(\ell)=\mathfrak{N}_{+|123\ldots p|-}(\ell{-}k_1)\,,\quad \mathfrak{N}_{+|34\ldots p12|-}(\ell)=\mathfrak{N}_{+|123\ldots p|-}(\ell{-}k_{12})\,,\text{ etc.}
\end{align}
However, there is in principle no need for the numerators with $p\leq n$ 
to obey (\ref{eq:Nshift}) -- they may depart from this relation through contact terms 
that cancel out in the full amplitude.

The appearance of linearized propagators from the ambitwistor-string prescription was firstly noticed
in \cite{Geyer:2015bja} and their equivalence with quadratic propagators has been actively discussed in the recent
literature including \cite{Gomez:2016cqb, Gomez:2017lhy, Gomez:2017cpe, Ahmadiniaz:2018nvr, Agerskov:2019ryp, Farrow:2020voh}. In this work, we will find the first instance of this phenomenon in matrix elements of higher-mass-dimension operators of the schematic form ${\rm Tr}\{D^{2k} F^n\}$ or $D^{2k} R^n$ in the low-energy effective action of open and closed superstrings to be reviewed in section \ref{sec:2.8}. Moreover, motivated by the five-point examples in SYM and supergravity, the possibility to recombine linearized to quadratic propagators is conjectured to 
descend from homology invariance in the chiral-splitting formalism, see section \ref{stringysec}.


\subsection{Disk and sphere integrals}
\label{sec:2.3}

The doubly-partial amplitudes $m(\alpha|\beta)$ furnish an important
building block in the explicit evaluation of moduli-space integrals
(\ref{rev.11}) of ambitwistor string theories.
Also in conventional string theories, the $m(\alpha|\beta)$ feature
prominently in the field-theory limit of integrals $Z$ over punctured
disks and $J$ over punctured spheres in open- and closed-string
tree-level amplitudes,
\begin{align}
\! \! \! \! Z(\gamma,n|\rho,n) &= (-1)^n (\ap)^{n-3} \! \int_{\mathfrak{D}(\gamma)} \! \! \! \! \dd \sigma_2 \, \dd \sigma_3\, \ldots \, \dd \sigma_{n-2} \! \prod_{1\leq i<j}^{n-1}  \! \! |\sigma_{ij}|^{ \alpha' s_{ij}} {\rm PT}(\rho(1,2,\ldots,n{-}1),n)
\label{rev.31} \\
\! \! \! \! J(\gamma,n|\rho,n) &= \bigg( \frac{\ap}{\pi} \bigg)^{n-3} \int_{\mathbb C^{n-3}} \dd^2 \sigma_2 \, \dd^2 \sigma_3\, \ldots \, \dd^2 \sigma_{n-2} \, \prod_{1\leq i<j}^{n-1} |\sigma_{ij}|^{ 2\alpha' s_{ij}} 
\notag \\
& \ \ \ \ \ \ \times \overline{ {\rm PT}(\gamma(1,2,\ldots,n{-}1),n)} \, {\rm PT}(\rho(1,2,\ldots,n{-}1),n) \label{rev.32}
\end{align}
indexed by two permutations $\gamma,\rho \in S_{n-1}$ of legs
$1,2,\ldots,n{-}1$. We have already picked ${\rm SL}_2$-frames with
$(\sigma_1,\sigma_{n-1},\sigma_n) \rightarrow (0,1,\infty)$ such that
the Parke-Taylor factors have the chain structure of (\ref{rev.3}),
and the integration domain $\mathfrak{D}(\gamma)$ for the punctures on
the disk boundary is given by
\beq
\mathfrak{D}(\gamma) = \{\sigma_{j=1,2,\ldots,n-1} \in \mathbb R, \ {-}\infty< \sigma_{\gamma(1)}<\sigma_{\gamma(2)}<\ldots <
\sigma_{\gamma(n-1)} < \infty \}\, .
\label{rev.33}
\eeq
Throughout this work, the normalization convention for $\alpha'$ is tailored to
open strings, and the accurate numerical factors in closed-string quantities 
can be restored by rescaling $\alpha' \rightarrow \alpha'/4$.

The low-energy expansion of the disk and sphere integrals yields a
Laurent series in the dimensionless Mandelstam invariants
$\alpha' s_{ij}$ whose coefficients are $\mathbb Q$-linear
combinations of multiple zeta values (MZVs) \cite{Terasoma,
  Brown:2009qja}
\beq
\zeta_{n_1,n_2,\ldots,n_r} = \sum_{0<k_1<k_2<\ldots <k_r}^\infty k_1^{-n_1}  k_2^{-n_2}\ldots  k_r^{-n_r}\, , \ \ \ \ 
n_1,\ldots,n_r \in \mathbb N \, , \ \ \ \ n_r \geq 2 \, .
\label{rev.34}
\eeq
By the normalization of (\ref{rev.31}) and (\ref{rev.32}), their
$\alpha'\rightarrow 0$ limit is free of MZVs and yields the doubly
partial amplitudes $m(\alpha|\beta)$ in case of both disk integrals
\cite{Cachazo:2013iea} and sphere integrals \cite{Stieberger:2014hba},
\beq
\lim_{\alpha' \rightarrow 0} Z(\gamma,n|\rho,n) = \lim_{\alpha' \rightarrow 0} J(\gamma,n|\rho,n) = m(\gamma,n|\rho,n) \,,\label{rev.35}
\eeq
so $Z$ and $J$ are natural uplifts of the $m(\alpha|\beta)$ in
SYM and supergravity amplitudes (\ref{rev.21}) and (\ref{rev.22}),
respectively. The subleading orders in $\ap$ beyond the field-theory
limit (\ref{rev.35}) comprise Laurent polynomials $p_d(s_{ij})$ of
homogeneity degree $d=3{-}n{+}w$ in $s_{ij\ldots p}$ along with
$\alpha'^w$. The weight $n_1{+}\ldots {+}n_r=w$ of the accompanying
MZVs $\zeta_{n_1,\ldots,n_r} $ matches the order of $\alpha'^w$, i.e.\
both $Z$ and $J$ are said to be uniformly transcendental.

On top of their identical field-theory limits (\ref{rev.35}), the
$\alpha'$-expansions of sphere integrals are completely determined by
those of the disk integrals through the single-valued map ``sv'' of
MZVs
\cite{Schnetz:2013hqa, Brown:2013gia},\footnote{Strictly speaking, the single-valued map is only well
	defined for the motivic versions
	$\zeta^{\mathfrak{m}}_{n_1,n_2,\ldots,n_r}$ of MZVs.} e.g. ($k\in \mathbb N$)
\beq
{\rm sv} \, \zeta_{2k} = 0\, , \ \ \ \ \ \ {\rm sv}\, \zeta_{2k+1} = 2 \zeta_{2k+1} \, , \ \ \ \ \ \
{\rm sv} \, \zeta_{3,5} = -10 \zeta_3 \zeta_5
\label{rev.36}
\eeq
which acts order by order in the $\alpha'$-expansion of
\cite{Schlotterer:2012ny,Stieberger:2013wea,Stieberger:2014hba,Schlotterer:2018abc,Vanhove:2018elu,Brown:2019wna}
\begin{align}
J(\gamma,n|\rho,n)  = {\rm sv} \, Z(\gamma,n|\rho,n) \, .
\label{rev.37}
\end{align}
At the lowest orders in the Laurent-expansion beyond the field-theory
limit (\ref{rev.35}), the single-valued map (\ref{rev.37}) produces
\begin{align}
Z&=  m  + \alpha'^2 \zeta_2 p_{5-n}(s_{ij}) + \alpha'^3 \zeta_3 p_{6-n}(s_{ij}) + \alpha'^4 \zeta_4 p_{7-n}(s_{ij}) \notag \\
& \ \ \ \ \ \ \, + \alpha'^5 \big[ \zeta_5 p_{8-n}(s_{ij}) +\zeta_2 \zeta_3 \hat p_{8-n}(s_{ij}) \big] + {\cal O}(\alpha'^6)\,, 
\label{rev.38} \\
J&=  m  +2 \alpha'^3 \zeta_3 p_{6-n}(s_{ij}) +2 \alpha'^5 \zeta_5 p_{8-n}(s_{ij}) + {\cal O}(\alpha'^6) \, .
\label{rev.39} 
\end{align}
We have suppressed the common labels $(\gamma,n|\rho,n) $ of the doubly partial 
amplitudes $m= p_{3-n}(s_{ij})$ and the degree-$d$ Laurent-polynomials $p_d(s_{ij})$. 
The single-valued map removes the appearance of $ \zeta_2,\, \zeta_4,\,\zeta_2 \zeta_3$ 
and doubles the coefficients of $\zeta_3,\, \zeta_5$ in passing from disk to sphere integrals at the orders $\ap^{\leq 5}$,
i.e.\ the polynomials $p_{6-n}(s_{ij})$ and $p_{8-n}(s_{ij})$ are universal to (\ref{rev.38}) and (\ref{rev.39}). 
The coefficients of $\zeta_w \ap^w$ determine those of MZVs at higher depth $r\geq 2$ and arbitrary products 
such as $\zeta_2 \zeta_3$ in (\ref{rev.38}) \cite{Schlotterer:2012ny} which can be attributed to the group-like 
behaviour of the $Z$-integrals under the motivic coaction \cite{Drummond:2013vz}.

The pattern of Laurent polynomials at the leading orders of (\ref{rev.38}) 
can be illustrated through the six-point examples
\begin{align}
&Z(1,2,3,4,5,6|1, 4, 2, 3,6,5) =
\frac{ 1}{s_{23} s_{234} s_{56}}  \notag \\
&\ \ \ \ + \ap^2 \zeta_2 \bigg( 
 {-} \frac{s_{16}}{s_{23} s_{234}} -\frac{ s_{123}}{s_{23} s_{56}} - \frac{s_{34}}{s_{234} s_{56}} + \frac{1}{s_{23}} + \frac{1}{s_{56}}
\bigg)
\label{rev.40}  \\
&\ \ \ \ + \ap^3 \zeta_3 \bigg( \frac{ s_{16}(s_{16}{+}s_{56})}{s_{23} s_{234}} 
+\frac{ s_{123}(s_{123}{+}s_{234})}{s_{23} s_{56}} 
+ \frac{ (s_{23}{+}s_{34}) s_{34}}{ s_{234} s_{56}} 
\notag \\
&\ \ \ \ \ \ \ \ \ \  - \frac{s_{234}}{s_{23}} - \frac{2 s_{45}}{s_{23}}  - \frac{ s_{56}}{s_{23}} 
 - \frac{2 s_{12}}{s_{56}}   - \frac{ s_{23}}{s_{56}} - \frac{ s_{234}}{s_{56}}\bigg) + {\cal O}(\ap^4)
 \notag
\end{align}
as well as
\begin{align}
&\sum_{\rho \in S_4} Z(1,2,\ldots,6|1,\rho(2,3,4,5),6) = \frac{1}{s_{12} s_{123} s_{56}}
- \ap^2 \zeta_2 \bigg( 
\frac{s_{45}}{s_{12} s_{123}} + \frac{s_{23}}{s_{123} s_{56}} + \frac{ s_{34}}{s_{12} s_{56}}
\bigg)
\notag \\
&\ \ \ \ + \ap^3 \zeta_3 \bigg( \frac{ s_{45}(s_{45}{+}s_{56})}{s_{12} s_{123}}  
 + \frac{ (s_{12}{+}s_{23}) s_{23}}{s_{123} s_{56}}
 + \frac{ (s_{123} {+}s_{34})s_{34}}{s_{12} s_{56}} 
 - \frac{ s_{35}}{s_{12}} 
  - \frac{ s_{24}}{s_{56}} 
  \bigg) + {\cal O}(\ap^4)\, .
  \label{rev.40a} 
\end{align}
The results in section \ref{sec:4} for one-loop matrix elements of
higher-mass-dimension operators crucially rely on the availability of
six- and seven-point $\ap$-expansions which we have generated by means
of the Berends-Giele recursion in \cite{Mafra:2016mcc,
  gitrep}. Alternative methods to determine the polynomial structure
of the $\ap$-expansions at all multiplicities are based on the
Drinfeld associator \cite{Broedel:2013aza} (also see
\cite{Drummond:2013vz, Kaderli:2019dny}) or polylogarithm
manipulations \cite{Zfunctions}, with machine-readable expressions
available for download on \cite{MZVWebsite}.
Another approach to generate $\alpha'$-expansions of disk integrals
(\ref{rev.31}) at $n\leq 7$ points is to exploit their connection with
hypergeometric functions \cite{Oprisa:2005wu, Stieberger:2006te,
  Stieberger:2009rr, Boels:2013jua, Puhlfuerst:2015gta}.
  

\subsection{Tree-level effective action of type I and II superstrings}
\label{sec:2.8}

The schematic form of the tree-level effective action of type I and II superstrings at leading orders
in $\alpha'$ has been previewed in (\ref{seffclosed}). In spite of all-multiplicity formulae including the 
exact tensor structure
of tree amplitudes of gauge and gravity multiplets \cite{Mafra:2011nv, Schlotterer:2012ny}, the 
detailed tensor structure of the effective operators with more than four powers of
the non-abelian field strength $F$ is only known up to the $\alpha'^4$ order. One can anticipate
from the bosonic components at the subleading order of
\begin{align}
S_{\rm eff}^{\rm open}= \int \dd^{D} x \, {\rm Tr}\bigg\{ \frac{1}{4} F_{\mu \nu} F^{\mu \nu}
+ \frac{ \ap^2 \zeta_2 }{4} \Big[&{-} 8F^\mu{}_\lambda F^\lambda{}_\nu F^{\rho}{}_{\mu}  F^{\nu}{}_{\rho} 
 - 4  F^\mu{}_\lambda F^\lambda{}_\nu F^{\nu}{}_{\rho} F^{\rho}{}_{\mu} \label{tensorseff} \\
 &+ 2 F^{\mu \nu} F_{\mu \nu} F^{\lambda \rho} F_{\lambda \rho}
 + F^{\mu \nu} F^{\lambda \rho} F_{\mu \nu} F_{\lambda \rho} \Big] + \ldots \bigg\}
\notag
\end{align}
that the number of terms grows drastically with the order in $\alpha'$.
A state-of-the-art method to determine 8-term and 96-term-expressions 
for the Lagrangians $\alpha'^3 \zeta_3 {\rm Tr} \{D^2F^4{+}F^5\}$ and 
$\alpha'^4 \zeta_2^2 {\rm Tr}\{D^4F^4{+}D^2F^5{+}F^6\}$ is described in \cite{Barreiro:2012aw}, 
see for instance \cite{Kitazawa:1987xj, Barreiro:2005hv} and
\cite{Koerber:2002zb, Oprisa:2005wu} for earlier results at the respective orders.
The supersymmetric completion of (\ref{tensorseff}) and
its gravitational counterpart is discussed from a variety of perspectives in \cite{Green:1998by, Cederwall:2001td, Koerber:2001uu, Collinucci:2002ac, Berkovits:2002ag, Drummond:2003ex,
  Policastro:2006vt, Howe:2010nu, Cederwall:2011vy, Wang:2015jna}
and references therein. 

The bosonic terms in (\ref{tensorseff}) and their gravitational
counterparts in (\ref{seffclosed}) are universal to any torus compactification 
of the initially ten-dimensional type-I and type-II theories to $D$
spacetime dimensions. Such compactifications preserve all supersymmetries
and lead to a straightforward dimensional reduction of the massless tree-level
interactions, since Kaluza-Klein and winding modes are only produced in pairs.
Accordingly, the one-loop matrix elements to be obtained from forward limits
of tree amplitudes in the massless multiplets have a dimension-agnostic loop integrand.

For abelian external open-string states, the leading order in the low-energy effective
action is described by the Born-Infeld theory \cite{Metsaev:1987qp}. In four dimensions, one-loop 
$n$-point amplitudes of Born-Infeld have been discussed in \cite{Shmakova:1999ai, Elvang:2019twd, 
Elvang:2020kuj}, and the abelian limit of our results contain the $D$-dimensional uplift of the
loop integrands in the references. A detailed comparison is left for the future.

Low-energy effective actions are usually determined by reverse-engineering the Feynman rules
that generate the $\alpha'$-expansions of massless string amplitudes.
When expressing the tree-level amplitudes of the open superstring in terms of $Z$-integrals
\cite{Mafra:2011nv, Zfunctions} in (\ref{rev.31}), the contributions of $\zeta_2$ and $\zeta_3$ to
their expansion (\ref{rev.38}) yield matrix elements of the effective operators ${\rm Tr}\{F^4\}$ and 
${\rm Tr}\{D^2F^4{+}F^5\}$, respectively. Similarly, the $\zeta_3$- and $\zeta_5$-terms in the 
expansion (\ref{rev.39}) of $J$-integrals (\ref{rev.32}) correspond
to $R^4$ and $D^4R^4{+}D^2R^5{+}R^6$ couplings of type-II superstrings, respectively.
In both cases, each $\mathbb Q$-independent combination of MZVs in the
expansion of the scalar genus-zero integrals is associated with a separate effective operator.

The gauge-multiplet operators due to weight-$w$ MZVs in the disk integral (\ref{rev.38}) have bosonic components
of the schematic form ${\rm Tr}\{D^{2(w-2)} F^4 {+} D^{2(w-3)} F^5{+}\ldots {+}F^{w+2}\}$. Similarly
the single-valued MZVs of weight $w$ in the sphere integral (\ref{rev.39}) lead to gravitational components
$D^{2(w-3)}R^4{+}D^{2(w-4)}R^5{+}\ldots{+}R^{w+1}$ in the type-II effective action. The maximal supersymmetry
of the type-I and type-II theories usually interlocks operators with different numbers of $F$ and $R$ while 
balancing their mass dimension through the number of covariant derivatives $D^{2} \leftrightarrow F$ and 
$D^{2} \leftrightarrow R$. In fact, relations of the schematic form $D^2 F \cong F^2$ and $D^2 R \cong R^2$ together
with field redefinitions introduce immense ambiguities into the form of the effective action. Up to and including 
weight $w=6$, these composite gauge and gravity operators can be represented as seen in Table~\ref{effops}.
\begin{table}
\begin{center}
\[
\begin{array}{| rl | rl | rl |}\hline
&\! \! \! \! {\rm gauge} \ {\rm interactions}
&&\! \! \! \! \! \! \! \!  \!{\rm gauge} \ {\rm interactions}
&&\! \! \! \! \! \!  \! \!  \! \! \! \! \!  {\rm gravitational} \ {\rm interactions}
\\\hline
1 \ &\leftrightarrow \ F^2
&\alpha'^5 \zeta_5 \ &\leftrightarrow \ D^6F^4{+}\ldots{+}F^7 
&1 \ &\leftrightarrow \ R
\\
\ \alpha'^2 \zeta_2 \ &\leftrightarrow \ F^4
&\ \alpha'^5 \zeta_2 \zeta_3\ &\leftrightarrow \ D^6F^4{+}\ldots{+}F^7 
&\alpha'^3 \zeta_3 \ &\leftrightarrow \ R^4
\\
\alpha'^3 \zeta_3 \ &\leftrightarrow \ D^2F^4{+}F^5
&\alpha'^6 \zeta_2^3\ &\leftrightarrow \ D^8F^4{+}\ldots{+}F^8 \ \
&\alpha'^5 \zeta_5 \ &\leftrightarrow \ D^4R^4{+}D^2 R^5{+}R^6\ \
\\
\alpha'^4 \zeta_2^2 \ &\leftrightarrow \ D^4F^4{+}D^2F^5{+}F^6 \ \
&\alpha'^6 \zeta_3^2 \ &\leftrightarrow \ D^8F^4{+}\ldots{+}F^8 
&\ \alpha'^6 \zeta_3^2 \ &\leftrightarrow \ D^6R^4{+}\ldots{+}R^7 \\\hline
\end{array}
\]
\end{center}
\caption{Schematic form of gauge and gravity operators at the orders of $\alpha'^{\leq 6}$ in the
tree-level effective action of type-I and type-II superstrings (suppressing the trace of ${\rm Tr} \{ D^{2k}F^n\}$
operators to avoid cluttering)}
\label{effops}
\end{table}

Starting from weight $w=8$, indecomposable MZVs $\zeta_{n_1,\ldots,n_r}$ of depth $r\geq 2$
occur in the conjectural $\mathbb Q$-bases (see e.g.\ \cite{Blumlein:2009cf}), for
instance $\zeta_{3,5}, \zeta_{3,7}$ and $\zeta_{3,3,5}$ at weight $8,10$ and $11$, respectively.
Since the four-point $Z$- and $J$-integrals are expressible in terms of depth-one $\zeta_w$,
one can avoid four-field interactions $D^{2k}F^4$ along with irreducible MZVs beyond depth one, i.e.\ eliminate 
$\alpha'^8\zeta_{3,5}D^{12}F^4, \alpha'^{10}\zeta_{3,7}D^{16}F^4$ and $\alpha'^{11}\zeta_{3,3,5}D^{18}F^4$. 
The ubiquity of indecomposable higher-depth MZVs in $(n\geq 5)$-point string 
amplitudes \cite{Schlotterer:2012ny} leads to combinations of $F^{\geq 5}$-operators such as 
$\alpha'^8\zeta_{3,5}(D^{10}F^5{+}D^{8}F^6{+}\ldots)$. For closed strings, $\zeta_{3,5}$ and $\zeta_{3,7}$ are
removed from the $\alpha'$-expansion through the single-valued map, see (\ref{rev.36}),
so the first appearance of irreducible MZVs is via $\alpha'^{11} \zeta_{3,3,5}(D^{14} R^5{+}D^{12} R^6{+}\ldots)$
without any four-point admixture $\sim D^{16} R^4$.

The appearance of $\zeta_{3,5}$ in $F^{\geq 5}$ operators and its dropout from $R^5$ and $R^6$ 
interactions was firstly reported in \cite{Stieberger:2009rr}, and the all-order systematics of MZVs in massless 
$n$-point amplitudes of open and closed superstrings is described in \cite{Schlotterer:2012ny}. In the same
way as irreducible MZVs of depth $\geq 2$ decouple at the four-point level, five-point amplitudes still feature
dropouts of MZVs at higher weight $\geq 18$ \cite{Schlotterer:2012ny}.\footnote{At weight $w=18$ for instance, 
the dropout of one MZV specified in \cite{Schlotterer:2012ny} can be used to bring one effective operator
into the form $\alpha'^{18} (D^{28} F^6+\ldots)$ without any admixtures of
$\alpha'^{18} D^{30} F^5$ and $\alpha'^{18} D^{32} F^4$.}
At sufficiently large multiplicity, any conjecturally $\mathbb Q$-independent combination
of MZVs is expected to occur in the $\alpha'$-expansion of disk integrals and therefore
the open-superstring effective action. Accordingly, closed-string tree-level interactions should
feature all $\mathbb Q$-independent single-valued MZVs.


\section{One-loop matrix elements for the superstring effective action}
\label{sec:3}

We shall now combine the ingredients reviewed in the previous section to propose
expressions for one-loop matrix elements with single and multiple insertions
of any operator ${\rm Tr}\{D^{2k}F^n\}$ and $D^{2k}R^n$ in the above tree-level effective actions.


\subsection{\texorpdfstring{$\ap$}{a'} uplifts for one-loop amplitudes}
\label{sec:3.1}

The key idea of this work can be phrased as systematically introducing $\ap$-corrections to one-loop 
ambitwistor-string amplitudes. For this purpose, the doubly-partial amplitudes $m$ in the
expressions (\ref{rev.21}) and (\ref{rev.22}) for SYM and supergravity amplitudes are 
promoted to disk and sphere integrals (\ref{rev.31}) and (\ref{rev.32}), respectively,\footnote{Throughout 
this work, we suppress the dependence on the gauge coupling and 
gravitational constant, given by global prefactors $(g_{\rm YM})^n$ in case of
$A^{\te{1-loop}}_{\rm eff}(1,\ldots,n)$ and $\kappa^n$ in case of 
$M^{\te{1-loop}}_{n,{\rm eff}} $, respectively. Normalization factors
including powers of $i$ and $\sqrt{2}$ that apply to the entire $\ap$-expansion are not tracked.}
\begin{align}
A^{\te{1-loop}}_{\rm eff}(1,2,\ldots,n) &=  \int \frac{ \dd^D \ell}{\ell^2} \lim_{k_{\pm} \rightarrow \pm \ell} 
\sum_{\substack{\gamma \in {\rm cyc}(1,2,\ldots,n) \\ \rho \in S_n}} Z(+,\gamma,-|+,\rho,-) N_{+|\rho|-}(\ell)\,,
\label{prop.1} \\
M^{\te{1-loop}}_{n,{\rm eff}} &=  \int \frac{ \dd^D \ell}{\ell^2} \lim_{k_{\pm} \rightarrow \pm \ell} 
\sum_{\gamma, \rho \in S_n} J(+,\gamma,-|+,\rho,-) N_{+|\gamma |-}(\ell) \bar N_{+|\rho|-}(\ell)\, .
\label{prop.2}
\end{align}
By the common field-theory limit (\ref{rev.35}) of $Z$ and $J$, the leading orders in $\alpha'$ are engineered
to comprise SYM and supergravity amplitudes,
\beq
A^{\te{1-loop}}_{\rm eff}(1,2,\ldots,n) = A^{\te{1-loop}}_{\rm SYM}(1,2,\ldots,n) + {\cal O}(\ap^2)\, , \ \ \ \ \ \
M^{\te{1-loop}}_{n,{\rm eff}}  = M^{\te{1-loop}}_{n,{\rm SUGRA}}  + {\cal O}(\ap^3)\, .
\label{ftlimit}
\eeq
The $\ap$-expansion of (\ref{prop.1}) and (\ref{prop.2}) is expected to 
reproduce the matrix elements for single and multiple insertions of the higher-mass-dimension
operators in the superstring effective actions of (\ref{seffclosed}) and section \ref{sec:2.8}.
However, the expressions for $A^{\te{1-loop}}_{\rm eff}(1,2,\ldots,n)$ and $M^{\te{1-loop}}_{n,{\rm eff}} $ 
are {\it not} claimed to be the full massless one-loop amplitudes of 
open and closed superstrings: as will be explained in section \ref{sec:3.2}, they are
lacking important contributions from the bulk of the moduli space of punctured genus-one surfaces.
Indeed, localizing to the boundary of the moduli space naturally induces a truncation of the superstring one-loop amplitudes where massive string excitations in the loop are dropped at any fixed order in the $\alpha'$-expansion.
By resumming the $\alpha'$-expansion of these boundary contributions, in turn, one may in principle 
recover the subset of the massive propagators which stem from genus-zero integrals $Z$ and $J$.

In the open-string case, (\ref{prop.1}) corresponds to the planar or single-trace amplitudes in the color decomposition
\begin{align}
A^{\te{1-loop}}_{n,{\rm eff}} &= {\rm Tr}\{ 1 \} \sum_{\gamma \in S_{n-1}} 
{\rm Tr} \{ t^{a_1} t^{a_{\gamma(2)}} \ldots t^{a_{\gamma(n)}} \} A^{\te{1-loop}}_{\rm eff}\big(1,\gamma(2,\ldots,n) \big) 
\notag \\
&\ \ + \sum_{k=2}^{\lfloor n/2 \rfloor} 
\Big[ {\rm Tr} \{ t^{a_1} t^{a_{2}} \ldots t^{a_{k}} \} 
{\rm Tr} \{ t^{a_{k+1}} t^{a_{k+2}} \ldots t^{a_{n}} \}  \label{prop.0}\\
& \ \ \ \ \ \ \ \ \ \ \times
A^{\te{1-loop}}_{\rm eff}(1, 2,\ldots,k| k{+}1,k{+}2,\ldots,n) + {\rm permutations}\Big]\, ,
\notag
\end{align}
and a proposal for their non-planar counterparts will be motivated below. 
The permutation sum is understood to involve all cyclically inequivalent 
double-trace configurations without double-counting of
$A^{\te{1-loop}}_{\rm eff}(1, \ldots,k| k{+}1,\ldots,n)=
A^{\te{1-loop}}_{\rm eff}( k{+}1,\ldots,n | 1, \ldots,k)$.

In order to further explore the structure and relations of the
one-loop matrix elements (\ref{prop.1}) and (\ref{prop.2}), we
introduce an important substructure of the single-trace
$A^{\te{1-loop}}_{\rm eff}$: the kinematic half integrand
${\cal I}_{n}(\ell)$ in the ambitwistor-string formula (\ref{rev.6})
for one-loop SYM amplitudes is gauge invariant on the support of the
scattering equations. Hence, ${\cal I}_{n}(\ell) $ dressed with any of
the individual Parke-Taylor factors in the cyclic sum of (\ref{rev.6})
integrates to gauge invariant objects dubbed {\it partial integrands}
\cite{He:2016mzd, He:2017spx},
\begin{align}
a_{\rm SYM}(\gamma(+,1,2,\ldots,n,-)) &=  \lim_{k_{\pm} \rightarrow \pm \ell} 
 \int\limits_{\mathbb C^{n-1}} \dd \mu_{n+2}^{\te{tree}} \,{\rm PT}(\gamma(+,1,2,\ldots,n,-)) \, {\cal I}_{n}(\ell)    \label{prop.3}\\
&= \lim_{k_{\pm} \rightarrow \pm \ell} 
\sum_{  \rho \in S_n} m\big(\gamma(+,1,2,\ldots,n,-)|+,\rho,-\big) N_{+|\rho|-}(\ell)\,,
\notag
\end{align}
with $\gamma \in S_{n+2}$ which compose single- and double-trace one-loop amplitudes via
\begin{align}
A^{\te{1-loop}}_{\rm SYM}(1,2,\ldots,n) &= \int \frac{ \dd^D \ell}{\ell^2} \, a_{\rm SYM}(+,1,2,\ldots,n,-)  + {\rm cyc}(1,2,\ldots,n) \,,
\label{prop.4} \\
A^{\te{1-loop}}_{\rm SYM}(1,2,\ldots,k|k{+}1,\ldots,n) &= \int \frac{ \dd^D \ell}{\ell^2} \! \! \!
\sum_{ \beta \in {\rm cyc}(1,2,\ldots ,k) \atop{ \gamma \in {\rm cyc}(k+1,\ldots,n) } } \! \! \! a_{\rm SYM}\big(+,\beta(1,2,\ldots,k),-,\gamma(k{+}1,\ldots,n)\big)
\, .\notag
\end{align}
In the same way as the promotion $m \rightarrow Z$ in $A^{\te{1-loop}}_{\rm SYM}$ led to (\ref{prop.1}),
we introduce $\ap$-corrections to the partial integrand (\ref{prop.3}) by inserting disk integrals into
\begin{align}
a_{\rm eff}\big(\gamma(+,1,2,\ldots,n,-) \big) &= 
 \lim_{k_{\pm} \rightarrow \pm \ell} 
\sum_{  \rho \in S_n} Z\big(\gamma(+,1,2,\ldots,n,-)|+,\rho,-\big) N_{+|\rho|-}(\ell)\, .
\label{prop.5}
\end{align}
The ambitwistor-string definition of partial SYM integrands in the first line of (\ref{prop.3}) 
yields forward limits of color-ordered tree-level amplitudes where tentative divergences
are regularized by the dropout of singular solutions. The transition to the second line of (\ref{prop.3})
requires supersymmetry in order to reliably evaluate the $\dd \mu_{n+2}^{\te{tree}}$ integrals in terms
of doubly-partial amplitudes which do not diverge in the forward limit. Accordingly, we only 
define $\alpha'$-dressed partial integrands
(\ref{prop.5}) in a supersymmetric setup for now such as the open superstring in toroidal,
K3 or Calabi-Yau compactifications to $D$-dimensional Minkowski spacetime.\footnote{Still, the treatment
of forward-limit divergences in deriving non-supersymmetric gauge-theory amplitudes from 
ambitwistor strings \cite{Geyer:2017ela} is expected to have smooth $\alpha'$-uplifts via $m\rightarrow Z$
or $m\rightarrow J$.} 
They are forward limits of color-ordered tree-level amplitudes with 
massless external states and therefore inherit their relations. In particular, (\ref{prop.5})
is the forward limit of the $n!$-term representation of $(n{+}2)$-point open-superstring tree amplitudes in terms of BCJ master numerators given in appendix B.3 of \cite{Zfunctions}. 

With the $\alpha'$-dressed partial integrands (\ref{prop.5}), the assembly (\ref{prop.4}) of 
SYM single- and double-trace amplitudes generalizes to
\begin{align}
A^{\te{1-loop}}_{\rm eff}(1,2,\ldots,n) &= \int \frac{ \dd^D \ell}{\ell^2} \, a_{\rm eff}(+,1,2,\ldots,n,-)  + {\rm cyc}(1,2,\ldots,n) \,,
\label{prop.6}
\\
A^{\te{1-loop}}_{\rm eff}(1,2,\ldots,k|k{+}1,\ldots,n) &= \int \frac{ \dd^D \ell}{\ell^2} \! \! \!
\sum_{\substack{ \beta \in {\rm cyc}(1,2,\ldots ,k) \\ \gamma \in {\rm cyc}(k+1,\ldots,n) } } \! \! \! a_{\rm eff}\big(+,\beta(1,2,\ldots,k),-,\gamma(k{+}1,\ldots,n)\big)\, .
\notag
\end{align}
The single-trace relation is equivalent to (\ref{prop.1}), and the second line defines our
proposal for the double-trace sector of the one-loop matrix elements in the color decomposition (\ref{prop.0}).


\subsection{Relations among one-loop matrix elements}
\label{sec:3.0}

The forward limit in the definition (\ref{prop.5}) of partial integrands for one-loop matrix elements will preserve the monodromy relations among disk integrals with different orderings $\gamma \in S_{n+2}$ of the punctures on a disk boundary \cite{Plahte:1970wy, BjerrumBohr:2009rd, Stieberger:2009hq}
\begin{align}
 \sum_{j=0}^{n} e^{2\pi i \alpha' \ell \cdot k_{12\ldots j} }  a_{\rm eff}(1,2,\ldots,j,+,j{+}1,\ldots,n,-)  &=0
\label{prop.7} \\
\sum_{j=1}^{n} e^{2\pi i \alpha' k_1\cdot k_{23\ldots j} } a_{\rm eff}(2,3,\ldots,j,1,j{+}1,\ldots,n,-,+) &=
e^{2\pi i \alpha' \ell \cdot k_1}  a_{\rm eff}(2,3,\ldots,n,-,1,+)   \, .\notag
\end{align}
In the field-theory limit, the exponentials can be replaced by their Taylor
expansion to the first order in $\alpha'$ and imply the generalization of Kleiss-Kuijf \cite{Kleiss:1988ne} and BCJ relations \cite{BCJ}
that apply to partial integrands \cite{He:2016mzd, He:2017spx}, e.g.
\begin{align}\label{prop.8}
 \sum_{j=1}^{n-1}  \ell \cdot k_{12\ldots j}  a_{\rm SYM}(1,2,\ldots,j,+,j{+}1,\ldots,n,-)  &=0\,, \\
\sum_{j=1}^{n} a_{\rm SYM}(2,3,\ldots,j,1,j{+}1,\ldots,n,-,+)  &=  a_{\rm SYM}(2,3,\ldots,n,-,1,+)\, . \nonumber
\end{align}
By combining suitable permutations of (\ref{prop.7}), one can express the constituents $a_{\rm eff}(+,\beta,-,\gamma)$ in the double-trace sector (\ref{prop.6}) in terms of the single-trace quantities $a_{\rm eff}(+,\ldots,-)$.
In the field-theory limit, the resulting relations among $a_{\rm SYM}(+,\beta,-,\gamma)$ and 
$a_{\rm SYM}(+,\ldots,-)$ imply those between single- and double-trace 
amplitudes (\ref{prop.4}) in general gauge theories \cite{Bern:1994zx}.
The monodromy relations (\ref{prop.7}) leave at most $(n{-}1)!$ independent orderings of $a_{\rm eff}(\ldots)$,
but spacetime supersymmetry in general implies additional relations: by the 16 supercharges of the type-I superstring, $(n\leq 3)$-point instances of $a_{\rm eff}$ vanish, and all permutations of $a_{\rm eff}\big(\gamma(+,1,2,3,4,-)\big)$ at four points have the same kinematic factor and only differ in their series-expansion in $k_j$ and $\ell$.
It would be interesting to relate the monodromy relations (\ref{prop.7}) among the one-loop matrix elements 
to those among the full-fledged one-loop open-string
amplitudes \cite{Tourkine:2016bak, 
Hohenegger:2017kqy, Ochirov:2017jby, Tourkine:2019ukp, Casali:2019ihm, Casali:2020knc, Stieberger:2021daa}.

In preparation for relations between one-loop matrix elements of
${\rm Tr}\{D^{2k}F^n\}$ and $D^{2k}R^n$ operators, we note that the
BCJ relations (\ref{prop.8}) of the partial SYM integrands qualify
them to enter a KLT construction.  The field-theory KLT formula
\cite{Kawai:1985xq} among tree amplitudes of SYM and supergravity
generalizes as follows to partial integrands (\ref{prop.3}) and
one-loop supergravity integrands \cite{He:2016mzd, He:2017spx}
\begin{align}\label{prop.9}
M^{\te{1-loop}}_{n,{\rm SUGRA}} =  \int \frac{ \dd^D \ell}{\ell^2} & \sum_{\gamma,\rho\in S_{n-1}} a_{\rm SYM}\big(+,\gamma(1,2,\ldots,n{-}1),n,-\big) \nonumber\\
&\times S_0(\gamma|\rho)_\ell  \, \bar a_{\rm SYM}\big(+,\rho(1,2,\ldots,n{-}1),-,n\big) \, .
\end{align}
The field-theory KLT matrix $S_0$ encoding the all-multiplicity
formula at tree level \cite{Bern:1998sv} is used in its representation
within the momentum-kernel formalism \cite{momentumKernel} and obeys
the recursion
\beq
S_0(A,j | B,j,C)_\ell = k_j \cdot (\ell+ k_B) S_0(A|B,C)_\ell \, , \ \ \ \ \ \ S_0(\emptyset | \emptyset)_\ell = 1\, ,
\label{prop.10}
\eeq
where $A=a_1,a_2,\ldots$ as well as $B=b_1,b_2,\ldots$ and
$C=c_1,c_2,\ldots$ are ordered sequences of external legs. In the
$\alpha'$-dressed supergravity integrand (\ref{prop.2}), the forward
limit preserves the relations between the sphere integrals $J$ and the
disk integrals $Z$ that drive the string-theory KLT relations at tree
level \cite{Kawai:1985xq}. As a consequence, we obtain the following
uplift of (\ref{prop.9}) to one-loop matrix elements,
\begin{align}\label{prop.11}
M^{\te{1-loop}}_{n,{\rm eff}} =  \int \frac{ \dd^D \ell}{\ell^2} &\sum_{\gamma,\rho\in S_{n-1}} a_{\rm eff}\big(+,\gamma(1,2,\ldots,n{-}1),n,-\big) \nonumber \\
& \times S_\ap(\gamma|\rho)_\ell  \, \bar a_{\rm eff}\big(+,\rho(1,2,\ldots,n{-}1),-,n\big) \,,
\end{align}
with a string-theory KLT kernel subject to
\beq
S_{\alpha'}(A,j | B,j,C)_\ell = \frac{ \sin\! \big(2\pi \alpha' k_j \cdot (\ell+ k_B)\big) }{2\pi \alpha'}S_{\alpha'}(A|B,C)_\ell \, , \ \ \ \ \ \ S_{\alpha'}(\emptyset | \emptyset)_\ell = 1\, .
\label{prop.12}
\eeq
At the same time, the string-theory KLT formula is equivalent to a field-theory
double-copy formula involving single-valued open-string amplitudes \cite{Schlotterer:2012ny,Stieberger:2013wea,Stieberger:2014hba,Schlotterer:2018abc,Vanhove:2018elu,Brown:2019wna}.
In the forward limit, this reasoning leads to the following rewriting of (\ref{prop.11})
\begin{align}\label{prop.13} 
M^{\te{1-loop}}_{n,{\rm eff}} =  \int \frac{ \dd^D \ell}{\ell^2} & \sum_{\gamma,\rho\in S_{n-1}} a_{\rm SYM}\big(+,\gamma(1,2,\ldots,n{-}1),n,-\big) \nonumber\\
& \times S_0(\gamma|\rho)_\ell  \, {\rm sv} \, \bar a_{\rm eff}\big(+,\rho(1,2,\ldots,n{-}1),-,n\big) \, .
\end{align}
As in the relation (\ref{rev.37}) between sphere and disk integrals, the single-valued map acts
on the (motivic) MZVs in the $\ap$-expansion according to (\ref{rev.36}), also see 
\cite{Schnetz:2013hqa, Brown:2013gia}
for ${\rm sv}(\zeta_{n_1,\ldots,n_r})$ at general depth $r$. One may wonder about an echo of the
KLT relations (\ref{prop.11}) among one-loop matrix elements for the full-fledged closed-string one-loop amplitudes.
As we will see in section \ref{sec:5}, the appearance of single-valued disk integrals in the expressions
 (\ref{prop.2}) and (\ref{prop.13}) for $M^{\te{1-loop}}_{n,{\rm eff}}$ resonates with
recent progress in relating the degeneration limits of genus-one integrals of open and closed 
strings \cite{Broedel:2018izr, Gerken:2018jrq, Zagier:2019eus, Vanhove:2020qtt, Gerken:2020xfv}.


\section{Examples of numerator extractions}
\label{sec:4}

This section is dedicated to four- and five-point examples of the one-loop matrix
elements (\ref{prop.1}) and (\ref{prop.2}) of massless SYM and supergravity states
and the associated effective tree-level interactions. The six- and seven-point tree-level disk 
integrals used in \cref{prop.1} and \cref{prop.2} are expanded by the {\tt BGap} 
package~\cite{gitrep}, and the kinematic one-loop numerators for SYM are well-known from 
\cite{Green:1982sw, Mafra:2014gja, He:2017spx, Edison:2020uzf}. 
As detailed in section \ref{sec:fromlintoquad}, the conversion from linearized to
quadratic propagators is facilitated by {\tt BGap}: at both four and five points, the kinematic 
numerators at the orders of $\ap^{\leq 7}$ turned out to readily 
satisfy the criteria \cref{eq:Nshift} under the forward limit.


\subsection{Four points, open strings}
\label{sec:4.1}

At four points, all the polarization dependence of the kinematic half integrand (\ref{rev.7}) 
is captured by the permutation invariant kinematic factor \cite{Green:1982sw}
\beq
t_8(f_1,f_2,f_3,f_4) = {\rm tr}(f_1f_2f_3f_4) - \frac{1}{4} {\rm tr}(f_1f_2){\rm tr}(f_3f_4) + {\rm cyc}(2,3,4)
\label{exsec.1}
\eeq
and its supersymmetrization \cite{Berkovits:2004px}. The traces are over the Lorentz indices of the linearized
field strength $f_j^{\mu \nu} = k_j^\mu \ep_j^\nu- k_j^\nu \ep_j^\mu$ involving transverse
polarization vectors $\ep_j^\mu$ of the external states $j=1,2,3,4$ (with $\mu,\nu = 0,1,\ldots,D{-}1$). The accompanying
Parke-Taylor factors also form a permutation invariant,
\beq
{\cal I}_{4}= t_8(f_1,f_2,f_3,f_4) \sum_{\rho \in S_4} {\rm PT}(+,\rho(1,2,3,4),-)\, ,
\label{exsec.2}
\eeq
which is the ambitwistor-string explanation for the absence of triangle and bubble diagrams in the 
loop integrand (\ref{rev.6}) of SYM. One can read off the master numerators
\beq
N_{+|\rho(1234)|-}= t_8(f_1,f_2,f_3,f_4) 
\label{exsec.3}
\eeq
from (\ref{exsec.2}) which do not depend on the permutation $\rho$
of legs 1,2,3,4 in the half-ladder diagram of Figure~\ref{figcutting}.
When assembling the four-point instances of the
partial integrands (\ref{prop.5}) for one-loop matrix elements, it is
convenient to peel off the polarization dependent $t_8$-factor,
\begin{align}
a_{\rm eff}\big(\gamma(+,1,2,3,4,-)\big) &= t_8(f_1,f_2,f_3,f_4) \widehat a_{\rm eff}\big(\gamma(+,1,2,3,4,-)\big)  \,,
\label{exsec.4}  \\
\widehat a_{\rm eff}\big(\gamma(+,1,2,3,4,-)\big) &=  \lim_{k_{\pm} \rightarrow \pm \ell} 
\sum_{  \rho \in S_4} Z\big(\gamma(+,1,2,3,4,-)|+,\rho(1,2,3,4),-\big) \, .\nonumber
\end{align}
With the leading orders in the $\ap$-expansion (\ref{rev.40a}) of the six-point disk integrals on the right-hand sides, 
the forward limit yields an $\ap$-dressed partial integrand of the form
\begin{align}
\widehat a_{\rm eff}(+,1,2,3,4,-) &=  \frac{1}{s_{1,\ell} s_{12,\ell} s_{123,\ell}}
- \ap^2 \zeta_2 \bigg( 
\frac{s_{34}}{s_{1,\ell} s_{12,\ell}} + \frac{s_{12}}{s_{12,\ell} s_{123,\ell}} + \frac{ s_{23}}{s_{1,\ell} s_{123,\ell}}
\bigg)
\label{exsec.5} \\
&\! \! \! \! \! \! \! \! \! \! \! \! \! \! \! \! \! \! \! \! \! \! \! \! \! \! \!  \! \!  \! \!+ \ap^3 \zeta_3 \bigg( \frac{ s_{34}(s_{34}{+}s_{123,\ell})}{s_{1,\ell} s_{12,\ell}}  
 + \frac{ (s_{1,\ell}{+}s_{12}) s_{12}}{s_{12,\ell} s_{123,\ell}}
 + \frac{ (s_{12,\ell} {+}s_{23})s_{23}}{s_{1,\ell} s_{123,\ell}} 
 - \frac{ s_{24}}{s_{1,\ell}} 
  - \frac{ s_{13}}{s_{123,\ell}} 
  \bigg) + {\cal O}(\ap^4)\, .
\notag
\end{align}
By the conventions (\ref{rev.23}) for $\ell$-dependent Mandelstam invariants, we have for instance 
$s_{1,\ell} = 2k_1 \cdot \ell$ and $s_{12,\ell}= s_{12}+2k_{12}\cdot \ell$. In the one-loop matrix elements
assembled via \cref{prop.6}, 
\begin{align}
A^{\te{1-loop}}_{\rm eff}(1,2,3,4)  &= t_8(f_1,f_2,f_3,f_4) \widehat A^{\te{1-loop}}_{\rm eff}(1,2,3,4) \,,
\label{exsec.6}\\
  \widehat A^{\te{1-loop}}_{\rm eff}(1,2,3,4) 
  &=\int \frac{ \dd^D\ell}{\ell^2}   \lim_{k_{\pm} \rightarrow \pm \ell}  \sum_{\gamma \in {\rm cyc}(1,2,3,4)}
\sum_{\rho \in S_4} Z(+,\gamma(1,2,3,4),-|+,\rho(1,2,3,4),-)\,, \nonumber
\end{align}
the cyclic permutations of $a_{\rm eff}(+,1,2,3,4,-) $ in the loop integrand conspire to
the quadratic propagators upon shifts of $\ell$ as in (\ref{ngonPF}). The resulting box-,
triangle- and bubble diagrams tie in with the Feynman vertices of the operators
$F^2$, $\ap^2 \zeta_2 F^4$, $\ap^3 \zeta_3 (D^2 F^4 +F^5),\ldots,$ in the low-energy 
effective action $S_{\rm eff}^{\rm open}$ in \eqref{seffclosed} and their supersymmetric completions. 
In particular, the field-theory limit $(s_{1,\ell} s_{12,\ell} s_{123,\ell})^{-1}$ of (\ref{exsec.5})
recombines to the box integral (\ref{rev.25}) with the well-known $t_8$-numerator
in maximally supersymmetric gauge theories \cite{Green:1982sw}.

\subsubsection{$\ap^2$ and $\ap^3$ corrections}

The $\ap$-corrections to the $Z$-integrals in (\ref{exsec.5}) augment
the box integrals of SYM by triangles, bubbles and tadpoles. At the subleading orders, their explicit form resulting from (\ref{exsec.6}) is
\begin{align}
  \widehat{A}^{\te{1-loop}}_{\te{eff}}(1,2,3,4)&=\int \dd^D\ell \frac{1}{\ell^2(\ell{+}k_1)^2(\ell{+}k_{12})^2(\ell{+}k_{123})^2}\nonumber\\
  &\quad-\ap^2\zeta_2\int \dd^D\ell \left[\frac{s_{12}}{\ell^2(\ell{+}k_1)^2(\ell{+}k_{12})^2}+\text{cyc}_{\ell}(1,2,3,4)\right]\label{exsec.7}\\
  &\quad+\ap^3\zeta_3\int \dd^D\ell \left[\frac{s_{34}(s_{3,\ell}-s_{4,\ell}+s_{34})}{2\ell^2(\ell{+}k_1)^2(\ell{+}k_{12})^2}
 - \frac{3s_{13}}{2\ell^2(\ell{+}k_1)^2} +\text{cyc}_{\ell}(1,2,3,4)\right]\notag\\
 &\quad+\mathcal{O}(\ap^4)\,, \notag
\end{align}
where the numerators are properly symmetrized with respect to the
graph automorphism groups of the underlying propagator topologies. The
``$+\,\text{cyc}_{\ell}(1,2,\ldots,n)$'' prescription involves \emph{a
  sum over the cyclic permutations of the external legs together with
  a proper shift of loop momentum such that it always points from leg
  $n$ to leg $1$.} For example, the scalar triangle at the
$\ap^2\zeta_2$-order and the vector integral along with $\ap^3\zeta_3$
are symmetrized as follows:
\begin{align}
\frac{s_{12}}{\ell^2(\ell{+}k_1)^2(\ell{+}k_{12})^2}+\text{cyc}_{\ell}(1,2,3,4)&=\frac{s_{12}}{\ell^2(\ell{+}k_1)^2(\ell{+}k_{12})^2}+\frac{s_{23}}{(\ell{+}k_1)^2(\ell{+}k_{12})^2(\ell{+}k_{123})^2}\nonumber\\
&\quad+\frac{s_{34}}{\ell^2(\ell{+}k_{12})^2(\ell{+}k_{123})^2}+\frac{s_{14}}{\ell^2(\ell{+}k_{1})^2(\ell{+}k_{123})^2}\,,
\label{cycellex} \\
\frac{s_{34} s_{3,\ell}}{\ell^2(\ell{+}k_1)^2(\ell{+}k_{12})^2}+\text{cyc}_{\ell}(1,2,3,4)&=
                                                                                            \frac{s_{34} s_{3,\ell}}{\ell^2(\ell{+}k_1)^2(\ell{+}k_{12})^2} + \frac{s_{14}(s_{4,\ell}+s_{14})}{(\ell{+}k_1)^2(\ell{+}k_{12})^2(\ell{+}k_{123})^2} \nonumber\\
  &\quad+\frac{s_{12}(s_{1,\ell}+s_{12})}{\ell^2(\ell{+}k_{12})^2(\ell{+}k_{123})^2}+\frac{s_{23}(s_{2,\ell}+s_{12}+s_{23})}{\ell^2(\ell{+}k_{1})^2(\ell{+}k_{123})^2}\,.
\end{align}
Our convention of fixing the position of the loop momentum corresponds to fixing one of the
punctures in one-loop string amplitudes by translation invariance on a genus-one surface.

\begin{figure}
\begin{center}
\subfloat[$N_{\text{triangle}}^{(\zeta_2)}(1,2,34)$]{\begin{tikzpicture}[scale=1.05, line width=0.30mm]
\draw [decoration={markings, 
	mark= at position 0.6 with \arrow{Stealth[length=3mm]}},postaction={decorate}] (0,0)--(2,1) node[pos=0.5,above=1pt]{$\ell$};
\draw [decoration={markings, 
	mark= at position 0.5 with \arrow{Stealth[length=3mm]}},postaction={decorate}] (2,-1)--(0,0) node[pos=0.5,below=1pt]{$\ell{+}k_{12}$};
\draw [decoration={markings, 
	mark= at position 0.55 with \arrow{Stealth[length=3mm]}},postaction={decorate}] (2,1)--(2,-1) node[pos=0.5,right=1pt]{$\ell{+}k_1$};
\draw (2,1) -- (2.5,1) node[right]{$1$};
\draw (2,-1) -- (2.5,-1) node[right]{$2$};
\draw (0,0)--(-1,-0.5)node[left]{$3$};
\draw (0,0)--(-1,0.5)node[left]{$4$};
\draw [fill=white] (0,0)circle (0.4cm);
\draw (0,0) node{$F^4$};
\end{tikzpicture}\label{fig:ffourdiaga}
}
\quad\quad\quad\quad\quad\quad\quad
\subfloat[$N_{\text{bubble}}^{(\zeta_3)}(1,234)$]{\begin{tikzpicture}[scale=1.05, line width=0.30mm]
\draw (0,0) .. controls (0.5,1) and (1.5,1) .. (2,0) node[pos=0.5,above=1pt]{$\ell$};
\draw (0,0) .. controls (0.5,-1) and (1.5,-1) .. (2,0) node[pos=0.5,below=1pt]{$\ell{+}k_1$};
\path[decoration={markings,mark=at position 1 with \arrow{Stealth[length=3mm]}}, decorate] (1.1,0.75);
\path[decoration={markings,mark=at position 1 with \arrowreversed{Stealth[length=3mm]}}, decorate] (0.9,-0.75);
\draw(2,0) -- (2.5,0)node[right]{$1$};
\draw(0,0)--(-1,-1)node[left]{$2$};
\draw(0,0)--(-1,0)node[left]{$3$};
\draw(0,0)--(-1,1)node[left]{$4$};
\draw [fill=white] (0,0)circle (0.5cm);
\draw(0,0)node{$D^2 F^4$};
\end{tikzpicture}\label{fig:ffourdiagb}
}
\end{center}
\caption{Diagrammatic representation of the terms $\sim \zeta_2$ in the left panel and the contribution
  $\sim \frac{s_{13}}{\ell^2 (\ell{+}k_1)^2}$ to the third line in
  the $\alpha'$-expansion (\ref{exsec.7}) in the right panel.}
\label{ffourdiag}
\end{figure}
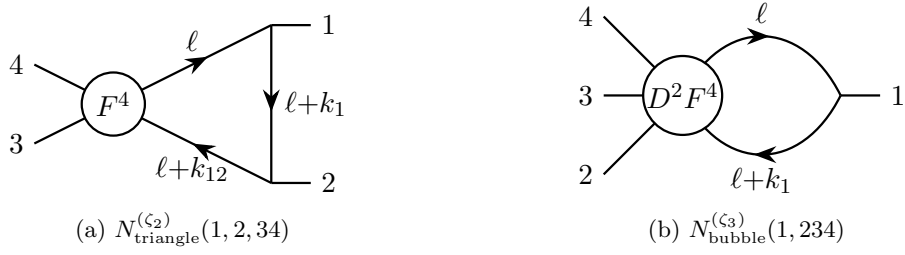

The term proportional to $\zeta_2$ in the second line of \cref{exsec.7}
corresponds to the triangle diagram in
Figure~\ref{fig:ffourdiaga} with a single-insertion of an $F^4$
operator. The propagators $\frac{1}{\ell^2(\ell{+}k_1)^{2}}$ in the third line 
of (\ref{exsec.7}) signal a bubble in external legs as
visualized in Figure~\ref{fig:ffourdiagb}, where the
quintic vertex can be attributed to the non-linear extensions of
$D^2 F^4$ beyond $\partial^2 (\partial A)^4$. Although such external
bubbles integrate to zero in dimensional regularization, we still keep
them in view of their contributions to UV divergences in four and fewer
dimensions, see section~\ref{sec:5}. Further triangle diagrams as in
Figure~\ref{ffourdiag} also arise with $F^4$ replaced by any
higher-derivative operator $D^{2k}F^4$ in the low-energy effective
action. While the four derivatives of the $F^4$ vertex in the second
line of \cref{exsec.7} exclusively contribute external momenta such as
the triangle numerator $s_{12}$, the $D^2 F^4$ vertex in the first
term of the third line yields up to one loop momentum in the numerator
$s_{34}(s_{3,\ell}-s_{4,\ell}+s_{34})$.

\subsubsection{Linearized versus quadratic propagators and bubbles on external legs}
\label{sec:fromlintoquad}

Given a linearized-propagator representation of generic one-loop integrands, 
the numerators usually do not have the properties \cref{eq:Nshift} under cyclic
shifts such that converting back to quadratic propagators may become challenging. 
In our setup where the numerators arise from $\alpha'$-expansions of the $Z$-integrals
in (\ref{prop.1}), the conversion can be greatly facilitated:
if we take forward limits on the output of the {\tt BGap} package \cite{gitrep}, the numerators of one-loop four- and five-point integrands are observed to automatically satisfy the criteria \cref{eq:Nshift}. 

First of all, {\tt BGap} uses different bases for the Mandelstam variables appearing in the numerators and 
denominators. For the disk ordering of $Z(+,i,i{+}1,\ldots,n,1,2,\ldots,i{-}1,-|\ldots)$, the denominators are 
expressed via region variable $s_{\ell,i,i+1,\ldots}$ respecting the ordering 
in the first slot, while the numerators are expressed in terms of two-particle $s_{ab}$ in which $k_{-}{=}{-}\ell$ and 
$s_{\ell,i-1}$ do not appear. One can check that both bases consist of $\tfrac{1}{2}(n{+}2)(n{-}1)$ elements, which are 
indeed minimal for $(n{+}2)$-point kinematics prior to the forward limit. 
The cyclic properties \cref{eq:Nshift} of generic numerators usually depend on the choice whether identities like
\beq
\frac{s_{2,\ell}}{s_{1,\ell}s_{12,\ell}} = \frac{1}{s_{1,\ell}} - \frac{1}{s_{12,\ell}} - \frac{s_{12}}{s_{1,\ell}s_{12,\ell}}
\label{bgoutput}
\eeq
are used to mix $p$-gon topologies with $(p{-}1)$-gons, i.e.\ triangles and bubbles in this case.
At least at four and five points, the output of {\tt BGap} singles out a scheme of
applying (\ref{bgoutput}): we have checked up 
to $\ap^7$, the highest order implemented in {\tt BGap}, that the criteria \cref{eq:Nshift} are satisfied by the numerators along with each combination of MZVs. Therefore, we can obtain quadratic-propagator 
representations without much effort up to and including $\ap^7$.

In contrary, the expressions for the six-point $Z$-integrals at $\ap^8$ and $\ap^9$ order 
given in~\cite{MZVWebsite} are not of this desired form. Consequently, the above shortcuts to get quadratic 
propagators from the forward limit do not apply here in an obvious way. We instead relied on the interpretation 
of forward limits as one-particle unitarity cuts to find a convenient management of the kinematic variables in 
the $\alpha'$-expansions from \cite{MZVWebsite}.

It is well known that scaleless integrals like scalar bubbles on an external leg are not visible to the 
linearized-propagator representation. In order to still specify their numerators in the $\ap$-expansion of $\widehat{A}^{\te{1-loop}}_{\te{eff}}$ in a systematic way, we first rewrite \cref{ngonPF} as
\begin{align}\label{eq:belint2}
  \int\frac{ \dd^D \ell \ \mathfrak{N}(\ell)}{\ell^2(\ell{+}k_1)^2}=\int\frac{\dd^D\ell}{\ell^2}\left(\frac{\mathfrak{N}(\ell)}{s_{1,\ell}}-\frac{\mathfrak{N}(\ell{-}k_1)}{s_{1,\ell}}\right)=\int\frac{\dd^D\ell}{\ell^2}\left(\frac{\mathfrak{N}(\ell)}{s_{1,\ell}}+\frac{\mathfrak{N}(\ell{-}k_1)}{s_{234,\ell}}\right)\,.
\end{align}
The forward limit \cref{exsec.6} taken on six-point disk integrals will naturally lead to the form in the rightmost step before we impose the consequence $s_{1,\ell} = - s_{234,\ell}$ of four-point kinematics. 
This is due to the above choice of Mandelstam basis in the six-point disk integrals:
one can systematically distinguish $s_{1,\ell}$ from $- s_{234,\ell}$ since the 
denominators in the $\alpha'$-expansion of $Z(+,i,i{+}1,\ldots|\ldots)$ from
\cite{Mafra:2016mcc} are expressed via region variables $s_{\ell,i,i{+}1\ldots}$.
 Then, if the numerators of $s_{1,\ell}$ and $s_{234,\ell}$ are related by the shift 
$\ell\rightarrow\ell{-}k_1$, we can recover the bubbles on external legs by reversing \cref{eq:belint2}
as done at the $\zeta_3$-order of (\ref{exsec.7}). As a 
consistency check, our prescription indeed produces the expected UV divergence of external bubbles 
in $D\leq 4$ dimensions, see section \ref{secMGF.3}.

\subsubsection{Higher orders}

\begin{figure}
\begin{center}
\subfloat[$N_{\text{bubble}}^{(\zeta_2^2)}(12,34)$]{\begin{tikzpicture} [scale=1.05, line width=0.30mm]
\draw(0,0) .. controls (0.5,1) and (1.5,1) .. (2,0);
\draw(0,0) .. controls (0.5,-1) and (1.5,-1) .. (2,0);
\path[decoration={markings,mark=at position 1 with \arrow{Stealth[length=3mm]}}, decorate] (1.1,0.75) node[above=1pt]{$\ell$};
\path[decoration={markings,mark=at position 1 with \arrowreversed{Stealth[length=3mm]}}, decorate] (0.9,-0.75) node[below=1pt]{$\ell{+}k_{12}$};
\draw(0,0)--(-1,-0.5)node[left]{$3$};
\draw(0,0)--(-1,0.5)node[left]{$4$};
\draw [fill=white] (0,0)circle (0.4cm);
\draw(0,0)node{$F^4$};
\draw(2,0)--(3,-0.5)node[right]{$2$};
\draw(2,0)--(3,0.5)node[right]{$1$};
\draw [fill=white] (2,0)circle (0.4cm);
\draw(2,0)node{$F^4$};
\end{tikzpicture}\label{fig:ffourbuba}}
\quad\quad\quad\quad\quad\quad\quad
\subfloat[$N_{\text{tadpole}}^{(\zeta_2^2)}(1234)$]{\begin{tikzpicture} [scale=1.05, line width=0.30mm]
\draw(0,0)--(-1.3,0)node[left]{$4$};
\draw(0,0)--(1.3,0)node[right]{$1$};
\draw(0,0)--(0.7,-1)node[right]{$2$};
\draw(0,0)--(-0.7,-1)node[left]{$3$};
\draw (0,0)  .. controls (-1.4,1.5) and (1.4,1.5) ..  (0,0);
 \path[decoration={markings,mark=at position 1 with \arrow{Stealth[length=3mm]}}, decorate] (0.1,1.12) node[above=1pt]{$\ell$};
\draw [fill=white] (0,0)circle (0.4cm);
\draw(0,0)node{$F^6$};
\end{tikzpicture}\label{fig:ffourbubb}}
\end{center}
\caption{Diagrammatic representation of the two-mass bubble in the left panel and the tadpole diagrams
in the right panel for the numerators in (\ref{exsec.10}).}
\label{ffourbub}
\end{figure}
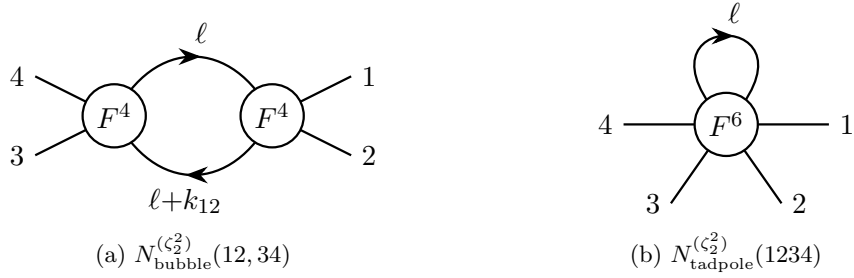

In order to compactly denote higher-order $\ap$-corrections to
(\ref{exsec.7}), we write\footnote{Note that the $1/2$ and $1/4$
  multiplying the two-mass bubble and tadpole diagrams are the symmetry factors that
  compensate the overcounting in the sum over cyclic permutations.}
\begin{align}\label{exsec.8}
  \widehat A^{\te{1-loop}}_{\te{eff}}&(1,2,3,4)=\int \dd^D\ell \; \Bigg\{\frac{1}{\ell^2(\ell{+}k_1)^2(\ell{+}k_{12})^2(\ell{+}k_{123})^2}+\left[\frac{N_{\text{triangle}}(1,2,34)}{\ell^2(\ell{+}k_1)^2(\ell{+}k_{12})^2}\right.\\*
  &\hspace{-1em} \left.+\frac{1}{2}\frac{N_{\text{bubble}}(12,34)}{\ell^2(\ell{+}k_{12})^2}+\frac{N_{\text{bubble}}(1,234)}{\ell^2(\ell{+}k_1)^2}+\frac{1}{4}\frac{N_{\text{tadpole}}(1234)}{\ell^2}+\text{cyc}_{\ell}(1,2,3,4)\right]\Bigg\}+\mathcal{O}(\ap^9)\, .\nonumber
\end{align}
In an ancillary file of the arXiv submission, we provide complete results for the numerators up to $\ap^8$ order, which are organized for each diagram $\Gamma \in \{ {\rm triangle},{\rm bubble}, {\rm tadpole} \}$ via
\begin{align}
N_\Gamma &= \ap^2 \zeta_2 N_\Gamma^{(\zeta_2)}
+\ap^3 \zeta_3 N_\Gamma^{(\zeta_3)}
+\ap^4 \zeta_2^2 N_\Gamma^{(\zeta_2^2)} 
+\ap^5 \Big[  \zeta_5 N_\Gamma^{(\zeta_5)} +  \zeta_2 \zeta_3 N_\Gamma^{(\zeta_2 \zeta_3)}\Big] \nonumber\\
&\quad + \ap^6 \Big[\zeta_3^2 N_{\Gamma}^{(\zeta_3^2)}+\zeta_2^3 N_{\Gamma}^{(\zeta_2^3)}\Big] + \ap^7 \Big[\zeta_7 N_{\Gamma}^{(\zeta_7)}+ \zeta_2\zeta_5 N_{\Gamma}^{(\zeta_2\zeta_5)} + \zeta_2^2\zeta_3 N_{\Gamma}^{(\zeta_2^2\zeta_3)}\Big]\nonumber\\
&\quad + \ap^8 \Big[\zeta_{3,5}N_{\Gamma}^{(\zeta_{3,5})}+\zeta_3\zeta_5 N_{\Gamma}^{(\zeta_3\zeta_5)}+\zeta_2\zeta_3^2 N_{\Gamma}^{(\zeta_2\zeta_3^2)}\Big]\,,
 \label{exsec.9}
\end{align}
with a separate numerator for each element in the tentative
$\mathbb Q$-bases of MZVs \cite{Blumlein:2009cf} at given weight. Note that up to $\ap^8$ order only the topologies given in \cref{exsec.8} show up. The
expressions in \cref{exsec.7} are equivalent to
\begin{align}
N^{(\zeta_2)}_{\text{triangle}}(1,2,34) &= -s_{12} \, , \ \ \ \ \ \  \ \ \ \ N^{(\zeta_2)}_{\text{bubble}}(12,34) = N^{(\zeta_2)}_{\text{bubble}}(123,4) = N^{(\zeta_2)}_{\text{tadpole}}(1234) = 0\,, \nonumber \\
N^{(\zeta_3)}_{\text{triangle}}(1,2,34) &= \frac{1}{2} s_{34}( s_{3,\ell}-s_{4,\ell} + s_{34})  \, , \ \ \ \ \ \ N^{(\zeta_3)}_{\text{bubble}}(12,34) =  0\,, 
 \label{exsec.9a} \\
N^{(\zeta_3)}_{\text{bubble}}(1,234) &= - \frac{3}{2} s_{13} \, , \ \ \ \ \ \  \ \ \ \ \ \ 
 \ \ \ \ \ \  \ \ \ \ \ \  \ \ \,  N^{(\zeta_3)}_{\text{tadpole}}(1234) = 0 \, ,
\nonumber
\end{align}
and the first higher-order correction beyond (\ref{exsec.7}) which
follows from the $\ap$-expansion of the six-point disk integrals in
(\ref{exsec.6}) is given by
\begin{align}
N^{(\zeta_2^2)}_{\text{triangle}}(1,2,34)&=-\frac{1}{20}s_{34}(4s_{3,\ell}^2+4s_{4,\ell}^2-7s_{3,\ell}s_{34}-s_{4,\ell}s_{34}+11s_{34}^2)\,, \nonumber \\
N^{(\zeta_2^2)}_{\text{bubble}}(12,34)&=s_{12}^2 \,,  \ \ \ \ \ \ N^{(\zeta_2^2)}_{\text{tadpole}}(1234)=\frac{24}{5}s_{13} \,,
 \label{exsec.10} \\
 N^{(\zeta_2^2)}_{\text{bubble}}(1,234)&=\frac{20s_{2,\ell}s_{24} {-}4s_{3,\ell}(4s_{34}{+}s_{23}) {-}13s_{34}^2{-}4s_{1,\ell}(2s_{34}{+}3s_{23}) {-}12s_{23}s_{34}{-}9s_{23}^2}{20} \, .
\nonumber
\end{align}
Apart from the triangle diagram (with a $\zeta_2^2 D^4 F^4$ vertex in the place of $\zeta_2F^4$ in Figure~\ref{fig:ffourdiaga}), the order of $\zeta_2^2= \frac{5}{2} \zeta_4$ is the first instance of a two-mass bubble diagram with a double insertion of $\zeta_2 F^4$ depicted in Figure~\ref{fig:ffourbuba}. Moreover, the tadpole numerator receives contributions from the $\zeta_2^2 F^6$ vertex in Figure~\ref{fig:ffourbubb}.

Starting from the order of $\ap^5$, the open-string effective action exhibits two combinations of
zeta values $\zeta_5$ and $\zeta_2 \zeta_3$ associated with numerators (see Table~\ref{effops}),
\begin{align}
N^{(\zeta_5)}_{\text{triangle}}(1,2,34)&=\frac{1}{2}s_{34}\Big[s_{3,\ell}^3-s_{4,\ell}^3-(s_{3,\ell}^2-2s_{4,\ell}^2)s_{34}+(s_{3,\ell}-2s_{4,\ell})s_{34}^2+s_{34}^3\Big]
\,,\nonumber \\
N^{(\zeta_5)}_{\text{bubble}}(12,34)&=0\, , \ \ \ \ \ \ N^{(\zeta_5)}_{\text{tadpole}}(1234)=2(s_{12}^2+4s_{12}s_{23}+s_{23}^2)\,,
\nonumber \\
N^{(\zeta_5)}_{\text{bubble}}(1,234)&=\frac{1}{4}\Big[s_{34}(8s_{2,\ell}^2+10s_{3,\ell}^2+6s_{4,\ell}^2+12s_{2,\ell}s_{3,\ell}+6s_{3,\ell}s_{4,\ell})  \label{exsec.11}\\*
&\qquad+s_{23}(6s_{2,\ell}^2+8s_{4,\ell}^2+10s_{3,\ell}^2+12s_{3,\ell}s_{4,\ell}+6s_{2,\ell}s_{3,\ell})\nonumber\\*
&\qquad+s_{34}^2(3s_{2,\ell}-9s_{3,\ell}-s_{4,\ell})-s_{23}^2(5s_{2,\ell}-s_{3,\ell}-s_{4,\ell})\nonumber\\*
&\qquad-s_{23}s_{34}(4s_{2,\ell}+14s_{3,\ell}+2s_{4,\ell})-2s_{24}(4s_{24}^2+3s_{34}^2)\Big]\,, \nonumber 
\end{align}
as well as
\begin{align}
N^{(\zeta_2\zeta_3)}_{\text{triangle}}(1,2,34)&=-\frac{1}{2}s_{34}^2\Big[s_{3,\ell}^2+s_{4,\ell}^2-(s_{3,\ell}+s_{4,\ell})s_{34}\Big]\,, \nonumber \\
N^{(\zeta_2\zeta_3)}_{\text{bubble}}(12,34)&=-s_{12}^2(s_{1,\ell}+s_{3,\ell}+s_{12}) \, , \ \ \ \ \ \ 
N^{(\zeta_2\zeta_3)}_{\text{tadpole}}(1234)=0\,,   \label{exsec.12}\\
N^{(\zeta_2\zeta_3)}_{\text{bubble}}(1,234)&=\frac{s_{34}^2(s_{2,\ell}{-}2s_{4,\ell}){-}s_{23}^2(s_{4,\ell}{-}2s_{2,\ell}){+}2s_{23}s_{34}(s_{2,\ell}{-}s_{4,\ell})  {-} s_{23}(s_{23}^2{-}6s_{34}s_{24})}{2} \, ,
\nonumber
\end{align}
and similar expressions for the numerators at order $\ap^6$ can be
found in appendix~\ref{app:4ptopen}.  Note that the two-mass bubble
numerator $N^{(\zeta_2\zeta_3)}_{\text{bubble}}(12,34)$ in
(\ref{exsec.12}) captures one insertion of both $\zeta_2 F^4$ and
$\zeta_3 D^2 F^4$ similar to Figure~\ref{fig:ffourbuba}.  

The simplest interaction
$\ap^8\zeta_{3,5}D^{10}F^5$ with an irreducible MZV beyond depth one gives
rise to the external-bubble and tadpole numerators in \cref{app:4ptopen}.
By unitarity, the absence of triangles and two-mass bubbles $\sim \zeta_{3,5}$
is a consequence of only having $\zeta_w$ of depth one in the four-point tree-level
amplitude. The contributions at $\ap^7$ and 
the remaining terms at $\ap^8$
are sufficiently complicated that we do not reproduce them in text,
but instead include them only in the ancillary files.


\subsection{Four points, closed strings}
\label{sec:4.8}

The double copy of the four-point half integrand (\ref{exsec.2}) leads to
the following simple form of the gravitational one-loop matrix element (\ref{prop.2}),
\begin{align}
M^{\te{1-loop}}_{4,{\rm eff}} &=  |t_8(f_1,f_2,f_3,f_4)|^2  \widehat M^{\te{1-loop}}_{4,{\rm eff}} \,,
\label{exsec.31} \\
 \widehat M^{\te{1-loop}}_{4,{\rm eff}} &= \int \frac{ \dd^D\ell}{\ell^2}   \lim_{k_{\pm} \rightarrow \pm \ell} \sum_{\gamma,\rho \in S_4} J(+,\gamma(1,2,3,4),-|+,\rho(1,2,3,4),-) \, ,\nonumber
\end{align}
where $|t_8(f_1,f_2,f_3,f_4)|^2$ instructs to replace $\ep_j^\mu \rightarrow \bar \ep_j^\mu$ in
one of the chiral halves (with analogous replacements for the fermionic completion). By the
single-valued map in (\ref{rev.37}), the leading $\ap$-orders of the six-point sphere integrals 
in (\ref{exsec.31}) can be obtained by setting $\zeta_2 \rightarrow 0$ and $\zeta_3 \rightarrow 2 \zeta_3$ 
in (\ref{exsec.5}). More specifically, by comparing the permutation sum over $\gamma$ in (\ref{exsec.31}) with the cyclic sums over $\gamma$ in (\ref{exsec.4}) and (\ref{exsec.6}), we are led to
\begin{align}
 \widehat M^{\te{1-loop}}_{4,{\rm eff}} &=  \sum_{\rho \in S_4} {\rm sv}\, \widehat A^{\te{1-loop}}_{\rm eff}\big(1,\rho(2,3,4)\big)\, ,
\label{exsec.32} \\
M^{\te{1-loop}}_{4,{\rm eff}} &= t_8(\bar f_1,\bar f_2,\bar f_3,\bar f_4) \sum_{\rho \in S_4} {\rm sv}\, A^{\te{1-loop}}_{\rm eff} \big(1,\rho(2,3,4) \big)\, . \notag
\end{align}
These relations holds to all orders in $\alpha'$, 
and one can readily assemble the orders of $\alpha'^{\leq 6}$ from (\ref{exsec.7}), (\ref{exsec.11}) and (\ref{ordz32}).
However, the permutation sum in (\ref{exsec.32}) leads to additional cancellations and improved power counting
as one can for instance see from the explicit form of the first orders of the $\ap$-expansion
\begin{align}
 \widehat M^{\te{1-loop}}_{4,{\rm eff}} 
= \int \dd^D\ell \,\bigg[&\,\frac{1}{4}\frac{1}{\ell^2(\ell{+}k_1)^2(\ell{+}k_{12})^2(\ell{+}k_{123})^2} 
+\frac{\ap^3 \zeta_3 s_{12}^2}{\ell^2(\ell{+}k_1)^2(\ell{+}k_{12})^2} 
\notag
\\
&+\frac{\ap^5 \zeta_5 s_{12}^2 [s_{3,\ell}^2+s_{4,\ell}^2-(s_{3,\ell}+s_{4,\ell})s_{12}+2s_{12}^2]}{2\ell^2(\ell{+}k_1)^2(\ell{+}k_{12})^2}  \label{exsec.33}  \\
&- \frac{ 5 \ap^5 \zeta_5  s_{12}s_{13}s_{14} }{ 2\ell^2 (\ell{+}k_1)^2}   + {\cal O}(\ap^6)\bigg]+\text{perm}(1,2,3,4)\, .
\notag
\end{align}
The coefficients of $\zeta_3$ and $\zeta_5$ are the one-loop matrix elements with a single insertion
of the operators $R^4$ and $D^4R^4$ in the low-energy effective action (\ref{seffclosed}), respectively.
Similar to the power counting of the triangle numerators from $F^4$ in (\ref{exsec.7}), the eight
derivatives of $R^4$ conspire to the external momenta in the numerator $\ap^3 \zeta_3 s_{12}^2$
of the first line of (\ref{exsec.33}). As a closed-string analogue of the $D^2 F^4$ numerator in (\ref{exsec.7}) which is
linear in loop momentum, the triangle numerator from $D^4 R^4$ in the second line of (\ref{exsec.33})
is quadratic in $\ell$ akin to the double-copy property of supergravity numerators. The diagrams 
in (\ref{exsec.33}) are identical to those in Figure~\ref{ffourdiag}
with $R^4$ and $D^4 R^4$ in the place of $F^4$ and $D^2 F^4$, respectively.

By analogy with (\ref{exsec.8}) and (\ref{exsec.9}) we organize higher orders in $\ap$
according to\footnote{The overall factor of 2 reconciles the
symmetry factors of the diagrams in the square brackets 
(compensating for an overcounting by the permutation sum)
with the normalization of the one-loop matrix 
element (\ref{prop.2}).}
\begin{align}
  \widehat M^{\te{1-loop}}_{4,{\rm eff}}  &=2 \int \dd^D\ell \left[\frac{1}{8}\frac{1}{\ell^2(\ell{+}k_1)^2(\ell{+}k_{12})^2(\ell{+}k_{123})^2}+\frac{1}{4}\frac{{\cal N}_{\text{triangle}}(1,2,34)}{\ell^2(\ell{+}k_1)^2(\ell{+}k_{12})^2}\right.\nonumber\\
&\quad \quad   \left.+\frac{1}{16}\frac{{\cal N}_{\text{bubble}}(12,34)}{\ell^2(\ell{+}k_{12})^2}+\frac{1}{12}\frac{{\cal N}_{\text{bubble}}(1,234)}{\ell^2(\ell{+}k_1)^2}+\mathcal{O}(\ap^9)\right]+\text{perm}(1,2,3,4)\,,\label{exsec.34}
\end{align}
and we provide complete results up to $\ap^8$ order,
\begin{align}
  {\cal N}_\Gamma &= \ap^3 \zeta_3 {\cal N}_\Gamma^{(\zeta_3)}
+\ap^5 \zeta_5 {\cal N}_\Gamma^{(\zeta_5)}
+\ap^6 \zeta_3^2 {\cal N}_\Gamma^{(\zeta_3^2)}
+\ap^7 \zeta_7 {\cal N}_\Gamma^{(\zeta_7)}
+\ap^8 \zeta_3 \zeta_5 {\cal N}_\Gamma^{(\zeta_3\zeta_5)}\, .
\end{align}
The subleading orders of (\ref{exsec.33}) translate into
\beq
{\cal N}^{(\zeta_3)}_{\text{triangle}}(1,2,34) = 2 s_{12}^2 \, , \quad\quad
{\cal N}^{(\zeta_3)}_{\text{bubble}}(12,34) = {\cal N}^{(\zeta_3)}_{\text{bubble}}(1,234) = 0 \,,
\label{exsec.35}
\eeq
as well as
\begin{align}
{\cal N}^{(\zeta_5)}_{\text{triangle}}(1,2,34)&=s_{12}^2\left[s_{3,\ell}^2+s_{4,\ell}^2-(s_{3,\ell}+s_{4,\ell})s_{12}+2s_{12}^2\right]\,,  \notag\\
{\cal N}^{(\zeta_5)}_{\text{bubble}}(1,234)&=-15s_{12}s_{13}s_{14} \, , \ \ \ \ \ \
{\cal N}^{(\zeta_5)}_{\text{bubble}}(12,34)=0 \, .
\label{exsec.36}
\end{align}
At higher orders, similar calculations yield 
\begin{align}
{\cal N}^{(\zeta_3^2)}_{\text{triangle}}(1,2,34)&=s_{12}^3\left[s_{3,\ell}^2+s_{4,\ell}^2-(s_{3,\ell}+s_{4,\ell})s_{12}\right], \nonumber \\
{\cal N}^{(\zeta_3^2)}_{\text{bubble}}(12,34)&=
-4s_{12}(s_{1,\ell}+s_{2,\ell})(s_{1,\ell}+s_{2,\ell}+2s_{12})\,,
\label{exsec.37}\\
{\cal N}^{(\zeta_3^2)}_{\text{bubble}}(1,234)&=
s_{12}^4+s_{13}^4+s_{14}^4+3s_{1,\ell}s_{12}s_{13}s_{14}
+2s_{\ell,1}\left[s_{2,\ell}s_{12}^2+s_{3,\ell}s_{13}^2+s_{4,\ell}s_{14}^2\right]\, ,
\notag
\end{align}
where the two-mass bubble
${\cal N}^{(\zeta_3^2)}_{\text{bubble}}(12,34)$ in the second line is
due to double insertion of $\zeta_3 R^4$ and can be viewed as a
closed-string analogue of the double insertion of $F^4$ in Figure~\ref{fig:ffourbuba}. Two-mass bubbles of this 
type do not occur along with single zeta values such as
\begin{align}
{\cal N}^{(\zeta_7)}_{\text{triangle}}(1,2,34)&=s_{12}^2\Big[s_{3,\ell}^4+s_{4,\ell}^4-2(s_{3,\ell}^3{+}s_{4,\ell}^3)s_{12}+3(s_{3,\ell}^2{+}s_{4,\ell}^2)s_{12}^2-2(s_{3,\ell}{+}s_{4,\ell})s_{12}^3+2s_{12}^4\Big],\nonumber \\
{\cal N}^{(\zeta_7)}_{\text{bubble}}(12,34)&=0\,, \label{exsec.38} \\
{\cal N}^{(\zeta_7)}_{\text{bubble}}(1,234)&=-\frac{13}{8}s_{1,\ell}^2s_{12}s_{13}s_{14}+4s_{1,\ell}(s_{12}^4+s_{13}^4+s_{14}^4) -14s_{12}s_{13}s_{14}(s_{12}^2+s_{13}^2+s_{14}^2)\nonumber\\
&\hspace{-2cm} -\Big[s_{2,\ell}^2(6s_{12}s_{13}s_{14}{+}5s_{12}^3)-8s_{2,\ell}s_{3,\ell}s_{14}^3+3s_{2,\ell}s_{12}^2(2s_{12}^2{-}3s_{13}^2{-}3s_{14}^2) +\text{cyc}(2,3,4)\Big]\,, \nonumber
\end{align}
and it is tedious but straightforward to obtain the next orders in
$\ap$ along the same lines.


\subsection{The five-point kinematic half integrand}
\label{sec:4.77}

The five-point worldsheet correlator in one-loop superstring amplitudes has been studied from
a variety of perspectives \cite{Tsuchiya:1988va, Stieberger:2002wk, Mafra:2009wi, Mafra:2012kh, Mafra:2018qqe},
and we shall now review the representation of~\cite{He:2017spx} for its analogue in the ambitwistor setup \cite{Adamo:2013tsa, Geyer:2015bja}. Its kinematic dependence can be compactly encoded in the two-particle polarization $\ep_{12}^{\mu}$ and field strength $f^{\mu \nu}_{12}$ \cite{Mafra:2014oia, Mafra:2015vca} given by
\begin{align}
\ep_{12}^\mu &= (k_2\cdot \ep_1) \ep_2^\mu - (k_1\cdot \ep_2) \ep_1^\mu + \frac{1}{2}(\ep_1\cdot \ep_2)(k_1^\mu - k_2^\mu) \notag \\
&= \frac{1}{2} \big[ (k_2\cdot \ep_1) \ep_2^\mu
+ (\ep_1)_\nu f_2^{\nu \mu}  -(1\leftrightarrow 2) \big] \,,
\label{eq5.1} \\
f_{12}^{\mu \nu} &=  (k_2\cdot \ep_1) f_2^{\mu \nu} - (k_1\cdot \ep_2) f_1^{\mu \nu} 
+ f_1^{\mu}{}_\la f_2^{\la \nu} -  f_2^{\mu}{}_\la f_1^{\la \nu} \notag \\
&= k_{12}^\mu \ep_{12}^\nu - k_{12}^\nu \ep_{12}^\mu - (k_1 \cdot k_2) (\ep_1^\mu \ep_2^\nu - \ep_1^\nu \ep_2^\mu )\,,\nonumber
\end{align}
and subject to
\beq
\ep_{12}^\mu=-\ep_{21}^\mu\, , \ \ \ \ \ \ f_{12}^{\mu \nu}=-f_{21}^{\mu \nu} = - f_{12}^{\nu \mu}\, .
\eeq
As a natural generalization of the four-point kinematic factor in (\ref{exsec.2}), 
the $t_8$-tensor (\ref{exsec.1}) in the five-point correlator either contracts the 
two-particle field strength or is dressed by additional polarization vectors\footnote{We are grateful to Carlos Mafra for discussions that led us to correct the
prefactor of the parity-odd terms to $\frac{1}{16}$ in place of
$\frac{i}{2}$ as in earlier versions. The factor of $i$ has been removed to account
for the minus sign in $\varepsilon_{10}^\mu(v_1,v_2,\ldots,v_9) \eta_{\mu \nu} \varepsilon_{10}^\nu(w_1,w_2,\ldots,w_9)
= - \det(v_i\cdot w_j)_{i,j=1,2,\ldots,9}$ from the Minkowskian signature.}
%
\begin{align}
t_8(12,3,4,5) &= t_8(f_{12},f_3,f_4,f_5)  = - t_8(21,3,4,5)
\label{t8defs}
\\
t_{\pm}^\mu(1,2,3,4,5) &= \big[ \ep_1^\mu t_8(f_2, f_3,f_4,f_5)  + (1\leftrightarrow 2,3,4,5) \big]
\pm 
 \frac{1}{16}  \varepsilon^\mu_{10}(\ep_1,f_2,f_3,f_4,f_5)  \, ,
\notag
\end{align}
where $t_8(12,3,4,5)$ and $t_{\pm}^\mu(1,2,3,4,5)$ are permutation-invariant in $\{3,4,5\}$ and
$\{1,2,3,4,5\}$, respectively.\footnote{Momentum conservation also implies the
symmetry $\varepsilon^\mu_{10}(\ep_1,f_2,f_3,f_4,f_5)=\varepsilon^\mu_{10}(\ep_2,f_1,f_3,f_4,f_5)$ of
the parity-odd contributions.} The $t_{\pm}$ notation for the vector accommodates two sign choices for the 
parity-odd term $\varepsilon^\mu_{10}(\ep_1,f_2,f_3,f_4,f_5)= (\varepsilon_{10})^{\mu}{}_{\nu \lambda_2 \rho_2
\lambda_3 \rho_3\lambda_4 \rho_4\lambda_5 \rho_5}\ep^\nu_1 f^{\lambda_2 \rho_2}_2f^{\lambda_3 \rho_3}_3f^{\lambda_4 \rho_4}_4 f^{\lambda_5 \rho_5}_5$ in order to later on unify expressions for 
type-IIA and type-IIB theories.

With the above building blocks, the five-point kinematic half integrand
can be compactly written as
\beq
{\cal I}_{5 }(\ell) = \frac{ \ell_\mu t_{+}^\mu(1,2,3,4,5) + \big[ G_{12} t_8({12},3,4,5)  + (1,2|1,2,3,4,5)\big]}{\sigma_1 \sigma_2 \sigma_3 \sigma_4 \sigma_5 }
\label{t8defs.2}
\eeq
with the following shorthand for the Green function on the nodal sphere
\beq
G_{ij} = \frac{\sigma_i {+} \sigma_j }{2(\sigma_i{-}\sigma_j)} \, .
\label{t8defs.3}
\eeq
The notation $+ (1,2|1,2,3,4,5)$ in (\ref{t8defs.2}) instructs to add the nine permutations of the term
$G_{12} t_8(f_{12},f_3,f_4,f_5) $ in the square bracket with $12$ replaced by any other pair $ij$ with
$1\leq i<j\leq 5$. By arranging the $\sigma_j$ dependence of (\ref{t8defs.2}) 
and (\ref{t8defs.3}) into Parke-Taylor
factors \cite{He:2017spx}, we arrive at the following
pentagon numerators in the expansion (\ref{rev.7}), 
\begin{align}
{\cal I}_{5}(\ell) &= \sum_{\rho \in S_5} 
N_{+|\rho(12345)| -}(\ell)  {\rm PT}(+,\rho(1,2,3,4,5),-) \,,
\notag \\
N_{+|12345|-}(\ell)&= \ell_\mu t_{+}^\mu(1,2,3,4,5) - \frac{1}{2} \Big\{
t_{8}({12},3,4,5)+t_8({13},2,4,5)  \label{eq5.5} \\
&\qquad
+t_8({14},2,3,5)+t_8({15},2,3,4)
+t_8({23},1,4,5)+t_8({24},1,3,5)\notag \\
&\qquad
+t_8({25},1,3,4)+t_8({34},1,2,5)
+t_8({35},1,2,4)+t_8({45},1,2,3)
\Big\} \, .  \nonumber
\end{align}
In contrast to the permutation-invariant four-point numerators $N_{+|\rho(1234)|-} {=}t_8(f_1,f_2,f_3,f_4)$, 
the five-point numerators depend on the ordering $\rho$ of the
pentagon. Its antisymmetric part 
\beq
N_{+|12345|-}- N_{+|21345|-}
= - t_8(12,3,4,5)
\label{t8defs.4}
\eeq
no longer depends on $\ell$, inherits the symmetry in $3,4,5$ of the
$t_8$-tensor\footnote{In other words, we also obtain $- t_8(12,3,4,5)$
  from $N_{+|31245|-}- N_{+|32145|-}$,
  $N_{+|34125|-}- N_{+|34215|-}$ and $N_{+|34512|-}- N_{+|34521|-} $
  as well as their permutations in $3,4,5$.}
and corresponds to a one-mass box numerator. The
structure of (\ref{eq5.5}) and its compatibility with the
color-kinematics duality was firstly discussed in
\cite{Mafra:2014gja}. The ancestors of the kinematic building blocks
(\ref{t8defs}) in pure-spinor superspace described in the reference
can be readily imported to supersymmetrize the $t_8$-tensors in the
$\alpha'$-expansions below. Color-kinematics dual five-point
numerators in terms of four-dimensional spinor-helicity variables can
be found in \cite{Carrasco:2011mn, He:2015wgf}.

For later reference, we note the linearized gauge variation
of the above building blocks which we shall implement via
\beq
\delta = \sum_{j=1}^5 \omega_j k_j^\mu \frac{ \partial }{\partial \epsilon_j^\mu}\, .
\label{t8defs.5}
\eeq
The formal bookkeeping variables $\omega_j$ track the transformation
$\ep_j \rightarrow k_j$ in the $j^{\rm th}$ leg, 
\begin{align}
\delta t_8(12,3,4,5) &= \frac{1}{2}s_{12} \big[ \omega_1 t_8(f_2,f_3,f_4,f_5) -  \omega_2 t_8(f_1,f_3,f_4,f_5) \big]
\notag
\\
\delta t_+^\mu(1,2,3,4,5) &= k_1^\mu \omega_1  t_8(f_2,f_3,f_4,f_5)  + (1\leftrightarrow 2,3,4,5) 
\notag \\
\delta N_{+|12345|-} &= \frac{\omega_1}{2} \big( \ell_1^2 - \ell^2 \big)t_8(f_2,f_3,f_4,f_5)
+ \frac{\omega_2}{2} \big( \ell_{12}^2 - \ell_1^2 \big)t_8(f_1,f_3,f_4,f_5) \label{t8defs.6} \\
&\quad + \frac{\omega_3}{2} \big( \ell_{123}^2 - \ell_{12}^2 \big)t_8(f_1,f_2,f_4,f_5) 
+ \frac{\omega_4}{2} \big( \ell_{1234}^2 - \ell_{123}^2 \big)t_8(f_1,f_2,f_3,f_5)\nonumber\\
&\quad + \frac{\omega_5}{2} \big( \ell^2 - \ell_{1234}^2 \big)t_8(f_1,f_2,f_3,f_4)\, ,
\notag
\end{align}
where we introduced the shorthand notations
\begin{equation}
\ell_1 = \ell + k_1\,, \qquad \ell_{12} = \ell + k_{12}\,,
\qquad \ell_{123} = \ell + k_{123} \,,\qquad \ell_{1234} = \ell + k_{1234} \,,
\label{t8defs.8}
\end{equation}
and we have used five-point momentum conservation to simplify the variation of $N_{+|12345|-}$.
The operator (\ref{t8defs.5}) implementing linearized gauge variations is 
engineered to mimic the BRST transformations of the supersymmetric
numerators in \cite{Mafra:2014gja}.


\subsection{Five points, open string}
\label{sec:4.78}

\begin{figure}
\begin{center}
\subfloat[$N_{\text{bubble}}^{(\zeta_3)}(1,2345)$]{\begin{tikzpicture} [scale=1.05, line width=0.30mm]
\draw(0,0) .. controls (0.5,1) and (1.5,1) .. (2,0);
\draw(0,0) .. controls (0.5,-1) and (1.5,-1) .. (2,0);
\path[decoration={markings,mark=at position 1 with \arrow{Stealth[length=3mm]}}, decorate] (1.1,0.75) node[above=1pt]{$\ell$};
\path[decoration={markings,mark=at position 1 with \arrowreversed{Stealth[length=3mm]}}, decorate] (0.9,-0.75) node[below=1pt]{$\ell_{1}$};
\draw(2,0) -- (2.5,0)node[right]{$1$};
\draw(0,0)--(-1,-1.2)node[left]{$2$};
\draw(0,0)--(-1,-0.4)node[left]{$3$};
\draw(0,0)--(-1,0.4)node[left]{$4$};
\draw(0,0)--(-1,1.2)node[left]{$5$};
\draw [fill=white] (0,0)circle (0.55cm);
\node at (0,0) [font=\fontsize{5pt}{5pt}\selectfont]{$D^2\!F^4{+}F^5$};
\end{tikzpicture}}
\quad\quad\quad
\subfloat[$N_{\text{bubble}}^{(\zeta_3)}({[12]},345)$]{\begin{tikzpicture} [scale=1.05, line width=0.30mm]
\draw(0,0) .. controls (0.5,1) and (1.5,1) .. (2,0);
\draw(0,0) .. controls (0.5,-1) and (1.5,-1) .. (2,0);
\path[decoration={markings,mark=at position 1 with \arrow{Stealth[length=3mm]}}, decorate] (1.1,0.75) node[above=1pt]{$\ell$};
\path[decoration={markings,mark=at position 1 with \arrowreversed{Stealth[length=3mm]}}, decorate] (0.9,-0.75) node[below=1pt]{$\ell_{12}$};
\draw(2,0) -- (2.4,0);
\draw(2.4,0)--(2.8,0.4)node[right]{$1$};
\draw(2.4,0)--(2.8,-0.4)node[right]{$2$};
\draw(0,0)--(-1,-1)node[left]{$3$};
\draw(0,0)--(-1,0)node[left]{$4$};
\draw(0,0)--(-1,1)node[left]{$5$};
\draw [fill=white] (0,0)circle (0.55cm);
\node at (0,0) [font=\fontsize{5pt}{5pt}\selectfont]{$D^2\!F^4{+}F^5$};
\end{tikzpicture}} \\
\subfloat[$N_{\text{bubble}}^{(\zeta_3)}(1,{[23]}45)$]{\begin{tikzpicture} [scale=1.05, line width=0.30mm]
\draw(0,0) .. controls (0.5,1) and (1.5,1) .. (2,0);
\draw(0,0) .. controls (0.5,-1) and (1.5,-1) .. (2,0);
\path[decoration={markings,mark=at position 1 with \arrow{Stealth[length=3mm]}}, decorate] (1.1,0.75) node[above=1pt]{$\ell$};
\path[decoration={markings,mark=at position 1 with \arrowreversed{Stealth[length=3mm]}}, decorate] (0.9,-0.75) node[below=1pt]{$\ell_{1}$};
\draw(2,0) -- (2.5,0)node[right]{$1$};
\draw(0,0)--(-0.6,-0.6);
\draw(-0.6,-0.6)--(-1.2,-0.6)node[left]{$3$};
\draw(-0.6,-0.6)--(-0.6,-1.2)node[left]{$2$};
\draw(0,0)--(-1,0)node[left]{$4$};
\draw(0,0)--(-1,1)node[left]{$5$};
\draw [fill=white] (0,0)circle (0.55cm);
\node at (0,0) [font=\fontsize{5pt}{5pt}\selectfont]{$D^2\!F^4{+}F^5$};
\end{tikzpicture}} 
\quad
\subfloat[$N_{\text{bubble}}^{(\zeta_3)}(1,2{[34]}5)$]{\begin{tikzpicture} [scale=1.05, line width=0.30mm]
\draw(0,0) .. controls (0.5,1) and (1.5,1) .. (2,0);
\draw(0,0) .. controls (0.5,-1) and (1.5,-1) .. (2,0);
\path[decoration={markings,mark=at position 1 with \arrow{Stealth[length=3mm]}}, decorate] (1.1,0.75) node[above=1pt]{$\ell$};
\path[decoration={markings,mark=at position 1 with \arrowreversed{Stealth[length=3mm]}}, decorate] (0.9,-0.75) node[below=1pt]{$\ell_{1}$};
\draw(2,0) -- (2.5,0)node[right]{$1$};
\draw(0,0)--(-1,-1)node[left]{$2$};
\draw(0,0)--(-0.8,0);
\draw(-0.8,0)--(-1.3,-0.35)node[left]{$3$};
\draw(-0.8,0)--(-1.3,0.35)node[left]{$4$};
\draw(0,0)--(-1,1)node[left]{$5$};
\draw [fill=white] (0,0)circle (0.55cm);
\node at (0,0) [font=\fontsize{5pt}{5pt}\selectfont]{$D^2\!F^4{+}F^5$};
\end{tikzpicture}}
\quad
\subfloat[$N_{\text{bubble}}^{(\zeta_3)}(1,23{[45]})$]{\begin{tikzpicture} [scale=1.05, line width=0.30mm]
\draw(0,0) .. controls (0.5,1) and (1.5,1) .. (2,0);
\draw(0,0) .. controls (0.5,-1) and (1.5,-1) .. (2,0);
\path[decoration={markings,mark=at position 1 with \arrow{Stealth[length=3mm]}}, decorate] (1.1,0.75) node[above=1pt]{$\ell$};
\path[decoration={markings,mark=at position 1 with \arrowreversed{Stealth[length=3mm]}}, decorate] (0.9,-0.75) node[below=1pt]{$\ell_{1}$};
\draw(2,0) -- (2.5,0)node[right]{$1$};
\draw(0,0)--(-0.6,0.6);
\draw(-0.6,0.6)--(-1.2,0.6)node[left]{$4$};
\draw(-0.6,0.6)--(-0.6,1.2)node[left]{$5$};
\draw(0,0)--(-1,0)node[left]{$3$};
\draw(0,0)--(-1,-1)node[left]{$2$};
\draw [fill=white] (0,0)circle (0.55cm);
\node at (0,0) [font=\fontsize{5pt}{5pt}\selectfont]{$D^2\!F^4{+}F^5$};
\end{tikzpicture}}
\end{center}
\caption{Bubble contributions to the $\alpha'^3 \zeta_3$-order of
  (\ref{t8defs.9}) with numerators given by (\ref{t8defs.12}). The
  blobs represent insertions of $D^2F^4+F^5$. Other $\zeta$ orders
  contain the same topologies but with different operator insertions.}
\label{ffivebubbs}
\end{figure}
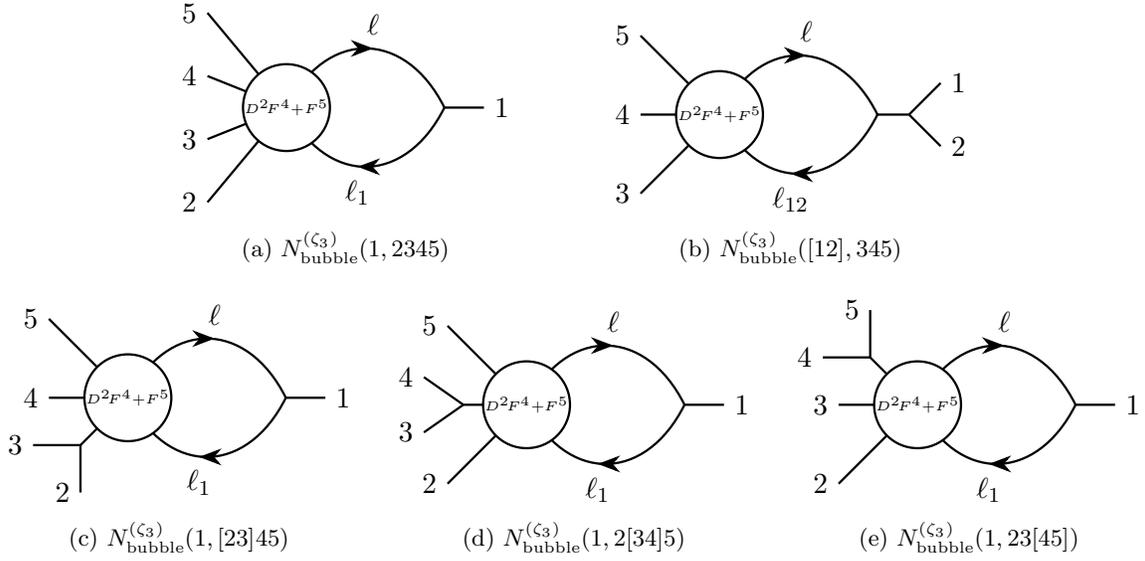

In contrast to the four-point case, already the field-theory limit of the one-loop
matrix elements (\ref{prop.1}) at five points is a combination of a pentagon 
and five massive boxes,
\begin{align}
A^{\te{1-loop}}_{\rm SYM}(1,2,3,4,5) &= \int  \dd^D \ell \, \bigg[
\frac{ N_{+|12345|-} }{\ell^2 \ell_1^2 \ell_{12}^2 \ell_{123}^2 \ell_{1234}^2 }
 - \frac{ t_8(12,3,4,5) }{ s_{12} \ell^2   \ell_{12}^2 \ell_{123}^2 \ell_{1234}^2 }
 - \frac{ t_8(23,1,4,5) }{ s_{23} \ell^2 \ell_1^2  \ell_{123}^2 \ell_{1234}^2 } 
\notag\\
 &\quad\quad
 - \frac{ t_8(34,1,2,5) }{ s_{34} \ell^2 \ell_1^2 \ell_{12}^2  \ell_{1234}^2 }
 - \frac{ t_8(45,1,2,3) }{ s_{12} \ell^2 \ell_1^2 \ell_{12}^2 \ell_{123}^2   }
- \frac{ t_8(51,2,3,4) }{ s_{12}  \ell_1^2 \ell_{12}^2 \ell_{123}^2 \ell_{1234}^2 } \bigg]\,.
 \label{t8defs.7}
\end{align}
In (\ref{t8defs.7}) and its $\ap$-corrections, the recombination of the linearized
propagators to quadratic ones relies on shifts of loop momenta in $\ell_\mu t_+^\mu(1,2,3,4,5)$
and the relation
\beq
(k_1)_\mu t_{\pm}^\mu(1,2,3,4,5) = - t_8(12,3,4,5)
- t_8(13,2,4,5)- t_8(14,2,3,5)- t_8(15,2,3,4)\, .
\label{keyid}
\eeq
Based on the $\ap$-expansion of the seven-point disk 
integrals \cite{Mafra:2016mcc} in the one-loop matrix element (\ref{prop.1}), we organize the corrections 
to \cref{t8defs.7} via
\begin{align}  \label{t8defs.9}
A^{\te{1-loop}}_{\rm eff}(1,2,3,4,5) &= A^{\te{1-loop}}_{\rm SYM}(1,2,3,4,5) + \int \dd^D \ell \, \bigg[\frac{N_{\text{box}}(1,2,3,45)}{\ell^2\ell_1^2\ell_{12}^2\ell_{123}^2}+\frac{N_{\text{triangle}}(1,2,345)}{\ell^2\ell_1^2\ell_{12}^2} \nonumber\\
&\quad+\frac{N_{\text{triangle}}(1,2,[34]5)}{\ell^2\ell_1^2\ell_{12}^2 s_{34}}+\frac{N_{\text{triangle}}(1,2,3[45])}{\ell^2\ell_1^2\ell_{12}^2 s_{45}}+\frac{N_{\text{triangle}}(1,23,45)}{\ell^2\ell_1^2\ell_{12}^2}\nonumber\\
&\quad+\frac{N_{\text{triangle}}(1,[23],45)}{\ell^2\ell_1^2\ell_{12}^2s_{23}}+\frac{N_{\text{triangle}}(1,23,[45])}{\ell^2\ell_1^2\ell_{12}^2s_{45}}+\frac{N_{\text{bubble}}(12,345)}{\ell^2\ell_{12}^2}\nonumber\\
&\quad+\frac{N_{\text{bubble}}([12],345)}{\ell^2\ell_{12}^2s_{12}}+\frac{N_{\text{bubble}}(12,[34]5)}{\ell^2\ell_{12}^2s_{34}}+\frac{N_{\text{bubble}}(12,3[45])}{\ell^2\ell_{12}^2s_{45}}\nonumber\\
&\quad +\frac{N_{\text{bubble}}(1,2345)}{\ell^2\ell_1^2}+\frac{N_{\text{bubble}}(1,[23]45)}{\ell^2\ell_1^2s_{23}}+\frac{N_{\text{bubble}}(1,2[34]5)}{\ell^2\ell_1^2s_{34}} \\
&\quad +\frac{N_{\text{bubble}}(1,23[45])}{\ell^2\ell_1^2s_{45}}+ \frac{1}{5}\frac{N_{\text{tadpole}}(12345)}{\ell^2} +\text{cyc}_{\ell}(1,2,3,4,5)\bigg] +\mathcal{O}(\ap^6) \, ,\nonumber
\end{align}
where square brackets as in $N_{\text{triangle}}(1,[23],45)$ or $N_{\text{bubble}}([12],345)$ indicate the
presence of propagators $s^{-1}_{ij}$ as opposed to the contact diagrams $N_{\text{triangle}}(1,23,45)$ and
$N_{\text{bubble}}(12,345)$ at different mass dimensions. The notation $\text{cyc}_{\ell}(1,2,3,4,5)$ is
again understood to include a shift of loop momentum such that $\ell$ points from leg 5 to leg 1
which may affect the numerators as exemplified in (\ref{cycellex}). We give complete results up to $\ap^5$ order,
\begin{align}
N_\Gamma &= \ap^2 \zeta_2 N_\Gamma^{(\zeta_2)}
+\ap^3 \zeta_3 N_\Gamma^{(\zeta_3)}
+\ap^4 \zeta_2^2 N_\Gamma^{(\zeta_2^2)} 
+\ap^5 \Big[  \zeta_5 N_\Gamma^{(\zeta_5)} +  \zeta_2 \zeta_3 N_\Gamma^{(\zeta_2 \zeta_3)}\Big] \,.
\end{align}
The $\ap$-expansion of the numerators in 
(\ref{t8defs.9}) is organized according to MZVs as in (\ref{exsec.9}): the coefficients of $\ap^2 \zeta_2$, 
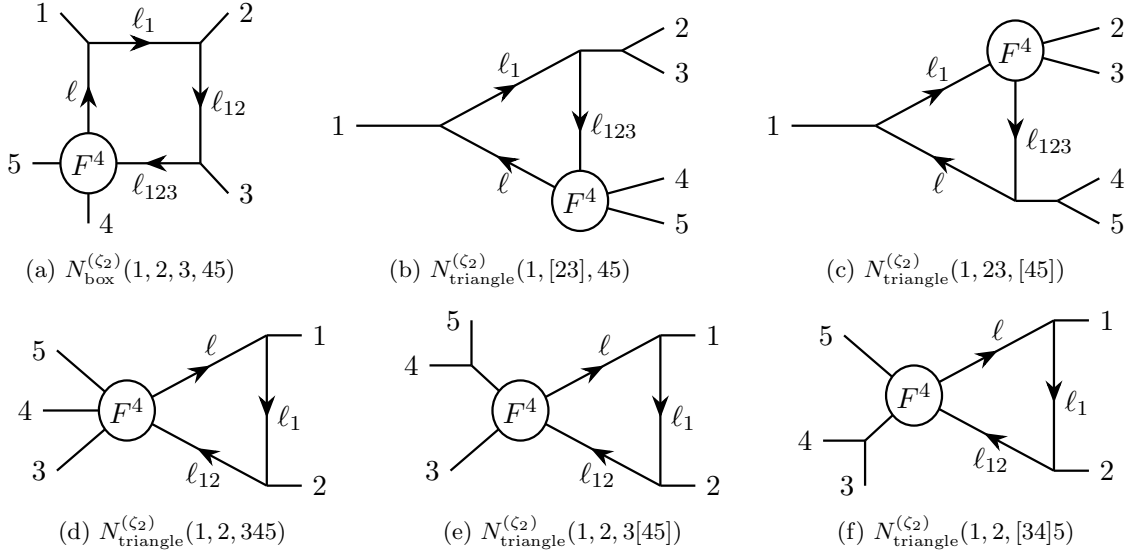
\begin{figure}
  \begin{center}
    \subfloat[$N_{\text{box}}^{(\zeta_2)}(1,2,3,45)$]{\begin{tikzpicture} [xscale=0.93, line width=0.30mm,scale=0.8]
    \draw [decoration={markings, 
      mark= at position 0.5 with \arrowreversed{Stealth[length=3mm]}},postaction={decorate}] (-1,-1)--(1,-1) node[pos=0.6,below=0pt]{$\ell_{123}$};
    \draw [decoration={markings, 
      mark= at position 0.55 with \arrow{Stealth[length=3mm]}},postaction={decorate}] (1,1)--(1,-1) node[pos=0.5,right=0pt]{$\ell_{12}$};
    \draw [decoration={markings, 
      mark= at position 0.43 with \arrowreversed{Stealth[length=3mm]}},postaction={decorate}] (1,1)--(-1,1) node[pos=0.5,above=0pt]{$\ell_1$};
    \draw [decoration={markings, 
      mark= at position 0.65 with \arrow{Stealth[length=3mm]}},postaction={decorate}] (-1,-1)--(-1,1) node[pos=0.6,left=0pt]{$\ell$};
    \draw(-1,1)--(-1.5,1.5)node[left]{$1$};
    \draw(1,1)--(1.5,1.5)node[right]{$2$};
    \draw(1,-1)--(1.5,-1.5)node[right]{$3$};
    \draw(-1,-1)--(-2,-1)node[left]{$5$};
    \draw(-1,-1)--(-1,-2)node[right]{$4$};
    \draw [fill=white] (-1,-1)circle (0.5cm);
    \draw(-1,-1)node{$F^4$};
    \end{tikzpicture}}\quad \quad
    \subfloat[$N_{\text{triangle}}^{(\zeta_2)}(1,{[23]},45)$]{\begin{tikzpicture} [xscale=0.93, line width=0.30mm]
    \draw [decoration={markings, 
      mark= at position 0.55 with \arrow{Stealth[length=3mm]}},postaction={decorate}] (0,0)--(2,1) node[pos=0.5,above=0pt]{$\ell_1$};
    \draw [decoration={markings, 
      mark= at position 0.4 with \arrowreversed{Stealth[length=3mm]}},postaction={decorate}] (0,0)--(2,-1) node[pos=0.45,below=1pt]{$\ell$};
    \draw [decoration={markings, 
      mark= at position 0.45 with \arrowreversed{Stealth[length=3mm]}},postaction={decorate}] (2,-1)--(2,1) node[pos=0.5,right=0pt]{$\ell_{123}$};
    \draw(0,0)--(-1.2,0)node[left]{1};
    \draw(2,-1)--(3.2,-1.3)node[right]{$5$};
    \draw(2,-1)--(3.2,-0.7)node[right]{$4$};
    \draw(2,1)--(2.6,1);
    \draw(2.6,1)--(3.2,0.7)node[right]{$3$};
    \draw(2.6,1)--(3.2,1.3)node[right]{$2$};
    \draw [fill=white] (2,-1)circle (0.4cm);
    \draw(2,-1)node{$F^4$};
    \end{tikzpicture}}\quad  \quad
    \subfloat[$N_{\text{triangle}}^{(\zeta_2)}(1,23,{[45]})$]{\begin{tikzpicture} [xscale=0.93, line width=0.30mm]
        \draw [decoration={markings, 
      mark= at position 0.5 with \arrow{Stealth[length=3mm]}},postaction={decorate}] (0,0)--(2,1) node[pos=0.45,above=0pt]{$\ell_1$};
    \draw [decoration={markings, 
      mark= at position 0.4 with \arrowreversed{Stealth[length=3mm]}},postaction={decorate}] (0,0)--(2,-1) node[pos=0.45,below=1pt]{$\ell$};
    \draw [decoration={markings, 
      mark= at position 0.4 with \arrowreversed{Stealth[length=3mm]}},postaction={decorate}] (2,-1)--(2,1) node[pos=0.4,right=0pt]{$\ell_{123}$};
    \draw(0,0)--(-1.2,0)node[left]{1};
    \draw(2,1)--(3.2,1.3)node[right]{$2$};
    \draw(2,1)--(3.2,0.7)node[right]{$3$};
    \draw(2,-1)--(2.6,-1);
    \draw(2.6,-1)--(3.2,-0.7)node[right]{$4$};
    \draw(2.6,-1)--(3.2,-1.3)node[right]{$5$};
    \draw [fill=white] (2,1)circle (0.4cm);
    \draw(2,1)node{$F^4$};
    \end{tikzpicture}} \\
  \subfloat[$N_{\text{triangle}}^{(\zeta_2)}(1,2,345)$]{\begin{tikzpicture} [xscale=0.93, line width=0.30mm]
      \draw [decoration={markings, 
        mark= at position 0.6 with \arrow{Stealth[length=3mm]}},postaction={decorate}] (0,0)--(2,1) node[pos=0.6,above=0pt]{$\ell$};
      \draw [decoration={markings, 
        mark= at position 0.5 with \arrowreversed{Stealth[length=3mm]}},postaction={decorate}] (0,0)--(2,-1) node[pos=0.55,below=0pt]{$\ell_{12}$};
      \draw [decoration={markings, 
        mark= at position 0.45 with \arrowreversed{Stealth[length=3mm]}},postaction={decorate}] (2,-1)--(2,1) node[pos=0.45,right=0pt]{$\ell_1$};
    \draw(2,1)--(2.5,1)node[right]{$1$};
    \draw(2,-1)--(2.5,-1)node[right]{$2$};
    \draw(0,0)--(-1,-0.8)node[left]{$3$};
    \draw(0,0)--(-1.2,0)node[left]{$4$};
    \draw(0,0)--(-1,0.8)node[left]{$5$};
    \draw [fill=white] (0,0)circle (0.4cm);
    \draw(0,0)node{$F^4$};
  \end{tikzpicture}}\quad \quad
    \subfloat[$N_{\text{triangle}}^{(\zeta_2)}(1,2,3{[45]})$]{\begin{tikzpicture} [xscale=0.93, line width=0.30mm]    
    \draw [decoration={markings, 
      mark= at position 0.6 with \arrow{Stealth[length=3mm]}},postaction={decorate}] (0,0)--(2,1) node[pos=0.6,above=0pt]{$\ell$};
    \draw [decoration={markings, 
      mark= at position 0.5 with \arrowreversed{Stealth[length=3mm]}},postaction={decorate}] (0,0)--(2,-1) node[pos=0.55,below=0pt]{$\ell_{12}$};
    \draw [decoration={markings, 
      mark= at position 0.45 with \arrowreversed{Stealth[length=3mm]}},postaction={decorate}] (2,-1)--(2,1) node[pos=0.45,right=0pt]{$\ell_1$};
    \draw(2,1)--(2.5,1)node[right]{$1$};
    \draw(2,-1)--(2.5,-1)node[right]{$2$};
    \draw(0,0) -- (-1,-0.8)node[left]{$3$};
    \draw(0,0) -- (-0.7,0.6);
    \draw(-0.7,0.6)--(-0.7,1.2)node[left]{$5$};
    \draw(-0.7,0.6)--(-1.3,0.6)node[left]{$4$};
    \draw [fill=white] (0,0)circle (0.4cm);
    \draw(0,0)node{$F^4$};
    \end{tikzpicture}}\quad \quad
    \subfloat[$N_{\text{triangle}}^{(\zeta_2)}(1,2,{[34]}5)$]{\begin{tikzpicture} [xscale=0.93, line width=0.30mm]    
        \draw [decoration={markings, 
      mark= at position 0.6 with \arrow{Stealth[length=3mm]}},postaction={decorate}] (0,0)--(2,1) node[pos=0.6,above=0pt]{$\ell$};
    \draw [decoration={markings, 
      mark= at position 0.5 with \arrowreversed{Stealth[length=3mm]}},postaction={decorate}] (0,0)--(2,-1) node[pos=0.55,below=0pt]{$\ell_{12}$};
    \draw [decoration={markings, 
      mark= at position 0.45 with \arrowreversed{Stealth[length=3mm]}},postaction={decorate}] (2,-1)--(2,1) node[pos=0.45,right=0pt]{$\ell_1$};
    \draw(2,1)--(2.5,1)node[right]{$1$};
    \draw(2,-1)--(2.5,-1)node[right]{$2$};
    \draw(0,0) -- (-1,0.8)node[left]{$5$};
    \draw(0,0) -- (-0.7,-0.6);
    \draw(-0.7,-0.6)--(-0.7,-1.2)node[left]{$3$};
    \draw(-0.7,-0.6)--(-1.3,-0.6)node[left]{$4$};
    \draw [fill=white] (0,0)circle (0.4cm);
    \draw(0,0)node{$F^4$};
    \end{tikzpicture}}
  \end{center}
  \caption{Diagrammatic representation of the contribution to the
    $ \zeta_2$-order of \cref{t8defs.9} with numerators in
    \cref{t8defs.9a}. Other $\zeta$ orders will have the same
    topologies with different operator insertions in place of $F^4$.}
  \label{fivefig.1}
\end{figure}
\begin{subequations}\label{t8defs.9a}
\begin{align}
N^{(\zeta_2)}_{\text{box}}(1,2,3,45)&=-s_{45}N_{+|12345|-} +\frac{1}{2}(s_{4,\ell}-s_{5,\ell}-s_{45})t_8(45,1,2,3)\,, \\
N^{(\zeta_2)}_{\text{triangle}}(1,[23],45)&=s_{45}t_8(23,1,4,5)\,,\\
N^{(\zeta_2)}_{\text{triangle}}(1,23,[45])&=s_{23}t_8(45,1,2,3)\,,  \\
N^{(\zeta_2)}_{\text{triangle}}(1,2,345)&=
\frac{1}{2}t_8(34,1,2,5)- t_8(35,1,2,4)+\frac{1}{2}t_8(45,1,2,3)\,,  \\
N^{(\zeta_2)}_{\text{triangle}}(1,2,[34]5)&=(s_{35}+s_{45})t_8(34,1,2,5)\,,\\
N^{(\zeta_2)}_{\text{triangle}}(1,2,3[45])&=(s_{34}+s_{35})t_8(45,1,2,3)\,,
\end{align}
\end{subequations}
comprise triangle and box diagrams with $F^4$-insertions depicted in Figure~\ref{fivefig.1}.
At the order of $\ap^3 \zeta_3 (D^2F^4+F^5)$, we have the same types of diagrams as
in (\ref{t8defs.9a}) with $D^2F^4$ in the place of $F^4$, 
\begin{subequations}\label{t8defs.11}
\begin{align}
N^{(\zeta_3)}_{\text{box}}(1,2,3,45)&=\frac{1}{2}s_{45}(s_{4,\ell}-s_{5,\ell}+s_{45})N_{+|12345|-} \nonumber\\
&\quad -\frac{1}{2}\Big[s_{4,\ell}^2+s_{5,\ell}^2-s_{45}(s_{4,\ell}+s_{5,\ell})\Big]t_8(45,1,2,3)\,, \\
N^{(\zeta_3)}_{\text{triangle}}(1,[23],45)&=-\frac{1}{2}s_{45}(s_{4,\ell}-s_{5,\ell}+s_{45})t_8(23,1,4,5)\,,\\
N^{(\zeta_3)}_{\text{triangle}}(1,23,[45])&=-\frac{1}{2}s_{23}(s_{2,\ell}-s_{3,\ell}+s_{12}+s_{23}-s_{13})t_8(45,1,2,3)\,, \\
N^{(\zeta_3)}_{\text{triangle}}(1,2,345)&=
\frac{1}{2} (s_{34}+s_{45}-2s_{35})N_{+|12345|-} \nonumber \\
&\quad+
\frac{1}{2}(2s_{4,\ell}+s_{5,\ell}-s_{34}-5s_{35}-3s_{45})t_8(34,1,2,5)\nonumber\\
&\quad-
\frac{1}{2}(s_{3,\ell}+2s_{4,\ell}-s_{45}+4s_{35})t_8(45,1,2,3)\nonumber\\
&\quad+
(s_{3,\ell}-s_{5,\ell}+s_{34}+2s_{45})t_8(35,1,2,4)\,, \\
N^{(\zeta_3)}_{\text{triangle}}(1,2,[34]5)&=-\frac{1}{2}(s_{35}+s_{45})(s_{3,\ell}+s_{4,\ell}-s_{5,\ell}+s_{12}-3s_{34})t_8(34,1,2,5)\,,\\
N^{(\zeta_3)}_{\text{triangle}}(1,2,3[45])&=-\frac{1}{2}(s_{34}+s_{35})(s_{\ell,3}-s_{4,\ell}-s_{5,\ell}+s_{34}+s_{35})t_8(45,1,2,3)\,,
\end{align}
\end{subequations}
and additionally the following types of bubbles depicted in Figure~\ref{ffivebubbs},
\begin{subequations}\label{t8defs.12}
\begin{align}
N^{(\zeta_3)}_{\text{bubble}}(1,2345)&=
-\frac{1}{2}t_8(23,1,4,5)+ t_8(24,1,3,5)- t_8(25,1,3,4)\nonumber\\
&\quad + t_8(35,1,2,4)-\frac{1}{2}t_8(45,1,2,3) \,, \\
N^{(\zeta_3)}_{\text{bubble}}([12],345)&=-
\frac{1}{2} (s_{34}+s_{45}-2s_{35})t_8(12,3,4,5)\,,\\
N^{(\zeta_3)}_{\text{bubble}}(1,[23]45)&=
\frac{1}{2} (s_{14}+2s_{25}+2s_{35})t_8(23,1,4,5) \,, \\
N^{(\zeta_3)}_{\text{bubble}}(1,2[34]5)&=\frac{1}{2} (s_{34}+3s_{25})t_8(34,1,2,5)\,,  \\
N^{(\zeta_3)}_{\text{bubble}}(1,23[45])&=
\frac{1}{2} (s_{13}+2s_{24}+2s_{25})t_8(45,1,2,3)\,.
\end{align}
\end{subequations}
The quintic vertex in Figure~\ref{fivefig.1} is due to the non-linear extensions of the field strengths in 
$\alpha'^2 \zeta_2 F^4$, and the vertices in Figure~\ref{ffivebubbs} of valence $>4$ can be attributed 
to both $\alpha'^3 \zeta_3D^2 F^4$ and $\alpha'^3 \zeta_3 F^5$.

The order of $\ap^4\zeta_2^2$ is again the onset of tadpoles and diagrams with double-insertions 
of higher-mass-dimension operators, see (\ref{exsec.10}) and Figure~\ref{ffourbub} for their four-point instances. 
At five points, particularly characteristic numerators in (\ref{t8defs.11}) are
\begin{subequations}\label{t8defs.13}
\begin{align}
N^{(\zeta_2^2)}_{\text{bubble}}(12,345)&=-\frac{1}{2}s_{12}\Big[t_8(34,1,2,5)-2t_8(35,1,2,4)+t_8(45,1,2,3)\Big]\,, \\
N^{(\zeta_2^2)}_{\text{bubble}}(12,[34]5)&=-
s_{12}(s_{12}-s_{34})t_8(34,1,2,5)\,,\\
N^{(\zeta_2^2)}_{\text{bubble}}(12,3[45])&=
s_{12}(s_{13}+s_{23})t_8(45,1,2,3)\,, \\
N^{(\zeta_2^2)}_{\text{tadpole}}(12345)&=\frac{12}{5}
\left[
\frac{s_{34}+s_{45}-s_{35}}{s_{12}}
t_8(12,3,4,5)-t_8(13,2,4,5)+\text{cyc}(1,2,3,4,5)\right],
\end{align}
\end{subequations}
see Figure~\ref{ffivebub} for the associated diagrams. The remaining numerators contributing to this order can
be found in appendix \ref{app:5ptzeta4}.

As a drawback of the tadpole numerator (\ref{t8defs.13}), gauge invariance of the $\alpha'^4$-order
of (\ref{t8defs.9}) can only be seen after shifts of loop momentum. One can attain gauge invariance 
pointwise in $\ell$ by employing the more lengthy alternative expression,
\begin{align}
  & N^{(\zeta_2^2)}_{\text{tadpole}}(12345)\rightarrow 2t_8(12,3,4,5)-3t_8(13,2,4,5)+3t_8(14,2,3,5)-2t_8(15,2,3,4) \nonumber\\
  &\quad +3t_8(23,1,4,5)+3t_8(25,1,3,4)+3t_8(34,1,2,5)-3t_8(35,1,2,4)+2t_8(45,1,2,3) \nonumber\\
  &\quad  - 6
\Big[\frac{s_{35}}{s_{12}}t_8(12,3,4,5)+\frac{s_{14}}{s_{23}}t_8(23,1,4,5)+\frac{s_{25}}{s_{34}}t_8(34,1,2,5) +\frac{s_{13}}{s_{45}}t_8(45,1,2,3)\Big]\,.
\label{alttadpole}
\end{align}

\begin{figure}
\begin{center}
\subfloat[$N_{\text{bubble}}^{(\zeta_2^2)}(12,345)$]{\begin{tikzpicture} [scale=1.05, line width=0.30mm]
\draw(0,0) .. controls (0.5,1) and (1.5,1) .. (2,0);
\draw(0,0) .. controls (0.5,-1) and (1.5,-1) .. (2,0);
\path[decoration={markings,mark=at position 1 with \arrow{Stealth[length=3mm]}}, decorate] (1.1,0.75) node[above=1pt]{$\ell_{12}$};
\path[decoration={markings,mark=at position 1 with \arrowreversed{Stealth[length=3mm]}}, decorate] (0.9,-0.75) node[below=1pt]{$\ell$};
\draw(0,0)--(-1,-0.5)node[left]{$1$};
\draw(0,0)--(-1,0.5)node[left]{$2$};
\draw [fill=white] (0,0)circle (0.4cm);
\draw(0,0)node{$F^4$};
\draw(2,0)--(3,-0.8)node[right]{$5$};
\draw(2,0)--(3.2,-0)node[right]{$4$};
\draw(2,0)--(3,0.8)node[right]{$3$};
\draw [fill=white] (2,0)circle (0.4cm);
\draw(2,0)node{$F^4$};
\end{tikzpicture}}\quad\quad\quad
\subfloat[$N_{\text{bubble}}^{(\zeta_2^2)}(12,{[34]}5)$]{\begin{tikzpicture} [scale=1.05, line width=0.30mm]
\draw(0,0) .. controls (0.5,1) and (1.5,1) .. (2,0);
\draw(0,0) .. controls (0.5,-1) and (1.5,-1) .. (2,0);
\path[decoration={markings,mark=at position 1 with \arrow{Stealth[length=3mm]}}, decorate] (1.1,0.75) node[above=1pt]{$\ell_{12}$};
\path[decoration={markings,mark=at position 1 with \arrowreversed{Stealth[length=3mm]}}, decorate] (0.9,-0.75) node[below=1pt]{$\ell$};
\draw(0,0)--(-1,-0.5)node[left]{$1$};
\draw(0,0)--(-1,0.5)node[left]{$2$};
\draw [fill=white] (0,0)circle (0.4cm);
\draw(0,0)node{$F^4$};
\draw(2,0)--(3,-0.8)node[right]{$5$};
\draw(2,0)--(2.7,0.6);
\draw(2.7,0.6)--(3.3,0.6)node[right]{$4$};
\draw(2.7,0.6)--(2.7,1.2)node[right]{$3$};
\draw [fill=white] (2,0)circle (0.4cm);
\draw(2,0)node{$F^4$};
\end{tikzpicture}} \\
\subfloat[$N_{\text{bubble}}^{(\zeta_2^2)}(12,3{[45]})$]{\begin{tikzpicture} [scale=1.05, line width=0.30mm]
\draw(0,0) .. controls (0.5,1) and (1.5,1) .. (2,0);
\draw(0,0) .. controls (0.5,-1) and (1.5,-1) .. (2,0);
\path[decoration={markings,mark=at position 1 with \arrow{Stealth[length=3mm]}}, decorate] (1.1,0.75) node[above=1pt]{$\ell_{12}$};
\path[decoration={markings,mark=at position 1 with \arrowreversed{Stealth[length=3mm]}}, decorate] (0.9,-0.75) node[below=1pt]{$\ell$};
\draw(0,0)--(-1,-0.5)node[left]{$1$};
\draw(0,0)--(-1,0.5)node[left]{$2$};
\draw [fill=white] (0,0)circle (0.4cm);
\draw(0,0)node{$F^4$};
\draw(2,0)--(3,0.8)node[right]{$3$};
\draw(2,0)--(2.7,-0.6);
\draw(2.7,-0.6)--(3.3,-0.6)node[right]{$4$};
\draw(2.7,-0.6)--(2.7,-1.2)node[right]{$5$};
\draw [fill=white] (2,0)circle (0.4cm);
\draw(2,0)node{$F^4$};
\end{tikzpicture}}\quad\quad\quad
\subfloat[$N_{\text{tadpole}}^{(\zeta_2^2)}(12345)$]{\begin{tikzpicture} [scale=1.05, line width=0.30mm]
    \draw [opacity=0] (-1,-0.5) node[left]{$1$} -- (3.3,-0.6) node[right]{$4$};
    \begin{scope}[xshift=1cm]
\draw(0,0)--(-1,0.3)node[left]{$5$};
\draw(0,0)--(-0.8,-0.6)node[left]{$4$};
\draw(0,0)--(1,0.3)node[right]{$1$};
\draw(0,0)--(0.8,-0.6)node[right]{$2$};
\draw(0,0)--(0,-1)node[below]{$3$};
 \draw (0,0)  .. controls (-1.4,1.5) and (1.4,1.5) ..  (0,0);
 \path[decoration={markings,mark=at position 1 with \arrow{Stealth[length=3mm]}}, decorate] (0.1,1.12) node[above=1pt]{$\ell$};
\draw [fill=white] (0,0)circle (0.4cm);
\draw(0,0)node{$F^6$};
\end{scope}
\end{tikzpicture}}
\end{center}
\caption{Diagrammatic representation of the two-mass bubbles and the tadpole diagrams at the
$\alpha'^4$ order of the five-point one-loop matrix element.}
\label{ffivebub}
\end{figure}
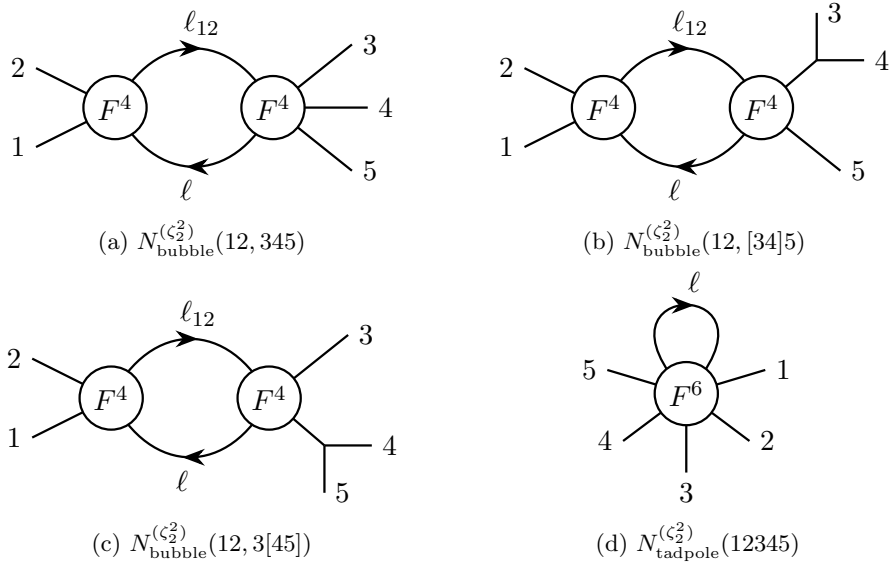


\subsection{Five points, closed string}
\label{sec:4.79}

For the one-loop matrix elements (\ref{prop.2}) of closed strings, the
ordering-dependence of the five-point numerators in (\ref{eq5.5}) does not
allow for a straightforward uplift of the four-point relation~(\ref{exsec.32}) to
single-valued open-string matrix elements (though $n$-point relations of this type 
are available in (\ref{prop.13}) at the level of linearized propagators).
Instead, we insert the five-point numerators into (\ref{prop.2}) and
assemble the $\ap$-expansion of the sphere integrals from
single-valued seven-point disk integrals. At the leading orders in
$\alpha'$, we have
\begin{align}
M^{\te{1-loop}}_{5,{\rm eff}}&=M^{\te{1-loop}}_{5,{\rm SUGRA}}+2 \int \dd^D \ell \, \bigg[\frac{1}{4}\frac{{\cal N}_{\text{box}}(1,2,3,45)}{\ell^2(\ell{+}k_1)^2(\ell{+}k_{12})^2(\ell{+}k_{123})^2}+\frac{1}{12}\frac{{\cal N}_{\text{triangle}}(1,2,345)}{\ell^2(\ell{+}k_1)^2(\ell{+}k_{12})^2}\notag \\
&\quad \!+\frac{1}{4}\frac{{\cal N}_{\text{triangle}}(1,2,[34]5)}{\ell^2(\ell{+}k_1)^2(\ell{+}k_{12})^2 s_{34}}+\frac{1}{4}\frac{{\cal N}_{\text{triangle}}(1,23,[45])}{\ell^2(\ell{+}k_1)^2(\ell{+}k_{123})^2s_{45}}+\frac{1}{48}\frac{\mathcal{N}_{\text{bubble}}(1,2345)}{\ell^2(\ell{+}k_1)^2}\nonumber\\
&\quad  \!+\frac{1}{8}\frac{\mathcal{N}_{\text{bubble}}(1,[23]45)}{\ell^2(\ell{+}k_1)^2s_{23}}+\frac{1}{24}\frac{\mathcal{N}_{\text{bubble}}([12],345)}{\ell^2(\ell{+}k_1)^2s_{12}}+\text{perm}(1{,}2{,}3{,}4{,}5)\bigg] + {\cal O}(\ap^6)\,.
\label{t8defs.21}
\end{align}
In this work, we have computed the numerators up to $\ap^5\zeta_5$ order,
\begin{align}
\mathcal{N}_{\Gamma}&=\ap^3 \zeta_3 \mathcal{N}^{(\zeta_3)}+\ap^5 \zeta_5 \mathcal{N}^{(\zeta_5)}\,.
\end{align}
The loop integrand $M^{\te{1-loop}}_{5,{\rm SUGRA}}$ of supergravity\footnote{The explicit form of $M^{\te{1-loop}}_{5,{\rm SUGRA}}$ can for instance be obtained from a cubic-graph expansion with the double-copy of the SYM numerators in (\ref{t8defs.7}) \cite{loopBCJ, Carrasco:2011mn, Mafra:2014gja} which follows from the ambitwistor-string formula (\ref{rev.1}) with half integrands ${\cal I}_L,{\cal I}_R \rightarrow {\cal I}_{5}$.} is followed by box- and triangle diagrams with a single insertion of $\alpha'^3 \zeta_3 R^4$ (in the place of $\alpha'^2 \zeta_2 F^4$ in Figure~\ref{fivefig.1}) and numerators 
\begin{subequations}\label{t8defs.22}
\begin{align}
{\cal N}_{\text{box}}^{(\zeta_3)}(1,2,3,45)&=s_{45}^2\big(\left|N_{+|12345|-}\right|^2+\left|N_{+|12354|-} \right|^2\big)\nonumber\\*
&\quad +(s_{\ell,4}-s_{\ell,5})s_{45}\big(\left|N_{+|12345|-} \right|^2-\left|N_{+|12354|-} \right|^2\big)\nonumber \\*
&\quad +\left[s_{\ell,4}^2+s_{\ell,5}^2-(s_{\ell,4}+s_{\ell,5})s_{45}\right]\left|t_8(45,1,2,3)\right|^2\,,\label{only1st} \\
{\cal N}_{\text{triangle}}^{(\zeta_3)}(1,2,345)&=\Big[
(s_{12}+3s_{35})\big(t_8(34,1,2,5)\bar{t}_8(45,1,2,3)+ (t_8 \leftrightarrow \bar t_8)\big)  \nonumber \\
&\quad
-3(s_{15}+s_{25})\left|t_8(34,1,2,5)\right|^2+\text{cyc}(3,4,5)\Big]\,, \\
{\cal N}_{\text{triangle}}^{(\zeta_3)}(1,2,[34]5)&=2(s_{35}+s_{45})(2s_{35}+2s_{45}-s_{12})\left|t_8(34,1,2,5)\right|^2\,, \\
{\cal N}_{\text{triangle}}^{(\zeta_3)}(1,23,[45])&=2s_{23}^2\left|t_8(45,1,2,3)\right|^2 \,,
\end{align}
\end{subequations}
and $\mathcal{N}_{\text{bubble}}^{(\zeta_3)}=0$ at this order. All the six topologies in \cref{t8defs.21} show up at $\ap^5\zeta_5$ order. The expressions for these numerators are lengthy and they are provided in an ancillary file.

\subsubsection{Type IIA versus type IIB}

Note that $M^{\te{1-loop}}_{5,{\rm SUGRA}}$ and (\ref{only1st}) feature
bilinears in the vector building block $t_{\pm}^\mu$ in (\ref{t8defs}) that 
differ between the type-IIA and IIB theories: the relative sign of the parity-odd
terms need to be chosen as $ t_{+}^\mu\bar t_{+}^\nu$ for the chiral type-IIB theory 
and as $ t_{+}^\mu\bar t_{-}^\nu$ in the non-chiral type-IIA theory. Accordingly, the
one-loop matrix elements in the two theories differ by
\begin{align} 
 M^{\te{1-loop}}_{5,{\rm eff}} \, \Big|_{\rm IIA} &=  M^{\te{1-loop}}_{5,{\rm eff}} \, \Big|_{\rm IIB}
 - \frac{1}{8}
 \int \frac{ \dd^D \ell}{\ell^2}  \, \ell_\mu \varepsilon_{10}^\mu(\bar \ep_1,\bar f_2,\bar f_3,\bar f_4,\bar f_5) 
 \label{IIBvsIIA} \\
 & \quad \times \lim_{k_{\pm} \rightarrow \pm \ell}  \sum_{\gamma,\rho \in S_5} 
 J\big(+,\gamma(1,2,3,4,5),-|+,\rho(1,2,3,4,5),-\big) N_{+|\rho(12345) |-}(\ell) \, ,\notag
  \end{align}
see (\ref{eq5.5}) for the numerators $N_{+|\rho(12345) |-}(\ell)$.
When the external states are chosen to be five gravitons $h^5$ 
with $\bar \ep_j \rightarrow \ep_j$ (rather than $B$-fields
or dilatons), the general formula (\ref{IIBvsIIA}) for the difference specializes as follows
\begin{align}
 M^{\te{1-loop}}_{5,{\rm eff}} \, \Big|^{h^5}_{\rm IIA} =  M^{\te{1-loop}}_{5,{\rm eff}} \, \Big|^{h^5}_{\rm IIB} &
 - \frac{1}{128}
 \int \frac{ \dd^D \ell}{\ell^2}\, \ell_\mu \varepsilon_{10}^\mu(\ep_1,f_2,f_3,f_4,f_5) \ell_\nu  \varepsilon_{10}^\nu(\ep_1, f_2, f_3,  f_4,  f_5)  \\
 &\quad\times
 \lim_{k_{\pm} \rightarrow \pm \ell} 
\sum_{\gamma, \rho \in S_5} J(+,\gamma(1,2,3,4,5),-|+,\rho(1,2,3,4,5),-) \, ,  \notag
\end{align}
where only the tensor structure $\ell_\mu \ell_\nu \sim \eta_{\mu \nu}$ contributes to the loop
integral. 

\subsubsection{Higher orders in $\alpha'$}

At the orders of $\alpha'^5 \zeta_5(D^4R^4+D^2 R^5+R^6)$ 
and beyond, expressions similar to (\ref{t8defs.22}) can be obtained from (\ref{prop.2}).
However, the computational effort grows drastically with the number of terms in the $\alpha'$-expansion
of sphere integrals which need to be combined to quadratic propagators as in (\ref{t8defs.21}). 
As will be detailed below, the reorganization (\ref{openEP.3}) of the five-point one-loop matrix element 
in terms of certain polarization-independent integrals ``$\langle E | E \rangle$'' with 
quadratic-propagator representations furnishes an 
improved starting point to obtain explicit results at higher orders in $\ap$.
By analogy with the four-point results in (\ref{exsec.36}) and (\ref{exsec.37}), the onset of 
external and two-mass bubble diagrams is expected at the orders of $\ap^5\zeta_5$ 
and $\ap^6 \zeta_3^2$, respectively.


\section{One-loop matrix elements and quadratic propagators from superstrings}
\label{stringysec}

In this section, we relate the one-loop matrix elements to superstring amplitudes and reorganize
the transition from linearized to quadratic propagators according to the homology invariance
of worldsheet correlators on the torus.


\subsection{Motivation from chiral splitting}
\label{sec:3.2}

In earlier sections, the proposal (\ref{prop.1}) and (\ref{prop.2}) for one-loop matrix elements of higher-mass-dimension
operators has been motivated from the ambitwistor-string perspective. We shall now provide a complementary
motivation from certain limits of one-loop amplitudes of conventional strings based on the chiral-splitting
formalism \cite{DHoker:1988pdl, DHoker:1989cxq}. 

The key idea of chiral splitting is to identify loop momenta as 
the joint zero modes of the worldsheet fields $\partial_z X^\mu$ and $\partial_{\bar z} X^\mu$ in
the left- and right-moving sector. By separating the loop-momentum integral
from the rest of the path integral that defines the string integrand, the contributions of the left- and 
right movers can be kept in a factorized form.
In particular, the evaluation and simplification of the correlation function of vertex operators that
determines the basis of kinematic factors can be performed separately for the two chiral halves
of the string amplitude. Accordingly, chiral splitting is the driving force in the recent construction
of the correlators in multiparticle one-loop \cite{Mafra:2018qqe}, five-point 
two-loop \cite{DHoker:2020prr} and three-loop four-point \cite{Geyer:2021oox} string amplitudes.

In the chiral-splitting approach to one-loop closed-string amplitudes, the integrand of\footnote{For ease of notation, the chiral-splitting representation (\ref{prop.41}) is adapted to the case of
real invariants $\ell^2$, $\ell\cdot k_i$ and $k_i\cdot k_j$. One can still accommodate complex kinematics
by employing $-k_j^\ast$, $-\ell^\ast$ in the place of $k_j$, $\ell$ in one of the chiral halves 
before complex conjugation of ${\cal K}_n(\ell,\tau) \, \KN_n(\ell,\tau)$. In this way, the real parts
of $\tau$ and $z_j$ drop out from the first and second term in the exponent of (\ref{prop.42}) when
combining the left- and right-movers.}
\beq
M^{\te{1-loop}}_{n,{\rm closed}} = \int \dd^D \ell \int_{\mathfrak{F}} \dd^2 \tau \int_{\mathfrak{T}_\tau^{n-1}}     \dd^2 z_2 \ldots \dd^2 z_n \, \big|{\cal K}_n(\ell,\tau) \, \KN_n(\ell,\tau)  \big|^2
\label{prop.41}
\eeq
is the absolute-value square of the chiral amplitude ${\cal K}_n \KN_n$ which is a
meromorphic function of the moduli $z_j$ and $\tau$ with $z_1=0$ by translation invariance. 
We have organized the chiral amplitude into the Koba-Nielsen factor
\beq
\KN_n(\ell,\tau) = \exp\bigg(2\pi i \alpha' \tau \ell^2
+ 4 \pi i  \alpha'\sum_{j=1}^n (\ell\cdot k_j) z_j 
+ 2\alpha' \sum_{1\leq i<j}^n k_{i}\cdot k_j \log \theta_1(z_{ij},\tau)  \bigg)
\label{prop.42}
\eeq
and a kinematic factor ${\cal K}_n $ that carries all the dependence on the external polarization.
While (\ref{prop.42}) is the correlator of the universal plane-wave parts $e^{i k_j\cdot X}$ of vertex operators,
the Wick contractions of the remaining worldsheet fields determine ${\cal K}_n $ in terms of derivatives
of the odd Jacobi theta function\footnote{More precisely, the chiral correlators ${\cal K}_n $ for massless
states in type-II, heterotic and bosonic string theories are 
polynomials in loop momenta, holomorphic Eisenstein series 
and expansion coefficients of a meromorphic Kronecker--Eisenstein series \cite{Dolan:2007eh, Broedel:2014vla, Gerken:2018jrq}.}
\beq
\theta_1(z,\tau) = 2iq^{1 / 8} \sin(\pi z) \prod_{j=1}^{\infty} (1-q^j) (1- e^{2\pi i z}q^j) (1- e^{-2\pi i z}q^j)  = - \theta_1(-z,\tau)  \, , \ \ \ \ q= e^{2\pi i \tau}\, .
\label{prop.43}
\eeq
In fact, the maximally supersymmetric ${\cal K}_n $ are degree-$(n{-}4)$ polynomials in loop momenta,
see (\ref{prop.51}) below for a relation to the half integrand ${\cal I}_{n}(\ell)$ in (\ref{rev.6}),
(\ref{exsec.2}) and (\ref{eq5.5}).

Finally, the integration domains for the moduli in (\ref{prop.41}) are the fundamental domain $\mathfrak{F} $ of the modular group ${\rm SL}_2(\ZZ)$ and the torus worldsheet $\mathfrak{T}_\tau = \frac{ \mathbb C}{\mathbb Z + \tau \mathbb Z}$ in its parametrization as a parallelogram.

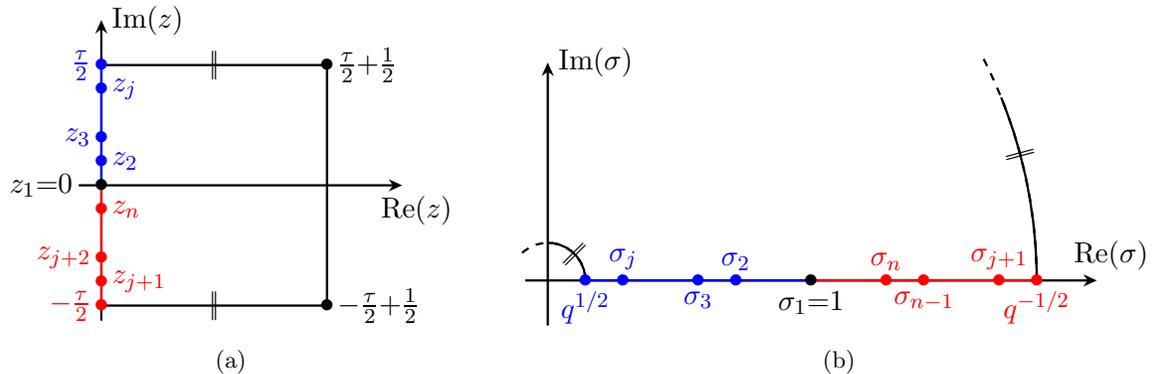
\begin{figure}
\begin{center}
\subfloat[]{\begin{tikzpicture}[line width=0.30mm]
  \draw [arrows={-Stealth[width=1.6mm, length=1.8mm]}] (-0.3,0) -- (4.0,0) node[below]{$\quad{\rm Re}(z)$};
  \draw [arrows={-Stealth[width=1.6mm, length=1.8mm]}] (0,-1.8) -- (0,2.2) node[right]{${\rm Im}(z)$};
\draw[red](0,-1.6) -- (0,0);
\draw[blue](0,0) -- (0,1.6);
\draw(3,-1.6) -- (3,1.6);
\draw(0,-1.6) -- (3,-1.6);
\draw(0,1.6) -- (3,1.6);
\draw(1.5,1.6)node{$| \! |$};
\draw(1.5,-1.6)node{$| \! |$};
\draw(0,0)node{$\bullet$};
\draw(-0.8,0)node{$z_1{=}0$};
\draw[blue](0,1.6)node{$\bullet$}node[left]{$\frac{\tau}{2}$};
\draw(3,1.6)node{$\bullet$}node[right]{$\frac{\tau}{2}{+}\frac{1}{2}$};
\draw(3,-1.6)node{$\bullet$}node[right]{${-}\frac{\tau}{2}{+}\frac{1}{2}$};
\draw[red](0,-1.6)node{$\bullet$}node[left]{${-}\frac{\tau}{2}$};
\draw[blue](0,0.32)node{$\bullet$}node[right]{$z_2$};
\draw[blue](0,0.64)node{$\bullet$}node[left]{$z_3$};
\draw[blue](0,1.28)node{$\bullet$}node[right]{$z_j$};
\draw[red](-0,-0.32)node{$\bullet$}node[right]{$z_{n}$};
\draw[red](-0,-0.96)node{$\bullet$}node[left]{$z_{j+2}$};
\draw[red](-0,-1.28)node{$\bullet$}node[right]{$z_{j+1}$};
\end{tikzpicture}\label{fig:bcyclparaa}}
\quad\quad
\subfloat[]{\begin{tikzpicture}[line width=0.30mm]
  \draw [arrows={-Stealth[width=1.6mm, length=1.8mm]}] (-0.3,0) -- (7.3,0) node[above]{$\quad{\rm Re}(\sigma)$};
  \draw [arrows={-Stealth[width=1.6mm, length=1.8mm]}] (0,-0.6) -- (0,2.9) node[right]{${\rm Im}(\sigma)$};
\draw[red] (6.5,0) -- (3.5,0);
\draw[blue] (0.5,0) -- (3.5,0);
\draw (0.5,0) arc (0:90:0.5cm);
\draw[dashed] (0.5,0) arc (0:135:0.5cm);
\draw (6.5,0) arc (0:22:6.5cm);
\draw[dashed] (6.5,0) arc (0:27:6.5cm);
\draw[rotate=45] (0.5,0)node[rotate=-45]{ $| \! |$};
\draw[rotate=15] (6.5,0)node[rotate=-75]{ $| \! |$};
  \draw[blue] (0.5,0)node{$\bullet$}  node[below]{$q^{1/2}$};
  \draw[blue] (1,0)node{$\bullet$}  node[above]{$\sigma_j$};
  \draw[blue] (2.0,0)node{$\bullet$}  node[below]{$\sigma_3$};
  \draw[blue] (2.5,0)node{$\bullet$}  node[above]{$\sigma_2$};
   \draw (3.5,0)node{$\bullet$}  node[below]{$\sigma_1{=}1$};
  \draw[red] (4.5,0)node{$\bullet$}  node[above]{$\sigma_n$};
  \draw[red] (5,0)node{$\bullet$}  node[below]{$\sigma_{n-1}$};
  \draw[red] (6,0)node{$\bullet$}  node[above]{$\sigma_{j+1}$};
  \draw[red] (6.5,0)node{$\bullet$}  node[below]{$q^{-1/2}$};
\end{tikzpicture}\label{fig:bcyclparab}}
\end{center}
\caption{Parametrization of the integration domain on the cylinder boundary in planar one-loop open-string amplitudes.}
\label{bcyclpara}
\end{figure}

The open-string counterpart of (\ref{prop.41}) in the chiral-splitting formalism is  
\beq
A^{\te{1-loop}}_{{\rm open}} (\gamma(1,2,\ldots,n)) = \int \dd^D \ell \int_{0}^{i\infty} \dd \tau   \int_{\mathfrak{C}_\tau(\gamma)}    \dd z_2 \ldots \dd z_n \, {\cal K}_n(\ell,\tau) \, |\KN_n(\ell, \tau)   |\, ,
\label{prop.46}
\eeq
where $z_1=0$ by translation invariance, the modular parameter $\tau$ is integrated over the positive 
imaginary axis instead of $\mathfrak{F}$, and the punctures $z_i=\tau u_i$ (with real $u_i$) 
are inserted on the cylinder boundary
\beq
\mathfrak{C}_\tau(\gamma)= \bigoplus_{j=1}^n \{   -\tfrac{1}{2} < u_{\gamma(j+1)}
< u_{\gamma(j+2)} < \ldots < u_{\gamma(n)} < 0 < u_{\gamma(2)} < u_{\gamma(3)}<\ldots < u_{\gamma(j)}< \tfrac{1}{2}
\} \, .
\label{prop.47}
\eeq
The integration domains in (\ref{prop.46}) are tailored to the amplitude contributions from planar 
cylinder diagrams and can be adapted to the non-planar cylinder or the M\"obius strip by shifting
$z_j \rightarrow z_j + \frac{1}{2}$ or $\tau \rightarrow \tau + \frac{1}{2}$, respectively \cite{Polchinski:1998rq}.
As visualized in Figure~\ref{fig:bcyclparaa}, the cylinder worldsheet in (\ref{prop.46}) 
and (\ref{prop.47}) arises from a rectangular torus whose homology cycle $\mathfrak{B}$ sets the cylinder boundary where all the state 
insertions occur. The second
cylinder boundary relevant for non-planar one-loop open-string amplitudes is displaced by half the
homology cycle $\mathfrak{A}$, i.e.\ by $z_j \rightarrow z_j + \frac{1}{2}$. As reflected by the appearance 
of negative $\Im (z_i)$ in (\ref{prop.47}), we have chosen the fundamental cell of the parental torus to be 
the rectangle with corners at $\pm \frac{\tau}{2}$ and $1\pm \frac{\tau}{2}$, and the cylinder covers its half 
bounded by $\pm \frac{\tau}{2}$ and $\frac{1}{2}\pm \frac{\tau}{2}$ seen in Figure~\ref{fig:bcyclparaa}.

\subsubsection{The $\tau \rightarrow i \infty$ degeneration}
\label{sec:3.3.1}

One-loop amplitudes of SYM and supergravity arise in the limit $\tau \rightarrow i \infty$
where the $\mathfrak{A}$-cycle pinches, and the torus degenerates to a nodal sphere. 
In other words, the field-theory limit $\ap \rightarrow 0$
is traditionally performed while keeping the dimensionful quantity $\ap \Im \tau$ finite and identifying it with
one of the Schwinger parameters of Feynman integrals 
\cite{Green:1982sw, Bern:1990cu, Bern:1990ux, Bern:1991aq, Bern:1991an, Bern:1993wt, Dunbar:1994bn, Schubert:2001he, Tourkine:2013rda}.

The $\tau \rightarrow i \infty$ limits of the constituents of the one-loop open-string amplitude (\ref{prop.46}) are most conveniently analyzed in terms of the coordinate $\sigma_j$ on the nodal sphere
\beq
 \sigma_j = e^{2\pi i z_j} \, , \ \ \ \ \ \ \dd z_j = \frac{ \dd \sigma_j }{2\pi i \sigma_j} \, , \ \ \ \ \ \ \dd^2 z_j = \frac{ \dd^2 \sigma_j }{4\pi^2 \sigma_j \bar \sigma_j} \, .
\label{rev.61}
\eeq
First, the cylinder ordering in the parametrization of (\ref{prop.47}) tends to a cyclic combination of disk
orderings (\ref{rev.33}) in the $\sigma_j$-variable, see Figure~\ref{fig:bcyclparab},
\beq
\lim_{\tau \rightarrow i \infty} \mathfrak{C}_{\tau}(1,2,\ldots,n) = \bigoplus_{j=1}^n
 \mathfrak{D}_j
\, , \ \ \ \ \ \  \mathfrak{D}_j= \mathfrak{D}(+,j,j{-}1,\ldots,2,1,n,n{-}1,\ldots,j{+}1,-)\, ,
\label{rev.62}
\eeq
where $\sigma_+=0$ and $\sigma_-\rightarrow \infty$ as in (\ref{rev.4}). Second, the chiral Koba-Nielsen factor
(\ref{prop.42}) turns into the forward limit of the tree-level one in (\ref{rev.31}) after removing the
scaling by $e^{2\pi i \alpha'  \tau \ell^2}$,
\beq
\lim_{\tau \rightarrow i \infty} e^{-2\pi i \alpha'  \tau \ell^2} \KN_n(\ell,\tau)
= \prod_{j=1}^n \sigma_{j+}^{2\ap \ell \cdot k_j} \prod_{1\leq i<j}^n |\sigma_{ij}|^{2\ap k_{i}\cdot k_{j}}\, ,
\label{rev.63}
\eeq
where we used $\sigma_1 =1$ and momentum conservation\footnote{More precisely, we start
from the degeneration $\log \theta_1(z_{ij},\tau) = \frac{i\pi \tau}{4} + \log( \frac{ \sigma_{ij} }{\sqrt{\sigma_i \sigma_j}}) + {\cal O}(e^{2\pi i\tau})$ and apply momentum conservation in the form of $\sum_{j\neq i}^n s_{ij}=0$ to cancel the contributions from $\frac{i\pi \tau}{4}$ and $-\frac{1}{2}(\log \sigma_i + \log \sigma_j)$ to
$\sum_{1\leq i<j}^n s_{ij} \log \theta_1(z_{ij},\tau) = \sum_{1\leq i<j}^n s_{ij} \log \sigma_{ij} + {\cal O}(e^{2\pi i\tau})$.}. 
Third, the polarization dependent part ${\cal K}_n$ of the chiral correlator yields the kinematic half integrand
of the ambitwistor string once its $z_{ij}$-dependence has been brought into logarithmic form via integration by parts \cite{He:2017spx},
\beq
{\cal I}_{n}(\ell) =  \lim_{\tau \rightarrow i \infty}   \frac{ {\cal K}_n(\ell, \tau) \, }{(2\pi i)^{n-4} \sigma_1 \sigma_2 
\ldots \sigma_n }  \, .
\label{prop.51}
\eeq
The inverse factors of $\sigma_j$ in (\ref{prop.51}) stem from the conversion of $\dd z_j$ into $\dd\sigma_j$ in (\ref{rev.61}). The four- and five-point instances of ${\cal I}_{n}(\ell)$ in (\ref{exsec.2}) and (\ref{eq5.5}) arise from
\begin{align}
{\cal K}_4(\ell, \tau) &=  t_8(f_1,f_2,f_3,f_4)\,, 
\label{cKexamples}
\\
{\cal K}_5(\ell, \tau) &= 2\pi i \ell_\mu t_+^\mu(1,2,3,4,5) + \big[ \partial_{z_1} \log \theta_1(z_{12},\tau) t_8({12},3,4,5)  + (1,2|1,2,3,4,5)\big]\,,
\notag
\end{align}
which can be checked by $\lim_{\tau \rightarrow i\infty}  \partial_{z_1} \log \theta_1(z_{12},\tau) = 2\pi i G_{12}$,
see (\ref{t8defs}) for the five-point kinematic factors and (\ref{t8defs.3}) for the nodal Green function $G_{12}$.
The $(n\leq 5)$-point examples in (\ref{cKexamples}) are inevitably in logarithmic form, while the analogous six-
and seven-point correlators \cite{Mafra:2018qqe} admit both logarithmic and non-logarithmic representations that
are related by integration-by-parts manipulations detailed in the reference. The vector building blocks in
the bilinear $|{\cal K}_5(\ell, \tau)|^2$ of the closed-string five-point amplitude (\ref{prop.41}) are $t_+^\mu \bar t_+^\nu$ for type IIB and $t_+^\mu \bar t_-^\nu$ for type IIA.

\subsubsection{\texorpdfstring{$\alpha'$}{a'}-expansions within the $\tau \rightarrow i\infty$ degeneration}
\label{sec:3.3.2}

We shall now argue that our proposal (\ref{prop.1}) and (\ref{prop.2}) for one-loop matrix
elements arises naturally when expanding the $\tau \rightarrow i\infty$ degeneration of one-loop
string amplitudes in $\alpha'$. In comparison to the full loop integrand of string amplitudes obtained
from performing the $z_j$- and $\tau$-integrals in (\ref{prop.41}) and (\ref{prop.46}), the
$\tau \rightarrow i\infty$ degeneration is insensitive to terms
without any massless propagators $\ell^{-2}$ or $(\ell + K)^{-2}$ which result in rational loop integrals. 
For open strings, the dominant contributions from the degeneration are most conveniently 
found by studying the $\tau$-integrand of (\ref{prop.46}) in terms of the quantity
\beq
F(\ell,\tau) = \int_{\mathfrak{C}_{\tau}(\gamma)}    \dd z_2 \ldots \dd z_n\, e^{-2\pi i \alpha'  \tau \ell^2} \,  \KN_n(\ell,\tau) \, {\cal K}_n(\ell,\tau)\, ,
\label{rev.64}
\eeq
and by inserting $0=  -F(\ell,i\infty) +F(\ell,i\infty)$ into
\begin{align}
A^{\te{1-loop}}_{{\rm open}} (1,2,\ldots,n) &= \int \dd^D \ell \int_{0}^{i\infty} \dd \tau  \, e^{2\pi i \alpha'  \tau \ell^2} F(\ell,\tau)  \label{rev.65} \\
&= \int \dd^D \ell \int_{0}^{i\infty} \dd \tau  \, e^{2\pi i \alpha'  \tau \ell^2} \Big\{ \underbrace{ F(\ell,\tau)  -F(\ell,i\infty) }_{(i)}+ \underbrace{ F(\ell,i\infty) }_{(ii)} \Big\} \, .
\notag
\end{align}
The factor of $e^{-2\pi i \alpha'  \tau \ell^2}$ in (\ref{rev.64}) is engineered to cancel the $\ell^2$-dependence of $\KN_n(\ell,\tau) $ in (\ref{rev.63}) and to ensure that $F(\ell,\tau)$ is independent on $\ell^2$.
The first part $(i)$ in (\ref{rev.65}) is suppressed at the cusp since $F(\ell,\tau)  -F(\ell,i\infty)$ vanishes
as $\tau \rightarrow i\infty$. In the second part $(ii)$, one can directly perform the $\tau$ integral
\beq
\int_{0}^{i\infty} \dd \tau  \, e^{2\pi i \alpha'  \tau \ell^2}  = \frac{i}{2\pi \alpha' \ell^2}
\label{rev.66}
\eeq
under the assumption that $\ell^2$ has a positive real part to drop the contribution from
the integration limit $\tau \rightarrow i \infty$. The accompanying limit $F(\ell,i\infty)$
can be simplified by means of the results in section \ref{sec:3.3.1}
\begin{align}
\lim_{\tau \rightarrow i\infty} F(\ell,\tau) &=  \frac{ 1}{(2\pi i)^{n-1}} \sum_{j=1}^n \int_{\mathfrak{D}_j} \frac{ \dd \sigma_2 \, \dd \sigma_3\ldots \dd \sigma_n  }{\sigma_2 \sigma_3 \ldots \sigma_n}
\lim_{\tau \rightarrow i\infty}  e^{-2\pi i \alpha'  \tau \ell^2} \, {\cal K}_n(\ell,\tau) \KN_n(\ell,\tau) \notag \\
&=  \frac{ 1}{(2\pi i)^{3}} \sum_{j=1}^n \int_{\mathfrak{D}_j} 
\dd \sigma_2 \, \ldots \, \dd \sigma_n \, {\cal I}_{n}(\ell) \prod_{j=1}^n \sigma_{j+}^{2\ap \ell \cdot k_j} \prod_{1\leq i<j}^n |\sigma_{ij}|^{2\ap k_{i}\cdot k_{j}}  \notag \\
&= \frac{ -1}{(2 \pi i)^{3} ( \alpha')^{n-1} }  \sum_{\rho \in S_n} N_{+|\rho(12\ldots n)|-}(\ell) \label{rev.67} \\
& \ \ \ \ \times \lim_{k_{\pm} \rightarrow \pm \ell}  \sum_{j=1}^n Z(+,j,j{-}1,\ldots,2,1,n,\ldots ,j{+}1,- | + ,\rho(1,2,\ldots,n),-)\notag \\
&= \frac{- 1}{(2 \pi i)^{3} ( \alpha')^{n-1} } \sum_{j=1}^n a_{\rm eff}(+,j,j{-}1,\ldots,2,1,n,\ldots ,j{+}1,- ) \, , \notag
\end{align}
based on the limits (\ref{rev.62}), (\ref{rev.63}) and (\ref{prop.51}).
The integration domains $\mathfrak{D}_j$ defined in (\ref{rev.62}) are the disk orderings 
of the $Z$-integrals identified in the third step, and we have expanded the kinematic
half integrand in terms of Parke-Taylor factors and $n$-gon numerators $N_{+|\rho|-}$
according to (\ref{rev.7}).

In summary, by combining (\ref{rev.65}), (\ref{rev.66}) and (\ref{rev.67}), the
part $(ii)$ of one-loop open-string amplitudes that carries the dominant contribution
at the cusp coincides with the proposal (\ref{prop.1}) for one-loop matrix elements,
\beq
A^{\te{1-loop}}_{{\rm open}} (1,2,\ldots,n) \, \big|_{(ii)}  
= \frac{A^{\te{1-loop}}_{\rm eff}(n,n{-}1,\ldots,2,1)}{16\pi^4 (\ap)^n}
=  \frac{A^{\te{1-loop}}_{\rm eff}(1,2,\ldots,n)}{16\pi^4 (-\ap)^n}
\, .
\label{rev.68}
\eeq
In particular, the $\alpha' \rightarrow 0$ limit of (\ref{rev.67}) reduces $Z \rightarrow m$
and extracts the representation (\ref{rev.21}) of SYM amplitudes from
conventional strings, complementing its earlier derivation from ambitwistor strings.

The other part
$(i)$ of the one-loop open-string amplitude in (\ref{rev.65}) does not
have the pole in $\ell^{2}$ which enters
$(ii)$ via (\ref{rev.66}). Since the quantity
$F(\ell,\tau)$ only depends on $\ell \cdot k_j$ but not on
$\ell^2$, there is no chance for
$A^{\te{1-loop}}_{{\rm open}} (1,2,\ldots,n) \,
\big|_{(i)}$ to have a massless quadratic propagator $(\ell +
K)^{-2}$. Upon loop integration, part
$(i)$ is therefore free of branch cuts in the Mandelstam invariants
corresponding to massless particles
and belongs to the analytic part of the one-loop open-string
amplitude.

A similar decomposition can be performed for the one-loop
closed-string amplitude (\ref{prop.41}). As detailed in appendix
\ref{app:closed}, the proposal (\ref{prop.2}) for gravitational
one-loop matrix elements can be motivated from the dominant
contributions at $\tau \rightarrow
i\infty$.  Note that the absolute normalization of the string
amplitudes (\ref{prop.41}) and (\ref{prop.46}) has not been tracked;
it can be determined from unitarity or S-duality of type-IIB
superstrings.

\subsubsection{Universal one-loop matrix elements from torus compactifications}
\label{sec:compact}

In preparation for one-loop string amplitudes in supersymmetry-preserving toroidal compactifications,
we have already adapted the loop integration measure $\dd^D \ell$ in (\ref{prop.41}) and (\ref{prop.46})
to a generic number $D$ of spacetime dimensions. Upon compactification on a $10{-}D$ torus,
open-string integrands additionally feature a $\tau$-dependent partition function to account for 
Kaluza-Klein modes, and closed-string integrands involve a similar partition function encoding
both Kaluza-Klein and winding modes \cite{Green:1982sw, Green:1987mn}. Both types of
partition functions reduce to a constant in the limit $\tau \rightarrow i\infty$ relevant to one-loop matrix
elements in (\ref{rev.68}) and (\ref{appcl.6}). By absorbing the $\tau \rightarrow i\infty$ degeneration
of the partition functions into the normalization, the loop integrands of the matrix elements
(\ref{prop.1}) and (\ref{prop.2}) turn out to be dimension agnostic under torus compactification
of the underlying string theory. This is consistent with the construction of the one-loop matrix
elements from forward limits in the massless states of tree amplitudes, without any Kaluza-Klein
or winding modes in the propagators.


\subsection{Quadratic propagators from homology invariance}
\label{sec:3.3}

The ambitwistor approach to SYM and supergravity amplitudes does
not manifest that the linearized propagators from the Parke-Taylor integrals
(\ref{rev.11}) eventually recombine to quadratic ones. In our proposal (\ref{prop.1})
and (\ref{prop.2}), the translation between
linearized and quadratic propagators
\cite{Gomez:2016cqb, Gomez:2017lhy, Gomez:2017cpe, Ahmadiniaz:2018nvr, Agerskov:2019ryp, Farrow:2020voh}
has to be performed order by order in the $\ap$-expansion of the $Z$- and $J$ integrals reviewed in section 
\ref{sec:2.3}. This procedure led to the four- and five-point results in section \ref{sec:4}, and there is no 
fundamental obstruction to higher-point extensions. Still, higher and higher orders in $\ap$ introduce a growing number of factors $\ell \cdot k_j$ into the numerators, so the efforts to find quadratic-propagator representations will grow accordingly by the shifts of loop momenta in the second step of (\ref{ngonPF}) or in (\ref{shiftnums}).

We shall now discuss a condition on amplitude building blocks to admit a quadratic-propagator representation. 
In particular, we pinpoint the smallest contributions to the half integrands ${\cal I}_{n}(\ell)$ which can integrate 
to quadratic propagators and thereby identify a particularly economic organization scheme to obtain them
from linearized propagators. On the one hand, having an integration measure $\dd \mu_{n+2}^{\rm tree}$ of 
forward-limit form in (\ref{rev.2}) is
tied to the degeneration $\tau \rightarrow i\infty$. On the other hand, the salient point for compatibility with
quadratic propagators will be the behaviour of the chiral integrands 
under translations $z_j \rightarrow z_j +\tau $ around the $\mathfrak{B}$-cycle which can only be meaningfully studied at finite $\tau$.

The torus integrals in one-loop closed-string amplitudes (\ref{prop.41}) are only well-defined for 
doubly-periodic integrands under $z_j \rightarrow z_j +1 $ and $z_j \rightarrow z_j +\tau $
for all of $j=1,2,\ldots,n$. By the integral over $\ell \in \mathbb R^{D}$,
it will be sufficient to attain double-periodicity up to shifts of the loop momentum. By momentum
conservation and the quasi-periodicity $\theta_1(z{+}\tau,\tau)=-e^{-i\pi \tau -2\pi i z} \theta_1(z,\tau)$,
one can easily check the chiral Koba-Nielsen factor (\ref{prop.42}) to exhibit the following $\mathfrak{A}$-
and $\mathfrak{B}$-cycle monodromies \cite{DHoker:1989cxq}
\begin{align}
{\cal J}_n (\ell,\tau)\big|_{z_j \rightarrow z_j + 1} &= e^{4\pi i \ap \ell \cdot k_j} {\cal J}_n (\ell,\tau)\,,
\notag \\
{\cal J}_n (\ell,\tau)\big|_{z_j \rightarrow z_j + \tau} &= {\cal J}_n (\ell{+}k_j,\tau)\, .
\label{prop.71}
\end{align}
The phases under the $\mathfrak{A}$-cycle shift $z_j \rightarrow z_j + 1$
drop out after combining the Koba-Nielsen factors (\ref{prop.42}) of both chiral halves,
so their contributions to the closed-string amplitude (\ref{prop.41}) have the 
required doubly-periodicity.\footnote{While the prescription (\ref{prop.41})
with $|{\cal J}_n (\ell,\tau)|^2$ in the integrand is tailored to real kinematic invariants 
$\ell^2,\ell\cdot k_j$ and $k_i\cdot k_j$,
one can reconcile homology invariance with complex kinematics by employing $-k_j^\ast,-\ell^\ast$
instead of $k_j, \ell$ in one of the chiral halves. Still, the loop integral in (\ref{prop.41}) only converges
for real kinematics after Wick rotating the zeroth component of both $\ell$ and $k_j$.}
In order to have the same doubly-periodicity properties for type-II and heterotic one-loop amplitudes,
the left- and right-moving chiral correlators ${\cal K}_n(\ell)$ in (\ref{prop.41}) need to separately satisfy \cite{DHoker:1989cxq}
\begin{align}
{\cal K}_n (\ell,\tau)\big|_{z_j \rightarrow z_j + 1} &= {\cal K}_n(\ell,\tau)\,,
\notag \\
{\cal K}_n (\ell,\tau)\big|_{z_j \rightarrow z_j + \tau} &= {\cal K}_n(\ell{+}k_j,\tau)\, .
\label{prop.72}
\end{align}
Functions of $z_j,\ell,k_j,\tau$ that obey (\ref{prop.72}) were dubbed {\it generalized elliptic integrands}
and systematically constructed in \cite{Mafra:2017ioj, Mafra:2018pll, Mafra:2018qqe}. We will attribute 
the shifts on both sides of (\ref{prop.72}) to the action of the homology group on $z_j,\ell$ and follow the terminology of \cite{DHoker:2020prr}:
\beq
E(z_j,\ell,\tau)=E(z_j{+}\tau,\ell {-}k_j,\tau)  \ \ \ \Longrightarrow \ \ \ E \ \text{is referred to as {\it homology invariant}}\, .
\label{prop.73}
\eeq
Clearly, elliptic (i.e.\ meromorphic and doubly-periodic) functions of $z_1,z_2,\ldots,z_n$ are automatically homology invariant.
Moreover, as becomes pressing for $(n\geq 6)$-point examples, the notion of homology invariance extends
to the situation where (\ref{prop.73}) holds up to total Koba-Nielsen derivatives in the $z_j$,\footnote{For instance,
the following quantity relevant to six-point correlators \cite{Mafra:2017ioj, Mafra:2018pll, Mafra:2018qqe} is 
homology invariant in the sense of (\ref{prop.73a}) but does not obey the stronger condition (\ref{prop.73}),
\[
E_{1|2|3,4,5,6} = - 2 s_{12} g^{(2)}(z_{12},\tau) + \bigg( 2\pi i \ell \cdot k_2 + \sum_{j=3}^6 s_{2j} \partial_{z_2} \log \theta_1(z_{2j},\tau) \bigg)\partial_{z_1} \log \theta_1(z_{12},\tau)\, .
\]
The first term features $g^{(2)}(z,\tau)= \frac{1}{2} ( \partial_{z} \log \theta_1(z,\tau))^2 -\frac{1}{2} \wp(z,\tau)$, a meromorphic Kronecker--Eisenstein coefficient ubiquitous to the
$(n\geq 6)$-point correlators in the references involving the Weierstra\ss{} function $ \wp(z,\tau)$.}
\begin{align}
&\int \bigg( \prod_{i=1}^n \dd z_i \bigg)  {\cal J}_n(\ell,\tau) E(z_j,\ell,\tau)= \int \bigg( \prod_{i=1}^n \dd z_i \bigg)  {\cal J}_n(\ell,\tau)E(z_j{+}\tau,\ell {-}k_j,\tau) 
\notag\\
& \ \ \ \ \ \ \ \ \ \ \ \ \ \ \ \Longrightarrow \ \ \ E \ \text{is referred to as {\it homology invariant}}\, .
\label{prop.73a} 
\end{align}
The same types of shifts $\ell \rightarrow \ell {-}k_j$ arise in the sufficient condition (\ref{eq:Nshift})
for numerators of generic $(p{\leq}n)$-gon diagrams in the $\alpha'$-expansions to admit a 
quadratic-propagator representation. Indeed, the associated
cyclic relabellings $\mathfrak{N}_{+|\ldots j|-} \leftrightarrow \mathfrak{N}_{+|j \ldots |-}$ of the numerators 
in (\ref{eq:Nshift}) descend from the 
shift $z_j \rightarrow z_j{+} \tau$ around the $\mathfrak{B}$-cycle. This is a first piece of evidence
that homology invariance has important implications for the appearance of quadratic propagators.

\subsubsection{Five-point examples}

The four-point correlator ${\cal K}_4$ in (\ref{cKexamples}) is independent on $z_j$ and $\ell$
and thereby trivially a homology invariant by (\ref{prop.73}).
The simplest non-trivial examples of homology invariants occur in
the one-loop five-point correlator \cite{Mafra:2017ioj,Mafra:2018qqe}
\begin{align}
E_{1|23,4,5} &= \partial_{z} \log \theta_1(z_{12},\tau)
+\partial_{z} \log \theta_1(z_{23},\tau)
+\partial_{z} \log \theta_1(z_{31},\tau) \label{prop.74}
\\
E^\mu_{1|2,3,4,5} &= 2\pi i \ell^\mu + \big[ k_2^\mu \partial_{z} \log \theta_1(z_{12},\tau) + (2\leftrightarrow 3,4,5)\big]\, .
\label{prop.75}
\end{align}
Given the transformations $\partial_z \log \theta_1(z{\pm}1,\tau)= \partial_z \log \theta_1(z,\tau)$
as well as $\partial_z \log \theta_1(z{\pm}\tau,\tau)= \partial_z \log \theta_1(z,\tau) \mp 2\pi i$, the
first example (\ref{prop.74}) is elliptic and therefore homology invariant by virtue of the
cyclic sum of theta functions. For the
vectorial function in (\ref{prop.75}), homology invariance w.r.t.\ $z_j$ is tied to the interplay of 
$\ell$ with the monodromies of the theta functions and in case of $z_1$ requires five-point
momentum conservation $k_{12345}=0$. Based on the kinematic identity (\ref{keyid}) for
$(k_j)_\mu t_+^\mu(1,2,3,4,5)$, the five-point correlator in (\ref{cKexamples}) can be written in manifestly
homology-invariant form as \cite{Mafra:2017ioj,Mafra:2018qqe}
\beq
{\cal K}_5(\ell) = (E_{1|2,3,4,5})_\mu t_+^\mu(1,2,3,4,5) + \big[ E_{1|23,4,5} t_8(23,1,4,5) + (2,3|2,3,4,5) \big]\, ,
\label{Eformk5}
\eeq
where $+ (2,3|2,3,4,5)$ instructs to add five inequivalent permutations of $E_{1|23,4,5} t_8(23,1,4,5)$ 
obtained from $23\leftrightarrow 24,25,34,35,45$.\footnote{This selection of permutations can be understood
from the properties $E_{1|23,4,5}=E_{1|23,5,4}=-E_{1|32,4,5}$. We also note that
$E_{1|23,4,5}=E_{2|31,4,5}$ and $E_{2|34,1,5}= E_{1|34,2,5}+E_{1|23,4,5}-E_{1|24,3,5}$. Moreover, $E^\mu_{1|2,3,4,5}$ is permutation symmetric in $2,3,4,5$, but $1\leftrightarrow 2$ acts via 
$E^\mu_{2|1,3,4,5}=E^\mu_{1|2,3,4,5} + [ k_3^\mu E_{1|23,4,5} + (3\leftrightarrow 4,5)]$.}

In extracting the kinematic half integrand via (\ref{prop.51}), we obtain the following contributions 
to ${\cal I}_{5}$ and $N_{+|\rho(12345)|-}$ from the $\tau \rightarrow i\infty$ degeneration of
(\ref{prop.74}) and (\ref{prop.75})
\begin{align}
e_{1|23,4,5} &= \lim_{\tau \rightarrow i \infty} \frac{ E_{1|23,4,5} }{2\pi i} = G_{12}+G_{23}+G_{31} \label{newprop.74}
\\
e^\mu_{1|2,3,4,5} &=   \lim_{\tau \rightarrow i \infty}\frac{ E^\mu_{1|2,3,4,5}}{2\pi i} = \ell^\mu + \big[ k_2^\mu G_{12}+ (2\leftrightarrow 3,4,5)\big]\, ,
\label{newprop.75}
\end{align}
see (\ref{t8defs.3}) for the nodal Green function $G_{ij}$.
As will be detailed in subsection \ref{sec:4.2} below, each of the 7 terms in the alternative form
\beq
{\cal I}_{5}(\ell) = \frac{ (e_{1|2,3,4,5})_\mu t_+^\mu(1,2,3,4,5) + \big[ e_{1|23,4,5} t_8(23,1,4,5) + (2,3|2,3,4,5) \big]}{\sigma_1\sigma_2\sigma_3\sigma_4\sigma_5}
\label{Eformi5}
\eeq
of (\ref{t8defs.2}) separately integrates to quadratic propagators order by order in the open-string $\alpha'$-expansion
of (\ref{prop.1}). Similarly, the double copy of (\ref{Eformi5}) organizes the gravitational one-loop matrix element (\ref{prop.2}) into $7\times 7$ terms with integrands $e_{\ldots} \bar e_{\ldots}$ that separately integrate
to quadratic propagators at each order in $\alpha'$. This refines the observations
of sections \ref{sec:4.78} and \ref{sec:4.79} that the {\rm overall} expressions for $A^{\te{1-loop}}_{\rm eff}(1,2,3,4,5) $ and $M^{\te{1-loop}}_{5,{\rm eff}}$ can be expanded in terms of conventional Feynman integrals with quadratic propagators. 

At the level of practicalities, the number of terms conspiring in the recombination of linearized propagators to quadratic ones is reduced by a factor of 49 on average for each bilinear of the $e_{\ldots}$ in (\ref{Eformi5}). 
At six points, by the 51 homology invariants contributing to ${\cal K}_6$ \cite{Mafra:2017ioj,Mafra:2018qqe},
the size of typical expressions that recombine to quadratic propagators is reduced by a factor of 2601 on average.
This applies to both the loop integrand in supergravity and the $\ap$-expansion of $M^{\te{1-loop}}_{6,{\rm eff}}$.

\subsubsection{General conjectures}

The above five-point observations and a variety of checks at six points\footnote{We are
grateful to Sebastian Mizera for testing (\ref{prop.78}) for all the six-point instances of $E_P$
seen in \cite{Mafra:2017ioj,Mafra:2018qqe}.} motivate us to formulate the following
general conjecture:
for any homology invariant in the sense of
(\ref{prop.73a}), the contribution to
$A^{\te{1-loop}}_{\rm eff}(1,2,\ldots,n) $ in (\ref{prop.1}) can be
rewritten in terms of quadratic propagators. The same is claimed to
hold for any pair of meromorphic and antimeromorphic
homology invariants contributing to $M^{\te{1-loop}}_{n,{\rm eff}}$ in
(\ref{prop.2}). In the field-theory limit $\alpha'\rightarrow 0$, this
conjecture can be formulated within the framework of ambitwistor
strings:

\paragraph{Conjecture 1:} (on quadratic propagators in field-theory amplitudes)
\begin{align}
&E_P(z_j,\ell,\tau) , \ E_Q(z_j,\ell,\tau)  \ \text{homology invariant} \notag \\
&\Longrightarrow \ \ \ \int \frac{ \dd^D \ell}{\ell^2}
\int\limits_{\mathbb C^{n-1}} \dd \mu_{n+2}^{\te{tree}} \,  \frac{ E_P(\sigma_j,\ell,\tau \rightarrow i\infty) }{\sigma_1 \sigma_2 \ldots \sigma_n} \! \!  \! \! \! \! \sum_{\gamma \in {\rm cyc}(1,2,\ldots,n)} \! \! \! \! \! \! {\rm PT}(+,\gamma(1,2,\ldots,n),-)  \notag \\
&\ \ \ \ \ \ \ \ \ \ \ \ \text{can be recombined to quadratic propagators} \label{prop.78} \\
&\Longrightarrow \ \ \ \int \frac{ \dd^D \ell}{\ell^2}
\int\limits_{\mathbb C^{n-1}} \dd \mu_{n+2}^{\te{tree}} \, \frac{ E_P(\sigma_j,\ell,\tau \rightarrow i\infty)\,
E_Q(\sigma_j,\ell,\tau \rightarrow i\infty) }{| \sigma_1 \sigma_2 \ldots \sigma_n |^2} \notag \\
&\ \ \ \ \ \ \ \ \ \ \ \ \text{can be recombined to quadratic propagators}  \, ,\label{prop.79}
\end{align}
where the labels $P$ and $Q$ of the homology invariants
may represent any combination of subscripts and vector 
indices such as $P \rightarrow 1|23,4,5$ or $P \rightarrow \genfrac{}{}{0pt}{}{\mu\hfill}{1|2,3,4,5\hfill}$in case of (\ref{prop.74}) or (\ref{prop.75}).

The $\alpha'$-corrections to the one-loop matrix elements in
(\ref{prop.1}) and (\ref{prop.2}) rely on the open- and closed-string
motivated measures
\begin{align}
\int_{12\ldots n} \dd \mu_{n+2}^{\rm open} &= \! \! \! \!   \! \! \! \! 
 \int \limits_{0=\sigma_{1}< \sigma_{2} <\ldots < \sigma_{n}<\infty} \! \! \! \!  \! \! \! \! \dd \sigma_2\ldots \dd \sigma_n \, \prod_{j=1}^n |\sigma_{j+}|^{2\ap \ell \cdot k_j} \prod_{1\leq i<j}^n |\sigma_{ij}|^{2\ap k_{i}\cdot k_{j}}
 + {\rm cyc}(1,2,\ldots,n) 
\notag \\
\int \dd \mu_{n+2}^{\rm closed} &= \int\limits_{\mathbb C^{n-1}} \dd^2 \sigma_2\ldots \dd^2 \sigma_n \prod_{j=1}^n |\sigma_{j+}|^{4\ap \ell \cdot k_j} \prod_{1\leq i<j}^n |\sigma_{ij}|^{4\ap k_{i}\cdot k_{j}} \, .
\label{stringmeas}
\end{align}
In this setting, we additionally claim that, order by order in the $\alpha'$-expansion,

\paragraph{Conjecture 2:} (on quadratic propagators in one-loop matrix elements)
\begin{align}
&E_P(z_j,\ell,\tau) , \ E_Q(z_j,\ell,\tau)  \ \text{homology invariant} \notag \\
&\Longrightarrow \ \ \ \int \frac{ \dd^D \ell}{\ell^2}\,
\int_{12\ldots n} \dd \mu_{n+2}^{\te{open}} \,  \frac{ E_P(\sigma_j,\ell,\tau \rightarrow i\infty) }{\sigma_1 \sigma_2 \ldots \sigma_n}  \notag \\
&\ \ \ \ \ \ \ \ \ \ \ \ \text{can be recombined to quadratic propagators}\,, \label{prop.80} \\
&\Longrightarrow \ \ \ \int \frac{ \dd^D \ell}{\ell^2}
\int\limits_{\mathbb C^{n-1}} \dd \mu_{n+2}^{\te{closed}} \, \frac{ E_P(\sigma_j,\ell,\tau \rightarrow i\infty)\,
\overline{E_Q(\sigma_j,\ell,\tau \rightarrow i\infty)} }{| \sigma_1 \sigma_2 \ldots \sigma_n |^2} \notag \\
&\ \ \ \ \ \ \ \ \ \ \ \ \text{can be recombined to quadratic propagators}  \, .\label{prop.81}
\end{align}
These conditions are sufficient (though not necessary) for the one-loop matrix elements at four 
and five points to admit quadratic-propagator representations at all orders in $\alpha'$ as exemplified in section~\ref{sec:4}. At the same time, (\ref{prop.78}) to (\ref{prop.81}) do not follow from the quadratic-propagator
representations of the overall $n$-point matrix elements with $n>4$ since kinematic relations (\ref{keyid}) may
in principle accommodate deviations at the level of the individual homology invariants in (\ref{Eformk5}).
For both the five-point examples in (\ref{prop.74}), (\ref{prop.75}) and the higher-point homology 
invariants $E_{P}$ in \cite{Mafra:2017ioj, Mafra:2018pll, Mafra:2018qqe}, the integrands 
of (\ref{prop.80}) and (\ref{prop.81}) can be brought into
Parke-Taylor form by the techniques in \cite{He:2017spx}.
Hence, (\ref{prop.80}) and (\ref{prop.81}) boil down to combinations of disk and sphere integrals
and line up with terms in the one-loop matrix elements (\ref{prop.1}) and (\ref{prop.2}), respectively.

The BCJ and KLT relations of section \ref{sec:3.0} and their
$\alpha'$-uplifts are formulated in terms of the linearized
propagators obtained from $m$ or the $\ap$-expansion of $Z$ and
$J$. It would be interesting to reformulate these relations in terms
of quadratic propagators, for instance on the basis of the conjectures
(\ref{prop.78}) to (\ref{prop.81}).  BCJ relations involving quadratic
propagators have been discussed in \cite{Boels:2011tp, Boels:2011mn,
  Du:2012mt, Primo:2016omk}, whereas quadratic-propagator formulations
of one-loop KLT relations are uncharted territory at the time of
writing.


\subsection{\texorpdfstring{$\ap$}{a'}-expansions due to individual homology invariants}
\label{sec:4.2}

We shall now illustrate the conjectures of the previous subsection by spelling
out various examples of $\alpha'$-expansions in terms of quadratic propagators.
At four points, the open-string expressions (\ref{exsec.7}), (\ref{exsec.8})
and closed-string expressions (\ref{exsec.33}), (\ref{exsec.34}) of this type can be
attributed to a single trivial homology invariant $E_{1|2,3,4}=1$. That is why in most
of our nontrivial examples of (\ref{prop.80}) and (\ref{prop.81}), the quantities
$E_P, E_Q$ are specialized to the five-point homology invariants in (\ref{prop.74}) and (\ref{prop.75}).

\subsubsection{Shorthands}

It will be convenient to introduce shorthand notations for the contributions of
homology invariants and their bilinears to open- and closed-string one-loop
matrix elements (\ref{prop.1}) and (\ref{prop.2}), respectively. For open strings,
the single-trace objects $A^{\te{1-loop}}_{\rm eff}(1,2,\ldots,n)$ are built from
dual pairings of cycles $a_1a_2\ldots a_n \cong a_2a_3\ldots a_na_1$ and a single homology invariant
\begin{align}
[ a_1a_2\ldots a_n \, |\,E_P \rangle &=   \alpha'^{n-1} \int \frac{ \dd^D \ell}{\ell^2} \,
\int_{a_1 a_2\ldots a_n} \dd \mu_{n+2}^{\rm open}\,
 \frac{ E_P(\sigma_j,\ell,\tau \rightarrow i\infty) }{(2\pi i)^{n-4}\sigma_1 \sigma_2 \ldots \sigma_n} \, ,
\label{openEP}
\end{align}
see (\ref{stringmeas}) for the definition of the $\alpha'$-dependent measures 
$\dd \mu_{n+2}^{\rm open},\dd \mu_{n+2}^{\rm closed}$.
One-loop matrix elements of closed strings in turn are built from sphere integrals
describing the pairing of meromorphic and antimeromorphic homology invariants
\begin{align}
\langle E_Q \, |\,E_P \rangle &= \bigg( \frac{\ap}{\pi} \bigg)^{n-1}
\int \frac{ \dd^D \ell}{\ell^2} \int \dd \mu_{n+2}^{\rm closed}\,
 \frac{ E_P(\sigma_j,\ell,\tau {\rightarrow} i\infty)\,
\overline{E_Q(\sigma_j,\ell,\tau {\rightarrow} i\infty)} }{ (2\pi )^{2n-8} | \sigma_1 \sigma_2 \ldots \sigma_n |^2}\, .
\label{closedEP}
\end{align}
By comparison with (\ref{prop.80}) and (\ref{prop.81}), the conjectures
in the previous section state that the $\ap$-expansions of any such $[ a_1a_2\ldots a_n | E_P \rangle $ 
and $\langle E_Q | E_P \rangle$ are expressible in terms of quadratic propagators. By
(\ref{prop.51}) and (\ref{Eformk5}), the five-point
examples below can be assembled to yield the one-loop matrix elements in section \ref{sec:4} via
\begin{align}
&A^{\te{1-loop}}_{\rm eff}(1,2,3,4,5) = [12345\, |\,E^\mu_{1|2,3,4,5}\rangle  t_{+\mu} (1,2,3,4,5) 
\label{openEP.2}  \\
& \ \ +[12345\, |\,E_{1|23,4,5} \rangle t_8(23,1,4,5)
+[12345\, |\,E_{1|24,3,5}\rangle t_8(24,1,3,5) \notag \\
& \ \ +[12345\, |\,E_{1|25,3,4}\rangle t_8(25,1,3,4) 
+[12345\, |\,E_{1|34,2,5}\rangle t_8(34,1,2,5)\notag \\
&\ \ +[12345\, |\,E_{1|35,2,4}\rangle t_8(35,1,2,4)
+[12345\, |\,E_{1|45,2,3}\rangle t_8(45,1,2,3)\,, \notag
\end{align}
as well as (with $\bar t_{\pm \nu} \rightarrow \bar t_{+ \nu}$ for type-IIB amplitudes
and $\bar t_{\pm \nu} \rightarrow \bar t_{-\nu}$ for those of type IIA)
\begin{align}
M^{\te{1-loop}}_{5,{\rm eff}} &= \bar t_{\pm\nu}(1,2,3,4,5) \langle E^\nu_{1|2,3,4,5} \, |\,E^\mu_{1|2,3,4,5}\rangle t_{+\mu}(1,2,3,4,5) \label{openEP.3}  \\
& \ \ +  \bigg[ \bar t_{8 }(23,1,4,5)  \Big(   \langle E_{1|23,4,5} \, |\,E^\mu_{1|2,3,4,5} \rangle t_{+\mu}(1,2,3,4,5) \notag \\
& \ \ \ \ \ \ + \langle E_{1|23,4,5} \, |\,E_{1|23,4,5} \rangle t_8(23,1,4,5)
+\langle E_{1|23,4,5} \, |\,E_{1|24,3,5}\rangle t_8(24,1,3,5) \notag \\
& \ \ \ \ \ \ +\langle E_{1|23,4,5}\, |\,E_{1|25,3,4}\rangle t_8(25,1,3,4) 
+ \langle E_{1|23,4,5}\, |\,E_{1|34,2,5}\rangle t_8(34,1,2,5)\notag \\
&\ \  \ \ \ \ + \langle E_{1|23,4,5}\, |\,E_{1|35,2,4}\rangle t_8(35,1,2,4)
+\langle E_{1|23,4,5}\, |\,E_{1|45,2,3}\rangle t_8(45,1,2,3)  \Big) \notag \\
&\ \  \ \ \ \  + \bar t_{\pm \nu}(1,2,3,4,5) \langle E^\nu_{1|2,3,4,5}\, |\,E_{1|23,4,5}\rangle  t_8(23,1,4,5)+ (2,3|2,3,4,5) \bigg] \, .  
\notag
\end{align}
At four points, in turn, the one-loop matrix elements (\ref{exsec.6}) and (\ref{exsec.31})
both result from a single pairing of the form (\ref{openEP}) and (\ref{closedEP}) with trivial homology invariants $E_P, E_Q\rightarrow 1$,
\begin{align}
A^{\te{1-loop}}_{\rm eff}(1,2,3,4) &= t_8(f_1,f_2,f_3,f_4) \,  [1234 \, | \, E{=}1 \rangle\,,
 \label{4ptbracket} \\
M^{\te{1-loop}}_{4,{\rm eff}} &= \big| t_8(f_1,f_2,f_3,f_4)   \big|^2 \, \langle E{=}1 \, | \, E{=}1 \rangle \, .
\notag
\end{align}
The results in sections \ref{sec:4.1} and \ref{sec:4.8} are then equivalent to the following leading orders
\begin{align}
 [1234 \, | \, E{=}1 \rangle &= \int \frac{ \dd^D\ell}{\ell^2}   \lim_{k_{\pm} \rightarrow \pm \ell}  \sum_{\gamma \in {\rm cyc}(1,2,3,4)}
\sum_{\rho \in S_4} Z(+,\gamma(1,2,3,4),-|+,\rho(1,2,3,4),-) \notag \\
&= \int  \dd^D\ell \,  \bigg\{  \frac{1}{\ell^2(\ell{+}k_1)^2(\ell{+}k_{12})^2(\ell{+}k_{123})^2} \label{4ptopex} \\
& \ \ \ \ \ \ \ \ \ \ \ \
-\ap^2 \zeta_2 \bigg[ \frac{s_{12}}{\ell^2(\ell{+}k_1)^2(\ell{+}k_{12})^2}
+ {\rm cyc}_{\ell}(1,2,3,4) \bigg]
+ {\cal O}(\ap^3) \bigg\}\,,
\notag \\
\langle E{=}1 \, | \, E{=}1 \rangle &=   \int \frac{ \dd^D\ell}{\ell^2}   \lim_{k_{\pm} \rightarrow \pm \ell} \sum_{\gamma,\rho \in S_4} J(+,\gamma(1,2,3,4),-|+,\rho(1,2,3,4),-)
\notag \\
&=  \int  \dd^D\ell \,  
\bigg\{ \bigg[  \frac{1}{\ell^2(\ell{+}k_1)^2(\ell{+}k_{12})^2(\ell{+}k_{123})^2}  + {\rm perm}(2,3,4) \bigg]
 \label{4ptclex} \\
 & \ \ \ \ \ \ \ \ \ \ \ \ + \ap^3 \zeta_3 \bigg[ \frac{s_{12}^2}{\ell^2(\ell{+}k_1)^2(\ell{+}k_{12})^2} + {\rm perm}(1,2,3,4) \bigg] + {\cal O}(\ap^5) \bigg\} \, ,
 \notag
\end{align}
see (\ref{exsec.7}) and (\ref{exsec.33}) for a further subleading order of both.

\subsubsection{Five-point open-string examples}

With the shorthands $\ell_{12\ldots j} = \ell {+}k_{12\ldots j} $ of (\ref{t8defs.8}) for the
momenta in the propagators of the planar $(1,2,3,4,5)$-ordering, the cyclically inequivalent $\ap$-expansions
relevant to (\ref{openEP.2}) are
\begin{align}
[12345 \, |\,E_{1|23,4,5} \rangle &= \int \dd^D \ell \, \bigg\{  \frac{-1}{2 \ell^2 \ell_{1}^2 \ell_{12}^2 \ell_{123}^2 \ell_{1234}^2} - \frac{1}{s_{12} \ell^2 \ell_{12}^2 \ell_{123}^2 \ell_{1234}^2} - \frac{1}{s_{23} \ell^2 \ell_{1}^2 \ell_{123}^2 \ell_{1234}^2} \label{openEP.5}\\
&\! \! \! \!   \! \! \! \! \! \! \! \!   \! \! \! \!  \! \! \! \!   \! \! \! \! \! \! \! \!   \! \! \! \!    +\ap^2 \zeta_2 \bigg[ 
\frac{s_{15} }{2 \ell_1^2 \ell_{12}^2 \ell_{123}^2 \ell_{1234}^2} 
+ \frac{s_{34}}{2 \ell^2 \ell_1^2 \ell_{12}^2  \ell_{1234}^2}
+\frac{ s_{45} }{2 \ell^2 \ell_1^2 \ell_{12}^2 \ell_{123}^2}
 - \frac{ s_{12}+2s_{ \ell ,2} }{2 \ell^2  \ell_{12}^2 \ell_{123}^2 \ell_{1234}^2} 
 \notag\\
 &\! \! \! \!   \! \! \! \!  \! \! \! \!   \! \! \! \! \! \! \! \!   \! \! \! \! \! \! \! \!   \! \! \! \!    \quad  
 - \frac{ s_{23}+2s_{13} + 2s_{\ell,3} }{2 \ell^2 \ell_1^2 \ell_{123}^2 \ell_{1234}^2} 
 + \frac{ s_{45}}{s_{23}  \ell^2 \ell_1^2   \ell_{123}^2 } + \frac{ s_{45} }{s_{12}  \ell^2   \ell_{12}^2 \ell_{123}^2 } 
 + \frac{ s_{15} }{s_{23}  \ell^2 \ell_1^2  \ell_{1234}^2}  + \frac{ s_{15}}{s_{23} \ell_1^2\ell_{123}^2 \ell_{1234}^2}\notag\\
 &\! \! \! \!   \! \! \! \!  \! \! \! \!   \! \! \! \!  \! \! \! \!   \! \! \! \! \! \! \! \!   \! \! \! \!   \quad 
+ \frac{ s_{34} }{s_{12}  \ell^2   \ell_{12}^2  \ell_{1234}^2}  
  +  \bigg( \frac{1}{s_{12} }+\frac{1}{s_{23}} \bigg)  \frac{ s_{45} }{\ell^2   \ell_{123}^2 \ell_{1234}^2} 
  + \bigg( \frac{s_{34} }{s_{12}}-1 \bigg)  \frac{1}{ \ell_{12}^2 \ell_{123}^2 \ell_{1234}^2}
  \bigg] + \mathcal{O}(\ap^3) \bigg\} \notag \\
[12345 \, |\,E_{1|24,3,5} \rangle  &= \int \dd^D \ell \, \bigg\{  \frac{-1}{2 \ell^2 \ell_{1}^2 \ell_{12}^2 \ell_{123}^2 \ell_{1234}^2} - \frac{1}{s_{12} \ell^2 \ell_{12}^2 \ell_{123}^2 \ell_{1234}^2} \label{openEP.6}\\*
  &\! \! \! \!   \! \! \! \! \! \! \! \!   \! \! \! \!  \! \! \! \!   \! \! \! \! \! \! \! \!   \! \! \! \!   +\ap^2 \zeta_2 \bigg[ 
  \frac{s_{15} }{2 \ell_1^2 \ell_{12}^2 \ell_{123}^2 \ell_{1234}^2} 
  + \frac{s_{23}}{2 \ell^2 \ell_1^2 \ell_{123}^2  \ell_{1234}^2}
+ \frac{s_{34}}{2 \ell^2 \ell_1^2 \ell_{12}^2  \ell_{1234}^2}
+\frac{ s_{45} }{2 \ell^2 \ell_1^2 \ell_{12}^2 \ell_{123}^2}
   \notag\\*
&\! \! \! \!   \! \! \! \! \! \! \! \!   \! \! \! \!   \! \! \! \!   \! \! \! \! \! \! \! \!   \! \! \! \!   \quad 
- \frac{ s_{12}+2s_{\ell,2} }{2  \ell^2  \ell_{12}^2 \ell_{123}^2 \ell_{1234}^2 }
- \frac{1}{  \ell^2 \ell_{1}^2  \ell_{1234}^2} 
- \frac{1}{   \ell_{1}^2 \ell_{12}^2 \ell_{123}^2 }
+ \frac{ s_{45} }{s_{12}  \ell^2   \ell_{12}^2 \ell_{123}^2 }
+ \frac{ s_{45} }{s_{12}   \ell^2  \ell_{123}^2 \ell_{1234}^2}
\notag\\*
&\! \! \! \!   \! \! \! \! \! \! \! \!   \! \! \! \!  \! \! \! \!   \! \! \! \! \! \! \! \!   \! \! \! \!    \quad 
+ \frac{ s_{34} }{s_{12}   \ell^2  \ell_{12}^2  \ell_{1234}^2}
+ \bigg( \frac{ s_{34} }{s_{12}}-1 \bigg) \frac{1}{  \ell_{12}^2 \ell_{123}^2 \ell_{1234}^2}
 \bigg]  + {\cal O}(\ap^3) \bigg\}\notag \\*
[12345 \, |\,E^\mu_{1|2,3,4,5} \rangle &= \int \dd^D \ell \, \bigg\{  \frac{\ell^\mu + \frac{1}{2} k_1^\mu}{\ell^2\ell_{1}^2\ell_{12}^2 \ell_{123}^2 \ell_{1234}^2} + \frac{k_{5}^\mu}{s_{15} \ell_{1}^2 \ell_{12}^2 \ell_{123}^2 \ell_{1234}^2} - \frac{k_2^\mu}{s_{12} \ell^2 \ell_{12}^2 \ell_{123}^2 \ell_{1234}^2} \label{openEP.7}\\*
&\! \! \! \!   \! \! \! \! \! \! \! \!   \! \! \! \!  \! \! \! \!   \! \! \! \! \! \! \! \!   \! \! \! \!  +\ap^2 \zeta_2 \bigg[{-} \bigg( \ell^\mu + \frac{1}{2} k_1^\mu \bigg) \bigg( \frac{s_{15} }{\ell_1^2 \ell_{12}^2 \ell_{123}^2 \ell_{1234}^2} 
  + \frac{s_{23}}{ \ell^2 \ell_1^2 \ell_{123}^2  \ell_{1234}^2}
+ \frac{s_{34}}{ \ell^2 \ell_1^2 \ell_{12}^2  \ell_{1234}^2}
+\frac{ s_{45} }{ \ell^2 \ell_1^2 \ell_{12}^2 \ell_{123}^2} \bigg)  \nonumber\\*
&\! \! \! \! \! \! \! \!   \! \! \! \!    \! \! \! \!  \! \! \! \!   \! \! \! \! \! \! \! \!   \! \! \! \!   \quad 
- \frac{ ( \ell^\mu + \tfrac{1}{2} k_1^\mu + k_2^\mu ) s_{12} }{ \ell^2 \ell_{12}^2 \ell_{123}^2  \ell_{1234}^2}
- \frac{ k_2^\mu s_{\ell,2} }{\ell^2 \ell_{12}^2 \ell_{123}^2 \ell_{1234}^2}
+ \frac{ k_5^\mu s_{\ell,1 }
}{\ell_{1}^2\ell_{12}^2 \ell_{123}^2 \ell_{1234}^2}
- \frac{ k_3^\mu }{\ell^2  \ell_{123}^2 \ell_{1234}^2}
+\frac{ k_4^\mu }{ \ell_{1}^2\ell_{12}^2 \ell_{123}^2}
 \notag\\
&\! \! \! \! \! \! \! \!   \! \! \! \!    \! \! \! \!  \! \! \! \!   \! \! \! \! \! \! \! \!   \! \! \! \!   \quad 
+ \frac{ k_2^\mu s_{45} }{s_{12} \ell^2 \ell_{12}^2 \ell_{123}^2 }
+ \frac{ k_2^\mu s_{45} }{ s_{12} \ell^2  \ell_{123}^2 \ell_{1234}^2 }
+ \frac{ k_2^\mu s_{34} }{s_{12} \ell^2 \ell_{12}^2  \ell_{1234}^2}
- \frac{ k_5^\mu s_{34} }{s_{15}  \ell_{1}^2\ell_{12}^2  \ell_{1234}^2}
- \frac{ k_5^\mu s_{34} }{s_{15} \ell_{12}^2 \ell_{123}^2 \ell_{1234}^2}
 \notag\\
&\! \! \! \! \! \! \! \!   \! \! \! \!    \! \! \! \! \! \! \! \!   \! \! \! \! \! \! \! \!   \! \! \! \!    \quad 
- \frac{ k_5^\mu s_{23} }{s_{15} \ell_{1}^2  \ell_{123}^2 \ell_{1234}^2}
+\bigg( 1 - \frac{ s_{23} }{s_{15} } \bigg) \frac{ k_5^\mu }{ \ell_{1}^2\ell_{12}^2 \ell_{123}^2 } 
+ \bigg( \frac{ s_{34} }{s_{12} } - 1 \bigg) \frac{ k_2^\mu }{\ell_{12}^2 \ell_{123}^2 \ell_{1234}^2} \bigg] + \mathcal{O}(\ap^3) \bigg\} \, .\notag
\end{align}
By inserting into (\ref{openEP.2}) and using the kinematic identity (\ref{keyid}), we reproduce
the SYM integrand (\ref{t8defs.7}) in the $\alpha' \rightarrow 0$ limit and the $\alpha'^2 \zeta_2 F^4$-
numerators (\ref{t8defs.9a}) from the subleading order.
Note that the expressions in (\ref{openEP.5}) to (\ref{openEP.7}) are compatible with the Koba-Nielsen dervatives
\beq
\frac{1}{\ap} \partial_{z_2} \log {\cal J}_5(\ell,\tau) = (k_2)_\mu E^\mu_{1|2,3,4,5} + k_2 \cdot k_3 E_{1|23,4,5}
+ k_2 \cdot k_4 E_{1|24,3,5} + k_2 \cdot k_5 E_{1|25,3,4}
\label{openEP.9}
\eeq
following from (\ref{prop.42}) which drop out from one-loop string amplitudes.
At the level of one-loop matrix elements and their building blocks
(\ref{openEP}) and (\ref{closedEP}), the total derivatives (\ref{openEP.9}) 
clearly lead to
\begin{align}
0 &= (k_2)_\mu [a_1a_2\ldots a_5\, |\, E^\mu_{1|2,3,4,5} \rangle + k_2 \cdot k_3 \, [a_1a_2\ldots a_5\, |\, E_{1|23,4,5} \rangle
\label{openEP.10} \\
&\ \ \ \
+ k_2 \cdot k_4 \, [a_1a_2\ldots a_5 \, | \, E_{1|24,3,5}  \rangle + k_2 \cdot k_5 \, [a_1a_2\ldots a_5\, | \, E_{1|25,3,4}  \rangle
\notag
\end{align}
and analogous integration-by-parts identities with $\langle E_Q |$ in the place of $ [a_1a_2\ldots a_5| $.

\subsubsection{Five-point closed-string examples}
\label{simcomb}

The gravitational five-point one-loop matrix element (\ref{openEP.3}) involves five 
pairings (\ref{closedEP}) of homology invariants that are inequivalent under permutations
of $2,3,4,5$:
\[
\begin{array}{ll}
\bullet \ \langle E_{1|23,4,5} \,|\,E_{1|23,4,5} \rangle \phantom{xxxxxxxxxxx}
&\bullet \ \langle E^\mu_{1|2,3,4,5}\,|\,E_{1|23,4,5}\rangle  \\ 
\bullet \ \langle E_{1|23,4,5}\,|\,E_{1|24,3,5}\rangle
&\bullet \ \langle E^\mu_{1|2,3,4,5}\,|\,E^\nu_{1|2,3,4,5}\rangle \\
\bullet \ \langle E_{1|23,4,5}\,|\,E_{1|45,2,3}\rangle
\end{array}
\]
Already the leading orders of their $\alpha'$-expansion are considerably longer than
their open-string counterparts and are therefore relegated to ancillary files. For instance,
the field-theory limit
\begin{align}
\langle E_{1|23,4,5}\,|\,E_{1|24,3,5}\rangle &= \int \dd^D \ell \, \bigg\{ 
\sum_{\rho \in S_3} \frac{1}{ s_{12} \ell^2 \ell_{12}^2 \ell_{12\rho(3)}^2 \ell_{12\rho(34)}^2} 
\label{closed5ex} \\
& \ \ + \sum_{\rho \in S_4} \frac{ {\rm sgn}_{23}^{\rho} {\rm sgn}_{24}^{\rho} }{4
\ell^2 \ell_1^2 \ell_{1\rho(2)}^2 \ell_{1\rho(23)}^2 \ell_{1\rho(234)}  }
+ {\cal O}(\ap^3) \bigg\} \notag
\end{align}
involves 6 boxes and 24 pentagons. The signs of the pentagons are governed by
\beq
{\rm sgn}_{ij}^{\rho} = \left\{ \begin{array}{rl}
+1 &: \ \te{$i$ is on the right of $j$ in $\rho(2,3,4,5)$} \\
-1 &: \ \te{$i$ is on the left of $j$ in $\rho(2,3,4,5)$} \\
\end{array} \right.
\eeq
introduced by the interplay of the nodal Green function (\ref{t8defs.3})
with Parke-Taylor factors, see section 3 of \cite{He:2017spx} for details.
Note that the integration-by-parts identity (\ref{openEP.10}) of the open string straightforwardly
carries over to
\begin{align}
0 &= (k_2)_\mu \, \langle E_P \, | \, E^\mu_{1|2,3,4,5} \rangle + k_2 \cdot k_3  \,\langle E_P \, | \, E_{1|23,4,5} \rangle
\label{cloEP.10} \\
&\ \ \ \
+ k_2 \cdot k_4\,  \langle E_P \, | \,E_{1|24,3,5}  \rangle + k_2 \cdot k_5\,  \langle E_P \, | \, E_{1|25,3,4}  \rangle\, .
\notag
\end{align}


\subsection{All-multiplicity formulae}
\label{sec:4.4}

While a detailed study of one-loop matrix elements at $n\geq 6$ points
is relegated to the future, we shall now present all-multiplicity
conjectures for scalar integrals over symmetrized Parke-Taylor
factors. As an $\alpha'$-uplift of the $n$-gon formulae in the ambitwistor
setup \cite{Geyer:2015bja, He:2015yua, Geyer:2015jch} (with
$\ell_{12\ldots j} = \ell+k_{12\ldots j}$),
\begin{align}
&\int \frac{ \dd^D\ell}{\ell^2} \int \frac{ \dd \mu_{n+2}^{\rm tree} }{\sigma_2\sigma_3\ldots \sigma_n}
 \sum_{\gamma \in {\rm cyc}(1,2,\ldots,n)} {\rm PT}(+,\gamma(1,2,\ldots,n),-)
 = \int \frac{(-1)^n \, \dd^D\ell}{\ell^2 \ell^2_{1} \ell_{12}^2 \ell_{123}^2 \ldots \ell_{12\ldots n-1}^2}\,, \notag \\
 &\int \frac{ \dd^D\ell}{\ell^2} \int \frac{ \dd \mu_{n+2}^{\rm tree} }{|\sigma_2\sigma_3\ldots \sigma_n|^2} = 
 \int \frac{ \dd^D\ell}{\ell^2 \ell^2_{1} \ell_{12}^2 \ell_{123}^2 \ldots \ell_{12\ldots n-1}^2} + {\rm perm}(2,3,\ldots,n)\,, 
 \label{ngons.0} 
\end{align}
we shall investigate the first string corrections to the symmetrized integrals
\begin{align}
Z^{\te{1-loop}}_{\rm symm}(1,2,\ldots,n) &= [ 12\ldots n \, |\,(2\pi i )^{n-4} \rangle 
= \alpha'^{n-1}  \int \frac{ \dd^D\ell}{\ell^2}  \int_{12\ldots n} \frac{ \dd \mu_{n+2}^{\rm open} }{\sigma_2 \sigma_3\ldots \sigma_n}  \label{ngons.3}\\
&=  \int \frac{ \dd^D\ell}{\ell^2}   \lim_{k_{\pm} \rightarrow \pm \ell}   \sum_{\substack{\gamma \in {\rm cyc}(1,2,\ldots,n) \\ \rho\in S_n}} Z(+,\gamma(1,2,\ldots,n),-|+,\rho(1,2,\ldots,n),-) \,,
\notag \\
J^{\te{1-loop}}_{n,{\rm symm}} &= \langle (2\pi i)^{n-4} \, |\,(2\pi i )^{n-4} \rangle 
= \bigg(\frac{  \alpha' }{\pi } \bigg)^{n-1}  \int \frac{ \dd^D\ell}{\ell^2}
\int \frac{ \dd \mu_{n+2}^{\rm closed} }{|\sigma_2 \sigma_3\ldots \sigma_n|^2}  \label{ngons.4}  \\
&=  \int \frac{ \dd^D\ell}{\ell^2}   \lim_{k_{\pm} \rightarrow \pm \ell} \sum_{\gamma,\rho \in S_n} J(+,\gamma(1,2,\ldots,n),-|+,\rho(1,2,\ldots,n),-) \, ,
\notag
\end{align}
see (\ref{stringmeas}) for the definition of the measures $\dd \mu_{n+2}^{\rm open}$ and 
$\dd \mu_{n+2}^{\rm closed} $. These all-multiplicity families of integrals are engineered to 
reproduce the $n$-gon integrals (\ref{ngons.0}) in their field-theory limits. The four- and five-point 
instances of the symmetrized open-string integrals,
\begin{align}
Z^{\te{1-loop}}_{\rm symm}(1,2,3,4) &=  \int   \dd^D\ell \,  \bigg\{ \frac{1}{\ell^2 \ell_1^2 \ell_{12}^2 \ell_{123}^2} -   \alpha'^2 \zeta_2\bigg[ \frac{ s_{12} }{\ell^2 \ell_{12}^2 \ell_{123}^2}+ {\rm cyc}_{\ell}(1,2,3,4) \bigg]  \label{ngons.5} \\
&\hspace{-1cm} + \frac{ \alpha'^3 \zeta_3}{2} \bigg[ \frac{ s_{12}^2 {+} s_{12}(s_{1,\ell}{-}s_{2,\ell}) }{ \ell^2 \ell_{12}^2 \ell_{123}^2}
+ \frac{s_{12}{-}2s_{13}{+}s_{23}}{\ell^2 \ell_{123}^2}+ {\rm cyc}_\ell(1,2,3,4) \bigg] +{\cal O}(\ap^4) \bigg\}\,,  \notag \\
Z^{\te{1-loop}}_{\rm symm}(1,2,3,4,5) &=  \int  \dd^D\ell \,  \bigg\{ \frac{1}{\ell^2 \ell_1^2 \ell_{12}^2 \ell_{123}^2 \ell_{1234}^2} - \alpha'^2 \zeta_2 \bigg[ \frac{   s_{12} }{\ell^2 \ell_{12}^2 \ell_{123}^2 \ell_{1234}^2}+ {\rm cyc}(1,2,3,4,5) \bigg]  \notag \\
&\hspace{-1cm} +  \frac{\alpha'^3 \zeta_3}{2} \bigg[ \frac{ s_{12}^2 {+} s_{12}(s_{1,\ell}{-}s_{2,\ell}) }{\ell^2 \ell_{12}^2 \ell_{123}^2 \ell_{1234}^2} + \frac{  s_{12}{-}2s_{13}{+}s_{23}}{\ell^2  \ell_{123}^2 \ell_{1234}^2}+ {\rm cyc}_\ell(1,2,3,4,5) \bigg]  +{\cal O}(\ap^4) \bigg\}\,, \notag
\end{align}
augment the $n$-gon integral in the prescribed color ordering by 
\begin{itemize}
\item a cyclic orbit of $(n{-}1)$ gons at
the order of the $\alpha'^2 \zeta_2 F^4$ and $\alpha'^3\zeta_3(D^2 F^4+F^5)$ interactions
\item by additional one-mass $(n{-}2)$ gons at
the $\alpha'^3\zeta_3(D^2 F^4+F^5)$-order.
\end{itemize}
The notation $+{\rm cyc}_\ell(1,2,\ldots,n)$ instructs to shift loop momenta in the
cyclic images of the numerators $\sim s_{1,\ell}{-}s_{2,\ell}$ as explained 
below (\ref{exsec.7}).

The same patterns of $(n{-}1)$ and $(n{-}2)$ gons are verified to arise 
at six points which motivates the $n$-point conjecture,
\begin{align}
&Z^{\te{1-loop}}_{\rm symm}(1,2,\ldots,n) =  \int   \dd^D\ell \,  \bigg\{ \frac{1}{\ell^2 \ell_1^2 \ell_{12}^2  \ldots  \ell_{12\ldots n-1}^2}
-  \bigg[ \frac{ \alpha'^2 \zeta_2  s_{12} }{\ell^2 \ell_{12}^2   \ldots \ell_{12\ldots n-1}^2}+ {\rm cyc}_{\ell}(1,2,\ldots,n) \bigg] \notag \\
&\ \ \  +\frac{  \alpha'^3 \zeta_3}{2}\bigg[ \frac{ s_{12}^2 {+} s_{12}(s_{1,\ell}{-}s_{2,\ell}) }{\ell^2 \ell_{12}^2 \ell_{123}^2 \ldots \ell_{12\ldots n-1}^2}
+ \frac{ s_{12}{-}2s_{13}{+}s_{23} }{\ell^2  \ell_{123}^2 \ldots \ell_{12\ldots n-1}^2}
+ {\rm cyc}_\ell(1,2,\ldots,n) \bigg] +{\cal O}(\ap^4) \bigg\} \, .\label{ngons.6}
\end{align}
Similarly, the closed-string results at four and five points,
\begin{align}
J^{\te{1-loop}}_{4,{\rm symm}} &=  \int   \dd^D\ell \,  \bigg\{  \bigg[  \frac{1}{\ell^2 \ell_1^2 \ell_{12}^2 \ell_{123}^2}  + {\rm perm}(2,3,4) \bigg] \notag  \\
& \ \ \ \  +  \alpha'^3 \zeta_3  \bigg[ \frac{s_{12}^2 }{\ell^2 \ell_{12}^2 \ell_{123}^2}+ {\rm perm}(1,2,3,4) \bigg] +{\cal O}(\ap^5) \bigg\}\,, \label{ngons.7} \\
J^{\te{1-loop}}_{5,{\rm symm}} &=  \int  \dd^D\ell \,  \bigg\{  \bigg[  \frac{1}{\ell^2 \ell_1^2 \ell_{12}^2 \ell_{123}^2 \ell_{1234}^2} + {\rm perm}(2,3,4,5) \bigg]  \notag \\
& \ \ \ \  +\alpha'^3 \zeta_3  \bigg[ \frac{    s^2_{12} }{\ell^2 \ell_{12}^2 \ell_{123}^2 \ell_{1234}^2}+ {\rm perm}(1,2,3,4,5) \bigg] +{\cal O}(\ap^5) \bigg\}\,, \notag
\end{align}
involve a characteristic permutation sum over scalar $(n{-}1)$-gons at the order of the $\alpha'^3 \zeta_3 R^4$ interaction.
We propose the $n$-point generalization,
\begin{align}
J^{\te{1-loop}}_{n,{\rm symm}} &=  \int  \dd^D\ell \,  \bigg\{  \bigg[  \frac{1}{\ell^2 \ell_1^2 \ell_{12}^2 \ldots \ell_{12 \ldots n-1}^2}  + {\rm perm}(2,3,\ldots,n) \bigg] \notag  \\
& \ \ \ \  + \alpha'^3 \zeta_3 \bigg[ \frac{ s_{12}^2 }{\ell^2 \ell_{12}^2 \ldots \ell_{12 \ldots n-1}^2}+ {\rm perm}(1,2,\ldots,n) \bigg] +{\cal O}(\ap^5) \bigg\}\,, \label{ngons.8} 
\end{align}
which has also been tested at six points. In fact, these expressions for $J^{\te{1-loop}}_{n,{\rm symm}} $
follow from our conjecture (\ref{ngons.6}) for $Z^{\te{1-loop}}_{\rm symm}(1,2,\ldots,n)$ since
the definitions (\ref{ngons.3}) and (\ref{ngons.4}) together with the tree-level relation $J = {\rm sv} \, Z$ imply that
\beq
J^{\te{1-loop}}_{n,{\rm symm}} = {\rm sv} \,Z^{\te{1-loop}}_{{\rm symm}}(1,2,\ldots,n)  + {\rm perm}(2,3,\ldots,n)\, .
 \label{ngons.10}
 \eeq 
The permutation sum removes the antisymmetric part $s_{1,\ell}{-}s_{2,\ell}$ from the $(n{-}1)$-gon
numerators of ${\rm sv}\, Z$ and furthermore cancels the $(n{-}2)$-gons since $s_{12}{-}2s_{13}{+}s_{23} +{\rm perm}(1,2,3)=0$.


\section{UV divergences versus degenerations of one-loop string amplitudes}
\label{sec:5}

In this section, we support our proposal for one-loop matrix elements
with insertions of $D^{2k}F^n$ and $D^{2k}R^n$ by comparing their
UV divergences with expectations from string
theory. More specifically, the behaviour of genus-one string
integrands at the boundary $\tau \rightarrow i\infty$ of moduli space
encodes the UV divergences of one-loop matrix elements computed from
low-energy effective actions \cite{Metsaev:1987ju,Green:2010sp, Pioline:2018pso}. We
will determine the UV divergences of the one-loop matrix elements
presented in earlier sections and verify their compatibility with
degenerate genus-one integrals of the corresponding string amplitude.

\subsection{Basics of degenerating one-loop closed-string amplitudes}
\label{secMGF.1}

In relating one-loop matrix elements to string amplitudes in section
\ref{sec:3.2} and appendix \ref{app:closed}, we have first performed
the integration over $\tau$ and then $z_j$ within a certain
approximation while leaving the loop momentum $\ell$ unintegrated. The
discussion of the present section will follow the more conventional
procedure \cite{Polchinski:1998rq, Green:1987mn} to first integrate
over $\ell$ which manifests modular invariance of closed-string genus-one amplitudes.

Based on properties of modular graph forms \cite{DHoker:2015wxz,
  DHoker:2016mwo}, we will investigate the integrals over $\ell$
followed by $z_j$ in the expansion around the cusp
$\tau \rightarrow i\infty$. The dominant behaviour of these integrals
at the cusp is governed by Laurent polynomials in $\pi \Im \tau$ with
$\mathbb Q$-linear combinations of MZVs (conjecturally single-valued
MZVs \cite{Zerbini:2015rss, DHoker:2015wxz}) in their coefficients.
Each term is in one-to-one correspondence with the UV divergences of
effective field theories in various spacetime dimensions
\cite{Green:2010sp, Pioline:2018pso}.

\subsubsection{Modular graph forms and their degeneration}

After performing the Gaussian loop integral, the closed-string 
four-point amplitude (\ref{prop.41}) is proportional to\footnote{By different 
conventions for the definition (\ref{rev.12}) of $s_{ij}$ in this work, the expressions in (\ref{MGFs.1}) and
(\ref{MGFs.4}) can be obtained from results of \cite{DHoker:2015gmr, DHoker:2015wxz} 
by setting $s_{ij} \rightarrow - \ap s_{ij}$ in the references.} 
\beq
M^{\te{1-loop}}_{4,{\rm closed}} \sim |t_8(f_1,f_2,f_3,f_4)|^2 
\int_{\mathfrak{F}} \frac{ \dd^2 \tau }{(\Im \tau)^2} \int_{\mathfrak{T}_\tau^{3}}  \bigg( \prod_{j=2}^4 \frac{ \dd^2 z_j  }{\Im \tau} \bigg) \, \exp \bigg(  {-}\ap\sum_{1\leq i<j}^4 s_{ij} {\cal G}(z_{ij},\tau) \bigg)
\label{MGFs.1}
\eeq
with modular invariant measures $ \frac{ \dd^2 \tau }{(\Im \tau)^2} $ and 
$\frac{ \dd^2 z_j  }{\Im \tau}$. The modular invariant closed-string Green function 
is given in terms of the $\theta_1$ function (\ref{prop.43}), the Dedekind eta 
function $\eta(\tau) = q^{1/24} \prod_{n=1}^{\infty}$ $(1-q^n)$ 
and the comoving coordinates $u,v \in \RR$ for $z=u \tau{+}v$:
\beq
{\cal G}(z,\tau) = - \log \bigg|  \frac{ \theta_1(z,\tau) }{\eta(\tau)}\bigg|^2 + \frac{ 2\pi (\Im z)^2 }{\Im \tau}
= \frac{ \Im \tau }{\pi} \sum_{m,n \in \mathbb Z \atop{ (m,n) \neq (0,0)} } \frac{ e^{2\pi i(nu-mv)} }{| m \tau + n |^2}\, .
\label{MGFs.2}
\eeq
The key information on the low-energy effective action can be obtained
by Taylor-expanding the integrand of (\ref{MGFs.1}) in $\ap$: upon
integration over $z_j$, this leads to a power series in $\ap s_{ij}$
whose coefficients are the modular invariant functions of $\tau$ that
result from integrating monomials in Green functions $ {\cal G}(z_{ij},\tau)$
over their first arguments. These $z_j$-integrations can be
systematically performed using the lattice-sum representation in
(\ref{MGFs.2}) and the resulting nested lattice sums are known as
modular graph forms \cite{DHoker:2015wxz, DHoker:2016mwo}.  The study
of modular graph forms became a vibrant research topic at the
interface of string theory, algebraic geometry and number theory, see
e.g.\ \cite{Brown:2017qwo, Brown:2017qwo2} for a conjectural
mathematical framework and the PhD thesis \cite{Gerken:review} for a
recent overview from a physics perspective.

In an expansion of modular graph forms around the cusp
$\tau \rightarrow i\infty$, the dominant contributions are Laurent
polynomials in
\beq
y = \pi  \Im \tau
\label{MGFs.3}
\eeq
up to terms that are exponentially suppressed with $q$ or $\bar q$,
i.e.\ $e^{-2y}$ in both cases. For the leading orders of the
$\tau$-integrand in (\ref{MGFs.1}), we have the degeneration
\begin{align}
m^{\te{1-loop}}_{4,{\rm closed}}(\tau) &=
(4\ap)^5
 (\Im \tau)^2  \int_{\mathfrak T_\tau^3} \prod_{j=2}^4 \dd^2 z_j \int \dd^{10}\ell \, \big| {\cal J}_4(z_j,\ell,\tau)\big|^2  \, \Big|_{q^0\bar q^0} 
\notag \\
&= \int_{\mathfrak{T}_\tau^{3}}  \bigg( \prod_{j=2}^4 \frac{ \dd^2 z_j  }{\Im \tau} \bigg) \, \exp \bigg(  {-}\ap\sum_{1\leq i<j}^4 s_{ij} {\cal G}(z_{ij},\tau) \bigg)  \, \Big|_{q^0\bar q^0} 
\notag \\
&=  1 +  2 \ap^2  (s_{12}^2 {-}  s_{13} s_{23}) \bigg( \frac{ y^2}{45} +\frac{ \zeta_3}{y} \bigg)  
-  \ap^3 s_{12} s_{13} s_{23} \bigg(\frac{ 2 y^3}{189} + \zeta_3 + \frac{15 \zeta_5}{4 y^2} \bigg) \label{MGFs.4} \\
& \ \ \ \ + \ap^4 (s_{12}^2 {-}  s_{13} s_{23})^2 \bigg(\frac{16 y^4}{14175} + \frac{8 y \zeta_3}{45} 
+ \frac{5 \zeta_5}{3 y}+\frac{ \zeta_3^2}{y^2}   + \frac{ \zeta_7}{y^3} \bigg)
\notag \\
& \ \ \ \ + \ap^5 s_{12} s_{13} s_{23}  (  s_{13} s_{23}{-} s_{12}^2) 
\bigg(
\frac{ 8 y^5}{22275} {+} \frac{2 y^2 \zeta_3}{21} {+} \frac{58 \zeta_5}{45}  {+} \frac{2 \zeta_3^2}{y} 
{+} \frac{49 \zeta_7}{8 y^2} {+} \frac{\zeta_3 \zeta_5}{ 2 y^3} {+} \frac{63 \zeta_9}{32 y^4}
\bigg)\notag \\
& \ \ \ \  + {\cal O}(\ap^6) \,, \notag
\end{align}
see for instance \cite{Zerbini:2015rss, DHoker:2019xef,
Zagier:2019eus, Gerken:2020yii, Vanhove:2020qtt} for the systematics
at higher order. The notation $|_{q^0\bar q^0}$ in the first two lines instructs to only keep the $m=n=0$ 
terms in the expansion $\sim (\Im \tau)^a q^m \bar q^n$ around the cusp. As will be detailed in
\cref{secMGF.4,sec:alt-div} below, each term in (\ref{MGFs.4}) accurately
matches the UV divergences of the closed-string one-loop matrix elements
in section~\ref{sec:4.8}, where the number of spacetime dimensions
correlates with the power of $y$.

\subsubsection{Analytic versus non-analytic sector of one-loop amplitudes}

The step of performing the $\ap$-expansion of (\ref{MGFs.1}) at the
level of the $\tau$-integrand is oblivious to the intricate branch
cuts of the closed-string one-loop amplitudes
\cite{DHoker:1994gnm}. Among other things, the power series in
(\ref{MGFs.4}) does not take the discontinuities of the form
$\log(\alpha' s_{ij})$ into account which can for instance be
determined from the Feynman integrals in section \ref{sec:4} (see
section~\ref{mainsec7} for four-point examples).  Instead,
(\ref{MGFs.4}) captures the rational functions in $s_{ij}$ obtained as
the residues of the UV divergences, and the ${\cal O}(e^{-2y})$ terms
which are accurately known to the order of $(\ap s_{ij})^6$ can be
used to infer the one-loop effective action. The interplay between the
power-series behaved terms (the ``analytic sector'') of closed-string
one-loop amplitudes and the branch cuts (the ``non-analytic sector'')
has for instance been discussed in \cite{Green:2008uj, DHoker:2015gmr,
 DHoker:2019blr}. In section \ref{mainsec7}, we will spell out the
discontinuities of one-loop four-point matrix elements which determine
the non-analytic sector of closed- and open-string amplitudes.

\subsubsection{Degenerate modular graph forms at five points}

The low-energy expansion of the closed-string five-point function can
again be organized into modular graph forms \cite{Green:2013bza} after
integrating $|{\cal K}_5 {\cal J}_5|^2$ over the loop momenta. Since the chiral
five-point correlator ${\cal K}_5$ in (\ref{cKexamples}) is a sum of seven homology
invariants $E_P$, it will be convenient to introduce a shorthand for
the five-point analogues of the four-point building block in
(\ref{MGFs.4}),
\beq
\langle \! \langle E_Q | E_P \rangle \! \rangle_\tau = 2^{10} (\ap)^6 (\Im \tau)^2  \int_{\mathfrak T_\tau^4} \prod_{j=2}^5 \dd^2 z_j \int \dd^{10}\ell \, E_P(z_j,\ell,\tau)\, \overline{E}_Q(z_j,\ell,\tau) \,\big| {\cal J}_5(z_j,\ell,\tau)\big|^2  \, \Big|_{q^0 \bar q^0} \, .
\label{MGFs.6}
\eeq
Similar to section \ref{simcomb}, the $7\times 7$ bilinears
$\overline{E}_Q E_P$ arising from $|{\cal K}_5 |^2$ can be organized
into five permutation inequivalent cases
\begin{align}
\langle \! \langle E_{1|23,4,5} | E_{1|23,4,5} \rangle \! \rangle_\tau &= -\frac{1}{s_{12}} -\frac{ 1}{s_{13}} -\frac{ 1}{s_{23}} + 
 2  \ap^2 \bigg(\frac{y^2}{45} + \frac{ \zeta_3}{y} \bigg) \bigg( \frac{ s_{34} s_{35}}{s_{12}} - \frac{ s_{45}^2}{s_{12}} 
 + \frac{    s_{24} s_{25}}{s_{13}} - \frac{  s_{45}^2}{s_{13} } \notag \\
 & \ \ \ \ + \frac{s_{14} s_{15}}{  s_{23} }
 - \frac{s_{45}^2}{s_{23}} + s_{45}
 \bigg)
+ {\cal O}(\ap^3)\,,  \notag\\
\langle \! \langle E_{1|23,4,5} | E_{1|24,3,5} \rangle \! \rangle_\tau &= - \frac{1}{s_{12}} + 
\ap^2 \bigg(\frac{y^2}{45} + \frac{ \zeta_3}{y} \bigg) \bigg(  \frac{2(s_{35} s_{45} - s_{34}^2)}{s_{12}} - s_{12} +   5  s_{34}\bigg) 
+ {\cal O}(\ap^3)\,,  \notag\\
\langle \! \langle E_{1|23,4,5} | E_{1|45,2,3} \rangle \! \rangle_\tau &= \ap^2 \bigg(\frac{y^2}{45} + \frac{ \zeta_3}{y} \bigg) (  s_{24}  + s_{35} - s_{25} - s_{34}) + {\cal O}(\ap^3)\,, \label{MGFs.7} \\
\langle \! \langle E_{1|23,4,5} | E^\mu_{1|2,3,4,5} \rangle \! \rangle_\tau &= - \frac{k_2^\mu}{s_{12}} + 
 \frac{k_3^\mu}{ s_{13}} + \ap^2 \bigg(\frac{y^2}{45} + \frac{ \zeta_3}{y} \bigg) \bigg( 
  k_3^\mu \frac{(s_{24}^2 {+} s_{25}^2 {+} s_{45}^2)}{s_{13} }
  -k_2^\mu \frac{(s_{34}^2 {+} s_{35}^2 {+} s_{45}^2)}{s_{12}}  \notag \\*
  &\ \ \ \ +(k_3^\mu-k_2^\mu) s_{45}+ 
    k_4^\mu  (s_{24} - s_{34}) + k_5^\mu  (s_{25} - s_{35})\bigg)
+ {\cal O}(\ap^3)\,, \notag \\
\langle \! \langle E^\nu_{1|2,3,4,5} | E^\mu_{1|2,3,4,5} \rangle \! \rangle_\tau &=
{-}\frac{k_2^\mu k_2^\nu }{s_{12}}
-\frac{k_3^\mu k_3^\nu }{s_{13}}
-\frac{k_4^\mu k_4^\nu }{s_{14}}
-\frac{k_5^\mu k_5^\nu }{s_{15}} \nonumber \\*
&\hspace{-3ex}  + 2\ap^2 \bigg(\frac{y^2}{45} + \frac{ \zeta_3}{y} \bigg) \bigg( k_2^\mu k_2^\nu \bigg[
\frac{(s_{34} s_{35} {+} s_{34} s_{45} {+}    s_{35} s_{45})}{s_{12}} - s_{12}\bigg]+(2\leftrightarrow 3,4,5)  \bigg)  \nonumber \\*
&\hspace{-3ex} +\ap^2 \bigg(\frac{y^2}{45} + \frac{ \zeta_3}{y} \bigg) \bigg[  (k_2^\mu k_3^\nu + k_2^\nu k_3^\mu) s_{23}  +(2,3|2,3,4,5) \bigg]
+ {\cal O}(\ap^3) \, . \nonumber
\end{align}
While the loop integrals in (\ref{MGFs.4}) and (\ref{MGFs.6}) have
been performed in $D=10$ dimensions, we will later on be interested in
UV divergences in arbitrary dimensions. The $D$-dimensional versions
of the closed-string loop integrals follow from inserting a
factor of $(2 \sqrt{\ap \Im \tau})^{10-D}$ due to the modified
Gaussian integral and a series in non-integer powers of $q,\bar q$ to accommodate
Kaluza-Klein and winding modes of toroidal compactifications \cite{Green:1982sw}. 
These $q,\bar q$ series do not contribute to the $\tau \rightarrow i \infty$ limit relevant for
our one-loop matrix elements, see section \ref{sec:compact}.

\subsection{Basics of degenerating one-loop open-string amplitudes}
\label{secMGF.open}

After integrating out the loop momentum in one-loop open-string
amplitudes (\ref{prop.46}), the integrand w.r.t.\ the modular
parameter $\tau$ admits similar expansions around the cusp as in the
closed-string case: when focusing on the analytic sector in
$\alpha' s_{ij}$, the integration over the punctures can be
systematically performed in terms of elliptic multiple zeta values
\cite{Enriquez:Emzv, Broedel:2014vla}.  The dominant contributions as
$\tau \rightarrow i\infty$ are Laurent polynomials in
\beq
T = \pi \tau
\label{MGFs.8}
\eeq
instead of the $y= \pi \Im \tau$ in the closed-string degenerations,
with $\mathbb Q$-linear combinations of MZVs in its coefficients. In
the four-point amplitude (\ref{prop.46}), we have
\cite{Broedel:2018izr, Gerken:2018jrq}\footnote{By the different
  conventions for the definition of $s_{ij}$, the following
  expressions can be obtained from the results of
  \cite{Broedel:2018izr, Gerken:2018jrq} by setting
  $s_{ij} \rightarrow - \ap s_{ij}$ in the references.}
\begin{align}
&a^{\te{1-loop}}_{{\rm open}}(1,2,3,4;\tau) = 
(-2i\ap)^5
 \tau^2 \int_{ \mathfrak C_\tau(1,2,3,4) } \prod_{j=2}^4 \dd z_j \int \dd^{10} \ell \, |{\cal J}_4(z_j,\ell,\tau)|  \, \Big|_{q^0 } 
\notag \\
& \ \ = \int_{0<u_2<u_3 <u_4<1} \bigg( \prod_{j=2}^4   \dd u_j   \bigg)
\exp\bigg(\ap \sum_{1\leq i<j}^4 s_{ij} \big[ \log \theta_1( |u_{ij}| \tau,\tau) + i \pi u_{ij}^2 \tau \big] \bigg)   \, \Big|_{q^0} 
\notag\\
& \ \ = 
\frac{1}{6} + \alpha' s_{13} \bigg({-}\frac{ i T}{60} + \frac{ i \zeta_2}{2 T} + \frac{3 \zeta_3}{2 T^2} 
- \frac{    3 i \zeta_4}{2 T^3} \bigg) + 
 \alpha'^2 (s_{12}^2 - s_{13} s_{23}) \bigg(   {-} \frac{T^2}{540} + \frac{ \zeta_2}{9} + \frac{i \zeta_3}{3 T} 
 + \frac{ \zeta_4}{2 T^2}  \bigg) \notag \\*
 &\ \ \ \ \ \ + 
 \alpha'^2 (s_{12}^2 + 4 s_{12} s_{23} + s_{23}^2) \bigg( \frac{T^2}{3780} - \frac{ \zeta_2}{ 36 } 
 + \frac{ \zeta_4}{4 T^2} - \frac{ i \zeta_5}{T^3} - \frac{5 \zeta_6}{4 T^4} \bigg)  \label{MGFs.9}  \\*
 &\ \ \ \ \ \ +  
 \alpha'^3  s_{12} s_{23} s_{13} \bigg({-} \frac{i T^3}{4536} + \frac{i T \zeta_2}{45} - \frac{ \zeta_3}{12} 
 - \frac{35 i \zeta_4}{ 24 T} - \frac{2 \zeta_2 \zeta_3}{T^2} + \frac{5 \zeta_5}{4 T^2} 
 + \frac{17 i \zeta_6}{ 12 T^3}\bigg)  \notag \\*
& \ \ \ \ \ \ + \alpha'^3 s_{13} (s_{12}^2 - s_{13} s_{23}) \bigg( \frac{ i T^3}{7560} - \frac{i T \zeta_2}{90} + 
   \frac{ \zeta_3}{20} + \frac{3 i \zeta_4}{4 T} + \frac{5 \zeta_5}{2 T^2}  - \frac{\zeta_2 \zeta_3}{2 T^2}
     \notag \\*
 &\ \ \ \ \ \ \ \ \ \ \ \ \ \ \ \ \  \ \ \ \ \ - \frac{29 i \zeta_6}{12 T^3} 
  - \frac{3 \zeta_7}{T^4}    - \frac{  3 \zeta_3 \zeta_4}{2 T^4} + \frac{21 i \zeta_8}{4 T^5} \bigg)+ {\cal O}(\ap^4)\, ,
\notag
\end{align}
where $z_j = u_j \tau$ and $|_{q^0}$ instructs to only keep the $n=0$ terms
in the expansion $\sim \tau^a q^n$ around the cusp. The methods of \cite{Broedel:2014vla,
  Broedel:2018izr, Gerken:2020xfv} can also be used to infer the
exponentially suppressed terms ${\cal O}(e^{2iT})$ at any order in
$\ap$ in terms of B-cycle elliptic multiple zeta values
\cite{Enriquez:Emzv, Zerbini:2018sox, Zerbini:2018hgs}. However, we
have only displayed the dominant terms in the expansion of the
$\tau$-integrand of string amplitudes around the cusp in
(\ref{MGFs.9}) since this is the regime relevant for one-loop matrix 
elements and their UV divergences.

\subsubsection{Open- versus closed-string Laurent polynomials}

A direct comparison between (\ref{MGFs.9}) and the closed-string
counterpart (\ref{MGFs.4}) can be made upon symmetrization over the
four-point color orderings. Mandelstam identities imply that for
instance the entire first $\ap$-order as well as the Laurent
coefficients $\ap^2 \zeta_5$ and $\ap^3 \zeta_7$ drop out from
(\ref{MGFs.9}) upon symmetrization,
\begin{align}
\sum_{\gamma\in S_3}&a^{\te{1-loop}}_{4,{\rm open}}(1,\gamma(2,3,4);\tau) = 
1 + \alpha'^2 (s_{12}^2 - s_{13} s_{23}) \bigg({-}\frac{T^2}{90} + \frac{2 \zeta_2}{3} + \frac{   2 i \zeta_3}{T} 
+ \frac{3 \zeta_4}{T^2} \bigg) \label{MGFs.10} \\
&\ + 
 \alpha'^3 s_{12} s_{23} s_{13} \bigg({-}  \frac{ i T^3}{756} + \frac{2 i T \zeta_2}{15} - \frac{ \zeta_3}{2}
  - \frac{35 i \zeta_4}{ 4 T} - \frac{12 \zeta_2 \zeta_3}{T^2} + \frac{15 \zeta_5}{2 T^2} + \frac{17 i \zeta_6}{ 2 T^3} \bigg)
     + {\cal O}(\ap^4)\, .
\notag
\end{align}
As firstly pointed out in \cite{Broedel:2018izr}, these symmetrized
open-string building blocks can be related to the closed-string
degerations (\ref{MGFs.4}) via
\beq
m^{\te{1-loop}}_{4,{\rm closed}}(\tau) = 
{\rm sv} \sum_{\gamma\in S_3}a^{\te{1-loop}}_{4,{\rm open}}(1,\gamma(2,3,4);\tau) 
\label{MGFs.11}
\eeq
if the action of the single-valued map (\ref{rev.36}) is extended to
\beq
{\rm sv} \, T = 2 i y\, .
\label{MGFs.12}
\eeq
The intimate relation between degenerate elliptic multiple zeta values
and modular graph forms has been proven for the two-point analogue of
(\ref{MGFs.11}) \cite{Zagier:2019eus}. Moreover, a conjectural
generalization of (\ref{MGFs.11}) to individual $n$-point open-string
integration cycles can be found in \cite{Gerken:2020xfv}.
As we will see in section \ref{secMGF.4}, the relation (\ref{MGFs.11})
between closed- and open-string Laurent polynomials can be understood
from the UV divergences in the relation (\ref{exsec.32}) between one-loop
matrix elements.

\subsubsection{Degenerate elliptic multiple zeta values at five points}

At five points, we introduce a shorthand similar to (\ref{MGFs.6}) for the contributions
of individual homology invariants to the $\tau$ integrand of (\ref{prop.46})
\beq
[ \! [ a_1 \ldots a_5\, |\, E_{P} \rangle \! \rangle_{\tau} = -32i (\ap)^6 \tau^2 \int_{ \mathfrak C_\tau(a_1,\ldots,a_5) } \prod_{j=2}^5 \dd z_j \int \dd^{10} \ell \, E_P(z_j,\ell,\tau) \, |{\cal J}_5(z_j,\ell,\tau)| \, \Big|_{q^0}\, .
\label{MGFs.14}
\eeq
For a given integration ordering $a_1a_2\ldots a_5 \rightarrow 12345$ (see (\ref{prop.47})
for the definition of $\mathfrak C_\tau$), there are three cyclically inequivalent 
cases to consider,
\begin{align}
[ \! [ 12345 \,|\, E_{1|23,4,5} \rangle \! \rangle_{\tau} &=  \frac{1}{6 s_{12}} + \frac{1}{6 s_{23}} + \ap \bigg( 3 - \frac{2 s_{14}}{s_{23}} - \frac{2 s_{35}}{s_{12}} \bigg)  \notag \\
&\ \ \ \ \ \ \ \ \ \ \times
\bigg({-}\frac{ i T}{120} {+} \frac{ i \zeta_2}{4 T} {+} \frac{3 \zeta_3}{4 T^2} 
{-} \frac{    3 i \zeta_4}{4 T^3} \bigg) + {\cal O}(\ap^2)\,,
\notag \\
[ \! [ 12345\, |\, E_{1|24,3,5} \rangle \! \rangle_{\tau} &= \frac{ 1}{6 s_{12}} + \ap \bigg( 1 + \frac{2 s_{35}}{s_{12}} \bigg)
\bigg({-}\frac{ i T}{120} {+} \frac{ i \zeta_2}{4 T} {+} \frac{3 \zeta_3}{4 T^2} 
{-} \frac{    3 i \zeta_4}{4 T^3} \bigg) + {\cal O}(\ap^2)\,,
\! \! \! \label{MGFs.15} \\
[ \! [ 12345\, |\, E^\mu_{1|2,3,4,5} \rangle \! \rangle_{\tau} &=  \frac{ k_2^\mu }{6 s_{12}} - \frac{ k_5^\mu }{6 s_{15}}
+ \ap \bigg(  k_{24}^\mu - k_{35}^\mu - \frac{2 s_{35}}{s_{12}}  k_2^\mu + \frac{2 s_{24}}{s_{15}} k_5^\mu\bigg)
 \notag \\
 &\ \ \ \ \ \ \ \ \ \ \times \bigg({-}\frac{ i T}{120} {+} \frac{ i \zeta_2}{4 T} {+} \frac{3 \zeta_3}{4 T^2} 
{-} \frac{    3 i \zeta_4}{4 T^3} \bigg)  + {\cal O}(\ap^2)\, .
\notag 
\end{align}
We have again displayed the ten-dimensional loop integrals in (\ref{MGFs.9}) and (\ref{MGFs.14}),
and the expressions in different dimensions follow by inserting an extra factor of $\sqrt{ -2i \ap \tau }^{10-D}$
and a series in $q$ to take the Kaluza-Klein modes from a toroidal compactification into account \cite{Green:1982sw}.
Again, the $q$-series of the Kaluza-Klein modes do not contribute to the $\tau \rightarrow i \infty$ limit
relevant for the one-loop matrix elements under investigation, see section \ref{sec:compact}.

\subsection{String Laurent polynomials, Schwinger parameters and UV divergences}
\label{secMGF.2}
In the field-theory limit of one-loop string amplitudes, the limit
$\alpha' \rightarrow 0$ is performed while degenerating the torus or
cylinder worldsheet to a worldline via $\Im \tau \rightarrow \infty$
\cite{Green:1982sw, Bern:1990cu, Bern:1990ux, Bern:1991an, Bern:1991aq, Bern:1993wt, Dunbar:1994bn, 
Schubert:2001he, Tourkine:2013rda, Mafra:2014gja}. In this process, Feynman integrals 
depending on momenta $K_j$ with possibly non-vanishing~$K_j^2$ are retained in their
Schwinger parametrization 
\begin{align}
&\int \frac{ \dd^D \ell }{\pi^{D/2} }  \, \frac{ (T_0  + T_1^\mu \ell_\mu + T_2^{\mu \nu} \ell_\mu \ell_\nu +\ldots)}{\ell^2 (\ell{+}K_1)^2
(\ell{+} K_{12})^2 \ldots (\ell{+}K_{12\ldots n-1})^2} =  \int^{\infty}_0 \frac{ \dd t}{t} \, t^{n-D/2}
\notag \\
&\ \ \times \! \! \! \! \! \! \! \!  \int \limits_{0<u_2<u_3<\ldots < u_n<1}
 \! \! \! \! \! \! \! \!  \dd u_2 \, \dd u_3\ldots \dd u_n\,
\exp\bigg[ {-} t \! \sum_{1\leq i<j}^{n} \! K_i \cdot K_j \big(u_{ij}^2 - |u_{ij}| \big) \bigg]
\label{WLUV.1} \\
&\ \ \ \ \ \ \ \ \ \ \ \ \ \ \ \ \ \ \ \ \times
 \bigg( T_0 + T_1^\mu L_\mu + T_2^{\mu \nu}\bigg[ L_\mu L_\nu + \frac{ \eta_{\mu \nu} }{2 t} \bigg]  + \ldots  \bigg)\, ,
 \notag
\end{align}
where the worldline proper times $u_j$ are the first comoving coordinate
of the worldsheet punctures $z_j = u_j \tau {+} v_j$ with $u_j,v_j\in \mathbb R$.
As will be detailed below, the worldline length $t$ is identified with $2\pi \alpha' \Im \tau$ for
open strings and $4\pi \alpha' \Im \tau$ for closed strings and remains a dynamical integration
variable in the simultaneous limit $\alpha' \rightarrow 0$ and $\tau \rightarrow i\infty$. 
Moreover, (\ref{WLUV.1}) exemplifies
the worldline treatment of loop momenta in the numerators, where
$T_0,T_1^\mu, T_2^{\mu \nu}$ are arbitrary $\ell$-independent tensors. The quantity
\beq
L^\mu = \sum_{j=1}^n K_j^\mu u_j
\label{WLUV.2}
\eeq
stems from the shift of loop momentum to complete the square in the Gaussian integral
over $\ell$, and higher powers $\ell^{N\geq 2}$ involve additional
Wick contractions $\ell_\mu \ell_\nu \rightarrow \frac{ \eta_{\mu \nu} }{2t}$.  
Finally, the exponential in the second line of
(\ref{WLUV.1}) is the worldline counterpart of the string-theory
Koba-Nielsen factors in (\ref{MGFs.1}) and (\ref{MGFs.9}) with
worldline Green function $u_{ij}^2 - |u_{ij}|$. We emphasize that
(\ref{WLUV.1}) also applies to any number of massive corners, i.e.\ momenta in
$K_{12\ldots p} = K_1{+}K_2{+}\ldots{+}K_p$ subject to $K_j^2\neq 0$.

In the Schwinger parametrization (\ref{WLUV.1}), UV divergences stem
from the integration region $t\rightarrow 0$ of small worldline
lengths, reflecting their origin from short-distance effects (see, for instance, 
\cite{Smirnov:2012gma}). Thus to study these UV divergences we first Taylor-expand the right-hand side of \cref{WLUV.1} as
\begin{align}
&\int \frac{ \dd^D \ell  }{  \pi^{D/2}} \, \frac{ (T_0  + T_1^\mu \ell_\mu + T_2^{\mu \nu} \ell_\mu \ell_\nu +\ldots)}{\ell^2 (\ell{+}K_1)^2
(\ell{+} K_{12})^2 \ldots (\ell{+}K_{12\ldots n-1})^2}\Bigg|_{\rm UV} = \int^{\tmax}_0 \dd t  \sum_{M=0}^\infty \frac{t^{n+M-1-D/2}}{M!} 
\label{WLUV.3} \\
&\ \ \times \! \! \! \! \! \! \! \!  \int \limits_{0<u_2<u_3<\ldots < u_n<1}
 \! \! \! \! \! \! \! \! \! \dd u_2 \ldots \dd u_n\,
\Bigg[   \! \sum_{1\leq i<j}^{n} \! \! K_i{\cdot} K_j \big(u_{ji} {-} u_{ji}^2 \big) \Bigg]^M  
\! \bigg( T_0 + T_1^\mu L_\mu + T_2^{\mu \nu}\bigg[ L_\mu L_\nu + \frac{ \eta_{\mu \nu} }{2 t} \bigg]  + \ldots  \bigg) 
 \,\, , 
 \notag
\end{align}
where the integration over the simplex $0{=}u_1{<}u_2{<}u_3{<}\ldots {<} u_n{<}1$ is straightforward for any $M$ and $n$, leading to a Laurent series in $t$ as the integrand. The 
series expansion of the exponential in (\ref{WLUV.1}) generically does not commute with the integration over $t$, and we employ the notation $\int \ldots \big|_{{\rm UV}}$ on the left-hand side of (\ref{WLUV.3}) to indicate that the
right-hand side does not provide an exact expression for the original Feynman integral. 
Still, the right-hand side of (\ref{WLUV.3}) accurately reproduces
the UV divergences in various spacetime dimensions while the series expansion in $K_i \cdot K_j$ does not 
capture the discontinuities such as the $\log (K_i \cdot K_j)$ in the $\epsilon$-expansions of appendix \ref{app:Feynints}. Moreover, we have introduced an infrared cutoff $\tmax$ into the integration domain for $t$ to decouple infrared divergences from the poles in the 
dimensional-regularization parameter $\epsilon$: if we view the Feynman integrals as analytic functions 
in $D=D_{\rm crit} {-} 2\epsilon$, then the UV divergences
of (\ref{WLUV.3}) in $D_{\rm crit}=2N$ are the residues at the $\frac{1}{\epsilon}$ pole of
\begin{align}
\int_{0}^{\tmax}\dd t \,t^{N-1-D/2}
=\frac{1}{\epsilon}+\mathcal{O}(\epsilon^0)\, ,
\end{align}
where the dependence on the cutoff $0<\tmax<\infty$ is relegated to the regular terms 
$\mathcal{O}(\epsilon^0)$. Upon comparison with (\ref{WLUV.3}), its $M^{\rm th}$ 
order in $K_i\cdot K_j$ captures the UV divergence of a scalar $n$-gon in the critical dimension $2n{+}2M$.
For tensor integrals, the loop momenta in the numerator lower the critical dimension through the 
Wick contractions $\ell_\mu \ell_\nu \rightarrow \frac{ \eta_{\mu \nu} }{2t}$. On these grounds,
the power series \cref{WLUV.3} in $t K_i\cdot K_j$ obtained from integrating out the $u_j$ 
order by order in $t$ serves as a generating function of higher-dimensional UV divergences.
In practice, we truncate the power series and only track terms up to a maximum order
of $t$ corresponding to the UV divergences up to some maximum spacetime dimension. 

In the cases of a massless scalar box with $(K_1,K_2,K_3, K_4) \rightarrow (k_1,k_2,k_3,k_4)$ and a one-mass scalar triangle with $(K_1,K_2,K_3) \rightarrow (k_{12},k_3,k_4)$, the series expansion in the integrand of (\ref{WLUV.3}) yields
\begin{align}\label{WLUV.5}
&\int \frac{\dd^D\ell }{\pi^{D/2} } \frac{1 }{\ell^2(\ell{+}k_1)^2(\ell{+}k_{12})^2(\ell{+}k_{123})^2}\Bigg|_{\rm UV}\! =\! \int_0^{\tmax} \dd t \, t^{3-D/2} \,
\bigg\{  \frac{1}{6} + \frac{ t    s_{13} }{120}
 \notag \\
& \hspace{2cm} 
+ \frac{ t^2  }{5040}(2 s_{12}^2 {+} s_{12}s_{23} {+} 2 s_{23}^2) 
+ \frac{ t^3 s_{13}  }{181440}  \,(3 s_{12}^2 {-} 2 s_{12}s_{23} {+} 3 s_{23}^2) + {\cal O}(t^4) 
\bigg\}\,,  \\
&\int  \frac{\dd^D\ell }{\pi^{D/2}} \frac{1}{\ell^2(\ell{+}k_{12})^2(\ell{+}k_{123})^2}\Bigg|_{\rm UV}\! =\! \int_0^{\tmax} \dd t \, t^{2-D/2} \,
\bigg\{ 
\frac{1}{2} - \frac{ t s_{12}}{24}  + \frac{ t^2 s_{12}^2 }{360}  - \frac{ t^3  s_{12}^3 }{6720}  + {\cal O}(t^4)
\bigg\} \, .\nonumber
\end{align}
The scalar box in particular already provides an opportunity to verify
the identifications between $t$ and the string moduli introduced
above.  The details of the mapping will be discussed as part of the
amplitude matchings below. 

An alternative method of constructing the Laurent polynomials directly
from logarithmic divergences in dimensional regularization is provided in \cref{sec:alt-div}.


\subsection{UV divergences from open-string effective actions}
\label{secMGF.3}

We start by matching the UV divergences of one-loop matrix elements of gauge multiplets
with the contributions to one-loop string amplitudes from the cusp $\tau \rightarrow i\infty$.
For type-I superstrings with gauge group $SO(32)$, UV divergences at one loop cancel 
between cylinder and Moebius-strip worldsheets in the single-trace sector and
are regulated by closed-string exchange in the double-trace sector \cite{Green:1987mn}. Hence, the 
Laurent polynomials (\ref{MGFs.9}) encoding the dominant contributions at the cusp may be 
viewed as the ``would-be''-UV-divergences of the effective field theory that drop out from the complete one-loop 
amplitude of the $SO(32)$ open superstring.

\subsubsection{Four points}

As in section \ref{sec:4.1}, we omit the overall factor of
$t_8(f_1,f_2,f_3,f_4)$ and study four-point UV divergences of the
scalar integral $ \widehat A^{\te{1-loop}}_{\te{eff}}(1,2,3,4)$ in (\ref{exsec.6}).
Our starting points are the Feynman integrals in
(\ref{exsec.7}) and (\ref{exsec.8}) as well as their higher-order counterparts 
in the $\alpha'$-expansion which will now be matched with Laurent polynomials in (\ref{MGFs.9}). It will be
natural to collect the results according to the zeta values, rather
than the $\ap$ striation of (\ref{MGFs.9}).

For the box integral and the scalar triangles in (\ref{exsec.7}), the UV divergences
are given by (\ref{WLUV.5}). At the order of $\alpha'^3 \zeta_3$, we
additionally encounter antisymmetrized vector triangles and
bubbles on external legs
with the following UV contributions according to (\ref{WLUV.3}):
\begin{align}
\int  \frac{\dd^D\ell }{\pi^{D/2}} \, \frac{ (s_{1,\ell} {-}s_{2,\ell})}{\ell^2(\ell{+}k_{12})^2(\ell{+}k_{123})^2}\Bigg|_{\rm UV} & \!=\!
 \int_0^{\tmax}\! \dd t \, t^{2-D/2} \,
(s_{23} {-} s_{13}) \,
\bigg[ 
\frac{1}{6} {-}\frac{ t s_{12}}{120}  {+} \frac{ t^2 s_{12}^2 }{2520} {-} \frac{ t^3 s_{12}^3}{ 60480 } {+} {\cal O}(t^4)
\bigg]
\notag \\
\int  \frac{\dd^D\ell }{\pi^{D/2}} \, \frac{ 1}{\ell^2(\ell{+}k_{1})^2}\Bigg|_{\rm UV} &\!=\!  \int_0^{\tmax}\! \dd t \, t^{1-D/2}\,.
\label{WLUV.23}
\end{align}
On these grounds, the Feynman integrals in (\ref{exsec.7}) have the combined UV behaviour
\begin{align}\label{WLUV.25}
\pi^{-D/2}   &\widehat A^{\te{1-loop}}_{\te{eff}}(1,2,3,4) \,\Big|_{\rm UV} = \int_0^{\tmax} \dd t \, t^{3-D/2} \bigg\{\bigg[
 \frac{1}{6  } + \frac{t s_{13}}{120 } + \frac{t^2(2 s_{12}^2 {+} s_{12} s_{23} {+} 2 s_{23}^2)}{5040} \\
& \ \ \ \ + \frac{t^3s_{13}(3 s_{12}^2 {-} 2 s_{12} s_{23} {+} 3 s_{23}^2)}{181440} + \ldots  \bigg]+ \ap^2 \zeta_2 \bigg[    \frac{s_{13}}{t} +
  \frac{s_{12}^2 {+} s_{23}^2}{12 } -
  \frac{t(s_{12}^3 {+} s_{23}^3) }{180 } + \ldots \bigg] \nonumber\\
& + \ap^3 \zeta_3 \bigg[
{-} \frac{6 s_{13}}{t^2} +
  \frac{2 (s_{12}^2 {+} s_{12} s_{23}{+} s_{23}^2)}{3t} 
  + \frac{s_{13}(3 s_{12}^2 -2 s_{12} s_{23} + 3 s_{23}^2)}{60}
 + \ldots \bigg] + {\cal O}(\ap^4) \bigg\}\, . \nonumber
\end{align}
Higher orders in $\alpha'$ additionally involve two-mass bubbles, tadpoles and multiple powers
of loop momenta in the numerators which can also be addressed by the techniques of
section \ref{secMGF.2}. At the order of $\alpha'^4 \zeta_4$ for instance,
the numerators in (\ref{exsec.10}) give rise to UV divergences
\begin{align}\label{WLUV.27}
\pi^{-D/2}\widehat A^{\te{1-loop}}_{\te{eff}}(1,2,3,4)  \,\Big|^{\rm  \ap^4 \zeta_4}_{\rm UV} 
= \int_0^{\tmax} &\dd t\, t^{3-D/2} \bigg\{ \frac{12 s_{13}}{ t^3}
  - \frac{3 s_{13}^2}{t^2}  \\
&+  \frac{  s_{13}(18 s_{12} ^2 {-} 17 s_{12} s_{23} {+} 18 s_{23}^2)}{12 t} + \ldots \bigg\}\, .
 \nonumber
\end{align}
All of these UV contributions are in perfect agreement in the Laurent polynomials in (\ref{MGFs.9}): 
with the dictionary
\beq
 \te{open strings:} \ \ \ t =-2\pi i \alpha' \tau = - 2i \alpha' T\, ,
 \label{dictopen}
\eeq
between the purely imaginary modular parameter $\tau$ of the cylinder
and the real Schwinger parameter $t$ on the worldline,\footnote{The
  dictionary (\ref{dictopen}) can be further validated by comparing
  the exponent $-t k_i \cdot k_j u_{ij}^2 $ in the Schwinger
  parametrization (\ref{WLUV.1}) of $n$-gons with the contributions
  $\sim e^{i\pi \alpha' \tau s_{ij}u_{ij}^2}$ to the open-string
  Koba-Nielsen factor in the second line of (\ref{MGFs.9}). Moreover,
  by rewriting (\ref{eq:os-eps-t}) as
  $ \widehat A^{\te{1-loop}}_{\rm eff}(1,2,3,4) \, \big|_{\rm UV} =
  \int^{\tmax}_0 \dd t\, t^3 ( \frac{ \pi }{t})^{D/2} \ldots$, the
  dimension-dependence $\sim ( \frac{ \pi }{t})^{D/2}$ on the
  right-hand side is consistent with the factor of
  $\sqrt{-2i\ap \tau}^{10-D}$ in $(D \neq 10)$-dimensional open-string
  amplitudes after loop integration.}  any power of $T$ at the
$\alpha'^{\leq 3}$-orders of (\ref{MGFs.9}) is accurately mapped to a
UV power divergence in the effective field theory, with precise
matching of the coefficients. In other words, we have confirmed up to
and including $\alpha'^{3}$ that
\begin{equation}
\pi^{-D/2}  \widehat A^{\te{1-loop}}_{\rm eff}(1,2,3,4) \, \Big|_{\rm UV} = 
\int^{\tmax}_0 \dd t\, t^{3-D/2}  a^{\te{1-loop}}_{{\rm open}}\Big(1,2,3,4;\tau {=} \frac{ i t }{2\pi \ap} \Big) 
  \label{eq:os-eps-t}
\end{equation}
reproduces the UV regime (\ref{WLUV.3}) of the Schwinger parametrized one-loop matrix elements.
Note that the leading orders $\alpha'^{\leq 3}$ in the Laurent polynomials are already
sensitive to the UV divergences due to higher $\alpha'$-orders of the one-loop matrix
elements. For instance, the coefficients of $\ap^2 \zeta_5,\, \ap^3\zeta_2 \zeta_3$ and $\ap^2\zeta_2^3$
in $a^{\te{1-loop}}_{{\rm open}}(1,2,3,4;\tau)$ already crosscheck the numerators of
$\widehat A^{\te{1-loop}}_{\rm eff}$ at $\alpha'^5$ (see (\ref{exsec.11}), (\ref{exsec.12})) and at $\alpha'^6$ 
(see appendix \ref{app:4ptopen}), respectively.
Our checks of (\ref{eq:os-eps-t}) therefore include the UV divergences 
\begin{align}
 \pi^{-D/2}  \afouruv{\rm  \ap^6 \zeta_6}
&=   \int^{\tmax}_0 \dd t\, t^{3-D/2} \bigg[ -\frac{20}{t^4}(s_{12}^2 + 4 s_{12} s_{23} + s_{23}^2) + \ldots\bigg]\,,
\notag \\
 \pi^{-D/2}   \afouruv{\rm  \ap^7 \zeta_7}
&=  \int^{\tmax}_0 \dd t\, t^{3-D/2} \bigg[- \frac{48}{t^4} \, s_{13} (s_{12}^2 + s_{12} s_{23} + s_{23}^2) + \ldots\bigg]\,,
\label{negdims} \\
 \pi^{-D/2}   \afouruv{\rm  \ap^8 \zeta_8}
&=  \int^{\tmax}_0 \dd t\, t^{3-D/2} \bigg[ \frac{168}{t^5} \,s_{13}(s_{12}^2 + s_{12} s_{23} + s_{23}^2) + \ldots\bigg] \, ,
\notag
\end{align}
which occur in formally vanishing or negative numbers of spacetime dimensions 
by the power of the Schwinger parameter $t$ in these examples, see the discussion 
around (\ref{WLUV.3}). \Cref{UVopen} summarizes all the UV divergences that we have matched with the $\alpha'^{\leq 3}$-orders 
of $a^{\te{1-loop}}_{{\rm open}}$.

\begin{table}
\begin{center}
\tikzpicture [scale=0.7, line width=0.30mm]
\draw(-4,0)--(14,0);
\draw(0,1)--(0,-10.5);
\draw(3,1)--(3,-10.5);
\draw(6,1)--(6,-10.5);
\draw(10,1)--(10,-10.5);
\draw(0,0) -- (-4,1);
\draw(-1,0.8)node{\footnotesize $\alpha'$-order};
\draw(-3,0.3)node{\footnotesize eff.\ op's};
\draw[dashed](8,-0.3)--(8,-10.5);
\draw[dashed](12,-0.3)--(12,-10.5);
\draw(1.5,0.6)node{$0$};
\draw(4.5,0.6)node{$1$};
\draw(8,0.6)node{$2$};
\draw(12,0.6)node{$3$};
\draw(-2,-0.8)node{$F^2$};
\draw(-2,-1.8)node{$\ap^2 \zeta_2   F^4$};
\draw(-2,-2.8)node{$\ap^3 \zeta_3  D^2 F^4$};
\draw(-2,-3.8)node{$\ap^4 \zeta_2^2  D^4 F^4$};
\draw(-2,-4.8)node{$\ap^5 \zeta_5 D^6 F^4$};
\draw(-2,-5.8)node{$\ap^5 \zeta_2 \zeta_3 D^6 F^4$};
\draw(-2,-6.8)node{$\ap^6 \zeta_2^3 D^8 F^4$};
\draw(-2,-7.8)node{$\ap^6 \zeta_3^2 D^8 F^4$};
\draw(-2,-8.8)node{$\ap^7 \zeta_7 D^{10} F^4$};
\draw(-2,-9.8)node{$\ap^8 \zeta_2^4 D^{12} F^4$};
\draw(1.5,-0.8)node{$8$};
\draw(4.5,-0.8)node{$10$};
\draw(4.5,-1.8)node{$6$};
\draw(4.5,-2.8)node{$4$};
\draw(4.5,-3.8)node{$2$};
\draw(7,-0.8)node{$12$};
\draw(7,-1.8)node{$8$};
\draw(7,-2.8)node{$6$};
\draw(7,-3.8)node{$4$};
\draw(9,-0.8)node{$12$};
\draw(9,-1.8)node{$8$};
\draw(9,-2.8)node{$$};
\draw(9,-3.8)node{$4$};
\draw(9,-4.8)node{$2$};
\draw(9,-5.8)node{$$};
\draw(9,-6.8)node{$0$};
\draw(11,-0.8)node{$14$};
\draw(11,-1.8)node{$10$};
\draw(11,-2.8)node{$8$};
\draw(11,-3.8)node{$6$};
\draw(11,-4.8)node{$4$};
\draw(11,-5.8)node{$4$};
\draw(11,-6.8)node{$2$};
\draw(13,-0.8)node{$14$};
\draw(13,-1.8)node{$10$};
\draw(13,-2.8)node{$8$};
\draw(13,-3.8)node{$6$};
\draw(13,-4.8)node{$4$};
\draw(13,-5.8)node{$4$};
\draw(13,-6.8)node{$2$};
\draw(13,-7.8)node{$$};
\draw(13,-8.8)node{$0$};
\draw(13,-9.8)node{$-2$};
\endtikzpicture
\end{center}
\caption{The critical dimensions where the UV divergences of the operators in the first column
appear in the $\ap$-expansion of the degeneration $a^{\te{1-loop}}_{{\rm open}}$ of the 
one-loop open-string amplitude, see (\ref{MGFs.9}). Dashed vertical lines separate the contributions that survive upon symmetrizing open-string amplitudes over the color orderings (left) from those that cancel upon symmetrization (right).}
\label{UVopen}
\end{table}

By imposing the well-tested correspondence (\ref{eq:os-eps-t}) to hold
for higher orders of $a^{\te{1-loop}}_{{\rm open}}$ beyond
$\alpha'^3$, one can efficiently predict the Laurent polynomials of
elliptic multiple zeta values  up to the transcendentality where
$\widehat A^{\te{1-loop}}_{\te{eff}}$ is available.  
We have determined the loop integrand of $\widehat A^{\te{1-loop}}_{\te{eff}}(1,2,3,4)$ up to and including the order of $\alpha'^8$, as well as
$\ap^9\zeta_9$, giving the following new predictions for the coefficients
of the leading orders in the Laurent polynomials
\begin{align}
  \afouruv{\ap^8 \zeta_3 \zeta_5} &\rightarrow \frac{-4}{t^4} \left(4 s_{12}^4+11 s_{12}^3 s_{23}+6 s_{12}^2 s_{23}^2+11 s_{12} s_{23}^3+4 s_{23}^4\right) + \dots \notag\\
  \afouruv{\ap^8 \zeta_{3,5}} &\rightarrow\frac{4}{5t^4} \left(10 s_{12}^4+17 s_{12}^3 s_{23}-6 s_{12}^2 s_{23}^2+17 s_{12} s_{23}^3+10 s_{23}^4\right)  \nonumber\\
  &\quad+\frac{4s_{13}}{5t^3}(2s_{12}^4+5s_{12}^3s_{23}+10s_{12}^2s_{23}^2+5s_{12}s_{23}^3+2s_{23}^4) 
  +\ldots 
  \label{eq:os-high-zeta}\\
  \afouruv{\ap^8 \zeta_2 \zeta_3^2} &\rightarrow \frac{2}{t^3} s_{12}  s_{13} s_{23}  \left(s_{12}^2+s_{23}^2\right) + \dots \notag\\
  \afouruv{\ap^9 \zeta_9} &\rightarrow \frac{4}{t^5} \left(94 s_{12}^4+227 s_{12}^3 s_{23}+78 s_{12}^2 s_{23}^2+227 s_{12} s_{23}^3+94 s_{23}^4\right) + \dots \, .\notag
\end{align}
The arrows indicate that the integration measure $\pi^{D/2}\int^{\tmax}_0 \dd t\, t^{3-D/2} $ 
has been suppressed on the right-hand sides. The methods in both \cref{secMGF.2} and 
\cref{sec:alt-div} are straightforward to extend such that combinatorics become the limiting
factor in extracting any desired order in $t$ and $s_{ij}$, which we plan to address in a future work. In this way,
integrating polynomials in proper times
$u_j$ can in principle give access to the MZVs of any weight for which
the expansion of six-point disk integrals is tractable, at arbitrary
$\alpha'$-orders of $a^{\te{1-loop}}_{{\rm
    open}}$, bypassing the degeneration limits of large numbers of
elliptic multiple zeta values.

\subsubsection{Five Points}

At five points, the UV expansion of the three series of Feynman integrals in (\ref{openEP.5}) 
to (\ref{openEP.7}) determines all other cases via permutations of the external legs. The Schwinger
parametrization of these momentum-space expressions in the UV regime
perfectly lines up with the Laurent polynomials in (\ref{MGFs.15}), 
\beq
\pi^{-D/2} \, [ a_1a_2a_3a_4a_5\,  | \, E_{P} \rangle \, \Big|_{\rm UV}  = - \int^{\tmax}_0 \dd t \, t^{3-D/2} \, [ \! [ a_1a_2a_3a_4 a_5\,  | \,  E_{P} \rangle \! \rangle_{\tau} \, \big|_{\tau = \frac{ i t}{2\pi \ap} }
\label{WLeq14} \, ,
\eeq
for instance
\begin{align}
\pi^{-D/2} \, [ 12345 | E_{1|23,4,5} \rangle \, \Big|_{\rm UV} &=  \int^{\tmax}_0 \dd t \, t^{3-D/2} \, \bigg\{ {-}\frac{1}{6 s_{12}} - \frac{1}{6 s_{23}}   \nonumber \\*
&\ \ \ \ \ \ \  +  \bigg(  \frac{2 s_{14}}{s_{23}} + \frac{2 s_{35}}{s_{12}} - 3  \bigg) 
\bigg( \frac{ t}{240} + \frac{ \ap^2 \zeta_2}{2 t} - \frac{ 3 \ap^3 \zeta_3}{t^2} + \frac{6 \ap^4 \zeta_4}{t^3} \bigg)\bigg\}
\notag \\
\pi^{-D/2} \, [ 12345 | E_{1|24,3,5} \rangle \, \Big|_{\rm UV}  &=  \int^{\tmax}_0 \dd t \, t^{3-D/2} \, \bigg\{{-} \frac{ 1}{6 s_{12}}  \label{WLeq15}  \\*
&\ \ \ \ \ \ \   -  \bigg( 1 + \frac{2 s_{35}}{s_{12}} \bigg)
\bigg( \frac{ t}{240} + \frac{ \ap^2 \zeta_2}{2 t} - \frac{ 3 \ap^3 \zeta_3}{t^2} + \frac{6 \ap^4 \zeta_4}{t^3} \bigg) + {\cal O}(s_{ij}) \bigg\}
\notag\\
\pi^{-D/2} \,  [ 12345 | E^\mu_{1|2,3,4,5} \rangle \, \Big|_{\rm UV} &=   \int^{\tmax}_0 \dd t \, t^{3-D/2} \, \bigg\{   \frac{ k_5^\mu }{6 s_{15}} {-} \frac{ k_2^\mu }{6 s_{12}} {+}  \bigg(   k_{35}^\mu -  k_{24}^\mu + \frac{2 s_{35}}{s_{12}}  k_2^\mu - \frac{2 s_{24}}{s_{15}} k_5^\mu\bigg)
 \notag \\
&\ \ \ \ \ \ \ \ \ \ \ \ \ \ \ \ \ \  \ \ \ \times \bigg( \frac{ t}{240} + \frac{ \ap^2 \zeta_2}{2 t} - \frac{ 3 \ap^3 \zeta_3}{t^2} + \frac{6 \ap^4 \zeta_4}{t^3} \bigg)  + {\cal O}(s_{ij}) \bigg\}\, .
\notag 
\end{align}
The full one-loop matrix element can be assembled from the loop
integrals in $[ 12345 | E_P \rangle $ together with the
$t_8$-tensors as prescribed in (\ref{openEP.2}).  For the scalars
$t_8(12,3,4,5)$ and vectors $t_+^\mu(1,2,3,4,5)$ defined in
(\ref{t8defs}), the linearized gauge transformations are detailed in
(\ref{t8defs.6}).  At each order in $t$ and $\alpha'$, the Schwinger
integrand resulting from (\ref{WLeq15}) is a gauge invariant
combination of $t_8$-tensors which can be expressed in terms of
$A_{\rm SYM}^{\rm tree}$ by repeated use of (\ref{keyid}) and
\begin{align}\label{WLeq16}
&\frac{ t_8(12,3,4,5) }{s_{12}} + \text{cyc}(1,2,3) = s_{45} \Big[
s_{34} A_{\rm SYM}^{\rm tree}(1,2,3,4,5)
- s_{24} A_{\rm SYM}^{\rm tree}(1,3,2,4,5)
\Big] \, . 
\end{align}
By furthermore reducing the SYM trees to 
the BCJ basis $\{ A_{\rm SYM}^{\rm tree}(1{,}2{,}3{,}4{,}5),A_{\rm SYM}^{\rm tree}(1{,}3{,}2{,}4{,}5)\}$, 
each order in the Schwinger integrands of $A^{\te{1-loop}}_{\rm eff}(1,2,3,4,5)$
can be characterized by a $2\times 2$ matrix with $s_{ij}$-dependent entries acting
on the two-component vector of $A_{\rm SYM}^{\rm tree}$, 
\begin{align}
&\left( \begin{array}{c}
A^{\te{1-loop}}_{\rm eff}(1,2,3,4,5) \\
A^{\te{1-loop}}_{\rm eff}(1,3,2,4,5) 
\end{array} \right) \Bigg|_{\rm UV} =
\int^{\tmax}_0 \dd t \, t^{3-D/2} \, \bigg\{  \frac{1}{6} P_2 \label{WLeq17}
\\
& \ \ \ \ \ \ +  \bigg( \frac{ t}{120} + \frac{ \ap^2 \zeta_2}{ t} - \frac{ 6 \ap^3 \zeta_3}{t^2} + \frac{12 \ap^4 \zeta_4}{t^3} \bigg)  M_3
 + {\cal O}(s_{ij}^4) \bigg\} 
\left( \begin{array}{c}
A^{\te{tree}}_{\rm SYM}(1,2,3,4,5) \\
A^{\te{tree}}_{\rm SYM}(1,3,2,4,5) 
\end{array} \right) \, .\notag
\end{align}
The entries of the $2\times 2$ matrices $P_w$ and $M_w$ are degree-$w$ polynomials in $s_{ij}$ with
rational coefficients \cite{Schlotterer:2012ny}, for instance
\beq
P_2 = \left( 
\begin{array}{cc} 
s_{12} s_{34} {-}s_{34}s_{45}{-}s_{15}s_{12} &s_{13}s_{24} \\
s_{12}s_{34} &s_{13} s_{24} {-}s_{24}s_{45}{-}s_{15}s_{13} 
\end{array}
\right) \, .
 \label{WLeq18}
 \eeq
The cases in (\ref{WLeq17}) are known from the $\alpha'$-expansion of five-point disk integrals,
and their explicit form at $w\leq 22$ can be downloaded from \cite{MZVWebsite}. The leading term
in (\ref{WLeq17}) arises from 
\beq
\frac{ t_8(12,3,4,5) }{s_{12}} + {\rm cyc}(1,2,3,4,5) = - \Big[(P_2)_{11} A^{\te{tree}}_{\rm SYM}(1,2,3,4,5)
+(P_2)_{12} A^{\te{tree}}_{\rm SYM}(1,3,2,4,5)\Big] \,,
 \label{WLeq19}
\eeq
which is the representation of \cite{Mafra:2011nv} for the color-ordered tree-level matrix element with 
a single-insertion of the effective operator $\alpha'^2  \zeta_2 {\rm Tr}(F^4)$. Higher orders
of (\ref{WLeq18}) will involve additional $2\times 2$ matrices beyond the $P_w$ and $M_w$
from genus-zero integrals, starting with $L_4$ in section 5.1 of \cite{Broedel:2014vla}.

\subsection{UV divergences from closed-string effective actions}
\label{secMGF.4}

Closed strings feature a similar correspondence between UV divergences in the effective field
theory and Laurent polynomials in the expansion of torus integrals around the cusp. The closed-string
amplitude is free of UV divergences as can be understood from the bound $\Im(\tau) \geq \sqrt{3}/2$ for 
the modular parameters $\tau \in \mathfrak{F}$ in (\ref{prop.41}): by the shape of the fundamental
domain of ${\rm SL}_2(\mathbb Z)$, modular invariance 
imposes an UV cutoff on the Schwinger parameter~$t$ which is absent for the Feynman integrals in
the $\alpha'$-expansion of the matrix elements~$M^{\te{1-loop}}_{n,{\rm eff}}$.

\subsubsection{Four points}

We again peel off a global factor of $|t_8(f_1,f_2,f_3,f_4)|^2$ and discuss four-point gravitational one-loop matrix
elements at the level of the scalar integrals $\widehat M^{\te{1-loop}}_{4,{\rm eff}}$ in (\ref{exsec.31}).
Based on standard Schwinger parametrizations of the Feynman integrals in (\ref{exsec.33}), we find
\begin{align}
\pi^{-D/2} &\widehat M^{\te{1-loop}}_{4,{\rm eff}} \, \Big|_{\rm UV} = \int^{\tmax}_0 \dd t \, t^{3-D/2}
\, \bigg\{
\bigg[1 + \frac{t^2 }{360} (s_{12}^2{-}s_{13}s_{23}) - \frac{ t^3 s_{12}s_{13}s_{23} }{6048} + \ldots\bigg]
\label{WLUV.41}\\
&\ \ + \ap^3 \zeta_3 \bigg[ \frac{ 8}{t} (s_{12}^2 {-}s_{13}s_{23}) - s_{12}s_{13}s_{23} + \ldots \bigg]
+ \ap^5 \zeta_5 \bigg[  {-} \frac{ 60 s_{12}s_{13}s_{23} }{t^2} + \ldots \bigg]
+ {\cal O}(\ap^6) \bigg\} \, . \notag 
\end{align}
At higher order $\ap^6\zeta_3^2$ and $\ap^7\zeta_7$, the numerators in (\ref{exsec.37}) and (\ref{exsec.38})
yield UV divergences
\begin{align}
\pi^{-D/2} \widehat M^{\te{1-loop}}_{4,{\rm eff}} \, \Big|^{\ap^6 \zeta_3^2}_{\rm UV}
&= \int^{\tmax}_0 \dd t \, t^{3-D/2}\, \bigg\{
\frac{32\sigma_4}{t^2}
-\frac{8\sigma_5}{t} 
+ \frac{12\sigma_6+\sigma_3^2}{15}
 -\frac{43\sigma_7}{630} \, t  + \ldots \bigg\}\,,
\notag \\
\pi^{-D/2} \widehat M^{\te{1-loop}}_{4,{\rm eff}} \, \Big|^{\ap^7 \zeta_7}_{\rm UV}
&= \int^{\tmax}_0 \dd t \, t^{3-D/2}\, \bigg\{
 \frac{128\sigma_4}{t^3}
 -\frac{98\sigma_5}{t^2}  - \frac{36\sigma_6-46\sigma_3^2}{5t}\label{WLUV.51} \\
 &\ \ \ \ \ \ \ \ \ \ \ \ \ \ \ \ \ \ \ \ \ \ \
  -\frac{89\sigma_7}{210}
 +\frac{1959\sigma_8+121\sigma_4^2}{18900} \, t   + \ldots \bigg\} \, , \notag
\end{align}
where the Mandelstam variables are organized in terms of the symmetric polynomials
\beq
\sigma_k = \frac{1}{k} \, (s_{12}^k + s_{23}^k + s_{13}^k) \, .
\label{WLUV.52}
\eeq
Similar to the open-string discussion in the previous section, we compare the $t$-integrand of
(\ref{WLUV.41}) and (\ref{WLUV.51}) with the
Laurent polynomials of the modular graph forms in (\ref{MGFs.4}). For closed strings, the dictionary between 
Schwinger parameter of the worldline and modular parameter of the worldsheet is\footnote{Similar to
(\ref{dictopen}), the dictionary (\ref{dictclosed}) can be validated by matching the contribution $ e^{- t s_{ij} u_{ij}^2/t}$
to the worldline Koba-Nielsen factor with the contribution $e^{-2\pi \ap \Im \tau  s_{ij} u_{ij}^2}$ to the closed-string
Koba-Nielsen factor in (\ref{MGFs.1}). Alternatively, one can compare the $D$-dependence on the right-hand side of $\widehat M^{\te{1-loop}}_{4,{\rm eff}} \, \big|_{\rm UV} = \int^{\tmax}_0 \dd t \, t^{3} ( \frac{ \pi }{t} )^{D/2}$ with the factor of $(2 \sqrt{ \ap \Im \tau})^{10-D}$ in $D\neq 10$-dimensional
closed-string amplitudes from the $D$-dependent loop integral.}
\beq
\te{closed strings:} \ \ \ t = 4 \pi \ap \Im( \tau ) = 4 \ap y \, .
\label{dictclosed}
\eeq
Indeed, (\ref{WLUV.41}) and higher-order terms in $t$ or $\alpha'$ are consistent with
\begin{equation} 
\pi^{-D/2}   \widehat M^{\te{1-loop}}_{4,{\rm eff}}   \, \Big|_{\rm UV} 
  = \int^{\tmax}_0 \dd t \, t^{3-D/2} \, m^{\te{1-loop}}_{4,{\rm closed}} \Big(\Im \tau = \frac{t}{4\pi \ap} \Big)  \,.
  \label{WLUV.44}
\end{equation}
The $\zeta_3 \zeta_5$ term is a further verification of this correspondence,
given by
\begin{equation}
  \pi^{-D/2} \widehat M^{\te{1-loop}}_{4,{\rm eff}} \, \Big|^{\ap^8 \zeta_3\zeta_5}_{\rm UV}
  = \int_0^{\tmax} \dd t\, t^{3-D/2} \, \bigg\{ 
   \frac{32 }{t^3}  s_{12} s_{23} s_{13}( s_{13} s_{23}{-}s_{12}^2)+ \dots \bigg\}
   \label{thez3z5term}
\end{equation}
which indeed matches the corresponding coefficient in \cref{MGFs.4}.
Moreover, the absence of $t^{-4}$ terms in (\ref{thez3z5term}) is consistent with the single-valued map (\ref{exsec.32}) and the vanishing of the $\zeta_3 \zeta_5t^{-4},\zeta_{3,5}t^{-4}$ terms in 
(\ref{eq:os-high-zeta}) upon symmetrization in $s_{ij}$.
Again, a fixed $\alpha'$-order of $m^{\te{1-loop}}_{4,{\rm closed}} $
is already sensitive to UV divergences from effective operators at
higher order in $\alpha'$. 

As a closed-string analogue of \cref{negdims,eq:os-high-zeta}, we also obtain UV divergences from
Schwinger integrands $\dd t \, t^{-D/2-1}$ and lower powers of $t$, 
which in the language of \cref{sec:alt-div} correspond to a divergence in formal spacetime
dimensions $\leq 0$. The simplest such example is a zero-dimensional UV divergence from
\begin{equation}
  \mfouruv{\ap^9 \zeta_9} = \int^{\tmax}_0 \dd t \, t^{3-D/2}  \, \bigg\{ 
 \frac{504}{t^4} s_{12} s_{13} s_{23} ( s_{13} s_{23} {-} s_{12}^2) 
  + \dots \bigg\}
  \label{eq:closed-h-zeta}
\end{equation}
which we have assembled from \cref{eq:os-high-zeta} and the single-valued map,
see (\ref{exsec.32}). The UV divergences of gravitational effective interactions that we have matched with Laurent
polynomials in (\ref{MGFs.4}) are summarized in \cref{UVclosed}.

One can again reverse the logic and rely on the well-tested loop integrands in (\ref{exsec.33})
and (\ref{exsec.34}) to determine Laurent polynomials at higher order in $\alpha'$. The numerators in
(\ref{exsec.35}) to (\ref{exsec.38}) together with the straightforward integrals over Schwinger proper
times in (\ref{WLUV.3}) can be used to determine the appearance of $\zeta_3,\zeta_5,\zeta_3^2$ and 
$\zeta_7$ in the Laurent polynomials at arbitrary $\alpha'$-orders of $m^{\te{1-loop}}_{4,{\rm closed}} $.
This settles a large number of terms in the degeneration of infinitely many four-point modular graph forms.
Since most of the closed-form results on Laurent polynomials concern the considerably simpler 
{\it two-point} modular graph forms \cite{DHoker:2019xef, Zagier:2019eus, Gerken:2020yii}, 
the correspondence (\ref{WLUV.44}) with UV divergences is a valuable source of mathematical information.

\begin{table}
\begin{center}
\tikzpicture [scale=0.7, line width=0.30mm]
\draw(-4,0)--(18,0);
\draw(0,1)--(0,-8.5);
\draw(3,1)--(3,-8.5);
\draw(6,1)--(6,-8.5);
\draw(9,1)--(9,-8.5);
\draw(12,1)--(12,-8.5);
\draw(15,1)--(15,-8.5);
\draw(0,0) -- (-4,1);
\draw(-1,0.8)node{\footnotesize $\alpha'$-order};
\draw(-3,0.3)node{\footnotesize eff.\ op's};
\draw(1.5,0.6)node{$0$};
\draw(4.5,0.6)node{$1$};
\draw(7.5,0.6)node{$2$};
\draw(10.5,0.6)node{$3$};
\draw(13.5,0.6)node{$4$};
\draw(16.5,0.6)node{$5$};
\draw(-2,-0.8)node{$R$};
\draw(-2,-1.8)node{$\ap^3 \zeta_3   R^4$};
\draw(-2,-2.8)node{$\ap^5 \zeta_5  D^4 R^4$};
\draw(-2,-3.8)node{$\ap^6 \zeta_3^2  D^6 R^4$};
\draw(-2,-4.8)node{$\ap^7 \zeta_7 D^8 R^4$};
\draw(-2,-5.8)node{$\ap^8 \zeta_3 \zeta_5 D^{10} R^4$};
\draw(-2,-6.8)node{$\ap^9 \zeta_9 D^{12} R^4$};
\draw(-2,-7.8)node{$\ap^9 \zeta_3^3 D^{12} R^4$};
\draw(1.5,-0.8)node{$8$};
\draw(7.5,-0.8)node{$12$};
\draw(7.5,-1.8)node{$6$};
\draw(10.5,-0.8)node{$14$};
\draw(10.5,-1.8)node{$8$};
\draw(10.5,-2.8)node{$4$};
\draw(13.5,-0.8)node{$16$};
\draw(13.5,-1.8)node{$10$};
\draw(13.5,-2.8)node{$6$};
\draw(13.5,-3.8)node{$4$};
\draw(13.5,-4.8)node{$2$};
\draw(16.5,-0.8)node{$18$};
\draw(16.5,-1.8)node{$12$};
\draw(16.5,-2.8)node{$8$};
\draw(16.5,-3.8)node{$6$};
\draw(16.5,-4.8)node{$4$};
\draw(16.5,-5.8)node{$2$};
\draw(16.5,-6.8)node{$0$};
\endtikzpicture
\end{center}
\caption{The critical dimensions where the UV divergences of the operators in the first column
appear in the $\ap$-expansion of genus-one amplitudes in their Laurent polynomials (\ref{MGFs.4}) at the cusp.}
\label{UVclosed}
\end{table}

Finally, the comparison of UV divergences of gauge and gravitational one-loop matrix elements offers
a new perspective on the single-valued map (\ref{MGFs.11}) between Laurent polynomials of elliptic
multiple zeta values and modular graph forms. Based on tree-level results, $\widehat M^{\te{1-loop}}_{4,{\rm eff}} $
and the symmetrized version of $ \widehat A^{\te{1-loop}}_{\te{eff}}(1,2,3,4)$ are related by the single-valued
map of MZVs as in (\ref{exsec.32}). Moreover, the respective dictionaries (\ref{dictopen}) and (\ref{dictclosed}) between 
UV divergences and Laurent polynomials preserve the single-valued map $\pi \tau \rightarrow 2\pi i \Im \tau$ 
in (\ref{MGFs.12}). On these grounds, the relation (\ref{exsec.32}) between $\widehat M^{\te{1-loop}}_{4,{\rm eff}} $
and $ \widehat A^{\te{1-loop}}_{\te{eff}}$ implies the connection
(\ref{MGFs.11}) between Laurent polynomials of elliptic
multiple zeta values and modular graph forms.

\subsubsection{Five Points}

We continue the matching at five points, analyzing the pairings
$\langle E_Q | E_P \rangle$ in (\ref{closedEP}).  We have verified that the
Schwinger parametrizations of the Feynman integrals match the Laurent
polynomials of the torus integrals
$\langle \! \langle E_Q | E_P \rangle \! \rangle_\tau$ in
(\ref{MGFs.6}). For instance, the leading orders in the UV expansions
of the results in section \ref{simcomb} such as
\begin{align}
 \pi^{-D/2} \, \langle E_{1|23,4,5} | E_{1|24,3,5} \rangle \, \Big|_{\rm UV} &= \int_0^{\tmax} \dd t\, t^{3-D/2} \, \bigg\{ {-} \frac{1}{s_{12}} \notag \\
 & + 
\bigg( \frac{t^2}{720} + \frac{4 \ap^3 \zeta_3}{t} \bigg) \bigg(  \frac{2(s_{35} s_{45} - s_{34}^2)}{s_{12}} - s_{12} +   5  s_{34}\bigg) 
+ \ldots  \bigg\} \notag\\
 \pi^{-D/2} \, \langle E_{1|23,4,5} | E_{1|45,2,3} \rangle  \, \Big|_{\rm UV} &=  \int_0^{\tmax} \dd t\, t^{3-D/2} \, \bigg\{
 \bigg(
 \frac{t^2}{720} {+} \frac{4 \ap^3 \zeta_3}{t}  \bigg) (  s_{24}  {+} s_{35} {-} s_{25} {-} s_{34}) + \ldots \bigg\} \notag \\
 \pi^{-D/2} \, \langle E_{1|23,4,5} | E^\mu_{1|2,3,4,5} \rangle  \, \Big|_{\rm UV} &=  \int_0^{\tmax} \dd t\, t^{3-D/2} \, \bigg\{
 \frac{k_3^\mu}{ s_{13}}   - \frac{k_2^\mu}{s_{12}}  +  \bigg( \frac{t^2}{720} + \frac{4 \ap^3 \zeta_3}{t}  \bigg)  \label{WLeq.7}\\
 &\ \ \ \ \ \times \bigg( 
  k_3^\mu \frac{(s_{24}^2 {+} s_{25}^2 {+} s_{45}^2)}{s_{13} }
  -k_2^\mu \frac{(s_{34}^2 {+} s_{35}^2 {+} s_{45}^2)}{s_{12}}  \notag \\
  &\ \ \ \ \ \ \ \ \ \  +(k_3^\mu{-}k_2^\mu) s_{45}+ 
    k_4^\mu  (s_{24} {-} s_{34}) + k_5^\mu  (s_{25} {-} s_{35})\bigg)
+\ldots \bigg\} \notag
\end{align}
are related to the degenerate torus integrals in (\ref{MGFs.7})
according to the general dictionary 
\beq
 \pi^{-D/2} \, \langle E_{Q} | E_{P} \rangle \, \Big|_{\rm UV} = - \int^{\tmax}_0  \dd t\, t^{3-D/2} 
\, \langle \! \langle E_Q | E_P \rangle \! \rangle_\tau \, \big|_{\Im \tau = \frac{ t} {4\pi \ap} } \, ,
\label{WLeq.8}
\eeq
see (\ref{WLeq14}) for its open-string analogue.

The one-loop matrix element $M^{\te{1-loop}}_{5,{\rm eff}} $ is given
in terms of $7\times 7$ pairings $ \langle E_{Q} | E_{P} \rangle$ and
bilinears of $t_8$-tensors specified in (\ref{openEP.3}).  In the UV
regime (\ref{WLUV.3}) of the Schwinger parametrization, gauge invariance again holds
order by order in $t$ and $\alpha'$. Similar to the open-string
five-point matrix element (\ref{WLeq17}), the tensor structures of
type-IIB components can be expressed in terms of
$A^{\rm tree}_{\rm SYM}$. The type-IIA components are related via
(\ref{IIBvsIIA}) by a sign-flip in one of the parity-odd terms in (\ref{t8defs}).
The coefficients of $A^{\rm tree}_{\rm SYM}$ bilinears
in $M^{\te{1-loop}}_{5,{\rm eff}} \, \big|_{\rm IIB} $ now depend on the total R-symmetry 
charges of the external legs in the type-IIB supergravity multiplet. For instance, 
the leading order in the UV expansion leads to the 
following $R^4$-type kinematic factor \cite{Green:2013bza, Mafra:2015mja} 
\begin{align}
&\bigg[ \frac{ t_8(12,3,4,5) \bar t_8(12,3,4,5) }{s_{12} } + (1{,}2|1{,}2{,}3{,}4{,}5) \bigg] + \frac{1}{2} t^\mu_+(1,2,3,4,5) \eta_{\mu \nu} \bar t^\nu_+(1,2,3,4,5) \, \big|_{\rm IIB} \nonumber \\
&\hspace{2cm} =  \left( \begin{array}{c}
A^{\te{tree}}_{\rm SYM}(1{,}2{,}3{,}5{,}4) \\
A^{\te{tree}}_{\rm SYM}(1{,}3{,}2{,}5{,}4) 
\end{array} \right)^{\! \! {\rm T}}
S_0\cdot M_3
\left( \begin{array}{c}
A^{\te{tree}}_{\rm SYM}(1{,}2{,}3{,}4{,}5) \\
A^{\te{tree}}_{\rm SYM}(1{,}3{,}2{,}4{,}5) 
\end{array} \right) \, \left\{ \begin{array}{cl}
-1 &: \ h^5 \\
\frac{D{-}8}{6} &: \ \phi_D h^4
\end{array} \right. \,, \label{WLeq.38}
\end{align}
where the $2\times 2$ matrix $S_0=\big(\begin{smallmatrix}
s_{12}(s_{13}{+}s_{23})  &s_{12}s_{13}  \\
s_{12}s_{13} &s_{13}(s_{12}{+}s_{23})
\end{smallmatrix} \big)$ is the five-point instance of the field-theory KLT kernel~(\ref{prop.10}). 
The extra factor of ${(D{-}8)}/{6}$ arises for four gravitons $h$ and one  
$D$-dimensional dilaton $\phi_D$ as well as any
other external-state configuration with the same overall R-symmetry charges
as $\phi_D h^4$. The type-IIA counterpart of the eight-dimensional UV divergence 
(\ref{WLeq.38}) has $\bar t^\nu_-$ in the place of $\bar t^\nu_+$.
The type-IIA components with four gravitons and one $B$-field are parity odd 
and proportional to $ \varepsilon_{10}(B_1,f_2,f_3,f_4,f_5)t_8(f_2,f_3,f_4,f_5) $ \cite{Green:2013bza}
whereas $B h^4$ interactions in the type-IIB theory vanish to all orders in $\alpha'$ \cite{Vafa:1995fj}.

In the complete five-point type-IIB matrix element, we find
bilinears in $A^{\rm tree}_{\rm SYM}$ at all orders of the UV expansion. For five external 
gravitons (or other R-symmetry-conserving state configurations), the coefficients at the leading
orders are
\begin{align}
M^{\te{1-loop}}_{5,{\rm eff}} \, \Big|_{\rm UV}^{{\rm IIB}, h^5} &=  \left( \begin{array}{c}
A^{\te{tree}}_{\rm SYM}(1{,}2{,}3{,}5{,}4) \\
A^{\te{tree}}_{\rm SYM}(1{,}3{,}2{,}5{,}4) 
\end{array} \right)^{\rm T} \cdot S_0 \int^{\tmax}_0  \dd t\, t^{3-D/2} 
\, \bigg[ 
{-}  M_3 -2  \bigg(  \frac{t^2}{720} {+} \frac{ 4 \ap^3 \zeta_3}{t}  \bigg)   M_5  \nonumber \\
&  \ \ \ \
- \bigg( \frac{t^3}{6048} + \ap^3 \zeta_3 + \frac{60 \ap^5 \zeta_5}{t^2} \bigg)   M_3^2 + {\cal O}(s_{ij}^7)
\bigg]
\,
\left( \begin{array}{c}
A^{\te{tree}}_{\rm SYM}(1{,}2{,}3{,}4{,}5) \\
A^{\te{tree}}_{\rm SYM}(1{,}3{,}2{,}4{,}5) 
\end{array} \right)\, ,\label{WLeq.39}
\end{align}
where higher orders in $s_{ij}$ involve $2\times 2$ matrices $M'_{w\geq 7}$ beyond the tree-level 
ones $M_w$ \cite{Green:2013bza}. For R-symmetry violating type-IIB states such as $\phi_D h^4$,
the numerical coefficients change to 
\begin{align}
M^{\te{1-loop}}_{5,{\rm eff}} &\Big|_{\rm UV}^{{\rm IIB}, \phi_D h^4} =  \left( \begin{array}{c}
A^{\te{tree}}_{\rm SYM}(1{,}2{,}3{,}5{,}4) \nonumber \\
A^{\te{tree}}_{\rm SYM}(1{,}3{,}2{,}5{,}4) 
\end{array} \right)^{\rm T} \cdot S_0 \int^{\tmax}_0  \dd t\, t^{3-D/2} \\*
&\times  \bigg[  
\frac{ D{-}8}{6} M_3 +  \frac{ D{-}12 }{5} \bigg( \frac{t^2}{720} + \frac{4 \ap^3 \zeta_3}{t}  \bigg)    M_5     \label{WLeq.40}  \\*
&\qquad + \frac{D{-}14}{12} 
\bigg( \frac{t^3}{6048} + \ap^3 \zeta_3 + \frac{60 \ap^5 \zeta_5}{t^2} \bigg)  M_3^2 + {\cal O}(s_{ij}^7)
\bigg]
\,
\left( \begin{array}{c}
A^{\te{tree}}_{\rm SYM}(1{,}2{,}3{,}4{,}5) \\
A^{\te{tree}}_{\rm SYM}(1{,}3{,}2{,}4{,}5) 
\end{array} \right) \, .\nonumber
\end{align} 
The $D$-dimensional relative factors in comparison to (\ref{WLeq.39})
play an important role for the S-duality properties of type-IIB string
amplitudes \cite{Green:2013bza, Mafra:2015mja, Green:2019rhz,
  DHoker:2020tcq}. From a supergravity point of view, the factors of
$D{-}8, \ D{-}12$ and $D{-}14$ in (\ref{WLeq.40}) prevent
R-symmetry violation through the respective UV divergence in $8,12$ and $14$
spacetime dimensions \cite{Mafra:2015mja, Pioline:2018pso, DHoker:2020tcq}.


\section{Comparison with non-analytic terms in string amplitudes}
\label{mainsec7}

On top of the analysis of UV divergences in the previous section, we can also
relate the discontinuities of one-loop matrix elements to string-theory computations.
Unitarity at the level of the $\alpha'$-expansion implies that the discontinuities of
one-loop string amplitudes in $s_{ij}$ agree with those of $A^{\te{1-loop}}_{{\rm eff}}(1,2,\ldots,n)$
and $M^{\te{1-loop}}_{n,{\rm eff}}$. We shall first confirm this for the four-point closed-string
results in the literature and then spell out predictions for the non-analytic sector of one-loop
four-point open-string amplitudes.


\subsection{Reproducing non-analytic terms of closed superstrings}
\label{subsec7.1}

The non-analytic part of the four-point one-loop amplitude of type-II
superstrings is known to all orders in $\alpha'$
\cite{DHoker:2019blr}, also see \cite{Green:1999pv, Green:2008uj,
  DHoker:2015gmr} for earlier results at the leading
$\alpha'$-orders. Based on the Feynman integrals in section
\ref{sec:4.8}, we find the following logarithmic behaviour in the
low-energy expansion 
\begin{align}
M^{\te{1-loop}}_{4,{\rm eff}}\, \big|_{\rm disc}&=  M^{\te{1-loop}}_{4,{\rm SUGRA}} \, \big|_{\rm disc}
+  |t_8(f_1,f_2,f_3,f_4)|^2 
\bigg\{{-} \frac{\alpha'^4\zeta_3}{45}s_{12}^4+\frac{\alpha'^6\zeta_5}{1260}s_{12}^4(s_{13}s_{23}{-}22s_{12}^2) \notag \\
&\! \! \!\! \! \!\! \! \! +\frac{\alpha'^7\zeta_3^2}{1260}s_{12}^5(12s_{12}^2+s_{13}s_{23})
-\frac{\alpha'^8\zeta_7}{18900}s_{12}^4(260s_{12}^4
-25  s_{12}^2 s_{13} s_{23} + 2  s_{13}^2 s_{23}^2)
\label{nonanalyticclosed}\\
&\! \! \!\! \! \!\! \! \!  + \frac{ \alpha'^9 \zeta_3 \zeta_5 }{4725} s_{12}^5
(70 s_{12}^4 + 5 s_{12}^2  s_{13} s_{23} - s_{13}^2 s_{23}^2)
 + {\cal O}(\alpha'^{10})\bigg\} \log\left(\frac{s_{12}}{\mu^2}\right)+\text{cyc}(2,3,4)\, ,
\notag
\end{align}
where $ \big|_{\rm disc}$ instructs to disregard any analytic dependence on $\alpha'$. In this way,
the polynomials in $s_{ij}$ introduced by a change of the scale $\mu$ do not affect the terms under
consideration, see \cref{app:Feynints} for the origin of the terms $\log(\frac{s_{ij}}{\mu^2})$. To the
orders shown, (\ref{nonanalyticclosed}) exactly matches the non-analytic terms of \cite{DHoker:2019blr}
obtained from the $\alpha'$-expansions in Theorem 4.1 in version 2 of the reference.\footnote{We are
grateful to Eric D'Hoker and Michael Green for correspondence on a factor of 4 that was
changed since the first version of \cite{DHoker:2019blr}.}

A naive power-series expansion of the integrals (\ref{MGFs.4}) over closed-string punctures
cannot detect the logarithmic dependence on $s_{ij}$. Still, the divergent $\tau$-integrals over
certain modular graph forms in the $\alpha'$-expansion of (\ref{MGFs.4}) have crucial information
on the non-analytic sector of one-loop closed-string amplitudes, see
\cite{Green:1999pv, Green:2008uj, DHoker:2015gmr, DHoker:2019blr} for further details.


\subsection{Predicting non-analytic terms of open superstrings}
\label{subsec7.2}

As an open-string counterpart of (\ref{nonanalyticclosed}), we can
assemble the discontinuities of the Feynman integrals in the
four-point one-loop matrix element of section \ref{sec:4.1},
\begin{align}
&A^{\te{1-loop}}_{{\rm eff}}(1,2,3,4)\, \big|_{\rm disc}=  A^{\te{1-loop}}_{{\rm SYM}}(1,2,3,4) \, \big|_{\rm disc} \\
&\ \ \ +  t_8(f_1,f_2,f_3,f_4) \bigg\{ f(s_{12},s_{23}) \log\left(\frac{s_{12}}{\mu^2}\right)
+f(s_{23},s_{12}) \log\left(\frac{s_{23}}{\mu^2}\right) \bigg\} \,,\notag
\end{align}
where\footnote{We are grateful to Lorenz Eberhardt and Sebastian Mizera for
pointing out typos in the exponents of $s_{12}$ and $s_{23}$ in the coefficient
of $\zeta_5$ which we corrected in the present version.}
\begin{align}
f(s_{12},&s_{23}) = \frac{\ap^3 \zeta_2 }{180} s_{12}^3
-\frac{\ap^4 \zeta_3}{1260} s_{12}^3(4s_{12}{+}s_{23})
+\frac{\ap^5 \zeta_2^2}{12600} s_{12}^3(46s_{12}^2{-}2s_{12}s_{23}{+}s_{23}^2)  \notag \\
&-\frac{\ap^6 \zeta_2\zeta_3}{5040} s_{12}^4(12s_{12}^2{+}s_{13}s_{23})
-\frac{\ap^6 \zeta_5}{15120} s_{12}^3(
 38s_{12}^3{+}12s_{12}^2s_{23}{+}3s_{12}s_{23}^2{+}
s_{23}^3) \nonumber\\
&+\frac{\ap^7\zeta_3^2}{30240} s_{12}^4(20s_{12}^3{+}2s_{12}^2s_{23}{-}3s_{12}s_{23}^2{-}s_{23}^3)
 \notag \\
&+\frac{\ap^7 \zeta_2^3}{5292000} s_{12}^3(12320s_{12}^4{-}275s_{12}^3s_{23}{+}357s_{12}^2s_{23}^2{-}36s_{12}s_{23}^3{+}32s_{23}^4)\label{opendisc}\\
&-\frac{\ap^8\zeta_7}{831600}  s_{12}^3(1660s_{12}^5{+}645s_{12}^4s_{23}{+}296s_{12}^3s_{23}^2{+}137s_{12}^2s_{23}^3{+}36s_{12}s_{23}^4{+}10s_{23}^5)\nonumber\\
&-\frac{\ap^8\zeta_2^2\zeta_3}{756000} s_{12}^4(1200s_{12}^4{-}115s_{12}^3s_{23}{-}48s_{12}^2s_{23}^2
{-}s_{12}s_{23}^3-8s_{23}^4)\nonumber\\
&-\frac{\ap^8\zeta_2\zeta_5}{37800} s_{12}^4(70s_{12}^4{-}5s_{12}^3s_{23}{-}6s_{12}^2s_{23}^2{-}2s_{12}s_{23}^3{-}s_{23}^4)+ {\cal O}(\ap^9) \, .\notag
\end{align}
This sets our prediction for the non-analytic terms in the planar-cylinder contribution to the four-point
one-loop open-string amplitude.
In addition to the above terms, we note that the $\zeta_{3,5}$
integrand only contains external bubbles (\ie Figure~\ref{fig:ffourdiagb}) and
tadpoles, which integrate to zero in dimensional reduction.  As such,
there are no terms of the form $\ap^9\zeta_{3,5} \log( \frac{s_{ij}}{\mu^2})$ as required
by unitarity (there is no $\zeta_{3,5}$ in the four-point tree-level amplitude).
Again, the logarithms in $s_{ij}$ can be anticipated from the divergent
$\tau$ integrals in a naive power-series expansion of the integrals
over the open-string punctures (with elliptic multiple zeta values in the $\tau$-integrand). 
With the notable exception at the subleading $\alpha'$-order discussed
in section 4.2 of \cite{Hohenegger:2017kqy}, the interplay between the analytic and non-analytic sector
of one-loop open-string amplitudes has not yet been studied at the level of explicit higher-order results.


\section{Conclusion and outlook}
\label{sec:out}

In this work, we have used a confluence of ideas from ambitwistor, conventional string and (effective) field
theories to obtain one-loop matrix elements of higher-mass-dimension operators 
$D^{2k} F^n$ and $D^{2k}R^n$. Our method applies to the supersymmetric
operators in the tree-level low-energy effective actions of type-I and type-II superstrings
and is driven by the forward limits of the disk and sphere integrals in massless tree
amplitudes. Both the one-loop matrix elements and their tree-level building blocks are
organized in a series expansion in the inverse string tension $\alpha'$. Moreover, one 
can further distinguish different operators at the same mass dimension from the 
conjecturally $\mathbb Q$-linearly independent multiple zeta values in their coefficients.

Our construction passes a variety of consistency checks including a matching of the matrix elements' UV 
divergences and discontinuities with expectations based on one-loop superstring amplitudes.
\begin{itemize}
\item The UV divergences of the loop integrals associated with various $\alpha'$-orders and 
spacetime dimensions line up with the degeneration limits of genus-one integrals over torus  or cylinder 
punctures. This was confirmed in all cases we considered, and by imposing consistency at higher orders, 
our proposal offers a new window into the expansion of modular graph forms
and elliptic multiple zeta values around the cusp.
\item The discontinuities of the one-loop matrix elements in this work via logarithmic dependence on Mandelstam invariants reproduce those of the one-loop four-graviton amplitude of type II superstrings (which is verified up to and including the order of $\alpha'^9$ \cite{DHoker:2019blr}). For open strings at $\geq 4$ points and closed strings at $\geq 5$ points, our one-loop matrix elements encode new results on the non-analytic sectors of one-loop string amplitudes.
\end{itemize}
By their construction from genus-zero integrals, the
one-loop matrix elements evidently enjoy an echo of the monodromy relations 
\cite{Plahte:1970wy, BjerrumBohr:2009rd, Stieberger:2009hq}
and KLT relations \cite{Kawai:1985xq} of open- and closed-string tree amplitudes, respectively.
These properties are spelt out for the loop integrand of $n$-point matrix elements 
and furnish an $\alpha'$-uplift of one-loop BCJ and KLT relations
among field-theory amplitudes in the formulation of \cite{He:2016mzd, He:2017spx}. The 
monodromy and KLT relations of one-loop matrix elements apply to the representation of their 
loop integrand where the inverse Feynman propagators are linearized
in loop momenta as familiar from ambitwistor strings \cite{Geyer:2015bja, Geyer:2015jch}.
It would be interesting to further explore the practical implications of the one-loop monodromy relations of
open-string amplitudes as given in \cite{Tourkine:2016bak, 
Hohenegger:2017kqy, Ochirov:2017jby, Tourkine:2019ukp, Casali:2019ihm, Casali:2020knc, Stieberger:2021daa}, as well as tentative KLT relations between one-loop open- and closed-string amplitudes that do not rely on linearized propagators, or that hold after integration.

Monodromy and KLT relations of one-loop matrix elements reflect the underlying
color-kinematics duality and double-copy properties of effective-field-theory operators,
order by order in $\alpha'$. This suggests an important line of follow-up work to manifest a one-loop cubic-diagram formulation of these properties. In particular, it would be very interesting to construct kinematic numerators at higher orders in $\alpha'$ that obey Jacobi identities on quadratic Feynman propagators. The numerators due to higher-derivative operators will involve 
large numbers of loop momenta and thereby serve as a useful laboratory for the rich 
loop-momentum dependence of field-theory numerators at higher loop orders. It is plausible that
one-loop matrix elements may even provide new input on the difficulties in finding local numerators
with kinematic Jacobi relations for the five-loop integrand of maximal supergravity
\cite{Bern:2017yxu, Bern:2017ucb}.

One-loop matrix elements at four and five points are spelt out to various orders in $\alpha'$,
with all-multiplicity conjectures for the $\alpha'^{\leq 3}$ orders of certain scalar toy integrals.
An obvious open problem for the future concerns explicit representations of higher-point matrix
elements. Similarly, our proposal for non-planar or double-trace matrix elements of gauge multiplets
calls for explicit examples and consistency checks.
Particularly interesting features can be expected at six points: first, our method qualifies for
explicit investigations of $\alpha'$-corrections to the one-loop hexagon anomaly of ten-dimensional
super-Yang-Mills \cite{Frampton:1983ah, Frampton:1983nr} and its cancellation for type-I 
superstrings \cite{Green:1984sg, Green:1984qs}. Second, 
one-loop six-point amplitudes in field theory pose additional challenges~\cite{Mafra:2014gja, Bridges:2021ebs} in manifesting the 
color-kinematics duality and double copy as compared 
to $\leq 5$ points, at least in $D=10$ dimensions and through
quadratic Feynman propagators \cite{He:2017spx}. 
More specifically, while a four-dimensional construction have been understood for some time~\cite{Bjerrum-Bohr:2013iza}, it remains to iron out details of the double copy for the SYM loop integrand recently obtained in~\cite{Bridges:2021ebs}.

The one-loop matrix elements in this work incorporate diagrams with single and multiple insertions
of higher-mass-dimension operators, see e.g.\ Figure~\ref{appetize}. In the open-string case, diagrams with single insertions
encode the zero-momentum limit of maximally supersymmetric one-loop form factors involving the respective ${\rm Tr}\{D^{2k} F^n\}$
operator. Supersymmetric form factors have prominently featured in the recent amplitudes literature, for instance
through their construction via unitarity-based methods \cite{Brandhuber:2010ad, Brandhuber:2011tv, Bork:2011cj, Bork:2012tt, Engelund:2012re, Nandan:2014oga, Bianchi:2018peu} or their color-kinematics duality \cite{Boels:2012ew, Yang:2016ear, Lin:2021kht}.

We expect fruitful connections between the one-loop matrix elements in this work
and stringy canonical forms \cite{Arkani-Hamed:2019mrd, Arkani-Hamed:2019plo, Arkani-Hamed:2020tuz}.
For instance, the $\alpha' \rightarrow 0$ limits of stringy canonical forms for finite-type cluster 
algebras produce the integrands for one-loop tadpole diagrams in case of type B, C and the 
one-loop planar $\phi^3$ diagrams in case of type D. It would be interesting to compare the
$\alpha'$-expansions of such stringy canonical forms with the forward limits of disk and sphere
integrals that carry the $\alpha'$-dependence in our results. Moreover, our results together with 
follow-up studies of analytic contributions to one-loop string amplitudes may lead to interesting
connections with the EFT-hedron \cite{Arkani-Hamed:2020blm}.
Relatedly, it will be rewarding to study a loop-level formulation of the recent advances 
in finding new double-copy constructions at higher mass dimension \cite{Carrasco:2019yyn, 
Carrasco:2021ptp, Chi:2021mio} and to compare with the one-loop matrix elements in this work.

Finally, it would also be rewarding to investigate similar proposals for
matrix elements at two loops and beyond. A promising starting point is
offered by the ambitwistor-string construction of two-loop
field-theory amplitudes via bi-nodal Riemann spheres
\cite{Geyer:2016wjx, Gomez:2016cqb, Geyer:2018xwu, Geyer:2019hnn},
also see \cite{Geyer:2021oox} for an impressive three-loop result in
this framework.  It is conceivable that suitable
$\alpha'$-uplifts via double-forward limits of disk and sphere
integrals yield two-loop matrix elements of $D^{2k} F^n$ and
$D^{2k}R^n$ that resonate with string-theory expectations. The main
crosschecks would concern the discontinuity structure of two-loop
string amplitudes \cite{Green:2010sp, Pioline:2018pso} and the
tropical degeneration of genus-two modular graph forms
\cite{Tourkine:2013rda, Pioline:2015qha, DHoker:2017pvk,
  DHoker:2018mys, Pioline:2018pso, DHoker:2020tcq}. Conversely, matrix
elements of effective operators beyond one loop have enormous
potential to infer physical and mathematical information on multiloop
string amplitudes.


\acknowledgments

We are grateful to Freddy~Cachazo, John~Joseph~Carrasco, Eric~D'Hoker, Paolo~Di~Vecchia, Michael~Green, 
Alfredo~Guevara, 
Song~He, Aidan~Herderschee, Carlos~Mafra, Sebastian~Mizera and Boris~Pioline for combinations of inspiring 
discussions, correspondence and collaboration on related topics. We especially thank Radu Roiban for reading through an early version of our manuscript and providing valuable comments. AE is supported by the Knut and Alice Wallenberg Foundation under 
KAW 2018.0116, {\it From Scattering Amplitudes to Gravitational Waves.}  
MG and OS are supported by the European Research Council under ERC-STG-804286 UNISCAMP.
HJ is supported by the Knut and Alice Wallenberg Foundation under grants KAW 2013.0235, KAW 2018.0116, KAW 2018.0162, the Swedish Research Council under grant 621-2014-5722, and the Ragnar S\"{o}derberg Foundation (Swedish Foundations' Starting Grant).
FT is supported by the US Department of Energy under Grant No.\ DE-SC0013699. 
Computational resources (project SNIC 2019/3-645) were provided by the Swedish National Infrastructure for Computing (SNIC) at UPPMAX, partially funded by the Swedish Research Council through grant no.\ 2018-05973.


\appendix


\section{Numerators at higher order in \texorpdfstring{$\ap$}{a'}}
\label{app:4pthigher}

This appendix gathers higher-order examples in the $\alpha'$-expansion of the
four- and five-point one-loop matrix elements in section \ref{sec:4}.

\subsection{Four-point open strings at higher orders in \texorpdfstring{$\ap$}{a'}}
\label{app:4ptopen}

In this appendix, we display the $\zeta_2^3 \alpha'^6, \zeta_3^2 \alpha'^6$ and $\zeta_{3,5} \alpha'^8$ 
contributions to the numerators in the four-point one-loop matrix elements (\ref{exsec.8}). The square 
of $\zeta_3$ is accompanied by 
\begin{align}
N^{(\zeta_3^2)}_{\text{triangle}}(1,2,34)&=\frac{1}{4}s_{34}^2\Big[s_{\ell,3}^3-s_{\ell,4}^3-s_{34}(s_{\ell,3}^2-2s_{\ell,4}^2)-s_{34}^2s_{\ell,4}\Big]\,, \\
N^{(\zeta_3^2)}_{\text{bubble}}(12,34)&=-\frac{1}{2}s_{12}^2\Big[s_{\ell,2}(s_{\ell,1}+s_{\ell,2})-s_{\ell,3}(s_{\ell,1}-s_{\ell,2})+s_{12}(s_{\ell,2}-s_{\ell,3})\Big]\,, \\
N^{(\zeta_3^2)}_{\text{bubble}}(1,234)&=\frac{1}{4}\Big[s_{34}^2(-2s_{\ell,2}^2+2s_{\ell,3}^2-3s_{\ell,4}^2-s_{\ell,2}s_{\ell,3}-2s_{\ell,3}s_{\ell,4})+(2\leftrightarrow 4)\nonumber\\
&\quad -2s_{23}s_{34}(2s_{\ell,2}^2+s_{\ell,3}^2+2s_{\ell,4}^2+2s_{\ell,2}s_{\ell,3}+2s_{\ell,3}s_{\ell,4}) \nonumber\\
&\quad -s_{34}^3(2s_{\ell,2}+2s_{\ell,3}-3s_{\ell,4})-s_{23}^3(s_{\ell,2}+3s_{\ell,3}+s_{\ell,4})\nonumber\\
&\quad +3s_{23}s_{34}s_{24}(s_{\ell,3}+s_{\ell,3})+3s_{34}^2s_{23}s_{\ell,4}\nonumber\\
&\quad -s_{23}s_{24}(8s_{34}^2+7s_{34}s_{23}+s_{23}^2)\Big]\,, \label{ordz32}\\
N^{(\zeta_3^2)}_{\text{tadpole}}(1234)&=-3s_{13}(s_{13}^2-s_{12}s_{23}) \,,
\end{align}
while even zeta values $\zeta_2^3=\frac{35}{8}\zeta_6$ introduce the more lengthy expressions
\begin{align}
N^{(\zeta_2^3)}_{\text{triangle}}(1,2,34)&=-\frac{1}{140}s_{34}\Big[16(s_{\ell,3}^4+s_{\ell,4}^4)-4s_{34}(13s_{\ell,3}^3+3s_{\ell,4}^3)+s_{34}^3(83s_{\ell,3}^2+23s_{\ell,4}^2)\nonumber\\
&\qquad\qquad\qquad-2s_{34}^3(31s_{\ell,3}+6s_{\ell,4})+47s_{34}^4\Big] \,, \\
N^{(\zeta_2^3)}_{\text{bubble}}(12,34)&=\frac{1}{10}s_{12}^2\Big[4(s_{\ell,1}^2+s_{\ell,2}^2+s_{\ell,3}^2+s_{\ell,1}s_{\ell,2}+s_{\ell,1}s_{\ell,3}+s_{\ell,2}s_{\ell,3})\nonumber\\
&\qquad\qquad+s_{12}(s_{\ell,1}+4s_{\ell,2}-3s_{\ell,3})+11s_{12}^2\Big] \,,  \\
N^{(\zeta_2^3)}_{\text{bubble}}(1,234)&=-\frac{124}{105} s_{12} s_{23} s_{\ell,1} s_{\ell,2}+\frac{58}{105} s_{12} s_{23} s_{\ell,1} s_{\ell,3}-\frac{271}{420}
s_{12}^2 s_{23} s_{\ell,1}-\frac{62}{105} s_{12} s_{23} s_{\ell,1}^2 \nonumber\\
&\quad -\frac{23}{420} s_{12} s_{23}^2
s_{\ell,1}+\frac{68}{105} s_{12} s_{23} s_{\ell,2} s_{\ell,3}-\frac{127}{210} s_{12}^2 s_{23} s_{\ell,2}-\frac{76}{105}
s_{12} s_{23} s_{\ell,2}^2 \nonumber\\
&\quad +\frac{31}{210} s_{12} s_{23}^2 s_{\ell,2}+\frac{79}{70} s_{12}^2 s_{23}
s_{\ell,3}-\frac{8}{21} s_{12} s_{23} s_{\ell,3}^2+\frac{23}{105} s_{12} s_{23}^2 s_{\ell,3} \nonumber\\
&\quad -\frac{8}{7} s_{12}^3
s_{23}-\frac{87}{70} s_{12}^2 s_{23}^2-\frac{47}{70} s_{12} s_{23}^3-\frac{48}{35} s_{12} s_{\ell,1} s_{\ell,2}
s_{\ell,3}-\frac{128}{105} s_{12}^2 s_{\ell,1} s_{\ell,2}\nonumber\\
&\quad -\frac{48}{35} s_{12} s_{\ell,1} s_{\ell,2}^2-\frac{48}{35}
s_{12} s_{\ell,1}^2 s_{\ell,2}-\frac{1}{5} s_{12}^2 s_{\ell,1} s_{\ell,3}-\frac{16}{35} s_{12} s_{\ell,1}
s_{\ell,3}^2 \nonumber\\
&\quad -\frac{24}{35} s_{12} s_{\ell,1}^2 s_{\ell,3}-\frac{64}{105} s_{12}^3 s_{\ell,1}-\frac{64}{105} s_{12}^2
s_{\ell,1}^2-\frac{16}{35} s_{12} s_{\ell,1}^3-\frac{1}{5} s_{12}^2 s_{\ell,2} s_{\ell,3} \nonumber\\
&\quad -\frac{8}{5} s_{12} s_{\ell,2}
s_{\ell,3}^2-\frac{64}{35} s_{12} s_{\ell,2}^2 s_{\ell,3}-\frac{593}{420} s_{12}^3 s_{\ell,2}-\frac{25}{21} s_{12}^2
s_{\ell,2}^2-\frac{36}{35} s_{12} s_{\ell,2}^3 \nonumber\\
&\quad -\frac{79}{210} s_{12}^3 s_{\ell,3}+\frac{1}{3} s_{12}^2
s_{\ell,3}^2-\frac{24}{35} s_{12} s_{\ell,3}^3-\frac{113}{140} s_{12}^4-\frac{8}{7} s_{23} s_{\ell,1} s_{\ell,2}
s_{\ell,3} \nonumber\\
&\quad -\frac{12}{7} s_{23} s_{\ell,1} s_{\ell,2}^2{-}\frac{12}{7} s_{23} s_{\ell,1}^2 s_{\ell,2}{-}\frac{2}{105}
s_{23}^2 s_{\ell,1} s_{\ell,2}{-}\frac{4}{7} s_{23} s_{\ell,1} s_{\ell,3}^2{-}\frac{4}{7} s_{23} s_{\ell,1}^2
s_{\ell,3} \nonumber\\
&\quad -\frac{2}{105} s_{23}^2 s_{\ell,1} s_{\ell,3}-\frac{4}{7} s_{23} s_{\ell,1}^3-\frac{1}{105} s_{23}^2
s_{\ell,1}{}^2-\frac{89}{420} s_{23}^3 s_{\ell,1}-\frac{36}{35} s_{23} s_{\ell,2} s_{\ell,3}^2 \nonumber\\
&\quad -\frac{44}{35} s_{23}
s_{\ell,2}^2 s_{\ell,3}-\frac{13}{105} s_{23}^2 s_{\ell,2} s_{\ell,3}-\frac{36}{35} s_{23} s_{\ell,2}^3+\frac{1}{15}
s_{23}^2 s_{\ell,2}^2-\frac{109}{420} s_{23}^3 s_{\ell,2} \nonumber\\
&\quad -\frac{4}{35} s_{23} s_{\ell,3}^3-\frac{11}{35} s_{23}^2
s_{\ell,3}{}^2-\frac{9}{140} s_{23}^3 s_{\ell,3}-\frac{11 }{35}s_{23}^4 \,, \\
N^{(\zeta_2^3)}_{\text{tadpole}}(1234)&=-\frac{104}{35} s_{12}^2 s_{23}-\frac{104}{35} s_{12} s_{23}^2-\frac{48}{7} s_{12} s_{\ell,1} s_{\ell,2}-\frac{16}{7} s_{12}
s_{\ell,1} s_{\ell,3}+\frac{152}{35} s_{12}^2 s_{\ell,1} \nonumber\\
&\quad -\frac{32}{7} s_{12} s_{\ell,1}^2-\frac{80}{7} s_{12} s_{\ell,2}
s_{\ell,3}-\frac{64}{7} s_{12} s_{\ell,2}^2+\frac{152}{35} s_{12}^2 s_{\ell,3}-\frac{48}{7} s_{12}
s_{\ell,3}^2-\frac{76}{35} s_{12}^3 \nonumber\\
&\quad -\frac{80}{7} s_{23} s_{\ell,1} s_{\ell,2}-\frac{16}{7} s_{23} s_{\ell,1}
s_{\ell,3}-\frac{48}{7} s_{23} s_{\ell,1}^2-\frac{152}{35} s_{23}^2 s_{\ell,1}-\frac{48}{7} s_{23} s_{\ell,2}
s_{\ell,3} \nonumber\\
&\quad -\frac{64}{7} s_{23} s_{\ell,2}^2-\frac{32}{7} s_{23} s_{\ell,3}^2-\frac{152}{35} s_{23}^2 s_{\ell,3}-\frac{76
	}{35}s_{23}^3 \, . 
\end{align}

Finally, as an example of the weight-8 contributions, we present the
$\zeta_{3,5}$ numerators.  Of note is that the triangle and two-mass bubble
do not contribute, namely
\begin{align}
  N^{(\zeta_{3,5})}_{\text{triangle}}(1,2,34) = 0 \, , \qquad   N^{(\zeta_{3,5})}_{\text{bubble}}(12,34)=0\, ,
\end{align}
while there is an external-bubble contribution
\begin{align}
  &N^{(\zeta_{3,5})}_{\text{bubble}}(1,234)=\frac{s_{13}}{2560}  \Bigg(s_{\ell,1}^3 \left[16 s_{\ell,2} (s_{12}+s_{23})+2 s_{\ell,3} (5 s_{12}+3 s_{23})+3
                                            \left(s_{12}^2-s_{23}^2\right)\right]  \notag\\
  &\quad +3 s_{\ell,1}^2 \bigg[s_{\ell,2} \left(2 s_{\ell,3} [5 s_{12}+3 s_{23}]+3
    \left[s_{12}^2-s_{23}^2\right]\right)+8 s_{\ell,2}^2 (s_{12}+s_{23})  \notag\\
  &\quad +s_{\ell,3} \left(5 s_{12} s_{23}+5 s_{12}^2+2
    s_{23}^2\right)+5 s_{12}^2 s_{23}+5 s_{12} s_{23}^2+s_{12}^3-2 s_{23} s_{\ell,3}^2+s_{23}^3\bigg]  \notag\\
  &\quad +s_{\ell,1} \bigg[3  s_{\ell,2}^2 \left(2 s_{\ell,3} [5 s_{12}+3 s_{23}]+3 \left[s_{12}^2-s_{23}^2\right]\right)+6 s_{\ell,2} \Big(s_{\ell,3}
    \left[5 s_{12} s_{23}+5 s_{12}^2+2 s_{23}^2\right]  \notag\\
    &\quad +5 s_{12}^2 s_{23}+5 s_{12} s_{23}^2+s_{12}^3-2 s_{23}s_{\ell,3}^2+s_{23}^3\Big)+16 s_{\ell,2}^3 (s_{12}+s_{23})-2 s_{\ell,3}^3 (5 s_{12}+2 s_{23})  \notag\\
    &\quad +3 s_{\ell,3}^2 \left(5 s_{12} s_{23}+5 s_{12}^2+2 s_{23}^2\right)+3 s_{23} s_{\ell,3} \left(5 s_{12} s_{23}+5 s_{12}^2+2 s_{23}^2\right)   \\
  &\quad +2 \left(5 s_{12}^3 s_{23}-5 s_{12} s_{23}^3+2 s_{12}^4-2 s_{23}^4\right)\bigg]+4 s_{\ell,1}^4 (s_{12}+s_{23})+2 \bigg[s_{\ell,2}^3
    \Big(2 s_{\ell,3} [5 s_{12}+3 s_{23}]  \notag \\
  &\quad +3 \left[s_{12}^2-s_{23}^2\right]\Big)+3 s_{\ell,2}^2 \Big(s_{\ell,3} \left[5 s_{12}
    s_{23}+5 s_{12}^2+2 s_{23}^2\right]+5 s_{12}^2 s_{23}+5 s_{12} s_{23}^2+s_{12}^3  \notag\\
  & \quad -2 s_{23}s_{\ell,3}^2+s_{23}^3\Big)+s_{\ell,2} \Big(-2 s_{\ell,3}^3 [5 s_{12}+2 s_{23}]+3 s_{\ell,3}^2 \left[5 s_{12} s_{23}+5
    s_{12}^2+2 s_{23}^2\right]  \notag\\
  & \quad +3 s_{23} s_{\ell,3} \left[5 s_{12} s_{23}+5 s_{12}^2+2 s_{23}^2\right]+2 \left[5 s_{12}^3
    s_{23}-5 s_{12} s_{23}^3+2 s_{12}^4-2 s_{23}^4\right]\Big)  \notag\\
  & \quad +4 s_{\ell,2}^4 [s_{12}+s_{23}]+s_{12} s_{\ell,3} \Big(3
    s_{\ell,3}^2 [s_{12}+2 s_{23}]+s_{\ell,3} \left[6 s_{12} s_{23}-3 s_{12}^2+6 s_{23}^2\right]  \notag\\
  & \quad +6 s_{12}^2 s_{23}-6 s_{12}s_{23}^2+4 s_{12}^3-4 s_{23}^3-4 s_{\ell,3}^3\Big)\bigg]\Bigg)
   \notag
\end{align}
and a tadpole
\begin{align}
  &N^{(\zeta_{3,5})}_{\text{tadpole}}(1234)=\frac{1}{640} \Bigg(s_{\ell,1} \bigg[s_{\ell,2} \left(81 s_{12}^2 s_{23}-9 s_{12} s_{23}^2+43 s_{12}^3-7
    s_{23}^3\right)  \notag\\
  &\quad +s_{13} \Big(10 \left[3 s_{12}^2 s_{23}-3 s_{12} s_{23}^2+2 s_{12}^3-2 s_{23}^3\right]-3 s_{\ell,3}
    \left[-4 s_{12} s_{23}+5 s_{12}^2+5 s_{23}^2\right]\Big)\bigg]  \notag\\
  &\quad +s_{\ell,1}^2 \left[42 s_{12}^2 s_{23}{-}3 s_{12} s_{23}^2{+}29
    s_{12}^3{+}4 s_{23}^3\right] +s_{\ell,2} s_{\ell,3} \left[81 s_{12} s_{23}^2{-}9 s_{12}^2 s_{23}{-}7 s_{12}^3{+}43
    s_{23}^3\right] \\
  &\quad +18 s_{\ell,2}^2 \left[2 s_{12}^2 s_{23}+2 s_{12} s_{23}^2+s_{12}^3+s_{23}^3\right]+s_{\ell,3} \bigg[s_{\ell,3} \left(42 s_{12} s_{23}^2-3 s_{12}^2 s_{23}+4 s_{12}^3+29 s_{23}^3\right)  \notag\\
  &\quad +10 \left(-5 s_{12}^3 s_{23}+5
   s_{12} s_{23}^3-2 s_{12}^4+2 s_{23}^4\right)\bigg]\Bigg)\,.
    \notag
\end{align}

\subsection{The five-point one-loop matrix element at the order of \texorpdfstring{$\ap^4$}{a'4}}
\label{app:5ptzeta4}

In the parametrization (\ref{t8defs.9}) of one-loop matrix elements at five points,
the $\alpha'^4 \zeta_2^2$-order of the numerators has been partially given in (\ref{t8defs.13}).
The remaining contributions to this order are given as follows (see (\ref{eq5.5}) for
the pentagon numerator $N_{+|12345|-}$ of SYM):
\begin{align}
N^{(\zeta_2^2)}_{\text{box}}(1,2,3,45)&=-\frac{1}{20}s_{45}\Big[4(s_{\ell,4}^2+s_{\ell,5}^2)-s_{45}(7s_{\ell,4}+s_{\ell,5})+11s_{45}^2\Big]N_{+|12345|-} \nonumber\\
&\quad+\frac{1}{20}\Big[4(s_{\ell,4}^3-s_{\ell,5}^3)-s_{45}(11s_{\ell,4}^2-s_{\ell,5}^2)\nonumber\\
&\qquad\qquad+2s_{45}^2(7s_{\ell,4}-2s_{\ell,5})-7s_{45}^3\Big]t_8(45,1,2,3)\,, \\
N^{(\zeta_2^2)}_{\text{triangle}}(1,2,[34]5)&=\frac{1}{20}(s_{35}+s_{45})\Big[4(s_{\ell,3}+s_{\ell,4})^2+4s_{\ell,5}^2-s_{\ell,5}(s_{35}+s_{45})\nonumber\\
&\quad-(s_{\ell,3}+s_{\ell,4})(16s_{34}+7s_{35}+7s_{45})+16s_{34}^2\nonumber\\
&\quad+14s_{34}(s_{35}+s_{45})+11(s_{35}+s_{45})^2\Big]t_8(34,1,2,5)\,, \\
N^{(\zeta_2^2)}_{\text{triangle}}(1,2,3[45])&=\frac{1}{20}(s_{34}+s_{35})\Big[4s_{\ell,3}^2+4(s_{\ell,4}+s_{\ell,5})^2-7s_{\ell,3}(s_{34}+s_{35})\nonumber\\
&\quad-(s_{\ell,4}+s_{\ell,5})(s_{34}+s_{35})+11(s_{34}+s_{35})^2\Big]t_8(45,1,2,3)\,, \\
N^{(\zeta_2^2)}_{\text{triangle}}(1,2,345)&=\frac{1}{20}
\Big[4s_{34}(2s_{\ell,4}+s_{\ell,5})-4s_{45}(s_{\ell,3}+2s_{\ell,4})+8s_{35}(s_{\ell,3}-s_{\ell,5})-s_{34}^2\nonumber\\
&\quad-6s_{35}^2+7s_{45}^2+2s_{34}s_{35}+16s_{34}s_{45}+14s_{35}s_{45}\Big]N_{+|12345|-} \nonumber\\
&\quad+\frac{1}{20}\Big[12s_{\ell,4}(s_{\ell,4}+s_{\ell,5})+4s_{\ell,5}^2+2s_{\ell,3}(5s_{35}-s_{45})\nonumber\\
&\quad+2s_{\ell,4}(8s_{35}-s_{34}-3s_{45})-s_{\ell,5}(s_{34}+2s_{35}+2s_{45})+4s_{34}^2+15s_{35}^2\nonumber\\
&\quad-5s_{45}^2+10s_{34}s_{35}+22s_{35}s_{45}\Big]t_8(34,1,2,5)\nonumber\\
&\quad+\frac{1}{20}\Big[12s_{\ell,4}(s_{\ell,3}+s_{\ell,4})+4s_{\ell,3}^2-s_{\ell,3}(18s_{34}+6s_{35}+11s_{45})\nonumber\\
&\quad-2s_{\ell,4}(15s_{34}+14s_{35}+11s_{45})+2s_{\ell,5}(s_{34}-5s_{35})+15s_{34}^2+27s_{35}^2\nonumber\\
&\quad+14s_{45}^2+25s_{34}s_{45}+52s_{34}s_{35}+47s_{35}s_{45}\Big]t_8(45,1,2,3)\nonumber\\
&\quad-\frac{1}{20}\Big[12s_{\ell,3}^2+8s_{\ell,4}^2+12s_{\ell,5}^2+12s_{\ell,4}(s_{\ell,3}+s_{\ell,5})\nonumber\\
&\quad-s_{\ell,3}(17s_{34}+22s_{35}+5s_{45})-s_{\ell,4}(9s_{34}+12s_{35}+19s_{45})\nonumber\\
&\quad-s_{\ell,5}(7s_{34}+2s_{35}+19s_{45})+29s_{34}^2+18s_{35}^2+35s_{45}^2+36s_{34}s_{45}\nonumber\\
&\quad+39s_{34}s_{35}+29s_{35}s_{45}\Big]t_8(35,1,2,4) \,, \\
N^{(\zeta_2^2)}_{\text{triangle}}(1,23,45)&=s_{23}s_{45}N_{+|12345|-} \nonumber\\*
&\quad-\frac{1}{2}s_{45}(s_{\ell,2}-s_{\ell,3}+s_{12}-s_{13}-s_{23})t_8(23,1,4,5)\nonumber\\*
&\quad+\frac{1}{2} s_{23}(s_{\ell,5}-s_{\ell,4}+s_{45})t_8(45,1,2,3)\,, \\
N^{(\zeta_2^2)}_{\text{triangle}}(1,[23],45)&=\frac{1}{20}s_{45}\Big[4(s_{\ell,4}^2+s_{\ell,5}^2)-(7s_{\ell,4}+s_{\ell,5})s_{45}+11s_{45}^2\Big]t_8(23,1,4,5)\, , \notag \\
N^{(\zeta_2^2)}_{\text{triangle}}(1,23,[45])&=\frac{1}{20}s_{23}\Big[4(s_{\ell,2}^2+s_{\ell,3}^2)+s_{\ell,2}(8s_{12}+s_{23})+s_{\ell,3}(8s_{13}+7s_{23})\nonumber\\
&\quad+4s_{12}^2+4s_{13}^2+11s_{23}^2+s_{12}s_{23}+7s_{13}s_{23}\Big]t_8(45,1,2,3)\,, \\
N^{(\zeta_2^2)}_{\text{bubble}}([12],345)&=
\frac{1}{20}\Big[4s_{\ell,3}(s_{45}-2s_{35}){+}8s_{\ell,4}(s_{45}-s_{34}){+}4s_{\ell,5}(2s_{35}-s_{34}){+}(s_{34}-s_{35})^2\nonumber\\
&\quad-7(s_{35}+s_{45})^2+4(3s_{35}^2-4s_{34}s_{45})\Big]t_8(12,3,4,5)\,, \\
N^{(\zeta_2^2)}_{\text{bubble}}(1,2345)&=\frac{1}{5}(3s_{24}-s_{25}+s_{34}+3s_{35})N_{+|12345|-} \nonumber\\
&\quad-\frac{1}{20}(12s_{\ell,3}+8s_{\ell,4}+4s_{\ell,5}+s_{23}+20s_{25}-2s_{35})t_8(23,1,4,5)\nonumber\\
&\quad+\frac{1}{20}(12s_{\ell,3}+24s_{\ell,4}+12s_{\ell,5}-5s_{23}+8s_{25}-17s_{34}-2s_{45})t_8(24,1,3,5)\nonumber\\
&\quad+\frac{1}{10} \Big[6(s_{\ell,2}-s_{\ell,5})+2(s_{\ell,3}-s_{\ell,4})-4(s_{23}-s_{45})-7s_{24}-8s_{25}\\
&\quad\quad-3s_{35}\Big]t_8(25,1,3,4)+\frac{1}{10}(6s_{24}-8s_{25}+s_{34}+6s_{35})t_8(34,1,2,5)\nonumber\\
&\quad-
\frac{1}{20}(12s_{\ell,2}{+}24s_{\ell,3}{+}12s_{\ell,4}{-}10s_{23}{+}4s_{25}{+}5s_{34}{+}17s_{45})t_8(35,1,2,4)\nonumber\\
&\quad+\frac{1}{20}(4s_{\ell,2}{+}8s_{\ell,3}{+}12s_{\ell,4}{-}2s_{24}{-}12s_{25}{-}8s_{34}{+}4s_{35}{-}s_{45})t_8(45,1,2,3)\,, \nonumber\\
N^{(\zeta_2^2)}_{\text{bubble}}(1,[23]45)&=\frac{1}{20}\Big[4(s_{\ell,2}+s_{\ell,3})(s_{45}-2s_{14}+2s_{23})+8s_{\ell,4}(2s_{45}+s_{14})\nonumber\\*
&\quad+4s_{\ell,5}(s_{45}+3s_{14}-2s_{23})+10(s_{12}+s_{13})^2+6s_{14}(s_{12}+s_{13})\nonumber\\*
&\quad+9s_{14}^2+24s_{23}(s_{12}+s_{13})+8s_{14}s_{23}+4s_{23}^2\Big]t_8(23,1,4,5)\,, \\
N^{(\zeta_2^2)}_{\text{bubble}}(1,2[34]5)&= \frac{1}{20}\Big[8(s_{\ell,3}+s_{\ell,4})(s_{12}-s_{15})-4s_{\ell,2}(s_{15}+3s_{25})+4s_{\ell,5}(s_{12}+3s_{25})\nonumber\\*
&\quad+46s_{12}^2+3(11s_{12}-3s_{15})(s_{13}+s_{14})\nonumber\\*
&\quad+12(3s_{12}-2s_{34})(s_{23}+s_{24})\Big]t_8(34,1,2,5)\,, \\
N^{(\zeta_2^2)}_{\text{bubble}}(1,23[45])&=\frac{1}{20}\Big[4s_{\ell,2}(2s_{12}{-}s_{13}{+}s_{23})-8s_{\ell,3}(s_{13}{+}2s_{23})-4(s_{\ell,4}{+}s_{\ell,5})(2s_{12}{+}3s_{23})\nonumber\\
&\quad+6s_{12}^2-7s_{13}^2+4s_{23}^2-14s_{12}s_{13}-12s_{13}s_{23}\Big]t_8(45,1,2,3)\,.
\end{align}


\section{Gravitational one-loop matrix element from closed strings}
\label{app:closed}

In this appendix, we sketch the closed-string analogue of the discussion in
section \ref{sec:3.2}, where one-loop matrix elements of gauge multiplets were recovered from the
$\tau \rightarrow i \infty$ contributions to open-string amplitudes. The key idea for the gravitational
counterparts is to decompose the expression (\ref{prop.41}) for closed-string one-loop amplitudes
in direct analogy with (\ref{rev.65}),
\beq
M^{\te{1-loop}}_{n,{\rm closed}} = \int \dd^D \ell  \int_{{\mathfrak F}}\dd^2 \tau \, e^{-4\pi \alpha' \Im \tau\, \ell^2} \,
 \Big\{ \underbrace{ {\cal F}(\ell,\tau )  - {\cal F}(\ell,i \infty ) }_{(i)}+ \underbrace{ {\cal F}(\ell, i \infty ) }_{(ii)} \Big\}\, ,
\label{appcl.1}
\eeq
where the following quantity no longer depends on $\ell^2$
\beq
{\cal F}(\ell,\tau ) = \int_{{\mathfrak T}^{n-1}_{\tau}} \dd^2 z_2 \ldots \dd^2 z_n \, e^{4\pi \alpha' \Im \tau \, \ell^2} 
 \, \big|{\cal K}_n(\ell,\tau) \, \KN_n(\ell,\tau)  \big|^2\, .
\label{appcl.2}
\eeq
We focus on the dominant contribution $(ii)$, where the worldsheet correlator and the integrals 
over the punctures are evaluated on a degenerate torus. From the leftover integral over
$\tau$ in the expression (\ref{appcl.1}) for $M^{\te{1-loop}}_{n,{\rm closed}} \, |_{(ii)}$,
one can extract
\beq
\int_{{\mathfrak F}}\dd^2 \tau \, e^{-4\pi \alpha' \Im \tau \ell^2} = \frac{1}{4\pi \ap \ell^2} + {\cal O}((\ell^2)^0)
\label{appcl.3}
\eeq
as the dominant contribution from the cusp, see (\ref{rev.66}) for the open-string counterpart.
By largely following the steps in (\ref{rev.67}),
the quantity ${\cal F}(\ell, i \infty )$ will now be related to the loop integrand in the proposal
(\ref{prop.2}) for gravitational one-loop matrix elements.

Based on the change of variables (\ref{rev.61}) and the relation (\ref{prop.51}) between chiral integrands
and half integrands of the ambitwistor string, we can rewrite
\begin{align}
\lim_{\tau \rightarrow i \infty} {\cal F}(\ell,\tau) &\rightarrow \frac{1}{(4\pi^2)^{n-1}} \int_{\mathbb C^{n-1}} \frac{ \dd^2 \sigma_2 \ldots \dd^2 \sigma_{n} }{| \sigma_2 \ldots \sigma_n |^2} \, \lim_{\tau \rightarrow i \infty}  e^{4\pi \alpha' \Im \tau \ell^2} 
 \, \big|{\cal K}_n(\ell,\tau) \, \KN_n(\ell,\tau)  \big|^2
\label{appcl.4} \\
&= \frac{1}{(4\pi^2)^{3}} \int_{\mathbb C^{n-1}} 
\dd^2 \sigma_2 \ldots \dd^2 \sigma_{n} \, | {\cal I}_n(\ell) |^2
\prod_{j=1}^n |\sigma_{j+}|^{4\ap \ell \cdot k_j} \prod_{1\leq i<j}^n |\sigma_{ij}|^{4\ap k_{i}\cdot k_{j}} \, . \notag
\end{align}
In the parametrization of the torus worldsheet through the parallelogram with corners $-\frac{\tau}{2}$ ,
$ -\frac{\tau}{2}{+}1,\, \frac{\tau}{2}{+}1$ and $\frac{\tau}{2}$, the limit $\tau \rightarrow i \infty$
recovers $\mathbb C^{n-1}$ from the integration domain in (\ref{appcl.2}).
  
As a next step, we insert the Parke-Taylor decomposition (\ref{rev.7}) of the ambitwistor-string correlators
into (\ref{appcl.4}) and identify the sphere integrals defined in (\ref{rev.32}),
\begin{align}
\lim_{\tau \rightarrow i \infty} {\cal F}(\ell,\tau) &= \frac{1}{(4\pi^2)^{3}} \bigg( \frac{\pi}{\ap} \bigg)^{n-1}
\sum_{\rho,\gamma\in S_{n-1}} N_{+|\rho|-} \bar N_{+|\gamma|-}
\lim_{k_{\pm} \rightarrow \pm \ell} J(+,\gamma,-|+,\rho,-)
\notag\\
&=\frac{1}{(4\pi^2)^{3}} \bigg( \frac{\pi}{\ap} \bigg)^{n-1}  M^{\te{1-loop}}_{n,{\rm eff}} \, . \label{appcl.5} 
\end{align}
In summary, the above assumptions and the restriction to the singular term $\ell^{-2}$
in (\ref{appcl.3}) lead to a tentative string origin
\beq
M^{\te{1-loop}}_{n,{\rm closed}} \, \big|_{(ii)} =
\frac{1}{(2\pi )^8}\bigg( \frac{\pi}{\ap} \bigg)^{n} M^{\te{1-loop}}_{n,{\rm eff}}
+ {\rm analytic}
\label{appcl.6}
\eeq
of gravitational one-loop matrix elements (\ref{prop.2}). Regular terms $(\ell^2)^{\geq 0}$
in (\ref{appcl.3}) may additionally introduce analytic contributions to (\ref{appcl.6}), i.e.\
without any branch cuts corresponding to massless particles similar to the ones in part $(i)$ of (\ref{appcl.1}).

Similar to the partitioning of open-string amplitudes in (\ref{rev.65}), gravitational one-loop matrix elements
have been formally recovered from the terms $(ii)$ in the closed-string amplitude (\ref{appcl.1})
which gather the dominant contributions from $\tau \rightarrow i\infty$. This is the boundary
of moduli space where the torus degenerates to a nodal sphere and the origin of all the poles in
$\ell^{2}$. Conversely, the one-loop matrix elements do
not account for the terms $(i)$ of closed-string amplitudes and the regular terms in (\ref{appcl.3}) which are free
of massless quadratic propagators $(\ell+K)^{-2}$ (with external momenta $K$) after 
integration over the punctures.
Hence, $M^{\te{1-loop}}_{n,{\rm eff}}$ is insensitive to the analytic  
dependence of closed-string one-loop amplitudes that arises from
$(i)$ and the regular terms in (\ref{appcl.3}) and plays a crucial role for the one-loop effective action of
type-II theories.

\section{Laurent series from logarithmic divergences}
\label{sec:alt-div}
In this appendix, we discuss an alternative approach to constructing the
Laurent polynomials in the Schwinger parameter $t$ from \cref{secMGF.2} via direct
momentum-space calculations of logarithmic UV divergences in dimensional reduction. The following integrals have been used for the one-loop four and five-point matrix elements, 
\begin{subequations}\label{defFeyn}
\begin{align}\label{eq:bub1}
  &\text{bubble:} &&I_2(k)=\mu^{2\epsilon}e^{\epsilon\gamma_{\rm E}}\int\frac{\dd^D\ell}{\pi^{D/2}}\frac{1}{\ell^2(\ell+k)^2}\,,
           \\ \label{eq:1tri}
  &\text{one-mass triangle:} && I_3(k_1,k_2)=\mu^{2\epsilon}e^{\epsilon\gamma_{\rm E}}\int\frac{\dd^D\ell}{\pi^{D/2}}\frac{1}{\ell^2(\ell+k_1)^2(\ell+k_{12})^2}\,,
                \\ \label{eq:2tri}
  &\text{two-mass triangle:} && I_3(k_1,k_2{+}k_3)=\mu^{2\epsilon}e^{\epsilon\gamma_{\rm E}}\int\frac{\dd^D\ell}{\pi^{D/2}}\frac{1}{\ell^2(\ell+k_1)^2(\ell+k_{123})^2}\,,
                     \\ \label{eq:1box}
  &\text{one-mass box:} && I_4(k_1,k_2,k_3)=\mu^{2\epsilon}e^{\epsilon\gamma_{\rm E}}\int\frac{\dd^D\ell}{\pi^{D/2}}\frac{1}{\ell^2(\ell+k_1)^2(\ell+k_{12})^2(\ell+k_{123})^2}\,,
\end{align}
\end{subequations}
where $\mu^2$ is an arbitrary mass scale that compensates the dimension shift $D=\mathcal{D}-2\epsilon$ such that the mass dimension of the integral is not changed, and $e^{\epsilon\gamma_{\rm E}}$ factor absorbs all the Euler constant $\gamma_{\rm E}$ in the $\epsilon$-expansion. All the $k_i$ in (\ref{eq:1tri}) to (\ref{eq:1box})
are lightlike, while the momentum in $I_2$ is in general off-shell, i.e.\ $k^2 \neq 0$. We recall that the 
signature of the spacetime metric is $(-,+,+,\ldots,+)$ and the integration measure 
$\dd^D \ell = -i \dd\ell^0 \,\dd\ell^1 \ldots \dd\ell^{D-1}$ 
is taken to absorb the factor of $i$ which appears in the more common conventions for Feynman integrals.
 
\subsection{Logarithmic in one dimension is power in another}
\label{sec:log-eq-pow}
The first point to make is what the section title says: UV power
divergences in a specific dimension, $\mathcal{D}$, can be probed
using logarithmic UV divergences in other dimensions.  This idea is
well known, if not often discussed, in the amplitudes and QFT
communities.  However, since the reader may not be familiar with this
idea, we present a brief argument for it here.

The idea is most obvious when considering the Schwinger representation
of \cref{WLUV.1} and \cref{WLUV.3} where, in $D_{\text{crit}} = \mathcal{D}$, an integrand that
carries $t^{-1}$ will produce the logarithm divergence in
$\mathcal{D}$ while all other powers of $t$ will contribute to power
divergences.  However, from the point of view of the $t$ integration,
there is nothing special about the choice of $D$ in integrand
prefactor of $t^{n-D/2}$.  If we look at the coefficient of $t^{-1+m}$
within the $t$ integration, in addition to interpreting the term as
sourcing a power divergence we could instead think of $m$ as shifting
the dimension away from $\mathcal{D}$.  Running this logic in reverse,
we can find the coefficient of $t^{-1+m}$ in $\mathcal{D}$ dimensions
by evaluating the logarithmic divergence of the same
momentum-space integrand in $D_{\text{crit}} = \mathcal{D} + 2 m$
dimensions instead.  Importantly, this should be thought of as an
integration trick and not an actual change to the underlying theory:
we are not manipulating the dimension the theory and its states live
in, merely exploring the analytic structure of the loop integral.

\subsection{Momentum-space perspective on UV divergences}

We now present an alternate method for directly extracting the
logarithmic UV divergence in a given dimension directly from the
momentum-space integral, which is the one-loop analogue of the methods
employed in many studies of multi-loop super-Yang-Mills and
supergravity (see
\cite{Bern:2013qca,Bern:2012uf,Johansson:2017bfl,Bern:2018jmv} for a
small sample of these studies).  We begin by scaling the loop momenta
via $ \ell \to \beta \ell$ and series expanding the integrand near
$\beta \to \infty$.  The resulting expressions will contain only powers
of $\ell^2$ in the denominator, with all of the $\ell \cdot k_i$
relegated to the numerator.  Such terms require only no-scale tensor
reduction, \eg
$ \ell^\mu \ell^\nu \to \frac{\ell^2}{D} \eta^{\mu \nu}$.  After the
reduction, the numerators $n_\lambda$ will be free of $\ell$ and the denominators
will be powers $\lambda \in \mathbb Z$ of $\ell^2$,
\begin{equation}
  \int \frac{\dd^D \ell}{\pi^{D/2}} I(\ell)\Bigg|_{\text{UV}} \to \int\frac{\dd^D\ell}{\pi^{D/2}} \sum_{\lambda} \frac{\bar{n}_{\lambda}}{\beta^{2 \lambda}(\ell^2)^\lambda}\Bigg|_{\text{UV}} \,,
  \label{eq:large-ell}
\end{equation}
where $I(\ell)$ refers to a generic loop integrand,
and the $\bar{n}_\lambda$ are polynomials in the external momenta.  In dimensional
regularization, integrals of this form are set to zero via a cancellation between UV
and IR divergences.  Since we are interested in the UV divergence
specifically, we can introduce a mass to regulate the IR divergence
separately from the UV divergence.  The logarithmic UV divergence in
$D=D_{\rm crit}-2\epsilon$ spacetime dimensions (given the mostly-plus metric convention and $s_{ij}>0$ region) is given by
\begin{equation}
\int\frac{\dd^D\ell}{\pi^{D/2}}\frac{1}{(\ell^2)^{D_{\text{crit}}/2}}\Bigg|_{\text{UV}}\!  
\rightarrow \!\int\frac{\dd^D\ell}{\pi^{D/2}} \frac{1}{(\ell^2 - m^2)^{D_{\rm crit}/2}} = 
  \frac{\Gamma(\epsilon)}{\Gamma(\frac{D_{\rm crit}}{2} ) (m^2)^{\epsilon}}  = 
  \frac{1}{\Gamma( \frac{D_{\rm crit}}{2}) \epsilon} + \mathcal{O}(m,\epsilon^0) \, .
  \label{eq:mass-reg}
\end{equation}
The UV divergence is the residue of the simple pole $\frac{1}{\epsilon}$, with the IR
regulator decoupled into the non-singular terms of order $\epsilon^{\geq 0}$. Thus, the various
orders of $\beta$ in the large-$\ell$ expansion of (\ref{eq:large-ell})
correspond to logarithmic UV divergences via the mapping
\begin{equation}
  \int\frac{\dd^D\ell}{\pi^{D/2}} I(\ell)\Bigg|_{\text{UV}} \rightarrow \bar{n}_{D_{\text{crit}}/2} \int\frac{\dd^D\ell}{\pi^{D/2}}\frac{1}{\beta^{D_{\rm crit}} (\ell^2)^{D_{\rm crit}/2}} \Bigg|_{\text{UV}} \to \frac{\bar{n}_{D_{\text{crit}}/2}}{\Gamma(\frac{D_{\rm crit}}{2}) \epsilon} \,.
  \label{eq:uv-eps}
\end{equation}
As discussed above, these UV divergences directly correspond to terms
in the expansion of the Schwinger exponent.  For example, we can apply
\cref{eq:uv-eps} to the scalar one-mass triangle~\cref{eq:1tri}, 
\begin{equation}  \label{eq:tri-eps}
\setlength{\arraycolsep}{3pt}
  I_3(k_1,k_2) \ \xrightarrow{{\rm UV}} \ \, \begin{array}{|*5{>{\displaystyle}c|}}
\hline
D_{\rm crit}=6 & D_{\rm crit}=8 & D_{\rm crit}=10 & D_{\rm crit}=12 & D_{\rm crit}=14  \\ \hline 
& & & & \\[-2.5ex]
\frac{1}{\epsilon} & -\frac{s_{12}}{24\epsilon} & \frac{s_{12}^2}{360\epsilon} & -\frac{s_{12}^3}{6720\epsilon} & \frac{s_{12}^4}{151200\epsilon} \\[1.5ex] \hline
\end{array}\;,
\end{equation}
which exactly matches the expansion in \cref{WLUV.5}.  The constructions in
\cref{secMGF.3} have all been reproduced using this method as well.
Then, as expected from the arguments in \cref{sec:log-eq-pow}, the
$D$-dimensional UV divergences at the residues of the simple poles in
$\epsilon=\frac{1}{2}(D_{\rm crit}-D)$ are related to the Laurent polynomials via
\begin{equation}
  \coef_{1/\epsilon} \pi^{-D/2}  \widehat A^{\te{1-loop}}_{\rm eff}(1,2,3,4) \, \big|_{\rm UV} = \coef_{ (-2 i \alpha' T)^{D_{\rm crit}/2-4}} a^{\te{1-loop}}_{{\rm open}}\Big(1,2,3,4;\tau=\frac{T}{\pi}\Big) \,,
  \label{eq:os-eps-tau}
\end{equation}
which is equivalent to \cref{eq:os-eps-t}.
The procedure can be just as easily applied to the closed-string
integrands which leads to the expected pattern of UV divergences
\begin{equation}
  \coef_{1/\epsilon}  \pi^{-D/2}  \widehat M^{\te{1-loop}}_{4,{\rm eff}}   \, \big|_{\rm UV} 
  = \coef_{(4\ap y)^{D_{\rm crit}/2-4}} m^{\te{1-loop}}_{4,{\rm closed}} \Big( \Im \tau=\frac{y}{\pi}\Big)  \,,
  \label{WLUV.45}
\end{equation}
equivalent to \cref{WLUV.44}.

\subsection{Momentum-space and \texorpdfstring{$D_{\text{crit}}\le 0$}{Dcrit<0}}

For the UV divergences along with higher $\zeta$ values, like those 
in \cref{negdims,eq:os-high-zeta,eq:closed-h-zeta}, we encounter
powers like $T^{-4}, T^{-5}, \dots$ in the associated Laurent polynomial~\cref{MGFs.9} for open string, and $y^{-4},\dots$ in \cref{MGFs.4} for closed string. By (\ref{eq:os-eps-tau}) and  
(\ref{WLUV.45}), these UV divergences occur in formally vanishing or
negative spacetime dimensions $D \le 0$.  There are two
points of interest in these cases.  The first is that the momentum-space tensor
reduction generates a prefactor $\frac{1}{D(D+2) \dots}$ that will
contribute a $\frac{1}{\epsilon}$ pole when $D_{\rm crit}=-2N$ is even and non-positive.  Second, the
$\Gamma$-function representation that \cref{eq:uv-eps} comes from,
\begin{equation}
  \frac{ \Gamma(\epsilon)}{ \Gamma\left( \frac{D_{\rm crit}}{2}\right)} \,,
\end{equation}
has an infinity-over-infinity ambiguity and thus needs to be properly regulated.
Since the tensor reduction already provides the $\frac{1}{\epsilon}$ pole,
the simplest regulation scheme should lead to divergence-free contributions.  To
this end, we propose the simple scheme,
\begin{equation}
  \frac{\Gamma(\epsilon)}{ \Gamma\left(\frac{D_{\rm crit}}{2} - \epsilon\right)} \,.
\end{equation}
For $D_{\rm crit}>0$, the additional $\epsilon$ regulator is irrelevant, reducing
the expression back to \cref{eq:uv-eps} in the leading-$\epsilon$
expansion.  However, for $D_{\rm crit}\le 0$, we instead get
\begin{equation}
  \frac{ \Gamma(\epsilon)}{ \Gamma\left(\frac{D_{\rm crit}}{2} - \epsilon\right)}
  \xrightarrow[D_{\rm crit}\text{ even and }D_{\rm crit}\le 0]{\epsilon\to 0}
   (-1)^{ \frac{D_{\rm crit}}{2}+1}\Gamma\left( \frac{|D_{\rm crit}|}{2} +1\right) \,.
  \label{eq:neg-d}
\end{equation}
Applying this regularization scheme to the leading terms of $\zeta_6$,
$\zeta_7$, and $\zeta_8$ in (\ref{negdims})
exactly lines up with the established correspondence of
\cref{eq:os-eps-tau} as desired.


\subsection{Integrals with closed formulas}  
\label{app:Feynints}

As yet another check, we can extract UV divergence and non-analytic terms from those integrals with closed formulas. The bubble integral~\cref{eq:bub1} and the one-mass triangle \cref{eq:1tri}
are given by
\begin{subequations}
\begin{align}
  I_2(k)&=\mu^{2\epsilon}e^{\epsilon\gamma_E}\frac{\Gamma(2-\frac{D}{2})\Gamma(\frac{D}{2}-1)^2}{\Gamma(D-2)}(k^2)^{\frac{D}{2}-2}\nonumber\\
  &\xrightarrow{D=10-2\epsilon}\;-\frac{k^6}{840}\left[\frac{1}{\epsilon}-\log\frac{k^2}{\mu^2}\right] - \frac{44k^6}{11025}+\mathcal{O}(\epsilon)\,,\\
  I_3(k_1,k_2)&=\mu^{2\epsilon}e^{\epsilon\gamma_E}\frac{\Gamma(3-\frac{D}{2})\Gamma(\frac{D}{2}-2)^2}{\Gamma(D-3)}(s_{12})^{\frac{D}{2}-3}\nonumber\\
  &\xrightarrow{D=10-2\epsilon}\,\frac{s_{12}^2}{360}\left[\frac{1}{\epsilon}-\log\frac{s_{12}}{\mu^2}\right] + \frac{17s_{12}^2}{1800}+\mathcal{O}(\epsilon)\,.
\label{eq:i3-exp}
\end{align}
\end{subequations}
Both of the formulas can be found in~\cite{Smirnov:2012gma}. On the
other hand, the two-mass triangle~\cref{eq:2tri} is not a master
integral. It can be reduced to a combination of two bubbles
\begin{equation}
I_3(k_1,k_2{+}k_3)=\frac{2(D-3)}{D-4}\frac{I_2(k_2{+}k_3)-I_2(k_1{+}k_2{+}k_3)}{s_{12}+s_{13}}\,.
\end{equation}
The one-mass box~\cref{eq:1box} also has a closed formula~\cite{Bern:1993kr,Kozlov:2015kol},
\begin{align}
I_4(k_1,k_2,k_3)&=\frac{\mu^{2\epsilon}e^{\epsilon\gamma_E}C_D}{s_{12}s_{23}}\Bigg[(s_{12})^{\frac{D}{2}-2}\mathcal{F}\Big(\frac{s_{13}}{s_{23}}\Big)+(s_{23})^{\frac{D}{2}-2}\mathcal{F}\Big(\frac{s_{13}}{s_{12}}\Big)-(s_{123})^{\frac{D}{2}-2}\mathcal{F}\Big(\frac{s_{13}s_{123}}{s_{12}s_{23}}\Big)\Bigg],
\end{align}
where
\begin{align}
C_D=\frac{8\Gamma(3-\frac{D}{2})\Gamma(\frac{D}{2}-1)^2}{(D-4)^2\Gamma(D-3)}\,,& &\mathcal{F}(x)=\tensor[_2]{F}{_1}\bigg[ \begin{array}{c} 1,\, \tfrac{D}{2}{-}2 \\ \tfrac{D}{2}{-}1 \end{array};-x\bigg]\,.
\end{align}
The UV divergence of the one-mass box at $D=10{-}2\epsilon$ is 
\begin{align}
I_4(k_1,k_2,k_3)\xrightarrow{D=10-2\epsilon}-\frac{s_{12}+s_{23}+s_{123}}{120\epsilon}+\mathcal{O}(\epsilon^0)\,.
\end{align}
The non-analytic terms at $\epsilon^0$ order contain logarithms and first-order derivatives of hypergeometric functions.
%



\providecommand{\href}[2]{#2}\begingroup\raggedright\endgroup

\end{document}